\documentclass[superscriptaddress,onecolumn,showpacs,a4paper,amssymb,amsmath,nobibnotes,aps,prd,showkeys,nofootinbib,notitlepage]{revtex4-1}

\usepackage{graphicx,subfigure,bm,color,psfrag}
\usepackage{amsfonts}
\usepackage{mathtools}
\usepackage{verbatim}
\usepackage[arrowdel]{physics}
\usepackage{accents}
\usepackage{xspace}
\usepackage[pscoord]{eso-pic}
\usepackage[normalem]{ulem}
\usepackage[dvipsnames]{xcolor}
\usepackage[colorlinks = true, linkcolor = purple, urlcolor = purple, citecolor = blue, anchorcolor = blue]{hyperref}
\newcommand\e{\mathrm{e}}
\newcommand{\bea}{\begin{align}}
\newcommand{\eea}{\end{align}}

\newcommand\be{\begin{align}}
\newcommand\ee{\end{align}}

\begin{document}


\title{Finite-time Cosmological Singularities and the Possible Fate of the Universe}

\author{Jaume~de~Haro}
\email{jaime.haro@upc.edu}
\affiliation{Departament de Matem\`{a}tica Aplicada, Universitat Polit\`{e}cnica de Catalunya, Diagonal 647, 08028 Barcelona, Spain}

\author{Shin'ichi~Nojiri}
\email{nojiri@gravity.phys.nagoya-u.ac.jp}
\affiliation{Department of Physics, Nagoya University, Nagoya 464-8602, Japan}
\affiliation{Kobayashi-Maskawa Institute for
the Origin of Particles and the Universe, Nagoya University, Nagoya 464-8602, Japan}

\author{S.~D.~Odintsov}
\email{odintsov@ice.csic.es}
\affiliation{ICREA, Passeig Luis Companys, 23, 08010 Barcelona, Spain}
\affiliation{Institute of Space Sciences (ICE, CSIC), C. Can Magrans, s/n, 08193 Barcelona, Spain}

\author{V.~K.~Oikonomou}
\email{voikonomou@gapps.auth.gr}
\affiliation{Department of Physics, Aristotle University of Thessaloniki, Thessaloniki 54124, Greece}

\author{Supriya~Pan}
\email{supriya.maths@presiuniv.ac.in}
\affiliation{Department of Mathematics, Presidency University, 86/1 College Street, Kolkata 700073, India}

\begin{abstract}
Singularities in any physical theory are either remarkable indicators of the unknown underlying fundamental theory, or indicate a change in the description of the physical reality. In General Relativity there are three fundamental kinds of singularities that might occur, firstly the black hole spacelike crushing singularities, e.g. in the Schwarzschild case and two cosmological spacelike singularities appearing in finite-time, namely, the Big Bang singularity and the Big Rip singularity. In the case of black hole and Big Bang singularity, the singularity indicates that the physics is no longer described by the classical gravity theory but some quantum version of gravity is probably needed. The Big Rip is a future singularity which appears in the context of General Relativity due to a phantom scalar field needed to describe the dark energy era. Apart from the Big Rip singularity, a variety of finite-time future singularities, such as, sudden singularity, Big Freeze singularity, generalized sudden singularity, $w$-singularity and so on, are allowed in various class of cosmological models irrespective of their origin. The occurrence of these finite-time singularities has been intensively investigated in the context of a variety of dark energy, modified gravity, and other alternative cosmological theories. These singularities suggest that the current cosmological scenario is probably an approximate version of a fundamental theory yet to be discovered. 
In this review we provide a concrete overview of the 
cosmological theories constructed in the context of Einstein's General Relativity and modified gravity theories that may lead to finite-time cosmological singularities. We also discuss various approaches suggested in the literature that could potentially prevent or mitigate finite-time singularities within the cosmological scenarios. 

\end{abstract}
\maketitle
\newpage

\tableofcontents{}
\newpage

\section{Introduction}

The dynamics of our Universe is one of the most intriguing
mysteries in modern theoretical physics. The physics of the
Universe at the early phase and the late phase is understood at a
certain level. With the increasing sensitivity in the astronomical
observations, modern cosmology has witnessed remarkable success
being now a precision science, but there are many fundamental
questions that are still unanswered. The finite-time cosmic
singularities constitute one of the mysteries in modern cosmology,
first pointed out in \cite{Caldwell:2003vq} (also see \cite{McInnes:2001zw}) motivated by the existence of some phantom fluid in the universe sector \cite{Caldwell:1999ew}. The theory of
inflation \cite{Liddle:2000cg} is able to describe consistently
the post-Planck classical early Universe and plays a crucial role
to answer some theoretical shortcomings of the standard Big Bang
model of cosmology. From the observational point of view, even
though the cosmic microwave background radiation goes in support
of inflation
\cite{COBE:1992syq,WMAP:2003ivt,WMAP:2003syu,Planck:2015sxf,Planck:2018jri},
however, inflation also has some limitations, and thus, at this
moment, it is very hard to conclude that the inflation is the only
theory for the early Universe. On the other hand, the discovery of
the late-time accelerating phase of the Universe
\cite{SupernovaSearchTeam:1998fmf,SupernovaCosmologyProject:1998vns}
has made compelling the presence of some hypothetical fluid with a
negative pressure in the Universe sector. However, the source of
such hypothetical fluid is not clearly known yet and this imposed
to the entire scientific community a difficult puzzle to solve.
The puzzle is basically how to model the late-time acceleration
era in an observationally viable way. One has to have in mind,
that describing nature in broken phases might not be a subtle way
to describe accurately nature, so the correct description must
also take into account the consistent description of all the
evolutionary eras of the Universe. The most basic approach to
explain this late-time cosmic acceleration is to add a
hypothetical fluid dubbed as Dark Energy (DE), with strong
negative pressure in the context of Einstein's gravitational
theory described by the General Relativity (GR). The simplest DE
candidate is Einstein's positive cosmological constant $\Lambda$,
which together with Cold Dark Matter (CDM), the so-called
$\Lambda$-Cold Dark Matter ($\Lambda$CDM) is constructed, and the
latter has been quite successful in describing a large span of
observational data related with the Cosmic Microwave Background
radiation polarization anisotropies. However, there are mainly two
potential issues within this simplest $\Lambda$CDM cosmology,
namely, the cosmological constant problem \cite{Weinberg:1988cp}
and the cosmic coincidence problem (also known as `why now?'
problem) \cite{Zlatev:1998tr}. Apart from that, the latest
observational data \cite{Planck:2018vyg} indicate that the DE
equation of state (EoS) parameter is allowed to take values that
cross the phantom divide line. Within the framework of simple GR,
the only way to model such an observationally allowed phantom
value is by using a phantom scalar field, which is not a
theoretically elegant description for any sensible physics model.
Apart from these issues, there is always the question whether the
DE can be dynamical or not. These shortcomings of the $\Lambda$CDM
description, point out towards a modification of the matter sector
of the universe within the context of Einstein's GR to describe
consistently the late-time era. Several modifications of the
matter sector are suggested in various forms in the literature,
and they have been investigated from theoretical perspectives and
further tested with the available observational data from diverse
astronomical sources, see for instance
\cite{Carroll:1998zi,Amendola:1999er,Barreiro:1999zs,Sahni:1999qe,delaMacorra:1999ff,Brax:1999yv,Urena-Lopez:2000ewq,Gonzalez-Diaz:2000glv,Matos:2000ng,Huterer:2000mj,Kamenshchik:2001cp,Bilic:2001cg,Zimdahl:2001ar,Baccigalupi:2001aa,Gu:2001tr,Bento:2001yv,Bento:2002ps,Padmanabhan:2002cp,Cardenas:2002np,Franca:2002iju,Huterer:2002hy,Corasaniti:2001mf,Freese:2002sq,Melchiorri:2002ux,Gerke:2002sx,Gorini:2002kf,Singh:2003vx,Huey:2004qv,Chae:2004jp,Chen:2004nqb,Li:2004rb,Scherrer:2004au,Alam:2004jy,Debnath:2004cd,Wetterich:2004pv,Jarv:2004uk,Bassett:2004wz,Capozziello:2005pa,Banerjee:2005ef,Banerjee:2005nv,Guo:2005ata,Cataldo:2005qh,Sahlen:2005zw,Kaloper:2005aj,Fabris:2005ts,Sola:2005et,Zhang:2004gc,Guo:2005qy,Banerjee:2006rp,Linder:2006sv,Sahlen:2006dn,Nojiri:2006zh,Zhang:2006qu,Huterer:2006mv,Zhang:2006av,Nojiri:2006jy,Wei:2007ty,Olivares:2007rt,Gao:2007ep,Linder:2007wa,Avelino:2008ph,Zhang:2008mb,Wu:2008jt,Dutta:2009yb,Zhang:2009un,Li:2009mf,Banerjee:2008rs,Gagnon:2011id,Adak:2012bv,Chen:2013vea,Velten:2013qna,Roy:2013wqa,Mahata:2013oza,Sharov:2014voa,Chakraborty:2014fia,Banerjee:2015kva,Sharov:2015ifa,Mahata:2015lja,Magana:2017nfs,Zhao:2017cud,Rezaei:2017yyj,Yang:2018qmz,Malekjani:2018qcz,Chakraborty:2018nrk,Yang:2019nhz,Perkovic:2019vxm,Yang:2019jwn,Rezaei:2019hvb,Pan:2019hac,Rezaei:2019xwo,Chakraborty:2019swx,Sinha:2020vob,Yang:2021flj,Yang:2021eud,Saridakis:2021qxb,Saridakis:2021xqy,Benaoum:2020qsi,Sharov:2022hxi,Liu:2022mpj,Cardona:2022pwm,Yao:2023ybs}
(also see the review articles in this direction
\cite{Peebles:2002gy,Padmanabhan:2002ji,Sahni:2002kh,Copeland:2006wr,Sahni:2006pa,Padmanabhan:2007xy,Frieman:2008sn,Martin:2008qp,Caldwell:2009ix,Silvestri:2009hh,Bamba:2012cp,Li:2012dt,Mortonson:2013zfa,Sami:2013ssa,Brevik:2017msy}
and the references therein). On the other hand, this late-time
accelerating era, can also be described if one modifies Einstein's
gravitational theory (without modifying the matter sector) or
introduce new gravitational theories. The resulting hypothetical
fluid in this modified or new gravitational sector is dubbed as
the Geometric Dark Energy (GDE) and this resulted in a large
number of modified gravity theories and models that have been
widely investigated from both theoretical and observational
perspectives, see for instance
\cite{Banerjee:2000gt,Banerjee:2000mj,Capozziello:2002rd,Capozziello:2003gx,Nojiri:2003ft,Nojiri:2003ni,Carroll:2003wy,Nojiri:2004bi,Elizalde:2004mq,Nojiri:2005vv,Capozziello:2005ku,Nojiri:2005jg,Amarzguioui:2005zq,Carter:2005fu,Das:2005bn,Cognola:2006eg,Brookfield:2006mq,Song:2006ej,Koivisto:2006xf,Koivisto:2006ai,Capozziello:2006dj,Li:2006ag,Li:2006vi,Nojiri:2006gh,Li:2007xn,Nojiri:2007uq,Bertolami:2007gv,Hu:2007nk,Li:2007xw,Li:2007jm,Starobinsky:2007hu,Nojiri:2007te,Nojiri:2007cq,Cognola:2007zu,Fay:2007uy,Pogosian:2007sw,Carloni:2007yv,Bengochea:2008gz,Dev:2008rx,Jhingan:2008ym,Capozziello:2008gu,Das:2008iq,Brax:2008hh,Frolov:2008uf,Oyaizu:2008sr,Oyaizu:2008tb,Schmidt:2008tn,Bamba:2008ja,Bamba:2008xa,Nojiri:2009kx,Appleby:2009uf,Thongkool:2009js,Thongkool:2009vf,Linder:2010py,Wu:2010mn,Bamba:2010wb,Capozziello:2010uv,Li:2010cg,Dunsby:2010wg,Geng:2011aj,Harko:2011kv,Koyama:2011wx,Gumrukcuoglu:2011ew,Cai:2011tc,Paliathanasis:2011jq,Olmo:2011uz,Li:2012by,deHaro:2012zt,Gannouji:2012iy,Cardone:2012xq,Sami:2012uh,Chakraborty:2012kj,Maluf:2013gaa,Odintsov:2013iba,Nesseris:2013jea,Brax:2013fda,vandeBruck:2012vq,Nojiri:2013zza,Tamanini:2013ltp,Basilakos:2013rua,Cai:2013lqa,Chakraborty:2012sd,Chakraborty:2013ywa,Cai:2013toa,Nojiri:2014zqa,Cai:2014upa,Bose:2014zba,Barreira:2014ija,Chakraborty:2014xla,Harko:2014aja,Leon:2014yua,He:2014eva,Ling:2014xoi,Thomas:2015dfa,Chakraborty:2015bja,Chakraborty:2015wma,Paliathanasis:2015arj,Chakraborty:2015taq,He:2015bua,Nunes:2016qyp,Paliathanasis:2016tch,Paliathanasis:2016vsw,Nunes:2016plz,Nunes:2016drj,Liu:2016xes,Cataneo:2016iav,Shirasaki:2016twn,Bose:2016wms,Rezazadeh:2017edd,Paliathanasis:2017efk,Li:2017xdi,Roy:2017mnz,Paliathanasis:2017htk,Akarsu:2018aro,He:2018oai,Mitchell:2018qrg,Nunes:2018xbm,Li:2018ixg,Leon:2018skk,Nunes:2018evm,Nojiri:2012zu,Nojiri:2012re,Nojiri:2015qyc,Gannouji:2018aaw,Papagiannopoulos:2018mez,Banerjee:2018yyi,Mitchell:2019qke,Paliathanasis:2019luv,Arnold:2019zup,Arnold:2019vpg,Leon:2019mbo,Pozdeeva:2019agu,Choudhury:2019zod,Paliathanasis:2019ega,Cai:2019bdh,Chen:2019ftv,Papagiannopoulos:2019kar,Yan:2019gbw,Paliathanasis:2020bgs,Mitchell:2020aep,Paliathanasis:2020axi,Alam:2020jdv,Paliathanasis:2020plf,Mitchell:2021aex,Mitchell:2021uzh,Mitchell:2021ter,Ren:2021tfi,Ren:2021uqb,Paliathanasis:2021qns,Paliathanasis:2021uvd,dosSantos:2021owt,Hernandez-Aguayo:2021kuh,
Ruan:2021wup,Paliathanasis:2022pgu,Ren:2022aeo,Leon:2022oyy,Paliathanasis:2022xvn,Dimakis:2022wkj,Santos:2022atq,Kumar:2023bqj,Qi:2023ncd,Millano:2023czt,Hu:2023juh}
(also see the review articles in this direction
\cite{Nojiri:2006ri,Nojiri:2017ncd,Capozziello:2007ec,Padmanabhan:2007xy,Sotiriou:2008rp,Silvestri:2009hh,DeFelice:2010aj,Nojiri:2010wj,Clifton:2011jh,Hinterbichler:2011tt,Capozziello:2011et,Sami:2013ssa,deRham:2014zqa,Cai:2015emx,Bahamonde:2021gfp}
and the references therein). Modified gravity theories can also
explain the inflationary phase of the Universe
\cite{Carter:2005fu,Ferraro:2006jd,Nojiri:2007as,Nojiri:2007uq,Bamba:2008ja,Bamba:2008xa,Bonanno:2010bt,Elizalde:2010ep,Nojiri:2014zqa,Johnson:2017pwo,Zhong:2018tqn,Fomin:2018blx,He:2018gyf,Cuzinatto:2018chu,Antoniadis:2018ywb,Chakraborty:2018scm,Canko:2019mud,Gialamas:2019nly,Pozdeeva:2020apf,Gamonal:2020itt,Oikonomou:2020oex,Dimopoulos:2020pas,Sangtawee:2021mhz,Zhang:2021ppy,Baffou:2021ycm,Bhattacharjee:2022lcs,Shiravand:2022ccb}
and in some cases, the unified description of inflation with the
DE era is achieved while in parallel with the presence of dark matter
(DM) is also accommodated
\cite{Oikonomou:2022tux,Oikonomou:2020qah,Odintsov:2020nwm} (see
also
\cite{Nojiri:2003ft,Nojiri:2010ny,Nojiri:2019fft,Houndjo:2022oxe}).
Apart from the DE and modified gravity models, modern cosmology
has witnessed the emergence of a cluster of proposals for
explaining different phases of our Universe. The cosmology with
gravitationally induced adiabatic matter creation rates is one of
them where one can explain the late-time accelerating phase of the
Universe
\cite{Lima:2008qy,Steigman:2008bc,Lima:2009ic,Basilakos:2010yp,Lima:2011hq,Jesus:2011ek,Lima:2012cm,Lima:2014qpa,Ramos:2014dba,Fabris:2014fda,Lima:2015xpa,Pan:2016jli,Nunes:2016aup}
and the early inflationary phase as well
\cite{Lima:1995xz,Abramo:1996ip,Gunzig:1997tk} for some specific
matter creation rate function. However, there exists a large
number of proposals aiming to explain the dynamics of the Universe
at high and low energy scales, nevertheless, no cosmological
model, irrespective of its fundamental origin, has been found to
be successful over others. With the increasing sensitivity in the
astronomical data and thanks to the large amount of the
astronomical data, currently we are able to offer precise
constraints on the cosmological models and test their soundness
and predictions by confronting them with the observational data.

The peculiar prediction of modified gravity models of DE is the
occurrence of finite-time cosmological singularities.
Interestingly enough, the GR description of a phantom DE era also
results in a future finite-time cosmological singularity. The
cosmological community is already very much familiar with
finite-time singularities in the past, for example, the Big Bang
singularity. However, DE studies commenced a new perspective for
cosmological singularities. The first well-known prediction of the
occurrence of a finite-time singularity was developed in
Ref.~\cite{Caldwell:2003vq} and initiated a new perspective in
cosmology. Specifically, it was shown in \cite{Caldwell:2003vq}
that a phantom DE dominated Universe may end up in a Big Rip
singularity \cite{Caldwell:2003vq}. This pioneering work initiated
the investigation of the possible occurrence of finite-time future
singularities in the context of various DE models originating from
various theoretical contexts. It was reported by a number of
investigators that cosmological models may develop a variety of
finite-time singularities in the future and the Big Rip
singularity is not alone in the list of finite-time future
singularities appearing in several cosmological models, there may
appear other types of finite-time singularities in the
cosmological models, such as the sudden singularity
\cite{Barrow:2004xh} (also see
\cite{Lake:2004fu,Fernandez-Jambrina:2004yjt,Dabrowski:2004bz,Stefancic:2004kb,Barrow:2004he,Dabrowski:2005fg,Barrow:2009df,Barrow:2011ub,Elizalde:2012zu,Denkiewicz:2012bz,deHaro:2012wv,Barrow:2013ria,Perivolaropoulos:2016nhp,Lymperis:2017ulc,Barrow:2019cuv,Barrow:2020rhh,Rosa:2021ish,Goncalves:2022ggq,Balcerzak:2023ynk}), the big freeze singularity \cite{Bouhmadi-Lopez:2006fwq}, the
generalized sudden singularity \cite{Barrow:2004hk}, $w$
singularity \cite{Dabrowski:2009kg,Fernandez-Jambrina:2010ngm}.
The appearance of such finite-time future cosmological
singularities became an important part of modern cosmology and
this got massive attention in the scientific community. The first
concrete classification of finite-time future cosmological
singularities was done in Ref.~\cite{Nojiri:2005sx} and for an
important stream of articles on finite-time cosmological
singularities, see Refs.
\cite{Nojiri:2003vn,Sami:2003xv,Bouhmadi-Lopez:2004mpi,Scherrer:2004eq,Curbelo:2005dh,Nojiri:2005sr,Chimento:2005au,JimenezMadrid:2005gd,Nojiri:2006ww,Zhang:2006ck,Faraoni:2007kx,Naskar:2007dn,Dabrowski:2007ci,Yurov:2007tw,Kamenshchik:2007zj,Bouhmadi-Lopez:2007xco,Fernandez-Jambrina:2008pwx,Cannata:2008xc,Keresztes:2009vc,
Capozziello:2009hc,Bouhmadi-Lopez:2009ggt,Yurov:2009gj,Nojiri:2009pf,Brevik:2010okp,
Antoniadis:2010ik,Carloni:2010nx,Bamba:2011sm,Lopez-Revelles:2012ine,Haro:2011zzb,Xi:2011uz,Denkiewicz:2011uz,Pavon:2012pt,Astashenok:2012tv,Meng:2012mb,Bamba:2012vg,deHaro:2012cj,deHaro:2012wv,Saez-Gomez:2012uwp,deHaro:2012xj,delaCruz-Dombriz:2012bni,Bamba:2012ka,Astashenok:2012iy,Capozziello:2012re,Balakin:2012ee,Kamenshchik:2013naa,AghaeiAbchouyeh:2013jkw,Myrzakul:2013qka,Bouhmadi-Lopez:2013tua,Bouhmadi-Lopez:2014tna,Fernandez-Jambrina:2014sga,Odintsov:2015zza,Shojai:2015dpa,BeltranJimenez:2016fuy,Fernandez-Jambrina:2016clh,Carlson:2016iuw,BeltranJimenez:2016dfc,Perivolaropoulos:2016nhp,Odintsov:2015ynk,Cataldo:2017nck,Albarran:2017swy,Bouhmadi-Lopez:2017ckh,Elizalde:2018ahd,Granda:2019aan,Granda:2019iwj,Heydarzade:2019dpf,Galkina:2020zle,Nojiri:2020sti,Rosa:2021ish,Cruz:2021knz,Fernandez-Jambrina:2021foi,Mousavi:2022puq,Odintsov:2022eqm,Trivedi:2022ngt,Trivedi:2022svr,Trivedi:2022jbu}
and the references therein. Note that the possibility of finite time singularities in the context of inflation was investigated and some models along this line were proposed \cite{Nojiri:2015wsa,Nojiri:2015fia}.

It is worth discussing what a physical spacetime singularity is.
For smooth singularities, like the pressure and Type IV
singularities, no geodesics incompleteness occurs, but for strong,
crushing type singularities, geodesics incompleteness always
occurs. For all the singularities, these are classified according
the singular or non-singular behavior of physical quantities and
higher order curvature invariants defined on the three dimensional
spacelike hypersurface defined by the time instance $t=t_s$, where
$t_s$ is the time instance where the singularity occurs. The
physical quantities that determine the type of a finite-time
physical singularity, according to the classification of
\cite{Nojiri:2005sx}, are the pressure, the energy density and the
scale factor. According to Penrose's definition of a physical
spacetime singularity, if the strong and weak energy conditions
are satisfied, a singularity in spacetime is accompanied by
geodesics incompleteness, which also covers the case that the
spacetime is spatially flat and a single point is removed from the
spacetime. The latter is geodesically incomplete, although the
curvature is null everywhere. The physical spacetime singularities
on smooth spacetime manifolds are accompanied by higher order
curvature divergences. Specifically certain integrals of higher
order curvature invariants calculated on geodesics are divergent,
when a finite-time singularity is developed. One example of this
sort is the following integral \cite{Fernandez-Jambrina:2004yjt},
\begin{align}\label{geodesicincompleteness}
\int_{0}^{t_s}d t'\int_0^{t'}d t''R^{i}_{0j0}(t'')\, ,
\end{align}
which depends on the Riemann tensor, and if a crushing type
singularity occurs at $t=t_s$, it strongly diverges. This is the
best way to perceive an actual crushing type finite-time
singularity, the fact that geodesics incompleteness occurs. The
geodesics incompleteness is always accompanied by singular
behaviors in higher order curvature invariants, but for singular
spacetimes, the definition of the curvature itself is ill defined,
in an open set in the spacetime surrounding the spacetime
singularity. In fact, a strong finite-time singularity in
spacetime maybe have exotic implications on the topology of
spacetime, since it is speculated that topology change might
actually occur \cite{Oikonomou:2018qsc}. Also the presence of a
singularity is accompanied by closed time-like curves and in all
cases the notion of differentiability is lost because the
smoothness of spacetime itself is lost. Also a finite-time
singularity may not actually be considered as an isolated point in
spacetime but as a three dimensional hypersurface defined by
$t=t_s$, on which physical observables and curvature invariants
strongly diverge. This is also Penrose's view of the Big Bang
singularity, who states that the Big Bang singularity cannot be
viewed as a singular isolated spacetime point but an initial
singular hypersurface \cite{1988QJRAS..29...61P}. This perspective
is a valid one because if the Big Bang was an isolated point in
spacetime, this would indicate that when it occurs, this would
lead to an infinite number of overlapping particle horizons, and
therefore at the next time instance we would end up with an
infinity of causally disconnected regions as the Universe evolves.

With this review, we aim to present all the different theories
that may lead to finite-time future singularities. We shall
include many different theoretical frameworks that lead to
cosmological singularities and we shall analyze in detail the
reasons why these theories lead to cosmic singularities. Our
purpose is two-fold: first to provide to the literature a unique
text that gathers all the theoretical frameworks that lead to
cosmic singularities, and second by using our initial aim, to
further inspire the academic society to seek fundamental physics
behind cosmic singularities. The latter is the most important
motivation, because singularities in physics always indicate an
effective theory behind a classical physics description. Indeed
the singularities of classical electrodynamics are resolved if one
considers the complete effective theory of quantum
electrodynamics, so it is tempting to consider this analogy and
seek for a fundamental theory behind an apparent future
cosmological singularity. However, the cosmological singularities
in a GR or modified GR framework are not similar to the
singularities in quantum systems. It is known most GR
singularities are well hidden behind cosmic horizons. So the
question is whether such a horizon exists for future or past
finite-time singularities.
What are the implications of such horizons for classical physics,
and to some extent, what does a crushing type singularity implies
for the spacetime itself? Does a crushing singularity indicate a
change in the topology, or equivalently, the shape of the
Universe? Is the very own fabric of the Universe ripped by a
future cosmic singularity? These are the fundamental questions and
the answers to these are not trivial. Several aspects of such
exotic scenarios regarding the effects of horizons and topology
changing Universe due to the occurrence of crushing type
singularities were developed in
Refs.~\cite{Nojiri:2022nmu,Oikonomou:2018qsc}, which may be
considered as starting points, among other works too. Nature
provides us with similar pictures of singularities indicating
changes in topology of the physical system in solid state physics,
like for example in the Hele-Shaw
systems~\cite{2009PhyD..238.1113L}, so with this review we aim to
bring forth all the different theories that lead to cosmic
singularities and to impose the question whether there is a
quantum or even classical resolution of cosmic singularities, and
further inspire work towards this research line.

The review is organized as follows:
In Section~\ref{sec-classification-singularities}, we discuss various
types of finite-time singularities in the past and future, where
in particular, Section~\ref{sec-bigbang} deals with the hot Big Bang
singularity in the past, Section~\ref{sec-future-singularities}
describes the variety of finite-time future singularities.
In Section~\ref{sec-singularities-DE}, we discuss the emergence of
finite-time singularities in various DE models where in
particular, section \ref{sec-singularities-viscous-cosmology}
deals with the singularities in viscous cosmologies, section
\ref{sec-singularities-IDE} describes the appearance of finite
time singularities in the context of interacting dark matter--dark
energy cosmologies. Then in Section~\ref{sec-singularities-MG}, we
discuss the finite-time singularities appearing in various
modified gravity theories. In particular, we organize Section
\ref{sec-singularities-MG} as follows: Section
\ref{sec-scalar-tensor} describes the Scalar-tensor gravity;
Section \ref{sec-Brans-Dicke} describes the Brans-Dicke gravity;
Section \ref{sec-k-essence} describes the $k-$essence model;
Section \ref{sec-Scalar-Einstein-Gauss-Bonnet} presents the
Scalar-Einstein--Gauss--Bonnet gravity; Section
\ref{sec-F(R)-gravity} describes the $F(R)$ gravity ($R$ is the
Ricci scalar) which further includes Section
\ref{sec-correspondence-singularities-FR} and Section
\ref{sec-correspondence-singularities-unimodularFR} describing the
correspondence of singularities in $F (R)$ and unimodular $F (R)$
gravity theories, respectively; Section \ref{sec-F(G)-gravity}
describes the $F (G)$ gravity ($G$ is the Gauss-Bonnet invariant);
Section \ref{sec-f(R,G)-gravity} describes the $F (R, G)$ gravity;
Section \ref{sec-singularities-F(T)} describes the $F (T)$ gravity
($T$ is the torsion scalar); Section
\ref{sec-singularities-nolocal} discusses the non-local gravity;
Section \ref{sec-Maxwell-Einstein-gravity} describes the
non-minimal Maxwell--Einstein gravity; Section
\ref{sec-singularities-semi-classical} describes the
semi-classical gravity. Then in
Section~\ref{sec-singularities-braneworld}, we describe the
finite-time singularities in the braneworld gravity.
Section~\ref{sec-singularities-matter-creation} describes the
finite-time singularities in matter creation models. In
Section~\ref{sec-singularities-LQG} we discuss the finite-time
singularities in the context of Loop Quantum Cosmology (LQC).
Further, in Section~\ref{sec-dyn-system-vs-singularities}, we make
a correspondence between the dynamical analysis and the finite
time singularities. Then in Section~\ref{sec-avoid-singularities},
we discuss the possibility to avoid the finite-time singularities
through some heuristic routes that include the quantum, thermal
and other non-standard effects. Finally, in
Section~\ref{sec-summary}, we present a brief summary of the
review by highlighting the important features that need to be
considered for upcoming works. We also discuss the fundamental
future perspectives of cosmological finite-time singularities that need to be addressed by the future researchers.

\section{Classification of Finite-time Singularities}
\label{sec-classification-singularities}

In this section, we describe the finite-time singularities
appearing in the past and future evolution of the Universe. In
agreement with the observational evidences in the large scale, our
Universe is almost homogeneous and isotropic and such geometrical
configuration of the Universe is well described by the
Friedmann--Lema\^{i}tre--Robertson--Walker (FLRW) line element:
\begin{align}
\label{FLRWk}
d{\rm s}^2 = -dt^2 + a^2 (t) \Bigg[\frac{dr^2}{1-kr^2} + r^2 \left(d\theta^2 + \sin^2 \theta d\phi^2\right) \Bigg]\, ,
\end{align}
where $a(t)$ is the expansion scale factor of the Universe, $(t,
r, \theta, \phi)$ are the co-moving coordinates and $k$ describes
three different geometries for three distinct values, namely,
spatially flat ($k =0$), closed ($k = +1$), open ($k =-1$). We
often use the case $k=0$ for which Eq. (\ref{FLRWk}) takes the
form
\begin{align}
\label{FLRWk0}
d{\rm s}^2 = -dt^2 + a^2 (t) \bigg[dr^2 + r^2 \left(d\theta^2 + \sin^2 \theta d\phi^2\right) \bigg]\, .
\end{align}

\subsection{Big Bang: What is it?}
\label{sec-bigbang}

The Big Bang is the simplest past singularity, and appears, for
example, when one studies a fluid with linear Equation of State (EoS) $p=w\rho$ where
the EoS parameter satisfy $w>-1$ (non-phantom fluid), and $p$ and
$\rho$ are the pressure and the energy density of the Universe,
respectively. Writing $w=\gamma-1 \rightarrow \gamma > 0$, and
dealing with the flat FLRW line element, after combining the
Friedmann and Raychaudhuri equations
\begin{align}
H^2=\frac{\rho\kappa^2}{3}, \qquad \dot{H}=-\frac{(\rho+p)\kappa^2}{2}\, ,
\label{FReqs}
\end{align}
where $H$ is the Hubble rate of the FLRW Universe and $\kappa^2 = 8 \pi G_N$ ($G_N$ is the Newton's gravitational constant) is the Einstein's gravitational constant, one gets
\begin{align}
\dot{H}=-\frac{3\gamma}{2}H^2\, ,
\end{align}
which could be integrated obtaining
\begin{align}\label{1}
H(t)=\frac{H_0}{\frac{3}{2}\gamma H_0(t-t_0)+1}=\frac{2}{3\gamma (t-t_s)}\, ,
\end{align}
where $t_0$ is present cosmic time, $H_0$ is the current value of the Hubble rate and $t_s=t_0-\frac{2}{3\gamma H_0}$.
Inserting this expression into the Friedmann equation, one gets the following value of the energy density:
\begin{align}\label{2}
\rho(t)=\frac{4}{3\gamma^2(t-t_s)^2\kappa^2}\, .
\end{align}
One can see that both the Hubble rate and the energy density
diverge at the finite past time $t_s<t_0$. In addition, since the
scale factor $a(t)$, at time $t_s$ is given by
$a(t_s)=a_0\e^{-\int_{t_s}^{t_0}H(t)dt}$, where $a_0$ is the current
value of the scale factor, one easily gets $a(t_s)=0$. This is the
well-known {\it Big Bang} singularity.

On the other hand, note that $t_s$ gives us an indication of the
age of the Universe. Effectively, for this model one gets
\begin{align}
t_0-t_s=\frac{2}{3\gamma H_0}\sim 14 \quad \mbox{billion years}\, ,
\end{align}
where we have used that the current value of the Hubble rate is
approximately $H_0\sim 70 \, \mathrm{Km/s/Mpc}$.

Here, it is very important to realize that the Big Bang
solution we have found is just a singular mathematical solution
of the comic equations. In addition, it is well-known that the
General Theory of Relativity is a viable theory that has been
proved to match the observational data at low energy densities,
but we still do not know what are the valid physical laws at very
high energy densities.

Taking this into account, and the fact that GR is of no use to
describe the physics in very small scales, e.g., the atomic scale, it
is accepted that we need to quantize gravity in order to depict
our Universe at very early times, at least up to the Planck
scales. But, for the moment, nobody knows how to obtain a quantum
theory of gravity, it even might be simply impossible. Maybe
gravity is a force of a very different nature, as compared to the
electromagnetic and the nuclear forces. However, as we will see in
this review, there are attempts to introduce the quantum effects to
understand the past evolution of our Universe.

\subsection{Finite-time Future Singularities}
\label{sec-future-singularities}

In this section, we outline the type of future singularities that appear in various cosmological models, which
 are classified as follows in \cite{Nojiri:2005sx}:
\begin{enumerate}
\item Type I (Big Rip) singularity: $t\rightarrow t_s$, $a\rightarrow \infty$,
$\rho\rightarrow\infty$ and $|p|\rightarrow\infty$.

\item Type II (Sudden) singularity: $t\rightarrow t_s$, $a\rightarrow a_s$,
$\rho\rightarrow\rho_s$ and $|p|\rightarrow\infty$.

\item Type III (Big Freeze) singularity: $t\rightarrow t_s$, $a\rightarrow a_s$,
$\rho\rightarrow\infty$ and $|p|\rightarrow\infty$.

 \item Type IV (Generalized Sudden) singularity: $t\rightarrow t_s$, $a\rightarrow a_s$,
 $\rho\rightarrow \rho_s$, $|p|\rightarrow p_s$ and some higher derivatives of $H$ diverge.

 \item Type V ($w$) singularity: $w \rightarrow \infty$ but $p$, $\rho$ are finite.

 \end{enumerate}

In the following, we shall give a brief illustration to each
finite-time cosmological singularity mentioned above.

\subsubsection{Type I Singularity}
\label{sec-future-singularities-type-I}

This kind of future singularity was introduced first by the
authors of Ref.~\cite{Caldwell:2003vq}. For a linear EoS, when
$w<-1\rightarrow \gamma<0$, that is, when one deals with a phantom
fluid \cite{Caldwell:1999ew}, as we can see from the Eqs. (\ref{1}) and (\ref{2}),
one obtains a future singularity because in that case $t_s>t_0$.
In this situation the Hubble rate and the energy density also
diverge at this moment, but now $\lim_{t\rightarrow
t_s}a(t)=\infty$. This is known as the Type I or {\it Big Rip}
singularity.

\subsubsection{Type II Singularity}
\label{sec-future-singularities-type-II}

In Ref.~\cite{Barrow:2004xh}, Barrow proposed a new kind of
finite-time future singularity appearing in an expanding FLRW
Universe. The singularity may appear without violating the strong
energy condition: $\rho> 0$ and $3 p+ \rho >0$. This kind of
singularity was named as the {\it Sudden singularity}. To deal
with this kind of singularities, we consider a nonlinear EoS of the form \cite{Nojiri:2005sr}
\begin{eqnarray}\label{sec-2-general-EoS}
p = -\rho-f(\rho),
\end{eqnarray}
 where $f$ is an analytic function of the energy density
$\rho$. In that case the conservation equation
\begin{align}
\label{conservation}
\dot{\rho}+3H(\rho+p)=0\, ,
\end{align}
becomes
$\dot{\rho}=3Hf(\rho)$, and using the Friedmann equation, one gets

\begin{align}\label{sec-2-conservation-eqn-for-general-eos}
\dot{\rho}=\sqrt{3}\kappa{\rho^{1/2}}f(\rho)\, .
\end{align}

Choosing, as in Ref. \cite{Nojiri:2005sr}
\begin{eqnarray}\label{sec-2-eos-general-f-rho}
f(\rho)=\frac{A} {\sqrt{3}\kappa}\rho^{\nu+\frac{1}{2}},
\end{eqnarray}
where $A$ and $\nu$ are two free parameters, from Eq. (\ref{sec-2-conservation-eqn-for-general-eos}), one obtains the first order differential equation
\begin{align}
\dot{\rho}=A\rho^{\nu+1}\, ,
\end{align}
whose solution is given by
\begin{align}
\rho(t)=\left\{\begin{array}{cc}
\bigg(\rho_0^{-\nu}-\nu A(t-t_0)\bigg)^{-1/\nu}, \quad & \mbox{for} \quad \nu\not= 0, \\
\rho_0\e^{A(t-t_0)}, \quad & \mbox{for} \quad \nu=0,
\end{array} \right.
\end{align}
where $\rho_0$ is the current value of the energy density.
Firstly, we consider the case $\nu<-1/2$. From the Friedmann equation the Hubble rate is given by
\begin{align}
\label{hubble}
H(t)= \frac{\kappa}{\sqrt{3}}\bigg(\rho_0^{-\nu}-\nu A(t-t_0)\bigg)^{-1/2\nu},
\end{align}
which introducing the time $t_s=t_0+\frac{\rho_0^{-\nu}}{ \nu A}$, could be written as
\begin{align}\label{10}
H(t)= \frac{\kappa}{\sqrt{3}}\bigg(\nu A(t_s-t)\bigg)^{-1/2\nu}, \end{align}
and thus, the scale factor is given by
\begin{align}
\label{scalefactor}
\ln\left(\frac{a(t)}{a_0} \right)=-\frac{2\kappa}{3A(2\nu-1)}
\bigg(\nu A(t_s-t)\bigg)^{(2\nu-1)/2\nu}+ \frac{2\kappa}{3A(2\nu-1)}
\bigg(\nu A(t_s-t_0)\bigg)^{(2\nu-1)/2\nu}.
\end{align}
Then, for a non phantom fluid, that is for $A<0$, the effective
EoS parameter $w\equiv p/\rho$ is given by $w
=-1-\frac{A}{\sqrt{3}\kappa}\rho^{\nu-\frac{1}{2}}>-1$. Since in
that case, $t_s>t_0$ and $\rho(t_s)$ vanishes, we find that the
pressure $p$ diverges at the instant $t_s$, obtaining a Sudden
singularity \cite{Barrow:2004xh,Barrow:2004hk}.

\subsubsection{Type III Singularity}
\label{sec-future-singularities-type-III}

We consider the same EoS (\ref{sec-2-general-EoS}) where $f(\rho)$
is given in Eq. (\ref{sec-2-eos-general-f-rho}), but here we
consider the case $\nu>1/2$ and $A>0$, that is, a phantom fluid,
which implies that $t_s>t_0$.
 From Eq. (\ref{scalefactor}), we see that $a(t)\rightarrow a_s$ (finite) when $t\rightarrow t_s$, and from Eq. (\ref{hubble}),
we deduce that both the energy density and the pressure diverge at that instant, which leads to a Big Freeze singularity.

\subsubsection{Type IV Singularity}
\label{sec-future-singularities-type-IV}

We continue with the same EoS (\ref{sec-2-general-EoS}) where
$f(\rho)$ given in Eq. (\ref{sec-2-eos-general-f-rho}), but with
$-1/2<\nu<0$ and $A<0$ (non-phantom fluid). In this situation,
$t_s>t_0$, and once again, the scale factor converges to $a_s$
(finite) when $t\rightarrow t_s$, but now the energy density and
the pressure go to zero when the cosmic time approaches to $t_s$.
In addition, looking at the Hubble rate obtained in Eq.
(\ref{hubble}) we easily conclude that when $-1/(2\nu)$ is not a
natural number, some higher order derivatives of the Hubble
parameter diverge at $t = t_s$, obtaining a Generalized Sudden
singularity.

\subsubsection{Type V ($w$) Singularity}
\label{sec-future-singularities-type-V}

For $t \rightarrow t_s$, $a \rightarrow \infty$, $\rho \rightarrow
0$, $|p| \rightarrow 0$ or to a finite value, but the EoS, $w
\rightarrow \infty$. This kind of singularity is known as Type V
or $w$ singularity \cite{Dabrowski:2009kg}. This future
singularity appears when the Hubble rate has the following
analytic form near the singularity
\cite{Fernandez-Jambrina:2010pep},
\begin{align}
H(t)=\sum_{n=1}^{\infty} H_n\left( \frac{t_s-t}{t_n} \right)^n,
\end{align}
where all $t_n$'s ($n = 1, 2, ...$) are positive numbers. In that
case, using the EoS
\begin{align}
w(t)=-1-\frac{2\dot{H}}{3H^2},
\end{align}
near $t_s$, one finds that
\begin{align}
w(t)\cong -1+\frac{2r}{H_r t_r}\left( \frac{t_s-t}{t_r} \right)^{-r-1},
\end{align}
where $H_r$ is the first non-vanishing term of the series. We can
see that the EoS parameter $w$ diverges at time $t_s$, but the
energy density and the pressure,
\begin{align}
\rho= \frac{3H^2}{\kappa^2}\, ,\quad p=-\frac{1}{\kappa^2}(3H^2+2\dot{H})\, ,
\end{align}
are finite at time $t_s$. In fact, the energy density vanishes and
the pressure is equal to $\frac{2H_1}{\kappa^2t_1}$, so it
vanishes only when $H_1=0$.

As an example we continue with the model (\ref{sec-2-general-EoS})
where $f(\rho)$ is given by Eq. (\ref{sec-2-eos-general-f-rho}).
And we consider the simple case $\nu=-\frac{1}{2}$ and $A<0$.
Then, we have
\begin{align}
w=-1-\frac{A}{\sqrt{3}\kappa \rho}\, .
\end{align}

Now, from Eq. (\ref{10}) the Hubble rate has the form
\begin{align}\label{23}
H(t)=-\frac{A\kappa}{2\sqrt{3}}(t_s-t)\, ,
\end{align}
and the energy density becomes
\begin{align}
\rho(t)=\frac{A^2}{2}(t_s-t)^2\, ,
\end{align}
which means that at $t=t_s$, the energy density vanishes and the
pressure has the finite value $\frac{A}{\sqrt{3}\kappa}$, and, as a consequence, the EoS parameter $w$ diverges at $t=t_s$.

To end this section, note that the case $0<\nu<1/2$ and $A<0$
(non-phantom fluid) corresponds to a Big Bang singularity and for
$0<\nu<1/2$ and $A>0$ (phantom fluid) one gets a Big Rip
singularity. In addition, when $\nu=0$ and $A>0$ one obtains the
so-called Little Rip \cite{Frampton:2011sp, Frampton:2011rh}\footnote{The Little Rip scenario has been discussed in detail in section~\ref{subsection-little-rip}. } where $w<-1$ and it asymptotically converges to $-1$. Effectively, in this
case the energy density is given by $\rho(t)=\rho_0\e^{A(t-t_0)}$
which diverges when $t\rightarrow \infty$. So, we find
\begin{align}
w=\frac{p}{\rho}=-1-\frac{A}{\sqrt{3\rho}\kappa}\rightarrow -1\, .
\end{align}

Finally, for the remaining case $\nu=1/2$ and $A>0$, the Hubble rate is given by
\begin{align}
 H=\frac{2\kappa}{\sqrt{3}A (t_s-t)}\, ,
\end{align}
and thus, the scale factor is given by
\begin{align}
 \ln\left(\frac{a(t)}{a_0} \right)=-\frac{2\kappa}{\sqrt{3}A}\ln\left(\frac{t_s-t}{t_s-t_0}\right)\, ,
\end{align}
which shows that the scale factor diverges when $t\rightarrow t_s$, and thus, a Big Rip singularity occurs.

Summarizing, for the EoS
$p=-\rho-\frac{A}{\sqrt{3}\kappa}\rho^{\nu+\frac{1}{2}}$, we have
the following classification for future singularities as a
function of $\nu$:
\begin{itemize}
 \item For $\nu<-1/2$ and $A<0$, one has a Type II singularity.
 \item For $\nu=-1/2$ and $A>0$, one has a Little Rip. 
 \item For $\nu=-1/2$ and $A<0$, one has a Type V singularity.
 \item For $-1/2<\nu<0$ and $A<0$, one has a Type IV singularity.
 \item For $\nu=0$ and $A>0$, one has a Little Rip. 
 \item For $0<\nu\leq 1/2$ and $A>0$, one has a Big Rip singularity.
 \item For $\nu>1/2$ and $A>0$, one has a Type III singularity.
\end{itemize}

\subsection{The Hamilton-Jacobi approach to cosmological singularities}

The Hamilton-Jacobi approach developed by Salopek and Bond \cite{Salopek:1990jq} can be used in the context of cosmological singularities. 
According to this approach \cite{Salopek:1990jq}, cosmological models
can be written as a function of a scalar field $\phi$ with a potential associated with any dependence of the Hubble rate.
Therefore, one can find the corresponding potential associated with the singular behavior studied above.

Starting with the Big Rip singularity, with a Hubble rate given by
$H(t)=- \frac{2}{3\gamma (t_s-t)}$ with $\gamma<0$, one can find the potential of a phantom field leading to this dynamical behavior. Denoting $H'$ as the derivative with respect to the scalar field $\phi$, we have $\dot{H}=H'\dot{\phi}$, and thus, the Raychaudhuri equation becomes $H'=\frac{\kappa^2}{2}\dot{\phi}$ (recall that we are dealing with a phantom field). Thus, the Friedmann equation will be 
\begin{eqnarray}
 H^2(\phi)=\frac{\kappa^2}{3}\left(-\frac{2}{\kappa^4}(H'(\phi))^2+V(\phi) \right)\Longrightarrow V(\phi)=\frac{3}{\kappa^2}H^2(\phi)+\frac{2}{\kappa^4}(H'(\phi))^2.
 \end{eqnarray}
Next, from the Raychaudhuri equation one can find the following relation between the field and the cosmic time $\phi=\frac{1}{\kappa}\int \sqrt{2\dot{H}}dt$, which for our Hubble rate leads to 
\begin{eqnarray}
 \phi(t)=-\frac{2}{\kappa\sqrt{-3\gamma}}\ln\left(\frac{t_s-t}{\kappa} \right),
\end{eqnarray}
where as an initial condition we have chosen $\phi(t_s-\kappa)=0$.
Then, we have $H(\phi)=-\frac{2}{3\kappa\gamma}e^{\sqrt{-3\gamma}\kappa \phi/2}$, and the corresponding potential is given by
\begin{eqnarray}
 V(\phi)=\frac{2}{3\kappa^4\gamma^2}(2-\gamma)e^{\sqrt{-3\gamma}\kappa \phi}.
 \end{eqnarray}
Dealing with the singularities of Type II-IV, we consider that the Hubble rate given in the equation (\ref{10}), which can be written as
\begin{eqnarray}
 H(t)=B(t_s-t)^{-1/2\nu},\qquad \mbox{with} \qquad B=\frac{\kappa}{\sqrt{3}}(\nu A)^{-1/2\nu}.
\end{eqnarray}

Here, it is important to realize that in order to realize this dynamics, since for the Type II and IV singularities one has $w>-1$, we need a non phantom field, and for the Type III, a phantom field is needed. Therefore, following the similar steps as in the Big Rip singularity, we find 
\begin{eqnarray}
 H_{\pm}(\phi)=C_{\pm}\phi^{\frac{2}{1-2\nu}} \qquad \mbox{with} \qquad
 C_{\pm}=B\left(\pm \frac{\kappa^2(1-2\nu)^2}{16B\nu}\right)^{\frac{1}{1-2\nu}},
\end{eqnarray}
where the $(+)$ sign refers to the Type III singularity because in that case $\nu>1/2$, and the $(-)$ sign refers to the singularities of Type II and IV, because in that case $\nu<0$.

Then, the corresponding potential is given by
\begin{eqnarray}
V_{\pm}(\phi)=\frac{C_{\pm}^2}{\kappa^2}\left(3\pm \frac{8}{\kappa^2 (1-2\nu)^2\phi^2}\right) \phi^{\frac{4}{1-2\nu}},
\end{eqnarray}
the shape of which depends on the value of the parameter $\nu$. Effectively, we have the following observations:
\begin{enumerate}
 \item For the Type II singularity where $\nu<-1/2$, one has $V(-\infty)=+\infty$ and $V(0)=-\infty$.
\item For the Type III singularity where $\nu>1/2$, one has $V(-\infty)=0$ and $V(0)=+\infty$. 
\item For the Type IV singularity where $-1/2<\nu<0$, one has $V(-\infty)=+\infty$ and $V(0)=0$.

\end{enumerate}

Finally, to discuss the Type V singularity one may choose the Hubble rate given in Eq. (\ref{23}) which corresponds to the case $\nu=-1/2$ and $A<0$, and thus, a non phantom fluid is represented. As we have already explained, to depict this behavior we need a non phantom field whose potential, for this Hubble rate, is given by
\begin{eqnarray}
 V(\phi)=-\frac{\sqrt{3}A\kappa}{4}\left(\phi^2-\frac{2}{3\kappa^2} \right).
\end{eqnarray}

\section{Finite-time Singularities in DE Models}
\label{sec-singularities-DE}

Within the context of GR, a hypothetical cosmic fluid is added to
explain the accelerating phase of the Universe, known as DE. The
nature and dynamics of DE are not known. Hence, over the last
several years, a plethora of DE models have been introduced in the
literature~
\cite{Carroll:1998zi,Amendola:1999er,Barreiro:1999zs,Sahni:1999qe,delaMacorra:1999ff,Brax:1999yv,Urena-Lopez:2000ewq,Gonzalez-Diaz:2000glv,Matos:2000ng,Huterer:2000mj,Kamenshchik:2001cp,Bilic:2001cg,Zimdahl:2001ar,Baccigalupi:2001aa,Gu:2001tr,Bento:2001yv,Bento:2002ps,Padmanabhan:2002cp,Cardenas:2002np,Franca:2002iju,Huterer:2002hy,Corasaniti:2001mf,Freese:2002sq,Melchiorri:2002ux,Gerke:2002sx,Gorini:2002kf,Singh:2003vx,Huey:2004qv,Chae:2004jp,Chen:2004nqb,Li:2004rb,Scherrer:2004au,Alam:2004jy,Debnath:2004cd,Wetterich:2004pv,Jarv:2004uk,Bassett:2004wz,Capozziello:2005pa,Banerjee:2005ef,Banerjee:2005nv,Guo:2005ata,Cataldo:2005qh,Sahlen:2005zw,Kaloper:2005aj,Fabris:2005ts,Sola:2005et,Zhang:2004gc,Guo:2005qy,Banerjee:2006rp,Linder:2006sv,Sahlen:2006dn,Nojiri:2006zh,Zhang:2006qu,Huterer:2006mv,Zhang:2006av,Nojiri:2006jy,Wei:2007ty,Olivares:2007rt,Gao:2007ep,Linder:2007wa,Avelino:2008ph,Zhang:2008mb,Wu:2008jt,Dutta:2009yb,Zhang:2009un,Li:2009mf,Banerjee:2008rs,Gagnon:2011id,Adak:2012bv,Chen:2013vea,Velten:2013qna,Roy:2013wqa,Mahata:2013oza,Sharov:2014voa,Chakraborty:2014fia,Banerjee:2015kva,Sharov:2015ifa,Mahata:2015lja,Magana:2017nfs,Zhao:2017cud,Rezaei:2017yyj,Yang:2018qmz,Malekjani:2018qcz,Chakraborty:2018nrk,Yang:2019nhz,Perkovic:2019vxm,Yang:2019jwn,Rezaei:2019hvb,Pan:2019hac,Rezaei:2019xwo,Chakraborty:2019swx,Sinha:2020vob,Yang:2021flj,Yang:2021eud,Saridakis:2021qxb,Saridakis:2021xqy,Benaoum:2020qsi,Sharov:2022hxi,Liu:2022mpj,Cardona:2022pwm,Yao:2023ybs}
(also see the review articles in this direction
\cite{Peebles:2002gy,Padmanabhan:2002ji,Sahni:2002kh,Copeland:2006wr,Sahni:2006pa,Padmanabhan:2007xy,Frieman:2008sn,Martin:2008qp,Caldwell:2009ix,Silvestri:2009hh,Bamba:2012cp,Li:2012dt,Mortonson:2013zfa,Sami:2013ssa,Brevik:2017msy}
and the references therein).

\subsection{Viscous Cosmologies}
\label{sec-singularities-viscous-cosmology}

A bulk viscous fluid is characterized by its energy density $\rho$ and pressure $p$, where the pressure component has two sub-components,
one is the conventional component $p_\mathrm{conv} = w \rho $ and other is the bulk viscosity component $p_\mathrm{vis} = - \xi u^{\mu}_{; \mu}$,
where $u^{\mu}_{; \mu}$ is the fluid expansion scalar,
and $\xi$ is the bulk viscous coefficient which could be either constant or dynamical.
Thus, the pressure of a bulk viscous fluid is represented as
\begin{align}\label{p-VC}
p = p_\mathrm{conv} + p_\mathrm{vis} = w \rho - \xi u^{\mu}_{; \mu}.
\end{align}

In the flat FLRW Universe, one has $u^{\mu}_{; \mu}=3H$, thus, in
pressure of the bulk viscous fluid, the pressure term given in
Eq.~(\ref{p-VC}) reduces to
\begin{align}\label{p-explicit-VC}
p = w \rho - 3 H \xi\, ,
\end{align}
where if $\xi$ is dynamical, then it could take one of the forms
such as, $\xi \equiv \xi (t)$, $\xi \equiv \xi (\rho)$, $\xi \equiv \xi (a)$, $\xi \equiv \xi (H)$, $\xi \equiv \xi\left(H, \dot{H}, \ddot{H},... \right)$,
 or it can take a more general form like
$\xi = \xi \left(t, a, \rho, H, \dot{H}, \ddot{H},... \right)$.
The microscopic dynamics may determine the dynamics of $\xi$.
Because we do not know the microscopic dynamics, we consider arbitrary cases.

We start once again with the simple choice of the EoS, $p = - \rho - f(\rho)$, given in
Eq. (\ref{sec-2-general-EoS}).
One can realize that Eq. (\ref{sec-2-general-EoS}) is a very
special case of Eq.~(\ref{p-explicit-VC}). For example, if $\xi$
does not depend on time, i.e., $\xi = \bar{\xi}$ (constant),
then from the Friedmann equation (\ref{FReqs}),
one can express the Hubble rate as, $H = \kappa\sqrt{\rho/3}$
(taking the expansion of the Universe, i.e., $H > 0$) and
consequently, Eq. (\ref{p-explicit-VC}) turns out to be $p = w \rho -
\kappa \bar{\xi} \sqrt{3 \rho} = -\rho - (k\bar{\xi} \sqrt{3
\rho} - \rho - w \rho)$ in which $f(\rho) = k\bar{\xi} \sqrt{3
\rho} - \rho - w \rho$. On the other hand, if $\xi$ is a function
of $H$, then similarly one can find a suitable $f (\rho)$.

Let us introduce a more generalized EoS of the bulk viscous fluid
as follows \cite{Nojiri:2005sr}:
\begin{align}\label{inhom-equation-of-state-2}
p=-\rho - f(\rho) - G(H),
\end{align}
where $G(H)$ is some arbitrary function of the Hubble parameter
$H$. However, $G$ can be any arbitrary function of $H$ and its
derivatives of any order and the scale factor of the Universe
\cite{Odintsov:2020voa,Nojiri:2021dze}. We start with the simple
case where $G$ is the function of the Hubble parameter only, that
means, we focus on Eq. (\ref{sec-2-general-EoS}). Let us note
that for a spatially flat Universe, the EoS given in Eqs. (\ref{sec-2-general-EoS}) and (\ref{inhom-equation-of-state-2}) are equivalent
since using the Friedmann equation,
one can express the Hubble parameter in terms of the energy density of the Universe, while for a non-flat case, Eq. (\ref{inhom-equation-of-state-2}) is not equivalent to Eq. (\ref{sec-2-general-EoS}).
Therefore, we also keep the EoS shown in Eq. (\ref{inhom-equation-of-state-2}) in our discussion.
Now plugging Eq. (\ref{inhom-equation-of-state-2}) into the conservation equation, i.e. Eq. (\ref{conservation}),
one gets,
\begin{align}
\dot \rho - 3H\left(f(\rho) + G(H)\right) = 0\, .
\end{align}

And using the Friedmann equation, i.e. Eq. (\ref{FReqs}) in an expanding Universe,
one derives that
\begin{align}
\label{inhom-equation-of-state-2.1}
\dot\rho = F (\rho) \equiv 3\kappa\sqrt{\frac{\rho}{3}}\left[ f(\rho)
+ G\left(\kappa\sqrt{\rho/3}\;\right)\right]\, .
\end{align}
We start with a simple EoS as follows
\begin{align}
\label{inhom-equation-of-state-3}
p= w_1\rho + w_2 \frac{3}{\kappa^2}H^2\,,
\end{align}
where $w_1$ and $w_2$ are constants.
Now, using the Friedmann equation, the above EoS (\ref{inhom-equation-of-state-3}), can be
rewritten as
\begin{align}
p=\left(w_1 + w_2\right)\rho\,,
\end{align}
which takes the form of $ p = w_\mathrm{eff} \rho$ in which $w_\mathrm{eff}$ has the following expression
\begin{align}
\label{inhom-equation-of-state-3-eff}
w_\mathrm{eff} = w_1+w_2\, .
\end{align}
Now from Eq. (\ref{inhom-equation-of-state-3-eff}), as long as
$w_\mathrm{eff}>-1$ even if $w_1<-1$, no Big Rip singularity occurs.
However, on the other hand, even if $w_1 > -1$, but one may obtain
$w_\mathrm{eff} < -1$ for sufficiently negative value of $w_2$, and
thus the Big Rip singularity may occur.

A more generalized EoS can be considered where $f(\rho)$
can be modified as
\begin{align}
\label{ppH9}
f(\rho)
\rightarrow f(\rho) + G(H).
\end{align}

For example, choosing $f(\rho)=A\rho^{\alpha}$ and
$G(H)=BH^{2\beta}$, and using the Friedmann equation (\ref{FReqs}), $f(\rho)$
gets modified as,
\begin{align}
\label{new-f}
f_\mathrm{eff}(\rho)
= f(\rho) + G(H) = A\rho^{\alpha} B \left(\frac{\kappa^2}{3}\right)^\beta \rho^\beta.
\end{align}
For $\beta>\alpha$, when $\rho$ is large, the second term in the
right hand side (r.h.s.) of Eq.~(\ref{new-f}) becomes dominant leading to
\begin{align}
\label{new-f-asymptotic}
f_\mathrm{eff}(\rho) \rightarrow B \left(\frac{\kappa^2}{3}\right)^\beta \rho^\beta\, .
\end{align}
On the other hand, if $\beta<\alpha$, the second term in the right
hand side of Eq.~(\ref{new-f}) becomes dominant and we obtain again
(\ref{new-f-asymptotic}) for $\rho\to 0$.

Following Ref.~\cite{Nojiri:2005sr}, we summarize the appearance
of singularities for this EoS considering various ranges of
$\alpha$ and $\beta$:

\begin{itemize}

\item When $\alpha>1$: for most of the values of $\beta$, there Type III singularity occurs.
Additionally, when $0<\beta<1/2$, Type IV
singularity appears and when $\beta<0$, Type II singularity appears.

\item When $\alpha=1$: if $\beta>1$, then Type III singularity
appears (for $\beta=1$, we get the EoS as in
Eq.~(\ref{inhom-equation-of-state-3})); if $\beta<1$ and $A>0$, Big Rip or Type I singularity occurs.
In addition to the Type I singularity, Type IV singularity appears for $0<\beta<1/2$ and
Type II singularity appears for $\beta<1$.

\item When $1/2<\alpha<1$: Type III singularity appears for $\beta>1$;
Type I singularity appears for $1/2\leq\beta<1$ (even for $\beta=1/2$) or $\beta=1$ and $B>0$;
Type IV singularity appears for $0<\beta<1/2$; and Type II singularity appears for $\beta<0$.

\item When $\alpha=1/2$: Type III singularity appears for $\beta>1$;
Type I singularity appears for $1/2<\beta<1$ or $\beta=1$ and $B>0$;
Type IV singularity appears for $0<\beta<1/2$;
Type II singularity occurs for $\beta<0$.
We note that for $\beta=1/2$ or $\beta=0$, no singularity appears.

\item For $0<\alpha<1/2$: Type IV singularity appears for $0<\beta<1/2$;
Type II singularity appears for $\beta<0$.
In addition to Type IV singularity, Type III singularity occurs
for $\beta>1$ and Type I singularity occurs for $1/2\leq\beta<1$ or $\beta=1$ $B>0$.

\item When $\alpha<0$: Type II singularity occurs.
In addition to Type II singularity, Type III singularity appears for $\beta>1$;
Type I singularity for $1/2\leq\beta<1$ or $\beta=1$ and $B>0$.
\end{itemize}

We also propose an implicit inhomogeneous EoS by generalizing the
expression $p=-\rho-f(\rho)$ as
\begin{align}
\mathcal{F} (p, \rho, H) = 0\,.
\label{eq:III-E-18}
\end{align}
In order to understand the cosmological consequences of the
implicit inhomogeneous EoS, we present the following example:
\begin{align}
\left( p + \rho \right)^2 - C_{\mathrm{c}} \rho^2
\left( 1 - \frac{H_{\mathrm{c}}}{H} \right) = 0\,,
\label{eq:III-E-19}
\end{align}
with $C_{\mathrm{c}}$ (dimensionless) and $H_{\mathrm{c}}$ as positive constants.
Plugging Eq.~(\ref{eq:III-E-19}) into the Raychaudhuri equation
$\dot{H}=-\frac{\kappa^2(p+\rho)}{2}$ and using the Friedmann equation
$H^2=\frac{\kappa^2\rho}{3}$ in Eq. (\ref{FReqs}), we acquire
\begin{align}
\dot{H}^2 = \frac{9}{4} C_{\mathrm{c}} H^4 \left( 1 - \frac{H_{\mathrm{c}}}{H}
\right)\,.
\label{eq:III-E-20_0}
\end{align}
We can integrate Eq.~(\ref{eq:III-E-20_0}) as
\begin{align}
H = \frac{16}{9 C_{\mathrm{c}}^2 H_{\mathrm{c}} \left( t - t_- \right)
\left( t_+ - t \right)}\,,
\label{eq:III-E-20}
\end{align}
with
\begin{align}
t_\pm \equiv t_{\mathrm{c}} \pm \frac{4}{3 C_{\mathrm{c}} H_{\mathrm{c}}}\,,
\label{eq:III-E-21}
\end{align}
where $t_{\mathrm{c}}$ is a constant of integration.
By substituting Eq.~(\ref{eq:III-E-20}) into
$p=-\rho \left( -1-\frac{2\dot{H}}{3H^2} \right)$ and $\rho=\frac{3H^2}{\kappa^2}$ we find
\begin{align}
p = - \rho \left[ 1 + \frac{3 C_{\mathrm{c}}^2}{4 H_{\mathrm{c}}}
\left( t - t_{\mathrm{c}} \right) \right]\quad \mbox{and}\quad
\rho = \frac{256}{27 C_{\mathrm{c}}^4 H_{\mathrm{c}}^2 \kappa^2
\left( t - t_- \right)^2 \left( t_+ - t \right)^2}\,.
\label{eq:III-E-23}
\end{align}
Thus, we have
\begin{align}
w = \frac{p}{\rho} = -1 - \frac{3 C_{\mathrm{c}}^2}{4 H_{\mathrm{c}}}
\left( t - t_{\mathrm{c}} \right)\,.
\label{eq:III-E-24}
\end{align}
 From Eq.~(\ref{eq:III-E-20}), we see that if $t_- < t < t_+$, $H > 0$.
At $t = t_{\mathrm{c}} = \left( t_- + t_+ \right)/2$, $H$ becomes
the minimum of $H =H_{\mathrm{c}}$. On the other hand, in the
limit of $t \to t_\pm$, $H \to \infty$. This may be interpreted
that at $t = t_-$, there exists a Big Bang singularity, whereas at
$t = t_-$, a Big Rip singularity appears. It is clearly seen from
Eq.~(\ref{eq:III-E-20}) that when $t_- < t < t_{\mathrm{c}}$, $w >
-1$ (the non-phantom (quintessence) phase), and when
$t_{\mathrm{c}} < t < t_+$, $w < -1$ (the phantom phase). At $t =
t_{\mathrm{c}}$, the crossing of the phantom divide line from the
non-phantom phase to the phantom phase occurs.
This is realized by an inhomogeneous term in the EoS.

\subsection{Interacting DE Models}
\label{sec-singularities-IDE}

Cosmological models allowing a non-gravitational interaction
between DM and DE are widely known as Interacting DE (IDE) or
Coupled DE (CDE) models. The theory of IDE has received massive
attention due to its ability to explain several cosmological
puzzles and for offering some interesting results
\cite{Amendola:1999er,Gondolo:2002fh,Farrar:2003uw,Huey:2004qv,Cai:2004dk,Guo:2004xx,Wang:2005jx,Pavon:2005yx,Das:2005yg,Barrow:2006hia,Amendola:2006dg,delCampo:2006vv,Wang:2006qw,Bertolami:2007zm,delCampo:2008sr,Valiviita:2008iv,delCampo:2008jx,He:2008tn,Boehmer:2008av,Gavela:2009cy,Jackson:2009mz,Koyama:2009gd,Majerotto:2009np,Boehmer:2009tk,Jamil:2009eb,He:2010im,Li:2010ju,Chimento:2011pk,Chimento:2012zz,Pettorino:2013oxa,Costa:2013sva,Chimento:2013rya,Chakraborty:2012gr,Yang:2014gza,yang:2014vza,Yang:2014hea,Pan:2012ki,Nunes:2016dlj,vandeBruck:2016jgg,Pourtsidou:2016ico,vandeBruck:2016hpz,Mukherjee:2016shl,An:2017crg,Sharov:2017iue,DiValentino:2017iww,Yang:2017yme,vandeBruck:2017idm,Cai:2017yww,Kumar:2017dnp,Yang:2017ccc,Yang:2017zjs,VanDeBruck:2017mua,Mifsud:2017fsy,Pan:2017ent,Yang:2018euj,Yang:2018xlt,Gonzalez:2018rop,Yang:2018qec,Pan:2019jqh,Paliathanasis:2019hbi,Yang:2019uog,Li:2019loh,Yang:2019vni,vonMarttens:2019ixw,Li:2019ajo,Pan:2019gop,Mifsud:2019fut,Yang:2020uga,Pan:2020bur,DiValentino:2019ffd,DiValentino:2019jae,Lucca:2020zjb,Yang:2020tax,Pan:2020mst,DiValentino:2020kpf,Wang:2021kxc,Anchordoqui:2021gji,Gao:2021xnk,Yang:2021hxg,Guo:2021rrz,Johnson:2021wou,Yang:2021oxc,Paliathanasis:2021egx,Lucca:2021dxo,Gariazzo:2021qtg,Bonilla:2021dql,Mukherjee:2021ggf,Nunes:2022bhn,Yao:2022kub,Alvarez-Ortega:2022vbo,Harko:2022unn,Yang:2022csz,Pan:2022qrr,Zhao:2022ycr,Zhai:2023yny}
(also see two review articles in this direction, e.g.
Refs.~ \cite{Bolotin:2013jpa,Wang:2016lxa}). In IDE theory, the
choice of the interaction function or the interaction rate is
usually made from the phenomenological ground. Even though there
are some attempts to derive the interaction function from the
fundamental principle \cite{vandeBruck:2015ida,
Boehmer:2015kta,Boehmer:2015sha,Gleyzes:2015pma,DAmico:2016jbm,Pan:2020zza,Chatzidakis:2022mpf},
however, this sector needs considerable progress. Therefore,
usually, in most of the IDE scenarios, the interaction function is
chosen by hand, therefore, the appearance of finite-time
singularities in those models is not unnatural. In this section, we
shall investigate whether the interacting models allow the
finite-time singularities in the future.

In a homogeneous and isotropic Universe characterized by the FLRW
line element, the conservation equations of the DM and DE sectors
in presence of an interaction between them take the forms
\begin{align}
\label{dark}
\left\{ \begin{array}{rl}
\dot{\rho}_\mathrm{DM}+3H\rho_\mathrm{DM} =& -Q(t), \\
\dot{\rho}_\mathrm{DE}+3H(1+w_\mathrm{DE})\rho_\mathrm{DE} =&Q(t),
\end{array}
\right.
\end{align}
where $\rho_\mathrm{DM}$, $\rho_\mathrm{DE}$ are the energy
density of DM and DE, respectively, $w_\mathrm{DE} =
p_\mathrm{DE}/\rho_\mathrm{DE}$ ($p_\mathrm{DE}$ is the pressure
of DE) refers to the EoS parameter of DE and DM has been assumed
to be pressure-less, and $Q(t)$ is the energy exchange rate
between the dark sectors. The sign of $Q(t)$ determines the direction of energy flow between the dark components. For $Q > 0$, energy flows from DM to DE and for $Q (t) <0$ the direction of flow is reversed, i.e. energy flows from DE to DM. 
The Friedmann and Raychaudhuri equations
(\ref{FReqs}) in the flat FLRW Universe (\ref{FLRWk0}) are
\begin{align}
H^2 = \frac{\kappa^2}{3} (\rho_\mathrm{DM} + \rho_\mathrm{DE}), \quad \mbox{and} \quad
\dot{H} = - \frac{\kappa^2}{2} (\rho_\mathrm{DM} + \rho_\mathrm{DE} + p_\mathrm{DE})\, ,
\end{align}
from which one could rewrite the energy densities of DE and DM as
\begin{eqnarray}
 &&\rho_\mathrm{DE} = -\frac{1}{\kappa^2 w_\mathrm{DE}}(3H^2+2\dot{H}), \label{eqn-rho-DE-sec-IDE-sp} \\
 &&\rho_\mathrm{DM} = \frac{1}{\kappa^2 w_\mathrm{DE}}\left[3(1+w_\mathrm{DE})H^2+2\dot{H}\right]. \label{eqn-rho-DM-sec-IDE-sp}
 \end{eqnarray}
Now if one combine this last expression (i.e. Eq.
(\ref{eqn-rho-DM-sec-IDE-sp})) with the conservation equation for
DM given in Eq.~(\ref{dark}), one gets the following expression
for $Q(t)$ \cite{BeltranJimenez:2016dfc}
\begin{align}\label{Q}
 Q(t)= - \frac{1}{\kappa^2w_\mathrm{DE}}\left[9(1+w_\mathrm{DE})H^3+6(2+w_\mathrm{DE})H\dot{H} + 2\ddot{H}-\frac{\dot{w}_\mathrm{DE}}{w_\mathrm{DE}}(3H^2+2\dot{H}) \right],
\end{align}
which means that given a dynamics, characterized by $H(t)$, one recovers the interaction rate $Q(t)$. Moreover, any singularity in
$H$, $\dot{H}$ or $\ddot{H}$ determines a singularity in $Q(t)$
and vice versa. In the following we shall discuss the possibility
of singularities for various choices for the DE EoS parameter.

\subsubsection{IDE with constant EoS in DE}

For the choice of constant EoS in DE, the contribution of
$\dot{w}_\mathrm{DE}$ from Eq.~(\ref{Q}) is removed. We remark
that for some specific dynamics, the interaction models may
encounter with finite-time future singularities. We consider the
following dynamics where the Hubble rate takes the form
\cite{BeltranJimenez:2016dfc}
\begin{align}\label{model}
 H(t)=H_s+\frac{2}{3t}+\frac{h_s}{(t_s-t)^{\beta}}~,
\end{align}
where $H_s\gg 1/t_s$ and $h_s$ are some positive constants and
$\beta \neq 0$ is a dimensionless parameter which determines the
type of singularity at time $t_s$. Let us note that the above
parametrization depicts a matter dominated early Universe while at
late-time, close to $t_s$, the Universe is dominated by a
cosmological constant when $\beta<0$, with a deviation from the
$\Lambda$CDM model controlled by the last term in (\ref{model}).
On the contrary, it has a singularity at $t_s$ for positive
values of $\beta$.

Inserting Eq.~(\ref{model}) into Eq.~(\ref{Q}) and retaining the
leading term near $t_s$, for $\beta<1$ one gets $Q(t)\sim
(t_s-t)^{-3\beta},$ and for $-2<\beta<1$ the leading term is
$Q(t)\sim (t_s-t)^{-\beta-2}$. Then, the following cases appear:
\begin{itemize}
\item When $\beta>1$: close to $t_s$, one has
\begin{align}
\label{BigRip}
H(t)\sim {h_s}{(t_s-t)^{-\beta}}\rightarrow\infty, \qquad \mbox{and} \qquad a(t)\sim \exp\left(-\frac{h_s}{1-\beta}(t_s-t)^{1-\beta} \right)\rightarrow\infty\, ,
\end{align}
which means that a Big Rip singularity appears.
\item When $\beta=1$: near $t_s$ one has
\begin{align}
\label{BigRipa}
H(t)\cong \frac{h_s}{t_s-t} \rightarrow \infty, \qquad \mbox{and} \qquad a(t)\sim (t_s-t)^{-h_s}
\rightarrow\infty\, ,
\end{align}
which also indicates a Big Rip singularity.
\item When $0<\beta<1$: at $t\sim t_s$ one has
\begin{align}
H(t)\sim {h_s}{(t_s-t)^{-\beta}}\rightarrow\infty, \qquad \mbox{and} \qquad a(t)\sim \exp\left(-\frac{B}{1-\beta}(t_s-t)^{1-\beta} \right)\rightarrow 1\, ,
\end{align}
which indicates a Type III singularity.
\item When $-1<\beta<0$: at $t\sim t_s$ one has
\begin{align}
H(t)\sim H_{s}~,
\qquad \mbox{and} \qquad
a(t)\cong \e^{H_s(t-t_s)}\sim 1\, ,
\end{align}
and thus both quantities are finite, but $\dot{H}\cong \frac{h_s}{(t_s-t)^{1+\beta}}\rightarrow \infty$, which means that, 
the pressure diverges, and hence, a Type II singularity is obtained.
\item When $-2<\beta<-1$: at $t\sim t_s$
the scale factor, the Hubble rate and its derivative are finite,
but the second derivative of the Hubble rate diverges, and thus, a Type IV ({\it Generalized Sudden}) singularity appears.
\end{itemize}

\subsubsection{IDE with dynamical EoS in DE}

In this Section we follow the approach of
\cite{BeltranJimenez:2016dfc}, and we consider the following EoS
parameter of DE
\begin{align}
w_\mathrm{DE} = w_s + \frac{k_s}{(t_s-t)^{\beta}}~,
\end{align}
where $w_s$, $k_s>0$ and $\beta$ are free parameters.
Note that when $-1<\beta<0$, the quantity $\dot{w}_\mathrm{DE}/w_\mathrm{DE}^2$ in (\ref{Q}) diverges when $t=t_s$, and thus, $Q$ diverges. Here, we analyze different situations: we start with positive values of $\beta$. Then,
$\frac{k_s}{(t_s-t)^{\beta}}\cong -\frac{2\dot{H}}{3H^2}$ and the dynamics is given by
\begin{align}
H(t)\cong \frac{2(1-\beta)}{3k_s}(t_s-t)^{\beta-1}\, ,
\end{align}
and thus,
\begin{align}
\ln\left(\frac{a(t)}{a_s} \right)\cong \frac{2(\beta-1)}{3k_s\beta}(t_s-t)^{\beta}, \quad \mbox{and}\quad
\dot{H}(t)\cong \frac{2(1-\beta)^2}{3k_s}(t_s-t)^{\beta-2}\, .
\end{align}

From this result, we have the following observations:
\begin{itemize}
\item For $\beta\geq 2$, there is a Type V singularity at $t=t_s$.
\item For $0<\beta<2$, there is a Type II singularity at $t=t_s$.
\end{itemize}

On the other hand, for negative values of $\beta$, $w_\mathrm{DE}(t)\rightarrow w_s$ as $t\rightarrow t_s$. Thus, the Hubble rate and its derivatives do not diverge at $t=t_s$.
The only divergence appears in the energy exchange rate $Q(t)$, which diverges for $-1<\beta<0$.

\subsubsection{Some Specific Interaction Models}

According to the existing records, a variety of interaction
models have been proposed in the literature, see Refs.
\cite{Amendola:1999er,Huey:2004qv,Cai:2004dk,Wang:2005jx,Pavon:2005yx,Das:2005yg,Barrow:2006hia,Amendola:2006dg,delCampo:2006vv,delCampo:2008sr,Valiviita:2008iv,delCampo:2008jx,Gavela:2009cy,Koyama:2009gd,Majerotto:2009np,Boehmer:2009tk,Chimento:2011pk,Chimento:2012zz,Pettorino:2013oxa,Chimento:2013rya,Chakraborty:2012gr,Yang:2014gza,yang:2014vza,Yang:2014hea,Pan:2012ki,Nunes:2016dlj,vandeBruck:2016jgg,Pourtsidou:2016ico,vandeBruck:2016hpz,Mukherjee:2016shl,An:2017crg,DiValentino:2017iww,Yang:2017yme,vandeBruck:2017idm,Cai:2017yww,Kumar:2017dnp,Yang:2017ccc,Yang:2017zjs,VanDeBruck:2017mua,Mifsud:2017fsy,Pan:2017ent,Yang:2018euj,Yang:2018xlt,Gonzalez:2018rop,Yang:2018qec,Pan:2019jqh,Paliathanasis:2019hbi,Yang:2019uog,Li:2019loh,Yang:2019vni,vonMarttens:2019ixw,Li:2019ajo,Pan:2019gop,Mifsud:2019fut,Yang:2020uga,Pan:2020bur,DiValentino:2019ffd,DiValentino:2019jae,Lucca:2020zjb,Yang:2020tax,Pan:2020mst,DiValentino:2020kpf,Wang:2021kxc,Anchordoqui:2021gji,Gao:2021xnk,Yang:2021hxg,Guo:2021rrz,Johnson:2021wou,Yang:2021oxc,Paliathanasis:2021egx,Lucca:2021dxo,Gariazzo:2021qtg,Bonilla:2021dql,Mukherjee:2021ggf,Nunes:2022bhn,Yao:2022kub,Alvarez-Ortega:2022vbo,Harko:2022unn,Yang:2022csz,Pan:2022qrr,Zhao:2022ycr,Zhai:2023yny}
(also see Refs.~ \cite{Bolotin:2013jpa,Wang:2016lxa}) and they
have been investigated by many authors. If one closely looks at
the conservation equations for DM and DE, i.e. Eq. (\ref{dark})
one may realize that the interaction rate $Q$, irrespective of its
form, modifies the EoS parameter of DM and DE. And thus, similar
to the non-interacting cosmological scenarios, the interacting
scenarios may lead to finite time singularities. In this section
we consider some well known interaction functions in order to
investigate the possibility of finite time future singularities.


\begin{center}
 {\it (i) Interaction model I}
\end{center}

The first interaction rate characterizing the energy transfer between DE and DM has the following form \cite{BeltranJimenez:2016dfc}
\begin{equation}\label{chapter-IDE-Q-model-I-01}
Q= \zeta H \rho_\mathrm{DE}~,
\end{equation}
where $\zeta$ is the coupling parameter of the interaction rate. Now, using the expression of $\rho_\mathrm{DE}$ from Eq. (\ref{eqn-rho-DE-sec-IDE-sp}), the interaction rate $Q$ of Eq. (\ref{chapter-IDE-Q-model-I-01}) can be recast as

\begin{equation}\label{chapter-IDE-Q-model-I-02}
Q=\zeta H \left(\frac{3H^2+2\dot{H}}{\widetilde{\kappa}^2} \right),
\end{equation}
where $\widetilde{\kappa}^2= -w_\mathrm{DE} \kappa^2$. Now, plugging \ref{chapter-IDE-Q-model-I-02} into Eq. (\ref{Q}),

\begin{equation}\label{chapter-IDE-Q-model-I-03}
2\ddot{H}+9(1+w_\mathrm{DE})H^3+6(2+w_\mathrm{DE})H\dot{H}
-\frac{\dot{w}_\mathrm{DE}}{w_\mathrm{DE}} \left( 3 H^2 + 2 \dot{H}\right)+\zeta H\left(3H^2+2\dot{H}\right)=0.
\end{equation}
Now, from this equation (i.e. Eq.
(\ref{chapter-IDE-Q-model-I-03})) one can analyze the behavior of
$H (t)$, and hence, the presence of finite time future
singularities, if any. We consider the simplest case, i.e. the
constant equation of state in DE for which $\dot{w}_\mathrm{DE}=0$
for which Eq.(\ref{chapter-IDE-Q-model-I-03}) leads to

\begin{equation}\label{chapter-IDE-Q-model-I-04}
2\ddot{H}+9(1+w_\mathrm{DE})H^3+6(2+w_\mathrm{DE})H\dot{H} +\zeta H\left(3H^2+2\dot{H}\right)=0.
\end{equation}
Notice that the only critical point of Eq. (\ref{chapter-IDE-Q-model-I-04}) is the Minkowski solution with $H=0$, which is unstable as discussed in \cite{BeltranJimenez:2016dfc}. However, even though Eq. (\ref{chapter-IDE-Q-model-I-04}) is a nonlinear differential equation, but by taking advantage of its time rescaling invariance, one can obtain its analytical solution \cite{BeltranJimenez:2016dfc}. Following \cite{BeltranJimenez:2016dfc},
we look for the solutions of the form $\dot{H}=\lambda^{-1} H^2 $, where $\lambda$ is some dimensionless parameter. For this choice, one can solve the Hubble function as
\begin{equation}\label{chapter-IDE-Q-model-I-05}
H=\frac{\lambda}{t_s-t}~,
\end{equation}
where $t_s$ denotes some reference time. One can notice that as $t \rightarrow t_s$, the Hubble parameter diverges. Now, inserting (\ref{chapter-IDE-Q-model-I-05}) into Eq. (\ref{chapter-IDE-Q-model-I-04}), one arrives at

\begin{equation}
\big(3+2\lambda^{-1}\big)\Big[3(1+w_\mathrm{DE})+\zeta+2\lambda^{-1}\Big]=0,
\end{equation}
which leads to two different branches of solutions, namely,
$\lambda=-2/3$ and $\lambda=-2/(3(1+w_\mathrm{DE})+\zeta)$. The
first branch corresponds to a matter dominated universe {which is
unstable in the presence of the DE component} (see the Figure $1$
of \cite{BeltranJimenez:2016dfc}), while on the other hand, the
second branch corresponds to a universe where either the DE or the
interaction term dominates. In the second branch, the effective
equation of state is given by $w^{\rm
eff}_\mathrm{DE}=w_\mathrm{DE} -\zeta/3$\footnote{Notice that the
conservation equation of DE in presence of this interaction can be
written as
$\dot{\rho}_\mathrm{DE}+3H\left(1+w_\mathrm{DE}+\zeta/3\right)\rho_\mathrm{DE}=0
\Longleftrightarrow{} \dot{\rho}_\mathrm{DE}+3H\left(1+ w^{\rm
eff}_\mathrm{DE} \right)\rho_\mathrm{DE}=0$ in which $w^{\rm
eff}_\mathrm{DE} = w_\mathrm{DE}+\zeta/3$. }. Thus, if the
coupling parameter satisfies $\zeta > 3(1+w_\mathrm{DE})$, then an
effective phantom behavior is realized which leads to a Big Rip
singularity even if DE satisfies the null energy condition.


\begin{center}
 {\it (ii) Interaction model II}
\end{center}

The second interaction rate between DE and DM has the following form \cite{BeltranJimenez:2016dfc}

\begin{equation}\label{chapter-IDE-Q-model-II-01}
 Q= \Gamma \rho_\mathrm{DE}^n~,
\end{equation}
 where $n$ is a dimensionless parameter and $\Gamma$ is the coupling parameter having dimension equal to the dimension of the Hubble parameter. Again using the expression of $\rho_\mathrm{DE}$ from Eq. (\ref{eqn-rho-DE-sec-IDE-sp}), the interaction rate $Q$ of Eq. (\ref{chapter-IDE-Q-model-II-01})

 \begin{equation}\label{chapter-IDE-Q-model-II-02}
Q=\Gamma\left(\frac{3H^2+2\dot{H}}{\widetilde{\kappa}^2}\right)^n~.
\end{equation}
Now, plugging the expression of the interaction function $Q$ of Eq. \ref{chapter-IDE-Q-model-II-02} into Eq. (\ref{Q}), one gets

\begin{equation}\label{chapter-IDE-Q-model-II-03}
2\ddot{H}+6(2+w_\mathrm{DE})H\dot{H}+9(1+w_{DE})H^3 - \frac{\dot{w}_\mathrm{DE}}{w_\mathrm{DE}} \left( 3 H^2 + 2 \dot{H}\right) +\widetilde{\kappa}^{2(1-n)}\Gamma\left(3H^2+2\dot{H}\right)^n=0.
\end{equation}
For simplicity we consider the constant equation of state in DE
for which $\dot{w}_\mathrm{DE}=0$ in Eq.
(\ref{chapter-IDE-Q-model-II-03}) which leads to

\begin{equation}\label{chapter-IDE-Q-model-II-04}
2\ddot{H}+6(2+w_\mathrm{DE})H\dot{H}+9(1+w_\mathrm{DE})H^3 +\widetilde{\kappa}^{2(1-n)}\Gamma\left(3H^2+2\dot{H}\right)^n=0.
\end{equation}
Now, looking carefully at (\ref{chapter-IDE-Q-model-II-04}), one can see that in addition to the Minkowski critical point $H = 0$, it also admits
de Sitter critical points \cite{BeltranJimenez:2016dfc}
\begin{equation}\label{chapter-IDE-Q-model-II-05}
H_{\rm dS}=\left[-\frac{9 (1+w_\mathrm{DE})}{3^n \widetilde{\kappa}^{1-n}\Gamma}\right]^\frac{1}{2n-3}~,
\end{equation}
which exist for $w_\mathrm{DE} \neq -1$. Within this interaction model, under certain restrictions, one can obtain other type of future singularities apart from the Big Rip singularity.
{We look for the solutions where $|\dot{H}| \gg H^2$ and additionally we focus on the solutions driven by the interaction function which requires $|H| \ll |\Gamma (\dot{H}/\widetilde{\kappa}^2)^{n-1}|$.} Under these conditions, Eq. (\ref{chapter-IDE-Q-model-II-05}) reduces to \cite{BeltranJimenez:2016dfc}

\begin{equation}\label{chapter-IDE-Q-model-II-06}
\ddot{H}+ \Gamma \left(\frac{\widetilde{\kappa}^2}{2}\right)^{1-n} \dot{H}^n \simeq 0.
\end{equation}
Note that Eq. (\ref{chapter-IDE-Q-model-II-06}) remains invariant under a constant shift of $H$ which is important to keep $H$ finite. Moreover, Eq. (\ref{chapter-IDE-Q-model-II-06}) can be solved leading to
\begin{equation}
H \simeq C (t_s-t)^p+H_s~,
\end{equation}
where $t_s$ and $H_s$ are integration constants and
\begin{equation}
C = \frac{(1-n)^p}{n-2}\left[\Gamma \left(\frac{\widetilde{\kappa}^2}{2}\right)^{1-n} \right]^{p-1} \quad \quad p=\frac{n-2}{n-1}.
\end{equation}
Now, if we take $n=3$, then from Eq.
(\ref{chapter-IDE-Q-model-II-06}) one finds that $H\simeq
C\sqrt{t_s-t}+H_s$ which gives $H(t_s)=H_s$ but for $t\rightarrow
t_s$, $\dot{H}\to \infty$, i.e. a type II or Sudden Singularity
is realized. This behavior remains general for values of $n$ for
which $0<p<1$. Additionally, this interaction model could lead to
a Big Freeze or a type III singularity for $p=-1/2$ which is
obtained for $n=5/3$. In this case, $ H\simeq C(t_s-t)^{-1/2}$, and
hence, $a \simeq a_s e^{-2C\sqrt{t_s-t}}$. Thus, for $t\rightarrow
t_s$, $H$ and its derivatives diverge but $a$ remains finite which
shows the occurrence of a type III singularity. In fact, for those
values of $n$ which restricts $p$ in the interval, $-1<p<0$ one
always has a type III singularity.

\newpage

\begin{center}
 {\it (iii) Interaction model III}
\end{center}

The interaction
function has the following form
\cite{Chimento:2003iea,Quartin:2008px,MohseniSadjadi:2008na,Cruz:2008er,Caldera-Cabral:2008yyo,Pan:2012ki,Pan:2016ngu}
\begin{eqnarray}
Q=3H \left (\mu\rho_\mathrm{DM}+\nu \rho_\mathrm{DE} \right),
\end{eqnarray}
where $\mu$ and $\nu$ are the coupling parameters of the
interaction function. For a perfect fluid with stress-tensor $
T^{\alpha}_{\beta, A}=p_A
\delta_{\beta}^{\alpha}+(\rho_A+p_A)u^{\alpha}_Au_{\beta, A}$,
where $A = \mathrm{DM}, \mathrm{DE}$
and
$u_A^{\alpha}=\frac{dx^{\alpha}}{\sqrt{-ds^2}}$ is the
four-velocity of the fluid. We consider the following covariant
interacting system
\begin{align}
\nabla_{\alpha} T^{\alpha}_{\ \beta, A}=Q_{\beta, A}\, ,
\end{align}
where
\begin{align}
Q_{\beta, \mathrm{DM}}=-Q_{\beta, \mathrm{DE}}=\nabla_{\eta}u^{\eta}_\mathrm{DM} \left(
\bar{\mu}T^{\alpha}_{\ \alpha, \mathrm{DM}}u_{\beta, \mathrm{DM}} + \bar{\nu}T^{\alpha}_{\ \alpha, \mathrm{DE}}u_{\beta, \mathrm{DE}} \right)\, .
\end{align}

At the background level we will have
$\nabla_{\eta}u^{\eta}_\mathrm{DM}=3H$, $T^{\alpha}_{\ \alpha,
A}=3p_A-\rho_A$ and $\nabla_{\alpha} T^{\alpha}_{\ 0,
A}=-\dot{\rho}_A-3H(\rho_A+p_A)$, then if we define
$\bar{\mu}=\mu$ and $\bar{\nu}=\frac{\nu}{1-3w_\mathrm{DE}}$, we
obtain the dynamical system (\ref{dark})
with $Q=Q_{0,DM}=3H(\mu\rho_\mathrm{DM}+\nu \rho_\mathrm{DE} )$.
Now introducing a new variable $N=\ln \left(\frac{a}{a_0}
\right)$, where $a_0$ denotes the present value of the scale
factor, and taking into account that $\dot{N}=H$, the system
(\ref{dark}) leads to the following linear first order autonomous
dynamical system:
\begin{align}
\left\{ \begin{array}{rcl}\label{modelB}
{\rho}'_\mathrm{DM}+3\rho_\mathrm{DM} &=&-3(\mu\rho_\mathrm{DM}+\nu \rho_\mathrm{DE} ), \\
{\rho}'_\mathrm{DE}+3(1+w_\mathrm{DE})\rho_\mathrm{DE} &=& 3(\mu\rho_\mathrm{DM}+\nu \rho_\mathrm{DE} )\, ,
\end{array} \right.
\end{align}
where now the prime denotes the derivative with respect the
variable $N$. The autonomous system (\ref{modelB}) can be
expressed in the matrix form as $X'=BX$, where
\begin{align}
X=\left(\begin{array}{c}
\rho_\mathrm{DM} \\
\rho_\mathrm{DE}
\end{array}\right)\, ,
\end{align}
and
\begin{align}
 B=\left(\begin{array}{cc}
 -3(1+\mu) & -3\nu \\
 +3\mu & -3(1+w_\mathrm{DE}-\nu)
 \end{array}\right).
\end{align}
Now in order to get singular behaviors, one of the eigenvalues of
the matrix $B$ should have a positive real part. In terms of the
trace and determinant of $B$, this means that there are two
different situations:
\begin{enumerate}
 \item $\mbox{Tr} B>0$ and $\mbox{Det} B>0$.
 \item $\mbox{Det} B<0$.
\end{enumerate}
In the former case both the eigenvalues are positive, and the
origin of coordinates is a repeller. In order to ensure that the
energy densities of both the fluids are always positive, one has
to impose that the origin is not a focus, because if so, then at
early times, the orbits would oscillate around the origin leading
to negative energy densities. To prevent this behavior one has to
impose that the discriminant $\Delta=(\mbox{Tr} B)^2- 4 \mbox{Det}
B$ is positive, that means, the origin is a node. In addition, for
a node, to ensure that the energy densities should be positive,
both the orbits following the respective eigenvectors of the
matrix $B$ ($X_+(N)=\e^{\lambda_+N}V_+$ and
$X_-(N)=\e^{\lambda_-N}V_{-}$ being $\lambda_+$ and $\lambda_-$
the eigenvalues of the matrix $B$ and, $V_+= \left(
V_{1,+},V_{2,+} \right)$ and $V_-= \left( V_{1,-},V_{2,-} \right)$
their corresponding eigenvectors) must belong to the first
quadrant, that is, the condition $V_{1,\pm}\geq 0$ and
$V_{2,\pm}\geq 0$ must be satisfied. As a consequence, all orbits
with an initial condition in the first quadrant, i.e., with
initial positive values of $\rho_\mathrm{DM}$ and
$\rho_\mathrm{DE}$, will remain in the first quadrant, which
ensures that the energy densities are always positive.

All these conditions lead to the following constraints that the parameters $\mu$ and $\nu$ need to satisfy:
\begin{align}\label{condition}
\left\{\begin{array}{rcl}
2+\mu-\nu+w_\mathrm{DE} &<& 0, \\
(1+\mu)(1+w_\mathrm{DE})-\nu &>&0,\\
(w_\mathrm{DE}-\mu-\nu)^2-4\mu\nu &>& 0,
\end{array}\right.
\end{align}
where the first equation is $\mbox{Tr} B>0$, the second is $\mbox{Det} B>0$ and the third is $\Delta>0$. On the other hand, the eigenvalues of $B$ are given by $\lambda_{\pm}=
(\mbox{Tr}B\pm
\sqrt{\Delta})/2$, and the corresponding eigenvectors are given by as follows:
\begin{enumerate}
\item For $\nu\not=0$,
\begin{align}
V_{\pm}=\left(1, -\frac{\mu+1+\lambda_{\pm}/3}{\nu}\right)\, .
\end{align}
\item For $\nu=0$, the eigenvalues are $\lambda_+=-3(1+w_\mathrm{DE})$ and
$\lambda_-=-3(1+\mu)$. This implies that $w_\mathrm{DE}<-1$ (phantom fluid) and also $\mu<-1$.
The eigenvectors are given by
\begin{align}
V_+=(0,1), \quad V_-=\left(1, \frac{\mu}{w_\mathrm{DE}-\mu} \right)\, .
\end{align}
\end{enumerate}
Then, the conditions $V_{1,\pm}\geq 0$ and $V_{2,\pm}\geq 0$ lead to the following restrictions:
\begin{align}
\label{a}
\left\{\begin{array}{ccc}
 -\frac{1}{\nu}\left(\mu+1+ \frac{\lambda_{\pm}}{3} \right) \geq 0, &\mbox{for} &\nu\not=0, \\
w_\mathrm{DE}< \mu\leq 0, & \mbox{for} & \nu=0\, .
\end{array}\right.
\end{align}

Thus, in order that the initial condition was in the region
defined by the orbits $X_+(N)$ and $X_-(N)$, and the energy
densities are always positive, is
\begin{align}\label{b}
\mbox{min} \left( V_{2,+}, V_{2,-} \right) \leq \frac{\rho_\mathrm{DE,0}}{\rho_\mathrm{DM,0}}\leq \mbox{max} \left( V_{2,+}, V_{2,-} \right)
\end{align}
for $\nu\not= 0$, and
\begin{align}\label{nu0}
\frac{\rho_\mathrm{DE,0}}{\rho_\mathrm{DM,0}}\geq
\frac{\mu}{w_\mathrm{DE}-\mu}\, ,
\end{align}
for $\nu=0$. Now, considering the simple case $\nu=0$, and taking into account that
$\rho_\mathrm{DE,0}/\rho_\mathrm{DM,0}=\Omega_\mathrm{DE,0}/\Omega_\mathrm{DM,0}$,
where, as usual, we denote by
$\Omega_A=\frac{\kappa^2\rho_A}{3H^2}$, and taking the observationally reliable values of the present day density parameters $\Omega_\mathrm{DM,0}\cong 0.262$ and
$\Omega_\mathrm{DE,0}\cong 0.69$, from Eq. (\ref{nu0}) we deduce that the parameter $\mu$ must satisfy
the condition
\begin{align}
0.72w_\mathrm{DE}<\mu\leq 0\, ,
\end{align}
and from Eq. (\ref{condition}), we conclude that the value of the parameter $\mu$ has to satisfy
\begin{align}
0.72w_\mathrm{DE}<\mu<-1, \quad \mbox{ with }\quad w_\mathrm{DE}<-1.38\, .
\end{align}

Finally, we consider the case where the origin is a saddle point, i.e., when $\mbox{Det} B<0$, which leads to the following constraint
\begin{align}
 (1+\mu)(1+w_\mathrm{DE})-\nu<0\, .
\end{align}
This constraint has to be added to the constraints in Eqs. (\ref{a}), (\ref{b}) and (\ref{nu0}).
And for the simple case with $\nu=0$, one gets the following range of the parameters
\begin{align}
 0.72w_\mathrm{DE}<\mu<0,\quad w_\mathrm{DE}<-1, \quad \mu>-1\, .
\end{align}

\section{Finite time singularities in Modified gravity theories}
\label{sec-singularities-MG}

Modified gravity theories are rich from both theoretical and
observational
perspectives~\cite{Banerjee:2000gt,Banerjee:2000mj,Capozziello:2002rd,Capozziello:2003gx,Nojiri:2003ft,Nojiri:2003ni,Carroll:2003wy,Nojiri:2004bi,Elizalde:2004mq,Nojiri:2005vv,Capozziello:2005ku,Nojiri:2005jg,Amarzguioui:2005zq,Carter:2005fu,Das:2005bn,Cognola:2006eg,Brookfield:2006mq,Song:2006ej,Koivisto:2006xf,Koivisto:2006ai,Capozziello:2006dj,Li:2006ag,Li:2006vi,Nojiri:2006gh,Li:2007xn,Nojiri:2007uq,Bertolami:2007gv,Hu:2007nk,Li:2007xw,Li:2007jm,Starobinsky:2007hu,Nojiri:2007te,Nojiri:2007cq,Cognola:2007zu,Fay:2007uy,Pogosian:2007sw,Carloni:2007yv,Bengochea:2008gz,Dev:2008rx,Jhingan:2008ym,Capozziello:2008gu,Das:2008iq,Brax:2008hh,Frolov:2008uf,Oyaizu:2008sr,Oyaizu:2008tb,Schmidt:2008tn,Bamba:2008ja,Bamba:2008xa,Nojiri:2009kx,Appleby:2009uf,Thongkool:2009js,Thongkool:2009vf,Linder:2010py,Wu:2010mn,Bamba:2010wb,Capozziello:2010uv,Li:2010cg,Dunsby:2010wg,Geng:2011aj,Harko:2011kv,Koyama:2011wx,Gumrukcuoglu:2011ew,Cai:2011tc,Paliathanasis:2011jq,Olmo:2011uz,Li:2012by,deHaro:2012zt,Gannouji:2012iy,Cardone:2012xq,Sami:2012uh,Chakraborty:2012kj,Maluf:2013gaa,Odintsov:2013iba,Nesseris:2013jea,Brax:2013fda,vandeBruck:2012vq,Nojiri:2013zza,Tamanini:2013ltp,Basilakos:2013rua,Cai:2013lqa,Chakraborty:2012sd,Chakraborty:2013ywa,Cai:2013toa,Nojiri:2014zqa,Cai:2014upa,Bose:2014zba,Barreira:2014ija,Chakraborty:2014xla,Harko:2014aja,Leon:2014yua,He:2014eva,Ling:2014xoi,Thomas:2015dfa,Chakraborty:2015bja,Chakraborty:2015wma,Paliathanasis:2015arj,Chakraborty:2015taq,He:2015bua,Nunes:2016qyp,Paliathanasis:2016tch,Paliathanasis:2016vsw,Nunes:2016plz,Nunes:2016drj,Liu:2016xes,Cataneo:2016iav,Shirasaki:2016twn,Bose:2016wms,Rezazadeh:2017edd,Paliathanasis:2017efk,Li:2017xdi,Roy:2017mnz,Paliathanasis:2017htk,Akarsu:2018aro,He:2018oai,Mitchell:2018qrg,Nunes:2018xbm,Li:2018ixg,Leon:2018skk,Nunes:2018evm,Nojiri:2012zu,Nojiri:2012re,Nojiri:2015qyc,Gannouji:2018aaw,Papagiannopoulos:2018mez,Banerjee:2018yyi,Mitchell:2019qke,Paliathanasis:2019luv,Arnold:2019zup,Arnold:2019vpg,Leon:2019mbo,Pozdeeva:2019agu,Choudhury:2019zod,Paliathanasis:2019ega,Cai:2019bdh,Chen:2019ftv,Papagiannopoulos:2019kar,Yan:2019gbw,Paliathanasis:2020bgs,Mitchell:2020aep,Paliathanasis:2020axi,Alam:2020jdv,Paliathanasis:2020plf,Mitchell:2021aex,Mitchell:2021uzh,Mitchell:2021ter,Ren:2021tfi,Ren:2021uqb,Paliathanasis:2021qns,Paliathanasis:2021uvd,dosSantos:2021owt,Hernandez-Aguayo:2021kuh,Ruan:2021wup,Paliathanasis:2022pgu,Ren:2022aeo,Leon:2022oyy,Paliathanasis:2022xvn,Dimakis:2022wkj,Santos:2022atq,Kumar:2023bqj,Qi:2023ncd,Millano:2023czt,Hu:2023juh}
(also see the review articles in this direction
\cite{Nojiri:2006ri,Nojiri:2017ncd,Capozziello:2007ec,Padmanabhan:2007xy,Sotiriou:2008rp,Silvestri:2009hh,DeFelice:2010aj,Nojiri:2010wj,Clifton:2011jh,Hinterbichler:2011tt,Capozziello:2011et,Sami:2013ssa,deRham:2014zqa,Cai:2015emx,Bahamonde:2021gfp}
and the references therein). It has been consistently observed
that the modified gravity models can successfully describe both
the late-time accelerating expansion of the Universe
\cite{Capozziello:2002rd,Nojiri:2003ni,Nojiri:2005jg,Capozziello:2005ku,Cognola:2006eg,Nojiri:2008nt,Nojiri:2017ncd}
as well as its evolution in the early phase, known as inflationary
era
\cite{Hwang:2002fp,Ferraro:2006jd,Cognola:2007zu,Nojiri:2008nt,Sebastiani:2013eqa,Bamba:2014jia,Bamba:2015uma,Nojiri:2017ncd}.
Additionally, it has been found that the modified gravity theories
can also unify both inflation and DE eras in a single picture
\cite{Nojiri:2003ft,Nojiri:2007as,Nojiri:2007cq,Cognola:2007zu,Fay:2007uy,Elizalde:2010ep,Artymowski:2014gea,Odintsov:2018nch,Nojiri:2019fft,Odintsov:2020nwm,Oikonomou:2020qah}.
Thus, undoubtedly, modified gravity theories are very appealing to
explain different phases of the Universe. However, as there is no
unique way to modify the gravitational sector of the Universe,
therefore, over the years several modified gravity models have
been introduced and confronted with the observational data. Based
on the qualitative nature of a specific modified gravity model,
this can lead to finite-time singularities in the future. In this
section we shall discuss how finite-time singularities appear in
different modified gravity theories.

In the case of GR, we have already seen that the cosmological equations can be written as
\begin{align}
\label{FLRWs}
\rho = \frac{3H^2}{\kappa^2} \, , \quad
p = - \frac{1}{\kappa^2} \left( 2 \dot H + 3 H^2 \right) \, .
\end{align}
Motivated by Eq.~(\ref{FLRWs}), we may define the effective energy density $\rho_\mathrm{eff}$
and the effective pressure $p_\mathrm{eff}$, as follows,
\begin{align}
\label{FLRWs2}
\rho_\mathrm{eff} \equiv \frac{3H^2}{\kappa^2} \, , \quad
p_\mathrm{eff} \equiv - \frac{1}{\kappa^2} \left( 2 \dot H + 3 H^2 \right) \, .
\end{align}
Note that $\rho_\mathrm{eff}$ and $p_\mathrm{eff}$ defined in Eq. (\ref{FLRWs2}) satisfy the conservation equation 
\begin{eqnarray}\label{eq-continuity-new-section-MG}
\dot{\rho}_\mathrm{eff} + 3 H (\rho_\mathrm{eff}+ p_\mathrm{eff}) = 0. 
\end{eqnarray} 
We now assume that the Hubble rate $H$ behaves as in (\ref{BigRip}),
\begin{align}
\label{Hsin}
H\sim \frac{h_s}{\left(t_s - t \right)^\beta} \, ,
\end{align}
with a constant $h_s>0$ when $t\lesssim t_s$.
When $\beta<1$ and different from zero, $\rho_\mathrm{eff}$ and $p_\mathrm{eff}$ behaves as $\rho_\mathrm{eff}\sim \left(t_s - t \right)^{-2\beta}$,
$p_\mathrm{eff}\sim \left(t_s - t \right)^{-\beta-1}$.
On the other hand, when $\beta\geq 1$, we find $\rho_\mathrm{eff} \sim p \sim \left(t_s - t \right)^{-2\beta}$.
Then $\beta\geq 1$ corresponds to Type I singularity, $0<\beta<1$ to Type III, $-1<\beta<0$ to Type II,
and the case that $\beta<-1$ and $\beta$ is not an integer corresponds to Type IV.
In the case of Type IV singularity, if $-n<\beta<1-n$ for a positive integer, $\frac{d^m H}{dt^m}$ $\left(m\geq n\right)$ diverges at $t=t_s$.
For $\frac{d^m H}{dt^m}$ $\left(0\leq m\leq n-1 \right)$, the behavior of (\ref{Hsin}) could be modified to be
\begin{align}
\label{HsinR}
H= H_s(t) + \frac{h_s}{\left(t_s - t \right)^\beta} \, .
\end{align}
Here $H_s(t)$ is a function which can be differentiated $n$-times
(that is, it is a $\mathcal{C}^n$ class function). Then for
$\frac{d^m H}{dt^m}$ $\left(0\leq m\leq n-1 \right)$, the first
term in (\ref{HsinR}) becomes dominant when $t\sim t_s$ and for
$\frac{d^m H}{dt^m}$ $\left(m\geq n\right)$, the second term
dominates when $t\sim t_s$.

In this section, by using the formalism of the reconstruction, we
construct models which realize the above singularities in
(\ref{Hsin}). Usually, we start from a theory, which is defined by
the action, and solve equations of motion to define the background
dynamics. We may consider, however, the inverse problem, i.e., the
cosmological reconstruction of gravitational theories, that is, by
using the fact that modified gravity is defined in terms of some
arbitrary function(s), we can show how the complicated background
cosmology, which complies with the observational data, may be
reconstructed. The general approach to reconstruction in modified
gravity and DE models was developed in
Refs.~\cite{Nojiri:2006gh,Capozziello:2006dj, Nojiri:2006be}.

\subsection{Scalar-tensor Gravity}
\label{sec-scalar-tensor}

We remind the reconstruction of the scalar-Einstein gravity (or scalar-tensor theory),
whose action can be written as
\begin{align}
\label{ma7}
S=\int d^4 x \sqrt{-g}\left\{
\frac{1}{2\kappa^2}R - \frac{1}{2}\omega(\phi)\partial_\mu \phi
\partial^\mu\phi - V(\phi) + L_\mathrm{matter} \right\}\, .
\end{align}
Here, $L_\mathrm{matter}$ is the matter Lagrangian, and $\omega(\phi)$ and $V(\phi)$ are functions of the scalar $\phi$.
The function $\omega(\phi)$ is not relevant and can be absorbed into the
redefinition of the scalar field $\phi$. In fact, if one redefines
the scalar field $\phi$ by
\begin{align}
\label{ma13}
\varphi \equiv \int^\phi d\phi \sqrt{\left|\omega(\phi)\right|} \, ,
\end{align}
the kinetic term of the scalar field in the action (\ref{ma7}) has the
following form:
\begin{align}
\label{ma13b}
 - \omega(\phi) \partial_\mu \phi \partial^\mu\phi
= \left\{ \begin{array}{ll}
 - \partial_\mu \varphi \partial^\mu\varphi, &
\mbox{when $\omega(\phi) > 0$}, \\
\partial_\mu \varphi \partial^\mu\varphi, & \mbox{when $\omega(\phi) < 0$}.
\end{array} \right.
\end{align}
The case of $\omega(\phi) > 0$ corresponds to the quintessence or
non-phantom scalar field, but the case of $\omega(\phi) < 0$ corresponds
to the phantom scalar.
Although $\omega(\phi)$ can be absorbed into the redefinition of the
scalar field, we keep $\omega(\phi)$ for later convenience.
The reconstruction of the scalar-tensor gravity is based on the Refs.
\cite{Nojiri:2005pu,Capozziello:2005tf,Vikman:2004dc}.

For the action (\ref{ma7}), in the flat FLRW spacetime (\ref{FLRWk0}), the
energy density and the pressure are as follows,
\begin{align}
\label{ma8}
\rho = \frac{1}{2}\omega(\phi){\dot \phi}^2 + V(\phi)\, ,\quad
p = \frac{1}{2}\omega(\phi){\dot \phi}^2 - V(\phi)\, .
\end{align}
The above equations in (\ref{ma8}) can be rewritten as
\begin{align}
\label{ma9}
\omega(\phi) {\dot \phi}^2 = - \frac{2}{\kappa^2}\dot H\, ,\quad
V(\phi)=\frac{1}{\kappa^2}\left(3H^2 + \dot H\right)\, .
\end{align}
Assuming $\omega(\phi)$ and $V(\phi)$ are given by a single function
$f(\phi)$, as follows,
\begin{align}
\label{ma10}
\omega(\phi)=- \frac{2}{\kappa^2}f'(\phi)\, ,\quad
V(\phi)=\frac{1}{\kappa^2}\left(3f(\phi)^2 + f'(\phi)\right)\, ,
\end{align}
we find that in the case where we neglect the contribution from the matter,
the exact solution of the Friedmann
and Raychaudhuri equations or the FLRW equations with (\ref{ma8})
has the following form:
\begin{align}
\label{ma11}
\phi=t\, ,\quad H=f(t)\, .
\end{align}
We also note that the equation given by the variation with respect to $\phi$
\begin{align}
\label{ma12}
0=\omega(\phi)\ddot \phi + \frac{1}{2}\omega'(\phi){\dot\phi}^2
+ 3H\omega(\phi)\dot\phi + V'(\phi)\, ,
\end{align}
is also satisfied by the solution (\ref{ma11}). Therefore, the
arbitrary Universe evolution expressed by $H=f(t)$ can be realized
by an appropriate choice of $\omega(\phi)$ and $V(\phi)$. In other
words, by defining the particular type of Universe evolution, the
corresponding scalar-Einstein gravity can be found. Especially in
the case of the singular behavior in (\ref{Hsin}), we find
\begin{align}
\label{stsing}
\omega(\phi)= -\frac{2\beta h_s}{\kappa^2
\left(t_s - \phi \right)^{\beta +1}}\, , \quad
V(\phi)=\frac{1}{\kappa^2}\left( \frac{3h_s^2}{ \left(t_s - \phi \right)^{2\beta}}
 -
 \frac{\beta h_s }{ \left(t_s - \phi \right)^{\beta+1}} \right)\, .
\end{align}
We should note that in the case that $\beta$ is positive, $\omega(\phi)$ becomes negative
when $\phi<t_s$, which means that the scalar field $\phi$ is a ghost.
The ghost generates negative norm states in the quantum theories and conflicts with the
so-called Copenhagen interpretation.
By using other kinds of modified gravity theories, we can realize the behavior of
(\ref{Hsin}) without generating the ghost.

Now, from Eq. (\ref{ma13}), one can see that the new field $\varphi$, is given by
\begin{align}
\label{ma13BB}
\varphi = \left\{
\begin{array}{ll}
\frac{2\sqrt{-\beta h_s}}{\kappa \left(1- \beta \right)} \left( t_s - \phi \right)^\frac{1-\beta}{2},
& \quad \mbox{when}\ \beta<0, \\
\frac{2\sqrt{\beta h_s}}{\kappa \left(1- \beta \right)} \left( t_s - \phi \right)^\frac{1-\beta}{2},
& \quad \mbox{when}\ \beta>0, \quad \mbox{but} \quad \beta \neq 1, \\
\frac{2\sqrt{h_s}}{\kappa} \ln \left( t_s - \phi \right),
& \quad \mbox{when}\ \beta = 1.
\end{array} \right. \, .
\end{align}

Therefore, the action (\ref{ma7}) has the following form when $\beta<0$,
\begin{align}
\label{ma7BB}
S= \int d^4 x \sqrt{-g}&\, \left\{
\frac{1}{2\kappa^2}R - \frac{1}{2}\partial_\mu \varphi \partial^\mu\varphi
 - \frac{1}{\kappa^2}\left(3h_s^2 \left( \frac{2\sqrt{-\beta h_s}}{\kappa \left(1- \beta \right)} \right)^{\frac{4\beta}{1-\beta}}
\varphi^\frac{-4\beta}{1-\beta} \right. \right. \nonumber \\
&\, \left. \left. + h_s \beta \left( \frac{2\sqrt{-\beta h_s}}{\kappa \left( 1-\beta \right)} \right)^{-\frac{\beta (\beta +1)}{1-\beta }}
\varphi^\frac{\beta (\beta +1)}{1-\beta } \right) + L_\mathrm{matter} \right\}\, ,
\end{align}
when $\beta>0$ and $\beta\neq 1$,
\begin{align}
\label{ma7BB2}
S= \int d^4 x \sqrt{-g}&\, \left\{
\frac{1}{2\kappa^2}R + \frac{1}{2}\partial_\mu \varphi \partial^\mu\varphi
 - \frac{1}{\kappa^2}\left(3h_s^2 \left( \frac{2\sqrt{\beta h_s}}{\kappa \left(1- \beta \right)} \right)^{\frac{4\beta}{1-\beta}}
\varphi^\frac{4\beta}{\beta+1} \right. \right. \nonumber \\
&\, \left. \left. + h_s \beta \left( \frac{2\sqrt{\beta h_s}}{\kappa \left(1- \beta \right)} \right)^{-\frac{\beta (\beta +1)}{1-\beta }}
\varphi^\frac{\beta (\beta +1)}{1-\beta } \right) + L_\mathrm{matter} \right\}\, .
\end{align}
and when $\beta=1$,
\begin{align}
\label{ma7BB3}
S= \int d^4 x \sqrt{-g}&\, \left\{
\frac{1}{2\kappa^2}R + \frac{1}{2}\partial_\mu \varphi \partial^\mu\varphi
 - \frac{1}{\kappa^2}\left(3h_s^2 + H_s \right)\e^{- \frac{\kappa \varphi}{\sqrt{h_s}}} + L_\mathrm{matter} \right\}\, .
\end{align}
In the case of Eqs.~(\ref{ma7BB2}) and (\ref{ma7BB3}), the signature in front of the kinetic term of the scalar field $\varphi$ is negative and therefore
the scalar field becomes a ghost. The
Eq.~(\ref{ma7BB2}) with $\beta>1$ and (\ref{ma7BB3}) correspond to a Type I singularity;
Eq.~(\ref{ma7BB2}) with $0<\beta<1$ corresponds to a Type III singularity; Eq.~(\ref{ma7BB}) with $-1<\beta<0$ corresponds to a Type II singularity;
and Eq.~(\ref{ma7BB}) where $\beta<-1$ and $\beta$ is not a negative integer corresponds to a Type IV singularity.

\subsection{Brans-Dicke Gravity}
\label{sec-Brans-Dicke}

The action of the original Brans-Dicke model is given by \cite{Brans:1961sx}
\begin{align}
\label{BD1}
S_\mathrm{BD}=\frac{1}{2\kappa^2}\int d^4 x \sqrt{-g}
\left( \phi R - \omega_0
\frac{\partial_\mu \phi \partial^\mu \phi}{\phi}\right)\, .
\end{align}
Here $\omega_0$ is a constant known as the Brans-Dicke coupling constant.
We now consider the following generalization,
\begin{align}
\label{BD2}
S=\int d^4x\sqrt{-g}\left[\frac{
\e^{\varphi(\phi)}R}{2\kappa^2} -\frac{1}{2}\omega(\phi)\partial_\mu
\phi\partial^\mu \phi-V(\phi)+ L_\mathrm{matter}\right]\, .
\end{align}
In the original Brans-Dicke model, there is a strong constraint on
the Brans-Dicke coupling constant $\omega_0>40,000$ but in the
generalized model (\ref{BD2}), by adjusting the parameters in the
potential $V(\phi)$, which includes the mass term, we can escape
from the constraint. The generalized model was applied to the DE
problem in \cite{Elizalde:2004mq}, and it was found that the
phantom Universe can be realized even if $\omega(\phi)>0$, that
is, the scalar field is canonical, which is contrary to the case
of the scalar-tensor model in Eq. (\ref{ma7}), where, as discussed
after Eq.~(\ref{stsing}), there appears the ghost when $\beta$ is
negative and $\beta<-1$ in the case of the phantom Universe.

If we re-scale the metric by $g_{\mu\nu}=\e^{-\varphi(\phi)}g^\mathrm{(E)}_{\mu\nu}$, the action (\ref{BD2}) has the following form,
\begin{align}
\label{BD2E}
S= \int d^4x\sqrt{-g^{(E)}} &\, \left[\frac{R}{2\kappa^2}
 - \frac{1}{2}\left( \e^{-\varphi(\phi)}\omega(\phi) + \frac{3}{2\kappa^2} \varphi'(\phi)^2 \right)
\partial_\mu \phi\partial^\mu \phi - \e^{-2\varphi(\phi)} V(\phi) \right. \nonumber \\
& \left. + \e^{-2\varphi(\phi)} \left. L_\mathrm{matter}\right|_{g_{\mu\nu}=\e^{-\varphi(\phi)}g^{(E)}_{\mu\nu}}\right]\, ,
\end{align}
The action in Eq. (\ref{BD2E}) is called the Einstein frame action.
Except for the matter part, the action can be regarded with the
action of the scalar-tensor theory given in Eq. (\ref{ma7}). In the Einstein
frame, there appears the coupling of the scalar field $\phi$ with
the matter due to the rescaling
$g_{\mu\nu}=\e^{-\varphi(\phi)}g^{(E)}_{\mu\nu}$. If the matter is
minimally coupled with the gravity in the action (\ref{BD2}), the
cosmological time observed by the observer made of the matter is
given by the original metric $g_{\mu\nu}$ but never
$g^{(E)}_{\mu\nu}$. Then even if the Universe is not accelerating
expanding in the Einstein frame, the expansion of the Universe may
be accelerated in the original frame given by the original metric
$g_{\mu\nu}$, which is called the Jordan frame. It is
straightforward, however, to discuss the existence of the ghost in
the Einstein frame because there is no direct coupling of the
scalar field $\phi$ with the scalar curvature $R$. The ghost can
be avoided if
\begin{align}
\label{BGghost}
\e^{-\varphi(\phi)}\omega(\phi) + 3 \varphi'(\phi)^2 > 0 \, .
\end{align}
The Friedmann and Raychaudhuri equations or the FLRW equations are given as follows,
\begin{align}
\label{BD3}
H^2 = \frac{\e^{-\varphi}\kappa^2}{3}\left[\frac{1}{2}\omega(\phi)\dot\phi^2 +V(\phi) \right] \, , \quad
\dot H = -\frac{\e^{-\varphi}\kappa^2}{2}\omega(\phi)
\dot\phi^2 -\frac{1}{2}\left(\ddot\varphi+\dot\varphi^2\right) \, ,
\end{align}
this can be rewritten as
\begin{align}
\label{BD4}
\omega(\phi)\dot\phi^2 = -\frac{2\e^{\varphi}}{\kappa^2}\left[ \dot H
+ \frac{1}{2}\left(\ddot\varphi+\dot\varphi^2\right) \right]\, , \quad
V(\phi) = \frac{2\e^{\varphi}}{\kappa^2}\left[ \dot H + 3 H^2
+ \frac{1}{2}\left(\ddot\varphi+\dot\varphi^2\right) \right] \, .
\end{align}
Then, if we take the following choice of scalar potentials using a function $f(\phi)$,
\begin{align}
\label{BD5}
\omega(\phi) =&\, -\frac{2\e^{\varphi(\phi)}}{\kappa^2}\left[ f'(\phi)
+ \frac{1}{2}\left(\varphi''(\phi) + (\varphi'(\phi))^2\right) \right]\, , \nonumber \\
V(\phi) =&\, \frac{2\e^{\varphi(\phi)}}{\kappa^2}\left[ f'(\phi) + 3
f(\phi)^2
+ \frac{1}{2}\left(\varphi''(\phi) + (\varphi'(\phi))^2\right) \right]\, ,
\end{align}
the explicit solution is given by Eq.~(\ref{ma11}).
Note that $\varphi(\phi)$ can be an arbitrary function of $\phi$.
In the case of the singular behavior in Eq.~(\ref{Hsin}), we find
\begin{align}
\label{BD5B}
\omega(\phi) =&\, \frac{2\e^{\varphi(\phi)}}{\kappa^2}\left[
-\frac{\beta h_s}{ \left(t_s - \phi \right)^{\beta +1}}
 - \frac{1}{2}\left(\varphi''(\phi) + (\varphi'(\phi))^2\right) \right]\, , \nonumber \\
V(\phi) =&\, \frac{2\e^{\varphi(\phi)}}{\kappa^2}\left[
\frac{\beta h_s}{\left(t_s - \phi \right)^{\beta+1}}
+ \frac{3 h_s^2}{ \left(t_s - \phi \right)^{2\beta}} + \frac{1}{2}\left(\varphi''(\phi) + (\varphi'(\phi))^2\right) \right]\, ,
\end{align}
By using Eq. (\ref{BD5B}), the action in Eq. (\ref{BD2}) has the following form,
\begin{align}
\label{BD2}
S= \int &\, d^4x \sqrt{-g} \left[\frac{
\e^{\varphi(\phi)}R}{2\kappa^2}
 - \frac{\e^{\varphi(\phi)}}{\kappa^2}\left\{
-\frac{\beta h_s}{ \left(t_s - \phi \right)^{\beta +1}}
 - \frac{1}{2}\left(\varphi''(\phi) + (\varphi'(\phi))^2\right) \right\} \partial_\mu \phi\partial^\mu \phi\right. \nonumber \\
& \left. -\frac{2\e^{\varphi(\phi)}}{\kappa^2}\left\{
\frac{\beta h_s}{\left(t_s - \phi \right)^{\beta+1}}
+ \frac{3 h_s^2}{ \left(t_s - \phi \right)^{2\beta}} + \frac{1}{2}\left(\varphi''(\phi) + (\varphi'(\phi))^2\right) \right\}
+ L_\mathrm{matter} \right]\, .
\end{align}
Different from the case of the scalar-tensor theory (\ref{stsing}),
by adjusting the function $\varphi(\phi)$, we can choose $\omega(\phi)$ to be positive
and avoid the ghost even if $\beta$ is positive.
In fact, now the left hand side (l.h.s.) of Eq.~(\ref{BGghost}) has the following form,
\begin{align}
\label{BGghost2}
\e^{-\varphi(\phi)}\omega(\phi) + \frac{3}{2\kappa^2} \varphi'(\phi)^2
= \frac{1}{\kappa^2}\left[
-2\frac{\beta h_s}{ \left(t_s - \phi \right)^{\beta +1}}
 - \varphi''(\phi) + \frac{1}{2} (\varphi'(\phi))^2 \right] \, .
\end{align}
Then, for example, when $\beta\neq 1$, if we choose
\begin{align}
\label{chphi}
\varphi(\phi) = \frac{2h_s}{1-\beta }\left(t_s - \phi \right)^{1-\beta}\, ,
\end{align}
we obtain,
\begin{align}
\label{BGghost3}
\e^{-\varphi(\phi)}\omega(\phi) + \frac{3}{2\kappa^2} \varphi'(\phi)^2
= \frac{2{h_s}^2}{\kappa^2
\left(t_s - \phi \right)^{2\beta}} > 0 \, .
\end{align}
Therefore, the condition~(\ref{BGghost}) is satisfied.
When $\beta=1$, we may choose,
\begin{align}
\label{chphi2}
\varphi(\phi) = 2h_s \ln \left(t_s - \phi \right)\, ,
\end{align}
we obtain,
\begin{align}
\label{BGghost4}
\e^{-\varphi(\phi)}\omega(\phi) + \frac{3}{2\kappa^2} \varphi'(\phi)^2
= \frac{2h_s^2}{\kappa^2\left(t_s - \phi \right)^2} > 0 \, ,
\end{align}
and we find that the condition~(\ref{BGghost}) is satisfied, again.

By using (\ref{chphi}), when $\beta\neq -1$, the action (\ref{BD2}) can be rewritten as
\begin{align}
\label{BD2neq-1}
S= \int d^4x \sqrt{-g} &\, \left[\frac{
\e^{\frac{2h_s}{1-\beta }\left(t_s - \phi \right)^{1-\beta }}R}{2\kappa^2}
+ \frac{2 h_s^2 \e^{\frac{2h_s}{1-\beta }\left(t_s - \phi \right)^{1-\beta }}}{\kappa^2
\left(t_s - \phi \right)^{2\beta}}\partial_\mu \phi\partial^\mu \phi
-\frac{10 h_s^2\e^{\frac{2h_s}{1-\beta }\left(t_s - \phi \right)^{1-\beta }}}{\kappa^2
\left(t_s - \phi \right)^{2\beta}} + L_\mathrm{matter} \right]\, .
\end{align}
Furthermore, if we define a new scalar field $\xi$ by
\begin{align}
\label{xi}
\xi = \frac{2}{\kappa}\e^{\frac{h_s}{1-\beta }\left(t_s - \phi \right)^{1-\beta }} \, ,
\end{align}
the action (\ref{BD2neq-1}) is further rewritten as
\begin{align}
\label{BD2neq-1B}
S= \int d^4x \sqrt{-g} \left[ \frac{\xi^2 R}{8} + \frac{1}{2} \partial_\mu \xi \partial^\mu \xi
 -\frac{5 h_s^2\xi^2}{2} \left( \frac{1-\beta}{2h_s} \ln \frac{\kappa\xi}{2} \right)^\frac{-2\beta}{1-\beta}
+ L_\mathrm{matter} \right]\, .
\end{align}
On the other hand, when $\beta= 1$, by using (\ref{chphi2}) the action (\ref{BD2}) is rewritten as
\begin{align}
\label{BD2eq-1}
S= \int d^4x \sqrt{-g} &\, \left[\frac{\left(t_s - \phi \right)^{2h_s}R}{2\kappa^2}
+ \frac{2 h_s^2 \left(t_s - \phi \right)^{2h_s -2}}{\kappa^2} \partial_\mu \phi\partial^\mu \phi
 -\frac{10 h_s^2 \left(t_s - \phi \right)^{2h_s - 2}}{\kappa^2} + L_\mathrm{matter} \right]\, .
\end{align}
In addition, if we define a new scalar field $\xi$ by
\begin{align}
\label{xi2}
\xi = \frac{2}{\kappa}\left(t_s - \phi \right)^{h_s} \, ,
\end{align}
the action (\ref{BD2neq-1}) is further rewritten as
\begin{align}
\label{BD2eq-1B}
S= \int d^4x \sqrt{-g} \left[ \frac{\xi^2 R}{8} + \frac{1}{2} \partial_\mu \xi \partial^\mu \xi
 -\frac{10 h_s^2}{\kappa^2}\left(\frac{\kappa\xi}{2} \right)^{2-\frac{2}{h_s}} + L_\mathrm{matter} \right]\, .
\end{align}

We should note that the signature in front of the kinetic term of $\xi$ in the action (\ref{BD2neq-1B})
or (\ref{BD2eq-1B}) might seem to tell that $\xi$ might be a ghost
but as clear in the Einstein frame action (\ref{BD2E}), because the condition~(\ref{BGghost}) is satisfied as in
(\ref{BGghost3}) or (\ref{BGghost4}), there does not appear the ghost in the model.

\subsection{The $k-$essence Model}
\label{sec-k-essence}

Here, based on \cite{Matsumoto:2010uv}, we review the reconstruction of
the $k$-essence model, which is a generalization of quintessence theory.
The $k$-essence model is a rather general model that includes only one scalar
field and the action is given by
\begin{align}
\label{KK1}
S= \int d^4 x \sqrt{-g} \left( \frac{R}{2\kappa^2} - K \left(
\phi, X \right) + L_\mathrm{matter}\right)\, ,\quad X \equiv \partial^\mu
\phi
\partial_\mu \phi \, .
\end{align}
Here, $\phi$ is a scalar field, again.
The Einstein equation has the following form:
\begin{align}
\label{Sch2}
\frac{1}{\kappa^2}\left( R_{\mu\nu} - \frac{1}{2}g_{\mu\nu} R \right)
= - K \left( \phi, X \right) g_{\mu\nu} + 2 K_X \left( \phi, X \right)
\partial_\mu \phi \partial_\nu \phi
+ T_{\mu\nu}\, .
\end{align}
Here, $K_X \left( \phi, X \right) \equiv \frac{\partial K \left( \phi, X \right)}{\partial X}$.
On the other hand, the variation of the action with respect to $\phi$ gives
\begin{align}
\label{Sch3}
0= - K_\phi \left( \phi, X \right) + 2 \nabla^\mu \left( K \left( \phi, X \right) \partial_\mu \phi \right)\, .
\end{align}
Here, $K_\phi \left( \phi, X \right) \equiv \frac{\partial K \left( \phi, X \right)}{\partial \phi}$ and it is
assumed that the scalar field $\phi$ does not directly couple with the matter.
When we neglect the contribution from the matter, the Friedmann
and Raychaudhuri equations or the FLRW equations are given by
\begin{align}
\label{KK2}
\frac{3}{\kappa^2} H^2 = 2 X \frac{\partial K\left( \phi, X \right)}
{\partial X} - K\left( \phi, X \right) \, ,\quad
 - \frac{1}{\kappa^2} \left(2 \dot H + 3 H^2 \right) = K\left( \phi, X \right) \, .
\end{align}
As in the previous model, if we consider the following model
\begin{align}
\label{k4}
K(\phi,X) = \sum_{n=0}^\infty \left(X+1\right)^n K^{(n)} (\phi)\, ,\quad
K^{(0)} (\phi) = - \frac{1}{\kappa^2}\left(2 f'(\phi) + 3 f(\phi)^2
\right)\, ,\quad K^{(1)} (\phi) = \frac{1}{\kappa^2} f'(\phi) \, ,
\end{align}
there exists a solution given by (\ref{ma11})., again.
Note that in (\ref{k4}), $K^{(n)} (\phi)$ $n\geq 2$ can be an arbitrary function but
$K^{(2)}$ is related with the stability of the solution and $K^{(3)}$ is related with the existence
of the Schwarzschild spacetime.
As in the previous models, we can realize the singular behavior in (\ref{Hsin}) by choosing
$f(\phi) = \frac{1}{\left(t_s - \phi \right)^\beta}$.
More explicitly, $K(\phi,X)$ in (\ref{k4}) is given by
\begin{align}
\label{k4ex}
K(\phi,X) = - \frac{1}{\kappa^2}\left( \frac{2 \beta} {\left(t_s - \phi \right)^{\beta+1} }
+ \frac{3}{ \left(t_s - \phi \right)^{2\beta}} \right)
 +\frac{\beta}{\kappa^2 \left(t_s - \phi \right)^{\beta+1} }
+ \sum_{n=2}^\infty \left(X+1\right)^n K^{(n)} (\phi) \, ,
\end{align}
Then, the model with $\beta\geq 1$ generates Type I singularity, the model $0<\beta<1$, Type III,
the model with $-1<\beta<0$ Type II,
and the model where $\beta<-1$ and $\beta$ is not a negative integer generates Type IV singularity.

\subsection{Scalar-Einstein--Gauss--Bonnet Gravity}
\label{sec-Scalar-Einstein-Gauss-Bonnet}

We now consider the reconstruction of the
scalar-Einstein--Gauss--Bonnet gravity (For pioneering work on the
scalar-Einstein--Gauss--Bonnet gravity, see \cite{Boulware:1986dr})
based on \cite{Nojiri:2005vv,Nojiri:2006je}. The
scalar-Einstein--Gauss--Bonnet gravity was first considered as a
candidate of the DE in \cite{Nojiri:2005vv}.

The action of the Einstein--Gauss--Bonnet gravity with a scalar field $\phi$ is
\begin{align}
\label{ma22}
S=\int d^4 x \sqrt{-g}\left[ \frac{R}{2\kappa^2} - \frac{1}{2}\partial_\mu \phi \partial^\mu \phi
 - V(\phi) - \xi(\phi) {G}\right]\, .
\end{align}
Here ${G}$ is the Gauss-Bonnet invariant defined by
\begin{align}
\label{GB}
{G}=R^2 -4 R_{\mu\nu} R^{\mu\nu} + R_{\mu\nu\xi\sigma}R^{\mu\nu\xi\sigma}\, .
\end{align}
The variation of the action (\ref{ma22}) with respect to the scalar field $\phi$ yields the following equation:
\begin{align}
\label{g3}
\nabla^2 \phi-V'(\phi)+ \xi(\phi){G}=0\, .
\end{align}
When we neglect the contribution from the matter, the variation of the action (\ref{ma22}) with respect to the metric $g_{\mu\nu}$
yields the following field equations,
\begin{align}
\label{gb4bD4}
0 = & \frac{1}{2\kappa^2}\left(- R^{\mu\nu} + \frac{1}2 g^{\mu\nu} R\right)
+ \left(\frac{1}2 \partial^\mu \phi \partial^\nu \phi
 - \frac{1}{4}g^{\mu\nu} \partial_\rho \phi \partial^\rho \phi \right)
 - \frac{1}2 g^{\mu\nu}V(\phi) \nonumber \\
& + 2 \left( \nabla^\mu \nabla^\nu \xi(\phi)\right)R
 - 2 g^{\mu\nu} \left( \nabla^2 \xi(\phi)\right)R
 - 4 \left( \nabla_\rho \nabla^\mu \xi(\phi)\right)R^{\nu\rho}
 - 4 \left( \nabla_\rho \nabla^\nu \xi(\phi)\right)R^{\mu\rho} \nonumber \\
& + 4 \left( \nabla^2 \xi(\phi) \right)R^{\mu\nu}
+ 4g^{\mu\nu} \left( \nabla_{\rho} \nabla_\sigma \xi(\phi) \right) R^{\rho\sigma}
- 4 \left(\nabla_\rho \nabla_\sigma \xi(\phi) \right) R^{\mu\rho\nu\sigma}.
\end{align}
We should note that the scalar field equation (\ref{g3}) can be obtained from Eq.~(\ref{gb4bD4}).
Therefore Eq.~(\ref{g3}) is not an independent equation and we forget the equation hereafter.

In the flat FLRW spacetime (\ref{FLRWk0}), Eq.~(\ref{gb4bD4}) has the following forms,
\begin{align}
\label{ma24}
0=& - \frac{3}{\kappa^2}H^2 + \frac{1}{2}{\dot\phi}^2 + V(\phi)
+ 24 H^3 \frac{d \xi(\phi(t))}{dt}\, ,\\
\label{GBany5}
0=& \frac{1}{\kappa^2}\left(2\dot H + 3 H^2 \right)
+ \frac{1}{2}{\dot\phi}^2 - V(\phi) - 8H^2 \frac{d^2 \xi(\phi(t))}
{dt^2} - 16H \dot H \frac{d\xi(\phi(t))}{dt} - 16 H^3
\frac{d \xi(\phi(t))}{dt} \, .
\end{align}
By combining (\ref{ma24}) and (\ref{GBany5}) and deleting $V(\phi)$, we obtain
\begin{align}
\label{ma25}
0 =&\, \frac{2}{\kappa^2}\dot H
+ {\dot\phi}^2 - 8H^2 \frac{d^2 \xi(\phi(t))}{dt^2}
 - 16 H\dot H \frac{d\xi(\phi(t))}{dt} + 8H^3 \frac{d\xi(\phi(t))}{dt} \nonumber \\
=&\, \frac{2}{\kappa^2}\dot H + {\dot\phi}^2
 - 8a\frac{d}{dt}\left(\frac{H^2}{a}\frac{d\xi(\phi(t))}{dt}\right)\, ,
\end{align}
and integrating Eq.~(\ref{ma25}) with respect to $\xi(\phi)$, we get
\begin{align}
\label{ma26}
\xi(\phi(t)) =&\, \frac{1}{8}\int^t dt_1 \frac{a(t_1)}{H(t_1)^2} \int^{t_1}
\frac{dt_2}{a(t_2)} \left(\frac{2}{\kappa^2}\dot H (t_2) + {\dot\phi^2(t_2)} \right)\, .
\end{align}

Finally, by substituting $\xi$ in (\ref{ma26}) into (\ref{ma24}), we have
\begin{align}
\label{GBV}
V(\phi(t)) =&\, \frac{3}{\kappa^2}H(t)^2 - \frac{1}{2}{\dot\phi (t)}^2 - 3a(t) H(t) \int^t \frac{dt_1}{a(t_1)}
\left(\frac{2}{\kappa^2}\dot H (t_1) + {\dot\phi^2(t_1)} \right)\, .
\end{align}

Therefore, for the model where $V(\phi)$ and $\xi(\phi)$ are given by using
adequate functions $g(t)$ and $f(\phi)$ in the following way,
\begin{align}
\label{ma27}
V(\phi) =&\, \frac{3}{\kappa^2}g'\left(f(\phi)\right)^2 - \frac{1}{2(f'(\phi))^2} \nonumber \\
&\, - 3g'\left(f(\phi)\right) \e^{g\left(f(\phi)\right)} \int^\phi d\phi_1 f'(\phi_1 ) \e^{-g\left(f(\phi_1)\right)}
\left(\frac{2}{\kappa^2}g''\left(f(\phi_1)\right)
+ \frac{1}{(f'(\phi_1 ))^2} \right)\, , \nonumber \\
\xi(\phi) =&\, \frac{1}{8}\int^\phi d\phi_1 \frac{f'(\phi_1)
\e^{g\left(f(\phi_1)\right)} }{(g'(\phi_1))^2} \int^{\phi_1} d\phi_2 f'(\phi_2) \e^{-g\left(f(\phi_2)\right)}
\left(\frac{2}{\kappa^2}g''\left(f(\phi_2)\right) + \frac{1}{(f'(\phi_2))^2} \right)\, ,
\end{align}
the solution of (\ref{ma27}) is given by
\begin{align}
\label{ma28}
\phi=f^{-1}(t)\quad \left(t=f(\phi)\right)\, ,\quad
a=a_s\e^{g(t)}\ \left(H= g'(t)\right)\, .
\end{align}
Then by choosing
\begin{align}
\label{gphi}
g'(t) = \frac{h_s}{ \left(t_s - t \right)^\beta} \, ,
\end{align}
we obtain the singular behavior in (\ref{Hsin}).
We should note that as clear from the action~(\ref{ma22}), the ghost does not appear in this model
even if $\beta>0$, that is, the case of Type I and Type III singularities.

It is difficult to execute the integrations in (\ref{ma27}) for general $\beta$ in (\ref{gphi}).
Then as an example, we consider the following case which corresponds to Type I singularity,
\begin{align}
\label{GBex}
g(t)=- h_s \ln \left(t_s - t \right)\, , \quad f = t_s - f_s \e^\phi \, .
\end{align}
Then, we find $g= h_s \left( \phi +\ln f_s \right)$ and
\begin{align}
\label{ma27ex}
V(\phi) =&\, \frac{1}{f_s^2\left( h_s + 1 \right)}\left\{ \frac{3h_s^2}{\kappa^2} \left( - h_s + 1 \right)
 - 4 h_s + 1 \right\} \e^{-2\phi} \, , \nonumber \\
\xi(\phi) =&\, \frac{1}{16h_s^2\left( h_s+1 \right)}\left(\frac{2h_s}{\kappa^2} + 1 \right) \e^{-2\phi} \, ,
\end{align}
which is the model proposed in \cite{Elizalde:2004mq}.

\subsection{$F(R)$ Theories of Gravity}
\label{sec-F(R)-gravity}

The theory of $F(R)$-gravity is the most simplest and
straightforward generalization of Einstein's General theory of
Relativity and it has received quite remarkable attention in the
cosmological community for two reasons: One is due to its simple
construction and the other is its ability to explain the late-time
accelerating expansion of the Universe and the early dynamics of
the Universe. Over the last several years, the theory of $F (R)$
gravity has been investigated by numerous investigators including
its successes and failures in different domains of astrophysics
and cosmology
\cite{Capozziello:2002rd,Nojiri:2006gh,Capozziello:2006dj,Song:2006ej,Nojiri:2006be,Li:2007xn,Sawicki:2007tf,Fay:2007uy,Boehmer:2007tr,deSouza:2007zpn,Starobinsky:2007hu,Nojiri:2007jr,Nojiri:2007as,Capozziello:2007ms,Nojiri:2007cq,Paliathanasis:2011jq,Paliathanasis:2016tch,Dimakis:2017tvb,Papagiannopoulos:2018mez}.
We refer to the review articles on $F (R)$ gravity for more
details in this direction
\cite{Nojiri:2006ri,Sotiriou:2008rp,DeFelice:2010aj,Nojiri:2017ncd}.

The action of $F(R)$ gravity is obtained by replacing the scalar curvature $R$ of the Einstein-Hilbert action by a suitable function $F(R)$ as follows \cite{Capozziello:2002rd,Nojiri:2006ri}
\begin{align}
\label{action-F(R)}
S_{F(R)}=\int d^4 x \sqrt{-g} \left[\frac{F(R)}{2\kappa^2} + {L}_\mathrm{matter}\right]\, .
\end{align}

One can view the modified part in $F (R)$ compared to the Einstein-Hilbert part as $F (R) = R + f(R)$.
Now, in the background of a spatially flat FLRW Universe (\ref{FLRWk0}),
one can derive the gravitational equations in this modified gravitational theory taking the forms:
\begin{align}
\label{sp-new-eqn-F(R)-EFE}
\rho_\mathrm{eff} = \frac{3}{\kappa^2} H^2\, , \quad p_\mathrm{eff} = - \frac{1}{\kappa^2} (2 \dot{H} + 3 H^2)\, ,
\end{align}
where $\rho_\mathrm{eff}$ and $p_\mathrm{eff}$ are given by
\begin{align}
\rho_\mathrm{eff} =&\, \frac{1}{\kappa^2}\left[-\frac{1}{2}f(R) + 3\left(H^2 + \dot H\right) f'(R) - 18 \left(4H^2 \dot H + H \ddot H\right)f''(R) \right] + \rho\, ,
\label{rhoeff-F(R)}\\
p_\mathrm{eff} =&\, \frac{1}{\kappa^2}\left[\frac{1}{2}f(R) - \left(3H^2 + \dot H \right)f'(R)
+ 6 \left(8H^2 \dot H + 4{\dot H}^2+ 6 H \ddot H + \dddot H \right)f''(R) + 36\left(4H\dot H + \ddot H\right)^2f'''(R) \right] \nonumber \\
&\, + p\, .
\label{peff-F(R)}
\end{align}
in which $\rho$, $p$ are the energy density and
pressure of the matter sector, respectively, and $R = 12H^2 + 6\dot{H}$. If the
matter sector has a constant EoS parameter $w= p/\rho$, then using
Eqs.~(\ref{rhoeff-F(R)}) and (\ref{peff-F(R)}), one can quickly
derive that
\begin{align}\label{effective-eos-F(R)-MG}
p_\mathrm{eff} - w \rho_\mathrm{eff} = G\left(H, \dot H, \ddot H, \cdots\right)\, ,
\end{align}
where
\begin{align}\label{Geff-F(R)}
G\left(H, \dot{H}, \ddot{H},...\right) = - \frac{1}{\kappa^2}\left(2\dot H + 3(1+w)H^2 \right)\, ,
\end{align}
and the explicit form of $G\left(H, \dot{H}, \ddot{H},...\right)$ is given by
\begin{align}
\label{Geff-explicit-F(R)}
G\left(H, \dot{H}, \ddot{H} \cdots\right) =&
\frac{1}{\kappa^2}\Bigg[\frac{1 + w}{2}f(R) - \left\{3\left(1+w\right)H^2
+ \left(1+3w\right) \dot H \right\}f'(R) \nonumber\\
& + 6 \left\{ \left(8 + 12w\right) H^2 \dot H + 4{\dot H}^2
+ \left(6 +3 w\right) H \ddot H + \dddot H \right\}f''(R)
+ 36\left(4H\dot H + \ddot H\right)^2f'''(R) \Bigg].
\end{align}
The above equations (\ref{Geff-F(R)}) and (\ref{Geff-explicit-F(R)}) have a very important consequence in cosmology.
For example, if a cosmology is given in terms of the Hubble rate $H (t)$, then the r.h.s. of Eq.~(\ref{Geff-F(R)})
can be expressed in terms of a function of time, say $f(t)$.
Now, if we can find a combination of $H$, $\dot{H}$, $\ddot{H},...$ in $G\left(H, \dot{H}, \ddot{H},... \right)$ that reproduces
the function $f(t)$, then the cosmology given by $H(t)$ can be realized by this reconstruction mechanism.
Let us illustrate the above by taking an example which will be essential to investigate the singularities in this context.
Assume that a cosmology is given by the following Hubble rate:
\begin{align}
\label{example-H-F(R)}
H=h_1 + \frac{h_2}{t}\, ,
\end{align}
where $h_1$ and $h_2$ are constants.
Thus, for the choice of $H(t)$ in Eq.~(\ref{example-H-F(R)}), one can derive its time
derivatives as follows:
\begin{align}
\dot{H} = - \frac{h_2}{t^2}\, ,\quad \ddot{H} = \frac{2h_2}{t^3}\, ,\quad
\cdots\, ,
\end{align}
As a consequence, the r.h.s. of Eq.~(\ref{Geff-F(R)})
turns out to be
\begin{align}
- \frac{1}{\kappa^2}\left(2\dot H + 3(1+w)H^2 \right)
= - \frac{1}{\kappa^2}\left( 3(1+w)h_1^2 + \frac{6(1+w)h_1 h_2}{t}
+ \frac{-2h_2 + 3(1+w) h_2^2}{t^2} \right)\, .
\end{align}
Now, if $G \left(H,\dot{H}, \ddot{H},.. \right)$ is given by the following function
\begin{align}
\label{choice-G-F(R)}
G\left(H,\dot{H}, \ddot{H},...\right) = \frac{1}{\kappa^2}
\left\{ - 3(1+w)h_1^2 + 6(1+w)h_1 H
+ \left[ 2 - 3\left(1+w\right) h_1 \right] \dot H \right\}\, ,
\end{align}
then Eq.~(\ref{example-H-F(R)}) is a solution of
Eq.~(\ref{Geff-F(R)}). However, the choice of $G \left(H,\dot H,\ddot{H}, \cdots \right)$
in Eq.~(\ref{choice-G-F(R)}), is indeed
not unique and there is a large freedom in its choice, but the
choice mainly depends on the underlying gravitational theory where
we are interested in, see for instance Refs.
\cite{Capozziello:2006dj,Nojiri:2006gh,Nojiri:2006be}.

Now, we shall investigate the appearance of finite-time future
singularities in some $F(R)$ gravity models using the
reconstruction technique. This phenomenon is not surprising
because the modified gravity can be represented as the Einstein
gravity with an effective ideal fluid having a phantom or
quintessence-like EoS (see the details in Ref.
\cite{Capozziello:2005mj}). It is known that in some cases, such
ideal fluid having a phantom or quintessence-like EoS may induce
the finite-time future singularities.

\begin{center}
\textit{(i) Big Rip singularity:}
\end{center}

We start with the Big Rip singularity which is characterized by the following evolution of the Hubble rate \cite{Bamba:2008ut}
\begin{align}
\label{H-F(R)-big-rip}
H (t) = \frac{h_s}{t_s -t}\, ,
\end{align}
where $h_s$ and $t_s$ are positive real numbers.
As one can notice, for $t \rightarrow t_s$, $H (t)$ of
Eq.~(\ref{H-F(R)-big-rip}) diverges. Now, we shall apply the
reconstruction technique. That means one can reconstruct the $F(R)$ gravity theory realizing the cosmology given
in terms of the Hubble rate of Eq.~(\ref{H-F(R)-big-rip}) which has a Big Rip singularity.

One can rewrite the action of Eq.~(\ref{action-F(R)}) with the use
of proper functions $P(\phi)$ and $Q(\phi)$ of a scalar field
$\phi$ as follows \cite{Nojiri:2006gh}:
\begin{align}
\label{action-F(R)-rewritten}
S=\int d^4 x \sqrt{-g} \bigg(P(\phi) R + Q(\phi) + {L}_\mathrm{matter} \bigg)\, .
\end{align}
As the scalar field $\phi$ does not have a kinetic term,
therefore, it can be considered to be an auxiliary field.
Now, varying the action (\ref{action-F(R)-rewritten}) with respect to
the scalar field $\phi$, one gets
\begin{align}
\label{first-order-F(R)}
0=P'(\phi)R + Q'(\phi)\, ,
\end{align}
where the prime denotes the derivative with respect to $\phi$ and this
can in principle be solved with respect to $\phi$ as $\phi=\phi(R)$.
Now plugging $\phi=\phi(R)$ into the action (\ref{action-F(R)-rewritten}), one can certainly express the action in terms of $F (R)$ given by
\begin{align}
\label{F(R)-derived}
F(R) = P\left(\phi(R)\right) R + Q \left(\phi(R) \right)\, .
\end{align}

Now by varying the action (\ref{action-F(R)-rewritten}) with respect to the
metric tensor $g_{\mu\nu}$, one can find
\begin{align}
\label{EFE-F(R)}
-\frac{1}{2}g_{\mu\nu}\left\{P(\phi) R + Q(\phi) \right\}
 - R_{\mu\nu} P(\phi) + \nabla_\mu \nabla_\nu P(\phi)
 - g_{\mu\nu} \nabla^2 P(\phi) + \frac{1}{2}T_{\mu\nu} = 0\, ,
\end{align}
where $T_{\mu\nu}$ is the energy momentum tensor of the matter
sector. In the background of a spatially flat FLRW Universe (\ref{FLRWk0}), the
gravitational field equations in Eq.~(\ref{EFE-F(R)}) reduce to
\begin{align}
-6 H^2 P(\phi) - Q(\phi) - 6H\frac{dP(\phi(t))}{dt} + \rho = 0\, ,
\label{efe1-flrw}\\
\left(4\dot H + 6H^2\right)P(\phi) + Q(\phi) + 2\frac{d^2 P(\phi(t))}{dt^2} + 4H\frac{d P(\phi(t))}{dt} + p = 0\, .
\label{efe2-flrw}
\end{align}
By combining Eqs.~(\ref{efe1-flrw}) and (\ref{efe2-flrw}) we get
\begin{align}
\label{2nd-order-ode-F(R)}
2\frac{d^2 P(\phi(t))}{dt^2} - 2 H \frac{dP(\phi(t))}{dt} + 4\dot H P(\phi) + p + \rho = 0\, .
\end{align}
As one can redefine the scalar field $\phi$, therefore, we may
choose $\phi = t$. Now for the Hubble rate in
Eq.~(\ref{H-F(R)-big-rip}), one can solve for the scale factor as
\begin{align}
\label{scale-factor-F(R)}
a (t) = \widetilde{a}_0 \exp\left( g(t) \right)\, ,
\end{align}
where $\widetilde{a}_0 >0$ is a constant and $\dot{g}(t)=H(t)$.
Concerning the matter sector we can assume that $\rho = \sum_{i}
\rho_i$ and $p = \sum_{i} p_i$ where $\rho_i$ and $p_i$
denotes the energy density and pressure of the $i$-th fluid, respectively.
If the fluid components do not interact with each other, then using the
usual conservation equation, one can find $\rho_i = \rho_{i0} a^{-3 (1+w_i)}$, where $\rho_{i0}$
is the current value of the energy density $\rho_i$ and $w_i = p_i/\rho_i$ denotes the EoS
parameter of the $i$-th fluid. With the above considerations, we
can now re-express Eq.~(\ref{2nd-order-ode-F(R)}) as
\begin{align}
\label{rank2-oder-F(R)-final}
2 \frac{d^2 P(\phi)}{d\phi^2} - 2 g'(\phi) \frac{dP(\phi)}{d\phi}+ 4g''(\phi) P(\phi)
+ \sum_i (1 + w_i) \rho_{i0} a_0^{-3(1+w_i)} \exp\left[ -3 \left( 1+w_i \right)g(\phi) \right] = 0\, .
\end{align}
If Eq.~(\ref{rank2-oder-F(R)-final}) is solved for $P (\phi)$,
then from Eq.~(\ref{efe1-flrw}), one can find $Q (\phi)$ as
\begin{align}
\label{Q-for-F(R)}
Q(\phi) = -6 \left(g'(\phi)\right)^2 P(\phi) - 6g'(\phi) \frac{dP(\phi)}{d\phi}
+ \sum_i \rho_{i0} a_0^{-3(1+w_i)} \exp\left[ -3 \left( 1+w_i \right)g(\phi) \right]\, .
\end{align}
Thus, we see that a given expansion history of the Universe
specified by the Hubble rate or the scale factor can be
realized by some specific $F(R)$ gravity model.
Now, neglecting the matter sector from this context, the general solution of
Eq.~(\ref{rank2-oder-F(R)-final}), can either be given by
\begin{align}
\label{solution1-F(R)-BR}
P(\phi) = P_+ \left(t_s - \phi\right)^{\alpha_+}
+ P_- \left(t_s - \phi\right)^{\alpha_-},\quad \alpha_\pm \equiv \frac{- h_s + 1 \pm \sqrt{h_s^2 - 10h_s +1}}{2}\, ,
\end{align}
when $h_s > 5 + 2\sqrt{6}$ or $h_s < 5 - 2\sqrt{6}$ or can be
given by
\begin{align}
\label{solution-F(R)-BR}
P(\phi) = \left(t_s - \phi \right)^{-(h_s + 1)/2}
\left( \widetilde{A} \cos \left( \left(t_s - \phi \right) \ln \frac{ - h_s^2 + 10 h_s -1}{2}\right)
+ \widetilde{B} \sin \left( \left(t_s - \phi \right) \ln \frac{ - h_s^2 + 10 h_s -1}{2}\right) \right)\, ,
\end{align}
when $5 - 2\sqrt{6}< h_s < 5 + 2\sqrt{6}$. Here $\widetilde{A}$ and $\widetilde{B}$ are arbitrary 
constants. 
Now, using Eqs.~(\ref{first-order-F(R)}), (\ref{F(R)-derived}), and
(\ref{Q-for-F(R)}) (without matter sector), the forms of $F(R)$
for large $R$ can be derived as follows:
\begin{align}
F(R) \propto&\, R^{1 - \frac{\alpha_{-}}{2}}~, \quad \mbox{when} \quad h_s > 5 + 2\sqrt{6}, \quad \mbox{or} \quad h_s < 5 - 2\sqrt{6}\, ,
\label{F(R)-BR-section-sp-new1}\\
F(R) \propto &\, R^{\left(h_s + 1\right)/4} \times \left(\text{oscillating parts}\right), \quad \mbox{when} \quad 5 - 2\sqrt{6}< h_s < 5 + 2\sqrt{6}\, .
\label{F(R)-BR-section-sp-new2}
\end{align}


\begin{center}
 \textit{ (ii) Other types of singularities}
\end{center}

Here we discuss more general singularities appearing in the
context of $F (R)$ gravity. In order to proceed with the general
singularities, we consider that the Hubble rate be given in (\ref{BigRip}) or
(\ref{Hsin}) as~\cite{Nojiri:2008fk,Bamba:2008ut},
where we assume $h_s~(>0)$, $\beta~(\neq 0, 1)$ are real
numbers,\footnote{The case $\beta =0$ corresponds to the de Sitter
space and $\beta =1$ corresponds to the Big Rip singularity as
discussed above, hence, we are interested to investigate the cases
that $\beta \neq 0, 1$. } and $t<t_s$ since we are living in an
expanding Universe. Note that for non-integer $\beta<0$, some
derivatives of $H$ and therefore the curvature may become
singular. For the above Hubble rate in
(\ref{Hsin}), one may realize the evolution of the scale factor as in (\ref{BigRip}).
Notice from Eq.~(\ref{BigRip}) that for
non-integer values of $\beta$, when $t_s < t$, the scale factor,
and therefore the metric tensor may become complex number which is unphysical.
This could hint towards the ending of our Universe
at $t=t_s$ even if $\beta$ could be negative or less than $-1$.
As we are interested to explore the general singularities, we
focus on $\beta \neq 1$ and examine its various ranges. When
$\beta>1$, the scalar curvature $R$ behaves as
\begin{align}\label{R-F(R)-general-singularities-beta-greater-1}
 R \sim 12 H^2 \sim 12h_s^2 \left( t_s - t \right)^{-2\beta}\;,
\end{align}
while for $\beta<1$, the scalar curvature $R$ behaves as
\begin{align}
 \label{R-F(R)-general-singularities-beta-less-1}
R \sim 6\dot H \sim 6h_s\beta \left( t_s - t \right)^{-\beta-1}\, .
\end{align}

Now, it is possible to trace the asymptotic solution for $P$ when $\phi\to t_s$ as follows:
\begin{enumerate}
\item For $\beta>1$, one can find the following asymptotic expression of
$P(\phi)$:
\begin{align}
\label{F(R)-P-beta-greater-1}
P(\phi) \sim&\, \e^{\left(h_s/2\left(\beta - 1\right)\right)\left(t_s - \phi\right)^{-\beta + 1}}
\left(t_s- \phi\right)^{\beta/2}
\left(\widetilde A \cos \left(\omega \left(t_s - \phi\right)^{-\beta + 1}\right)
+ \widetilde B \sin \left(\omega \left(t_s - \phi\right)^{-\beta + 1}\right) \right)\, , \\
\omega \equiv&\, \frac{h_s}{2\left(\beta - 1\right)}\, . \nonumber
\end{align}
When $\phi\to t_s$, $P(\phi)$ tends to vanish.
Using (\ref{first-order-F(R)}), (\ref{F(R)-derived}), and (\ref{Q-for-F(R)}), at large $R$, $F(R)$ can be derived as
\begin{align}
\label{F(R)-form-beta-greater-1}
F(R) \propto \e^{\left(h_s/2\left(\beta - 1\right)\right)
\left(\frac{R}{12h_s}\right)^{(\beta - 1)/2\beta}}
R^{-1/4}\times\left( \mbox{oscillating part} \right)\, .
\end{align}
\item For $0 < \beta < 1$, one gets the asymptotic
expression of $P(\phi)$ as follows:
\begin{align}
\label{F(R)-P-beta-between-0-1}
P(\phi) \sim B \e^{-\left(h_s/2\left(1 - \beta\right)\right)
\left(t_s - \phi\right)^{1-\beta}}\left(t_s - \phi\right)^{\left(\beta + 1\right)/8},
\end{align}
and $F(R)$ is given by
\begin{align}
\label{F(R)-form-beta-between-0-1}
F(R) \sim \e^{-\left(h_s/2\left(1-\beta\right)\right)
\left( - 6\beta h_s R \right)^{(\beta - 1)/(\beta + 1)} } R^{7/8}\, .
\end{align}
\item For $\beta<0$, the asymptotic expression of $P(\phi)$ is given by:
\begin{align}
\label{F(R)-P-beta-less-0}
P(\phi) \sim A \e^{-\left(h_s/2\left(1 - \beta\right)\right)
\left(t_s - \phi\right)^{1-\beta}}
\left(t_s - \phi\right)^{- \left(\beta^2 - 6\beta + 1\right)/8},
\end{align}
and consequently, $F(R)$ is given by
\begin{align}
\label{F(R)-form-beta-less-0}
F(R) \sim
\left( -6h_s \beta R \right)^{\left(\beta^2 + 2\beta + 9\right)/8\left(\beta +1\right)}
\e^{-\left(h_s/2\left(1 - \beta\right)\right)
\left( -6h_s \beta R \right)^{\left(\beta-1\right)/\left(\beta + 1\right)}}\;,
\end{align}
where note that $-6h_s \beta R >0$ for real solution.
\end{enumerate}
Alternatively, one can trace the behavior of $H$ from the behavior
of $R$. Let us consider the case when $R$ behaves as,
\begin{align}
\label{form-R-sp} R \sim 6\dot H \sim R_s \left( t_s - t \right)^{-\gamma}\, .
\end{align}
Which corresponds to $\gamma=\beta+1<2$. In this situation: If $1<\gamma < 2$,
which corresponds to $ 0< \beta~(= \gamma -1) < 1$, $H$ is given by
\begin{align}
\label{H2-F(R)-sp}
H \sim
\frac{R_s}{6\left( \gamma - 1\right)} \left( t_s - t \right)^{-\gamma + 1}\, .
\end{align}
And if $\gamma<1$, which corresponds to $\beta = \gamma -1 <0$, then $H$ follows
\begin{align}
\label{H3-F(R)-sp} H
\sim H_s + \frac{R_s}{6\left( \gamma - 1\right)} \left( t_s - t
\right)^{-\gamma + 1}\, ,
\end{align}
where $H_s$ is an arbitrary constant
and it does not affect the behavior of $R$. Note that $H_s$ has
been chosen to vanish in Eq.~(\ref{Hsin}).
On the contrary,
 if $\gamma>2$, which corresponds to $\beta = \gamma/2 >1$, one has $R\sim 12 H^2$ and
$H$ behaves as
\begin{align}
\label{H1-F(R)-sp}
H \sim \sqrt{\frac{R_s}{12}} \left( t_s - t \right)^{-\gamma/2}.
\end{align}

Now, for the above expressions of the Hubble rate, one can
find the evolution of the scale factor. If $\gamma>2$, we find
that the scale factor evolves as
\begin{align}
\label{scale-factor1-F(R)-sp}
a(t) \propto \exp \left( \left(\frac{2}{\gamma} -1 \right)
\sqrt{\frac{R_s}{12}} \left( t_s - t \right)^{-\gamma/2 +1}\right)\, .
\end{align}
When $1 < \gamma < 2$, $a(t)$ evolves as
\begin{align}
\label{scale-factor2-F(R)-sp}
a(t) \propto \exp \left(
\frac{R_s}{6\gamma\left( \gamma - 1\right)} \left( t_s - t
\right)^{-\gamma}\right)\, .
\end{align}
And if $\gamma<1$, we get
\begin{align}
\label{scale-factor3-F(R)-sp}
a(t) \propto \exp \left( H_s t +
\frac{R_s}{6\gamma\left( \gamma - 1\right)} \left( t_s - t
\right)^{-\gamma}\right)\, .
\end{align}

However, we see that a sudden future singularity appears at
$t=t_s$~\cite{Barrow:2004xh,Nojiri:2004pf,Barrow:2004he,Fernandez-Jambrina:2004yjt}
when $\gamma<2$. Now, dealing with the case $\gamma<1$, as the
second term in Eq.~(\ref{H3-F(R)-sp}) is smaller than the first
term, one may solve the differential
Eq.~(\ref{rank2-oder-F(R)-final}) asymptotically in the following
way
\begin{align}
\label{eqn-P-F(R)-sp}
P\sim P_s \left( 1 + \frac{R_s}{3\beta(1-\beta)}\left(t_s -
\phi\right)^{1-\beta}\right)\, ,
\end{align}
where $P_s$ is a constant,
which finally leads to
\begin{align}
\label{sec-F(R)-expression-F(R)-sp}
F(R) \sim F_0 R + F_1 R^{2\beta/\left(\beta + 1\right)} \, .
\end{align}

Now, since for $F (R)$ gravity, one can introduce the effective
energy density and effective pressure, see Eqs.~(\ref{sp-new-eqn-F(R)-EFE}), (\ref{rhoeff-F(R)}),
(\ref{peff-F(R)}), thus, for different values of $\beta$ of the
Hubble rate (\ref{Hsin}), the nature
of the singularities can be studied as follows. When $\beta> 1$,
as $t\to t_s$, we see that
\begin{align}
a \sim \exp( h_s\left( t_s - t \right)^{1-\beta}/\left( \beta -1 \right) )
\to \infty,\;\; \mbox{and as a consequence},\; \rho_\mathrm{eff} \to \infty,\, |p_\mathrm{eff}| \to \infty,
\end{align}
which means that a Big Rip (Type I) singularity occurs. If
$0<\beta<1$, $a$ goes to a constant but $\rho_\mathrm{eff} \to
\infty,\, |p_\mathrm{eff}| \to \infty$, that means a Type III
singularity occurs. If $-1<\beta<0$, then $a$ and $\rho_\mathrm{eff}$
vanish but $|p_\mathrm{eff}| \to \infty$ which means that a Type II
singularity occurs. When $\beta<0$, instead of
Eq.~(\ref{Hsin}), as in
Eq.~(\ref{H3-F(R)-sp}), one may assume, as in (\ref{HsinR}) with $H_s(t)$ is a constant,
\begin{align}
\label{H3a-F(R)-sp}
H \sim H_s + h_s \left(t_s - t\right)^{-\beta}\, .
\end{align}
Now, if $-1<\beta<0$, for $t\to t_s$, $\rho_\mathrm{eff}$ approaches to finite value $3H_s^2/\kappa^2$
but $|p_\mathrm{eff}|$ diverges, so we have a Sudden singularity.
If $\beta<-1$ but $\beta$ is not an integer, then $a$ remains
finite and $\rho_\mathrm{eff}$, $p_\mathrm{eff}$ vanish if $H_s=0$ or
$\rho_\mathrm{eff}$ and $p_\mathrm{eff}$ are finite if $H_s\neq 0$,
however, higher derivatives of $H$ diverge. That means in this
case a Type IV singularity occurs. Thus, we see that $F(R)$
gravity may allow various type of finite-time future
singularities. This is not unnatural because in the context of
modified gravity, one can find an effective phantom/quintessence
phase~\cite{Nojiri:2006ri} and it is well-known that a
phantom/quintessence-dominated Universe may end up with
finite-time future singularities of various types. Hence, with the
reconstruction of the $F (R)$ gravity from a given cosmology one
can find the possible functional forms of $F (R)$ that may lead to
finite-time future singularities. For example, from the present
discussions, one can see that $F(R) = R + \widetilde{\alpha} R^n$
with $n>2$ leads to a Type I singularity and $F(R) = R -
\widetilde{\beta} R^{-n}$ with $n>0$ leads to a Type III
singularity where $\widetilde{\alpha}$ and $\widetilde{\beta}$ are
any real numbers.

\subsubsection{Occurrence of Singularities in Different Frames: $F(R)$ Gravity}
\label{sec-correspondence-singularities-FR}

The choice of the physical frame in the context of $F (R)$ gravity
is an important point because cosmology of an $F(R)$ gravity in
one frame could be different in other frame. The accelerating
phase of the Universe in one frame may correspond to its
decelerating phase \cite{Bahamonde:2017kbs}. Also, the type of a
singularity changes from one frame to other frame
\cite{Briscese:2006xu,Bahamonde:2016wmz}. In this section, we shall
discuss the second possibility in detail, that means, how the
choice of the frames affects the type of finite-time future
singularities appearing in $F (R)$ gravity theory.
We start with the vacuum $F(R)$ gravity in the Jordan frame whose action is
given by (\ref{action-F(R)}) when we neglect the contribution from matter by putting ${L}_\mathrm{matter}=0$
\cite{Bahamonde:2016wmz}.
Varying the action (\ref{action-F(R)}) with ${L}_\mathrm{matter}=0$ with
respect to the metric $g_{\mu\nu}$ representing a spatially flat
FLRW geometry (\ref{FLRWk0}), one can obtain the following gravitational field
equations, which is equivalent to (\ref{sp-new-eqn-F(R)-EFE}) with (\ref{rhoeff-F(R)}) and (\ref{peff-F(R)})
when we neglect the matter,
\begin{align}
0=&\, -\frac{F(R)}{2} + 3\left(H^2 + \dot H\right) F'(R)
- 18 \left( 4H^2 \dot H + H \ddot H\right) F''(R) \, ,
\label{different-frames-eq-2} \\
0=&\, \frac{F(R)}{2} - \left(\dot H + 3H^2\right)F'(R)
+ 6 \left( 8H^2 \dot H + 4 {\dot H}^2 + 6 H \ddot H + \dddot H\right) F''(R)
+ 36\left( 4H\dot H + \ddot H\right)^2 F'''(R)\, ,
\label{different-frames-eq-3}
\end{align}
where an overhead dot represents the differentiation with respect
to the cosmic time $t$ in (\ref{FLRWk0}) and the prime corresponds to the
differentiation with respect to the Ricci scalar $R$.

Now introducing the auxiliary fields $A$ and $B$, the action of
Eq.~(\ref{action-F(R)}) with ${L}_\mathrm{matter}=0$ written in the Jordan frame can
be expressed into an equivalent action 
as follows \cite{Maeda:1988ab} (see also \cite{Chiba:2003ir}) 
\begin{align}
S= \frac{ 1}{2\kappa^2} \int d^4 x \, \sqrt{-g} \Bigg[B(R-A)+F(A) \Bigg] \,.
\label{different-frames-eq-4}
\end{align}
The action (\ref{different-frames-eq-4}) can be varied with respect to the auxiliary scalar $B$ leading to the condition $A=R$,
and hence, one can recover the action~(\ref{action-F(R)}) with ${L}_\mathrm{matter}=0$.
Moreover, varying the action (\ref{different-frames-eq-4}) with respect to $A$, one can eliminate the auxiliary field $B$ from it ending up $B=F'(A)$.
Hence, the action (\ref{different-frames-eq-4}) takes the equivalent form
\begin{align}
S= \frac{1}{2\kappa^2} \int d^4 x \, \sqrt{-g} \Bigg[F'(A)(R-A)+F(A) \Bigg] \, .
\label{different-frames-eq-5}
\end{align}
Taking a conformal transformation of the metric tensor, one may
obtain a minimally coupled scalar-tensor theory, called the
Einstein frame scalar-tensor theory. We use a particular conformal
factor taking the following expression \cite{Bahamonde:2016wmz}
\begin{align}
\label{different-frames-eq-6}
\hat{g}_{\mu\nu}=\frac{1}{ F'(A)} g_{\mu\nu}\, ,
\end{align}
which modifies the Ricci scalar as $R\rightarrow \hat{R}$. With
the conformal transformation (\ref{different-frames-eq-6}), and
then by defining a new scalar field $\sigma$ in terms of the
auxiliary scalar field $A$,
\begin{align}
\sigma=-\ln F'(A)\,,
\label{different-frames-eq-7}
\end{align}
the action~(\ref{different-frames-eq-5}) takes the form
\begin{align}
\label{different-frames-eq-8}
S = \frac{1}{2\kappa^2}\int d^4 x \sqrt{-\hat{g}} \left\{\hat{R} - \frac{3}{2}\hat{g}^{\mu\nu} \partial_{\mu} \sigma \partial_{\nu} \sigma - V(\sigma)\right\} \, ,
\end{align}
where the potential $V(\sigma)$ takes the form
\begin{align}
\label{different-frames-eq-9}
V(\sigma)=\frac{A}{F'(A)}-\frac{F(A)}{F'(A)^2}\,.
\end{align}

Notice that with the use of Eq.~(\ref{different-frames-eq-7}), the
potential of Eq.~(\ref{different-frames-eq-9}) can be expressed in
terms of the scalar field $\sigma$. Thus, corresponding to the
Jordan frame $F(R)$ gravity characterized by the action of
Eq.~(\ref{action-F(R)}) with ${L}_\mathrm{matter}=0$, the Einstein frame scalar-tensor
theory can be found as given by the action of
Eq.~(\ref{different-frames-eq-8}). Finally, we note that with the
use of the following transformation
\begin{align}
\label{different-frames-eq-10}
\varphi= \sqrt{\frac{3}{2\kappa^2}}\sigma\, ,
\end{align}
the action (\ref{different-frames-eq-8}) can be expressed into the canonical form \cite{Bahamonde:2016wmz}
\begin{align}
\label{different-frames-eq-8A}
S = \int d^4 x \sqrt{-g} \left\{\frac{R}{2\kappa^2} - \frac{1}{2} \partial_{\mu} \varphi \partial^{\mu} \varphi - V(\varphi)\right\} \, ,
\end{align}
Alternatively, starting with the scalar-tensor canonical scalar field action given by \cite{Bahamonde:2016wmz}
\begin{align}
\label{different-frames-eq-11}
S = \int d^4 x \sqrt{-\hat{g}} \left\{\frac{\hat{R}}{2\kappa^2} - \frac{1}{2} \hat{\partial}_{\mu} \varphi \hat{\partial}^{\mu} \varphi - V(\varphi)\right\} \, ,
\end{align}
one can find its equivalent Jordan frame $F(R)$ gravity action.
Assuming a spatially flat FLRW line element (\ref{FLRWk0}), the FLRW equations corresponding to
the action (\ref{different-frames-eq-11}) can be written as
\begin{align}
3\widetilde{H}^2=\frac{1}{2}\dot{\varphi}^2+V\, , \quad
3\widetilde{H}^2+2\dot{\widetilde{H}}=-\frac{1}{2}\dot{\varphi}^2+V\, .
\end{align}
where we have used `{\it tilde}' in the physical quantities in
this frame (i.e., the Einstein frame) in order to differ the
corresponding physical quantities in the Jordan frame.

Now one can map the above action (\ref{different-frames-eq-11}) to
a modified $F(R)$ gravity theory. In order to do this, we need to
perform the conformal transformation given by
$g_{\mu\nu}\rightarrow \e^{\pm\sqrt{\frac{2}{3}}\kappa\varphi}\hat{g}_{\mu\nu}$, and as a
result, the FLRW metric becomes,
\begin{align}
ds_{F(R)}^2=\e^{\pm\sqrt{\frac{2}{3}}\kappa\varphi}\left(-d\widetilde{t}^2
+ \widetilde{a}^2 \left(~\widetilde{t}~\right)\sum_{i=1,2,3} \left(dx^i\right)^2 \right) ,
\end{align}
where we have introduced a new time coordinate $\widetilde{t}$, given by
$dt=\e^{\pm\frac{1}{2}\sqrt{\frac{2}{3}}\kappa\varphi} d\widetilde{t}$,
the solution of which, denoted by $t=f\left(~\widetilde{t}~\right)$, is an
increasing function. Let us note that if the function
$f\left(~\widetilde{t}~\right)$ allows singularities, then we may have some
problems. As we are considering two different frames, i.e., the
Jordan frame and the Einstein frame, therefore, following the
relation between $t$ and $\widetilde{t}$, the range of the values
of $\widetilde{t}$ could be mapped to a different region in the
$t$ coordinate. Let us consider an interval
$[\widetilde{t}_1,\widetilde{t}_2]$ in the Einstein frame, with
the scale factor at $\widetilde{t}=\widetilde{t}_1$ being equal to
$\widetilde{a}\left(~\widetilde{t}_1~\right)=0$ and at
$\widetilde{t}=\widetilde{t}_2$, the scale factor is equal to,
$\widetilde{a}\left(~\widetilde{t}_2~\right)=0$ or
$\widetilde{a}\left(~\widetilde{t}_2~\right)=\infty$. As one can identify
that $\widetilde{t}=\widetilde{t}_2$, corresponds to a Big Crunch
or to a Big Rip singularity, respectively, in which case with
potentially $\widetilde{t}_1=-\infty$ and/or
$\widetilde{t}_2=\infty$, so that these singularities effectively
do not occur, a fact which crucially depends on the particular
form of the scale factor at hand. The new range of $t$ coordinate
will be $[f\left( \widetilde{t}_1 \right)\,,f\left( \widetilde{t}_2 \right)]$, assuming
$\phi\left(~\widetilde{t}~\right)$ is regular everywhere in the interval
$[\widetilde{t}_1,\widetilde{t}_2]$.

In the following we shall discuss the correspondence of the
finite-time singularities in the Jordan and Einstein frames using
some simple examples. We note that while discussing the
singularities below, we have followed the same unit system as in
Ref.~\cite{Bahamonde:2016wmz}, that means, we have considered the
same unit system so that the gravitational coupling constant
becomes unity, i.e., $\kappa=1$.


\begin{center}
 {\it (i) Power law cosmology}
\end{center}

We consider the power law cosmology which is described by the following scale factor,
\begin{align} \label{singularity-correspondence-power-law}
\widetilde{a}\left( \widetilde{t} \right)=\widetilde{a}_c \left(\frac{\widetilde{t}}{\widetilde{t}_c} \right)^p \, ,
\end{align}
where $\widetilde{t}_c$ is a fiducial value of the cosmic time,
$p$ is a positive real number and from the above relation we
identify that $\widetilde{a}(~\widetilde{t}_c~)=\widetilde{a}_c$.
Note that such a power law cosmology described in
Eq.~(\ref{singularity-correspondence-power-law}) is a solution of
the Friedmann equation in the Einstein frame scalar-tensor theory
when the potential has the exponential form.
In this case the scalar field evolves as
\begin{align}
\phi=\pm \sqrt{2p}\; \ln \left(\frac{\widetilde{t}}{\widetilde{t}_c} \right)\,.
\end{align}
In this case $\widetilde{t}_1=0$ and $\widetilde{t}_2=\infty$ and in this model the Hubble rate $\widetilde{H}$ diverges at $\widetilde{t}=0$, so we have a
Big Bang singularity in the Einstein frame.

Now in order to understand the behavior of the singularities in
the Jordan frame, we can follow the procedure as described in
Section~\ref{sec-correspondence-singularities-FR}. The new time
variable $t$ in the Jordan frame can be found from the differential equation
\begin{align}
\frac{dt}{d\widetilde{t}}=\left(\frac{\widetilde{t}}{\widetilde{t}_c} \right)^{\pm\sqrt{\frac{p}{3}}}\,,
\end{align}
having the following solution,
\begin{align}
\label{ccc}
t= \frac{3 }{3\pm \sqrt{3p}} \widetilde{t} \left(\frac{\widetilde{t}}{\widetilde{t}_c} \right)^{\pm2\sqrt{\frac{p}{3}}}\,.
\end{align}
The corresponding scale factor as a function of the cosmic time $t$ takes the form
\begin{align}\label{correspondence-scale-factor-jordan}
a(t)\sim t^w \quad \mbox{where}\quad w =\frac{\sqrt{3p}\pm3 p}{\sqrt{3p}\pm 3}\, .
\end{align}
Now we have the following observations. When the minus sign is
chosen in the conformal factor, the cosmological
evolution~(\ref{correspondence-scale-factor-jordan}) has a Big
Bang singularity at $t=0$ if the power law parameter $p$ lies in
the interval $1/3\leq p<3/4$. If $p=1/3$, the Jordan frame metric
becomes static and we do not have any finite-time singularity.
When $0<p<1/3$, the Big Bang singularity at $\widetilde{t}=0$ in
the Einstein frame becomes the beginning of a contracting Universe
in the Jordan frame. In the case $3/4<p<3$, from (\ref{ccc}) we
can see the direction of time is reversed, meaning that this case
must be disregarded. Finally, for $p>3$ the time $\widetilde{t}=0$
correspond to $t=-\infty$, so the Big Bang singularity disappears
in the Jordan frame.

\begin{center}
 {\it (ii) Cosmology generated by $R^{-n}$ gravity in the Jordan Frame}
\end{center}

We now consider the cosmology driven by $R^{-n}$ gravity in the
Jordan frame (see~\cite{Briscese:2006xu} for details). So, for
$F(R)\sim R^{-n}$, from Eqs.~(\ref{different-frames-eq-2})
and~(\ref{different-frames-eq-3}) one can see that the
corresponding scale factor takes the power law form
\begin{align}
a \sim \left(t_s - t\right)^{\frac{(n+1)(2n+1)}{n+2}}\, .
\end{align}
Therefore, if either $n<-2$ or $-1<n<-1/2$, a Big Rip Type I singularity appears at $t=t_s$ in the Jordan frame,
and in the remaining cases, a Type III Big Crunch singularity is present at this point.
In this case, the corresponding Einstein frame canonical scalar field takes the form
\begin{align}
\sigma\sim (n+1)\ln R \sim -2(n+1)\ln (t_s - t)\, ,
\end{align}
and the Ricci scalar takes the following expression
\begin{align}
R\sim \frac{6(n+1)(2n+1)(4n+5)n}{(n+2)^2(t_s - t)^2}\, .
\end{align}
Now the time coordinate $\widetilde{t}$ in the corresponding Einstein frame scalar-tensor theory is given by
\begin{align}
d\widetilde t = \pm \e^{\frac{1}{2}\sigma}dt \;\; \sim \pm (t_s - t)^{-(n+1)}dt\, ,
\end{align}
which gives $\widetilde{t}=\pm (t_s-t)^{-n}$. Therefore, for $n >
0$, $t$ approaches $t \rightarrow t_s$ in the Jordan frame
corresponds to $\widetilde{t} \rightarrow \pm \infty$ in the
Einstein frame. As a consequence, we observe that the singularity
changes its structure abruptly, that means it does not appear in
finite-time in the Einstein frame scalar-tensor theory.
Nevertheless, a new additional singularity may be present, as when
$t$ approaches infinity in the Einstein frame, it corresponds to
the new time coordinate $\widetilde{t} \to 0$, and thus any
singularities at infinity can be brought back to a finite-time. On
the other hand, when $n<0$, the limit $t\to t_s$ in the Jordan
frame corresponds to $\widetilde{t}\to 0$ in the Einstein frame.
Finally, we note that the metric in the Einstein frame
scalar-tensor theory behaves as
\begin{align}
ds^2=\e^{\sigma} \left(-dt^2 + a^2(t)\sum_{i=1,2,3}(dx^i)^2\right)
\sim -d\widetilde{t}^2 + \widetilde{a}^2\left(~\widetilde{t}~\right) \sum_{i=1,2,3}(dx^i)^2\, ,\quad
\widetilde{a}^2\left(~\widetilde{t}~\right) \sim a_s^2\; \widetilde{t}^{~\frac{2n(n^2-1)}{n+2}}\, ,
\end{align}
where the constant $a_s$ is a real number. In this case, the power of the scale factor is negative only when $-2<n<-1$ or $0<n<1$, and thus a Big Rip singularity is present. {Thus, for the Big Rip singularity in the Jordan frame, the scale factor in the Einstein frame behaves as $\widetilde{a}^2\left(~\widetilde{t}~\right)\to 0$ when $\widetilde{t}\to 0$, thus, it becomes a Type III Big Crunch singularity in the Einstein frame. }


\begin{center}
 {\it (iii) A singular cosmological evolution}
\end{center}

In this section, we examine how the simplest singular cosmology behaves in different frames.
We consider the following Hubble rate which describes the simplest singular cosmology
as in (\ref{Hsin}),
\begin{align}
\label{correspondence-Hubble-singular}
H\left(~\widetilde{t}~\right)=h_s \left(~\widetilde{t}-\widetilde{t_s}~\right)^{-\beta}\,,
\end{align}
where $h_s$ is any positive real number and $\beta$ is a real number. From the values of $\beta$ one can determine the singularity type. In particular, we realize the following type of singularities based on the values of $\alpha$:
\begin{itemize}
\item For $\beta>1$, we realize a Type I singularity.
\item For $0<\beta<1$, a Type III singularity is found.
\item When $-1<\beta<0$, a Type II singularity occurs.
\item When $\beta<-1$, a Type IV singularity is realized.
\end{itemize}

Let us assume that the Hubble rate of
Eq.~(\ref{correspondence-Hubble-singular}) is given in the
Einstein frame and we aim to investigate the singular behavior
captured in the Hubble rate of
Eq.~(\ref{correspondence-Hubble-singular}) in the context of Jordan
frame $F(R)$ theory. We use the conformal factor given by
$\e^{\sqrt{\frac{2}{3}}\varphi}$ which results in the
transformation of the metric as
\begin{align}
ds_{F(R)}^2=\e^{\sqrt{\frac{2}{3}}\varphi}\left(-d\widetilde{t}^2+ \widetilde{a}^2(~\widetilde{t}~)\sum_{i=1,2,3} \left(dx^i \right)^2 \right) \, ,
\end{align}
and the scale factor now becomes,
\begin{align}
a(t)=\e^{\frac{1}{2}\sqrt{\frac{2}{3}}\varphi}\;\widetilde{a}\left(~\widetilde{t}~\right)\, ,
\end{align}
where the Jordan frame time parameter $t$ is defined as 
\begin{align}
dt=\e^{\frac{1}{2}\sqrt{\frac{2}{3}}\varphi} d\widetilde{t},
\end{align}
the solution of which is an incomplete gamma function. However, for the Hubble rate as given in
Eq.~(\ref{correspondence-Hubble-singular}), using the corresponding equations of motion, the Einstein frame scalar field can be found to be
\begin{align}
\varphi=\frac{2 \sqrt{2h_{s}\beta } \left(~\widetilde{t}-\widetilde{t}_s~\right)^{\frac{1-\beta}{2}}}{1-\beta}\,,
\end{align}
where $h_s \beta>0 $ is an essential criterion in order for the scalar field to be canonical.
We further notice that the transformation blows up at $\widetilde{t}_s$ if $\beta>1$, hence, extra caution is required in this case.
in effect, the new Hubble rate in terms of our initial time coordinate $\widetilde{t}$, i.e.,
$H(\tilde{t})\equiv \frac{1}{a}\frac{da}{d\tilde{t}}$, is equal to,
\begin{align}
H(\widetilde{t})=\frac{ \sqrt{h_s\beta}\;
(~\widetilde{t}-\widetilde{t}_s~)^{-\frac{\beta +1}{2}}}{\sqrt{3}}+h_s (~\widetilde{t}-\widetilde{t}_s~)^{-\beta }\, ,
\end{align}
and the question is what is the effect on the time coordinate $t$.
We suppose $t=f(~\widetilde{t}~)$. Now, when, $f(~\widetilde{t}~)\neq 0$, we have,
\begin{align}
f'(~\widetilde{t}~)=\e^{\frac{1}{2}\sqrt{\frac{2}{3}}\varphi}\, ,
\end{align}
and consequently the following relation
\begin{align}
\frac{dH}{dt}=\frac{d\widetilde{t}}{dt} \frac{d\widetilde{H}}{d\widetilde{t}}=\e^{-\frac{1}{2}\sqrt{\frac{2}{3}}\varphi}\frac{d\widetilde{H}}{d\widetilde{t}}\, .
\end{align}
holds, that means, the expression $dH/dt$ diverges if and only if
$d\widetilde{H}/d\widetilde{t}$ diverges for $\beta<1$, as the
conformal factor is finite at $\widetilde{t}_s$. Now the Big Rip
(Type I) singularity occurs when the Hubble rate $H(t)$ diverges
at a finite-time thus, we need to examine whether
$f(~\widetilde{t}_s~)$ is finite or not. One can easily derive
$t_s$ given by \cite{Bahamonde:2016wmz}
\begin{align}
t_s=f(~\widetilde{t}_s~)=c_1-c_2 \Gamma \left(\frac{2}{\alpha +1}\right)\, ,
\end{align}
which is finite, provided $2/(1-\beta)$ is not a negative integer.
This readily implies that the singularity appears in the Hubble
rate $H$ at a finite-time, as long as $\beta \neq -(2/n) +1$ where
$n\geq 2$ is an integer. Accordingly, a Type II singularity is
found if $dH/dt$ diverges, but $H$ does not diverge. Combining
together, these imply that $-3<\beta<-1$. On the other hand, for
$\beta<-3$, a Type IV singularity occurs. Now we investigate the
evolution of the scale factor when it is conformally transformed.
The evolution of the scale factor in the Einstein frame reads, as in (\ref{BigRip}),
\begin{align}
\widetilde{a}\left(~\widetilde{t}~\right)=\widetilde{a}_s \exp\left(\frac{h_s\left(~\widetilde{t}-\widetilde{t}_s ~\right)^{1-\beta}}{1-\beta}\right)\, ,
\end{align}
which in the Jordan frame reads,
\begin{align}
\label{scale-factor-singular-evolution-Jordan-frame}
a\left(~\widetilde{t}~\right)=a_{s} \exp\left(\frac{3 h_{s} \left(~\widetilde{t} -\widetilde{t}_{s}~\right)^{1-\beta}
\pm 2 \sqrt{3\beta h_{s}} \left(~\widetilde{t} -\widetilde{t}_{s}~\right)^{\frac{1-\beta}{2}}}{3 (1-\beta)}\right)\,.
\end{align}
Now, depending on the values of $\beta$, from the scale factor (\ref{scale-factor-singular-evolution-Jordan-frame}),
one can witness the following type of singularities:
\begin{enumerate}
 \item For $\beta>1$, a Type I or no singularity occurs.
 \item For $-1<\beta<1$, a Type III singularity occurs.
 \item For $-3<\beta<-1$, a Type II singularity occurs.
 \item For $\beta<-3$, a Type IV singularity occurs.
\end{enumerate}
In TABLE~\ref{tab:correspondence-F(R)}, we show the correspondence
of the finite-time singularities between the Jordan frame and the
Einstein frame. TABLE~\ref{tab:correspondence-F(R)} clearly
shows that the singularity in one frame may not be same in the
other frame. For example, as displayed in TABLE~\ref{tab:correspondence-F(R)}, the Type I (Big Rip) singularity in
the Einstein frame may correspond to a non-singular evolution in
the Jordan frame. Further, the Type II singularity in the Einstein
frame could be modified to a more severe Type III singularity in
the Jordan frame, and the Type IV singularity in the Einstein
frame may correspond to a Type II singularity in the Jordan frame.

\begin{table}
\caption{The table shows the correspondence of finite-time
singularities in the Einstein and Jordan frames for the
cosmological evolution given in terms of the Hubble rate
(\ref{correspondence-Hubble-singular}) in the Einstein frame. }
 \label{tab:correspondence-F(R)}
\begin{center}
\renewcommand{\arraystretch}{1.4}
\begin{tabular}{|c@{\hspace{1 cm}}|@{\hspace{1 cm}} c|}
\hline
\textbf{Singularity in the Einstein Frame} & \textbf{Singularity in the Jordan Frame}\\
\hline\hline
 Type I & Type I or No Singularity \\

Type II & Type III \\

Type III & Type III \\

Type IV & Type IV or Type II \\

\hline
\end{tabular}
\end{center}
\end{table}

We close this section with a
special case of the singular evolution (\ref{correspondence-Hubble-singular}) describing the singular bounce cosmology,
a special case of the symmetric bounce \cite{Nojiri:2015sfd,Nojiri:2016ppu,Nojiri:2016ygo}.
One can rewrite the scale factor of the cosmological evolution (\ref{correspondence-Hubble-singular}) as follows,
\begin{align}\label{scale-factor-bounce-F(R)-correspondence}
a(t)=a_s\; \exp\left(h_s(t-t_s)^{2(1+\epsilon)}\right)\, ,
\end{align}
for which the Hubble rate becomes
\begin{align}\label{Hubble-bounce-F(R)-correspondence}
H(t)=2(1+\epsilon)h_s (t-t_s)^{2\epsilon+1},
\end{align}
where $\epsilon>0$ and it has been chosen in such a way so that all the quantities remain real. Now, in order to realize a bouncing scenario, for $t<t_s$ the Hubble rate must become negative (i.e., $H<0$) and additionally, in order for the bounce (\ref{scale-factor-bounce-F(R)-correspondence}) to be a deformation of the symmetric bounce described by $a(t)\sim \exp(\beta t^2)$, the parameter $\epsilon$ must lie in the interval $0<\epsilon<1$ having the following form
\begin{align}\label{epsilon-correspondence-F(R)}
\epsilon=\frac{2n}{2m+1}\, ,
\end{align}
where $m$ and $n$ are integers and they are chosen in such a way so that $\epsilon<1$ is satisfied. Now, for this choice of $\epsilon$, the cosmology described by the scale factor (\ref{scale-factor-bounce-F(R)-correspondence}) and the Hubble rate (\ref{Hubble-bounce-F(R)-correspondence}) clearly depicts a Type IV singular cosmology, in which case, the Hubble rate and its first derivative with respect to the cosmic time, i.e., $\dot{H}$ remain finite, but its second derivative with respect to the cosmic time, i.e., $\ddot{H}$ diverges. As demonstrated in Refs.~\cite{Odintsov:2015jca,Odintsov:2015gba} with the use of reconstruction techniques, the pure $F(R)$ gravity realizing the cosmological evolution (\ref{Hubble-bounce-F(R)-correspondence}) can approximately be given by
\begin{align}\label{F(R)-form-correspondence-bouncing}
F(R)= R+ \frac{R^2}{4C}+\Lambda\,,
\end{align}
where $C$ is a positive real number, near the bouncing point, which is $t\simeq t_s$. Let us define a new parameter $x=t-t_{s}$ for simplicity. Thus, the limit near the bouncing point, i.e., $t\simeq t_s$, corresponds to the limit $x\rightarrow 0$. Now, in order to transform the theory in the Einstein frame, we consider the following conformal transformation
\begin{align}
g_{\mu\nu}=\e^{-\sigma} \hat{g}_{\mu\nu}\, ,
\end{align}
where the scalar field $\sigma$ is equal to $\sigma= - \ln F'(A)$.
In terms of the new parameter $x$, the Ricci scalar reads
\begin{align}
R=12 h_s (\epsilon +1) x^{2 \epsilon } \Bigg[4 h_s (\epsilon +1) x^{2 \epsilon +2}+2 \epsilon +1 \Bigg],
\end{align}
and if we are close to the singularity where $x$ is small, then the Ricci scalar becomes,
\begin{align}
\label{Ricci-correspondence-singular}
R\approx 12 h_s (\epsilon +1)(2\epsilon+1) x^{2 \epsilon } \, .
\end{align}
Consequently, by combining Eqs.~(\ref{F(R)-form-correspondence-bouncing}) and (\ref{Ricci-correspondence-singular}), one obtains
\begin{align}
F'(R)\approx 1+\frac{6 h_s (\epsilon +1)(2\epsilon+1) x^{2 \epsilon } }{C}\equiv 1+D x^{2\epsilon}\, ,
\end{align}
where $D$ is a positive real number.
This means that the new time coordinate of the Einstein frame FLRW metric will be given in terms of $x$ as follows
\begin{align}
\label{diffeq1}
d \widetilde{t}=(1+D x^{2\epsilon})^{\frac{1}{2}\sqrt{\frac{3}{2}}}d x\, ,
\end{align}
the solution of which is a hypergeometric function. The new scale factor in terms of $x$ reads
\begin{align}
a(~\widetilde{t}~)=(1+D x^{2\epsilon})^{\frac{1}{2}\sqrt{\frac{3}{2}}} \; \widetilde{a}(x)\, ,
\end{align}
and subsequently, the derivative of the scale factor is given by
\begin{align}
\frac{d a}{d\widetilde{t}}= \frac{dx}{d \widetilde{t}} \frac{d a}{dx}= \frac{dx}{d\widetilde{t}}\Bigg[(1+D x^{2\epsilon})^{\frac{1}{2}\sqrt{\frac{3}{2}}} \frac{d\widetilde{a}}{dx}+2\epsilon x^{2\epsilon}(1+D x^{2\epsilon})^{\frac{1}{2}\sqrt{\frac{3}{2}}-1} \widetilde{a}(x)\Bigg]\, .
\end{align}
As $d t=d x$ at $x\simeq 0$, thus, looking at the second derivative of the scale factor $a(t)$, one may conclude that the scale factor diverges at $x=0$ provided $\epsilon<1/2$. Hence, the Type IV singularity in the Jordan frame becomes a Type II singularity in the Einstein frame, also reflected from TABLE~\ref{tab:correspondence-F(R)}.

\subsubsection{Occurrence of Singularities in Different Frames: Unimodular $F(R)$ Gravity}
\label{sec-correspondence-singularities-unimodularFR}

In this section, we discuss the correspondence between the frames
in the context of unimodular $F(R)$ gravity
\cite{Nojiri:2015sfd,Nojiri:2016ppu,Nojiri:2016ygo,Odintsov:2016imq}.
The Jordan frame unimodular $F(R)$ gravity is described by the
action \cite{Nojiri:2015sfd}
\begin{align}
\label{correspondence-action-unimodular-FR-1}
S = \int d^4 x \bigg[\sqrt{-g}\; \bigg( F(R) - \lambda \bigg) + \lambda \bigg]
+ S_\mathrm{matter} \,,
\end{align}
where we assume that $F(R)$ is a smooth function of the Ricci scalar $R$,
$\lambda$ denotes the Lagrange multiplier function and
$S_\mathrm{matter}$ stands for the action of the matter fluids
present in the Universe sector. Notice that the variation of the
action (\ref{correspondence-action-unimodular-FR-1}) with respect
to $\lambda$ leads to the unimodular constraint
\begin{align}
\sqrt{-g}&=1\,.
\label{correspondence-unimodular-FR-2}
\end{align}
The unimodular constraint given in
Eq.~(\ref{correspondence-unimodular-FR-2}) is the key point of
unimodular $F(R)$ gravity. Now if we vary the unimodular $F(R)$
gravity action (\ref{correspondence-action-unimodular-FR-1}) with
respect to the metric tensor $g_{\mu \nu}$, we obtain the
unimodular $F(R)$ gravity field equations as follows
\begin{align}
0= \frac{1}{2}g_{\mu\nu} \left( F(R) - \lambda \right) - R_{\mu\nu} F'(R)
+ \nabla_\mu \nabla_\nu F'(R) - g_{\mu\nu}\nabla^2 F'(R) + \frac{1}{2} T_{\mu\nu} \, .
\label{F(R)2}
\end{align}
Now, in order to proceed with the cosmological evolution, one needs to be very careful
because the flat standard FLRW metric (\ref{FLRWk0})
does not satisfy the unimodular constraint given in
Eq.~(\ref{correspondence-unimodular-FR-2}). Nevertheless, taking
the following coordinate transformation
\begin{align}
d\tau=a^3(t) dt\, ,
\label{correspondence-unimodular-FR-3}
\end{align}
one can verify that the resulting metric satisfies the unimodular condition (\ref{correspondence-unimodular-FR-2}).
Now, with the use of the transformation (\ref{correspondence-unimodular-FR-3}), the FLRW metric in (\ref{FLRWk0}) can be written as
\begin{align}
\label{Unimodular-FLRW2}
ds^2 = -a^{-6}\left(t\left(\tau\right)\right) d\tau^2 + a^2\left(t\left(\tau\right)\right) \left(dx^2+dy^2+dz^2 \right),
\end{align}
which for the sake of brevity we call the unimodular FLRW metric.
Using this unimodular FLRW metric, the vacuum field equations become \cite{Nojiri:2015sfd,Nojiri:2016ppu,Nojiri:2016ygo,Odintsov:2016imq},
\begin{align}
\label{correspondence-unimodular-FR-4}
0 =&\, - \frac{a^{-6}}{2} \left( F(R) - \lambda \right) + \left( 3 \dot K + 12 K^2 \right) F'(R)
 - 3 K \frac{d F'(R)}{d\tau} \, , \\
\label{correspondence-unimodular-FR-5}
0=&\, \frac{a^{-6}}{2} \left( F(R) - \lambda \right) - \left( \dot K + 6 K^2 \right) F'(R)
+ 5 K \frac{d F'(R)}{d\tau} + \frac{d^2 F' (R)}{d\tau^2} \, ,
\end{align}
where the function $K(\tau)$ is defined as the corresponding Hubble rate in the ``$\tau$'' coordinate, that is,
\begin{align}
K=\frac{1}{a(\tau)}\frac{d a(\tau)}{d\tau}\, .
\end{align}
By using the unimodular FLRW metric of
Eq.~(\ref{Unimodular-FLRW2}), the corresponding Ricci scalar now
becomes
\begin{align}
R &=a^{6} \left( 6\dot{K}+30K^2 \right)\, .
\end{align}

Now, in the following sections we describe the correspondence of the Jordan frame unimodular $F(R)$ gravity in the Einstein frame.

We begin with the unimodular $F(R)$ gravity action in the Jordan
frame, i.e., Eq.~(\ref{correspondence-action-unimodular-FR-1})
without the matter sector, and following the same approach as in
the ordinary $F(R)$ gravity, that means, introducing the
auxiliary field $A$, we rewrite the action
(\ref{correspondence-action-unimodular-FR-1}) as follows
\begin{align}
\label{action-auxiliary-field-unimodular-FR-6}
S = \int d^4 x \Bigg[ \sqrt{-g} \Bigg( F'(A)(R-A)+F(A) - \lambda \Bigg) + \lambda \Bigg]\, .
\end{align}
We note that the last term of the action
(\ref{action-auxiliary-field-unimodular-FR-6}) will remain
unaffected by the conformal transformations. Now to obtain a
minimally coupled scalar-tensor theory, we perform the conformal
transformation as in the standard $F(R)$ gravity case, i.e.,
$\hat{g}_{\mu\nu}= \e^\sigma g_{\mu\nu}$,
where $\hat{g}$ denotes the metric in the Einstein frame, and $g$
is the metric in the Jordan frame. Further, the scalar field
$\sigma$ in terms of the auxiliary field $A$ is given by,
$\sigma=-\ln F'(A)$, hence, the action
(\ref{action-auxiliary-field-unimodular-FR-6}) becomes,
\begin{align}
S = \int d^4 x \left\lbrace \sqrt{-\hat{g}} \left( \hat{R}-\frac{3}{2}\hat{g}^{\mu\nu}\partial_{\mu}\sigma \partial_\nu \sigma
 -V(\sigma) - \lambda \e^{-2\sigma}\right) + \lambda \right\rbrace\, ,
\end{align}
which describes a canonical scalar field action, in the absence of any matter sector.
We note that the unimodular constraint is not unaffected by the conformal transformation, so in the case it becomes
\begin{align}
\label{cons-unimodular-FR-7}
\sqrt{-\hat{g}}=\e^{2\sigma}\, .
\end{align}
In effect, the FLRW metric does not satisfy this constraint
identically, so in order to overcome this issue, similar to the
earlier approach, we introduce a new time coordinate $\tilde\tau$, which
is related to the cosmic time $\tilde t$ as follows
\begin{align}
d\widetilde{\tau}=\widetilde{a}^3\left(~\widetilde{t} ~\right)\e^{2\sigma\left(~\widetilde{t} ~\right)}d\widetilde{t}\, ,
\end{align}
so that the conformally transformed unimodular constraint of Eq.
(\ref{cons-unimodular-FR-7}) is satisfied and the corresponding
Einstein frame unimodular FLRW metric takes the form
\begin{align}
\label{FLRW-unimodular-FR-8}
ds^2=-\frac{\e^{4\sigma\left( \widetilde{\tau} \right)}}{\widetilde{a}^6\left(~\widetilde{\tau} ~\right)}d\widetilde{\tau}^2+\widetilde{a}^2\left( \widetilde{\tau} \right)\sum_{i=1}^3dx_i^2 \, .
\end{align}

In summary, in order to transform from one frame to other at the
level of metric, we have the following transformations. The scale
factor transforms as
\begin{align}
\label{trans-unimodular-FR-9}
\widetilde{a}\left(~\widetilde{\tau}~\right)&=\e^{\sigma/2} a(\tau)\,,
\end{align}
where the parameter $\widetilde{\tau}$ is related to $\tau $ as follows,
\begin{align}
\label{trans-unimodular-FR-10}
\left(\frac{\e^{4\sigma\left( \widetilde{\tau} \right)}}{\widetilde{a}^6\left(~\widetilde{\tau}~\right)}\right) d\widetilde{\tau}^2
= \left(\frac{\e^\sigma}{a^6(\tau)} \right)d\tau^2\,.
\end{align}
Combining Eqs.~(\ref{trans-unimodular-FR-9}) and
(\ref{trans-unimodular-FR-10}), one can easily observe that the
new time coordinate $\widetilde{\tau}$ is the same as the
coordinate $\tau$, i.e., $\widetilde{\tau}=\tau$.

So far we discuss the unimodular $F(R)$ gravity scenario, however,
we also need to discuss the scalar-tensor unimodular gravity. To
proceed, we begin with the following minimally coupled
scalar-tensor action
\begin{align}\label{action-unimodular-FR-11}
S = \int d^4 x \left\lbrace \sqrt{-\hat{g}} \left( \frac{\hat{R}}{2\kappa^2}-\frac{1}{2}\hat{g}^{\mu\nu}\partial_{\mu}\varphi \partial_\nu \varphi-V(\varphi)
 - \lambda h(\varphi)\right) + \lambda \right\rbrace\,,
\end{align}
where $\lambda$ is a constant and in addition, for the time being
we have assumed that the determinant of the metric is given by an
arbitrary function of the scalar field but this will be determined
later by the requirement that the action has a Jordan frame. Now,
considering the flat unimodular FLRW metric, for the above action
(\ref{action-unimodular-FR-11}), one can write down the
gravitational equations which are nothing but the standard scalar
field cosmological equations with a modified scalar potential
having the form $V(\varphi) +\lambda h(\varphi)$. In particular,
the gravitational equations read
\begin{align}
3\widetilde{K}^{2}=&\, \frac{1}{2}\dot{\varphi}^{2}+\Big(V(\varphi) +\lambda h(\varphi)\Big)\widetilde{a}^{-6}(~\widetilde{\tau}~)\,,
\label{EQ1-unimodular-FR-11}\\
9\widetilde{K}^2+2\dot{\widetilde{K}}=&\, -\frac{1}{2}\dot{\varphi}^{2}+\Big(V(\varphi) +\lambda h(\varphi)\Big)\widetilde{a}^{-6}(\widetilde{\tau})\,,
\label{EQ2-unimodular-FR-12}
\end{align}
where $\widetilde{K}(\widetilde\tau)$ denotes the unimodular Hubble parameter in the Einstein frame, explicitly given by
\begin{align}
\label{Hubble-unimodular-FR-13}
\widetilde{K}\left( \widetilde{\tau} \right)&=\widetilde{K}(\tau)=\frac{1}{\widetilde{a}(\tau)}\frac{d\widetilde{a}(\tau)}{d\tau}\,.
\end{align}

Now in order to conformally transform this action (\ref{action-unimodular-FR-11}) to a Jordan frame, we apply the following conformal transformation
\begin{align}
\label{conf-unimodular-FR-14}
g_{\mu\nu}=\e^{\pm\kappa\sqrt{\frac{2}{3}}\varphi}\hat{g}_{\mu\nu}\, .
\end{align}
This rescaling eliminates the kinetic term from the action (\ref{action-unimodular-FR-11}) and recast into the following form
\begin{align}
\label{action-unimodular-FR-14}
S = \int d^4 x \Bigg[ \sqrt{-g} \bigg\{ \e^{\pm\kappa\sqrt{\frac{2}{3}}\varphi}{2\kappa^2}R-\e^{\pm2\kappa\sqrt{\frac{2}{3}}\varphi} \bigg(V(\varphi)
+ \lambda h(\varphi) \bigg)\bigg\} + \lambda \Bigg]\,.
\end{align}
Now, varying the action (\ref{action-unimodular-FR-14}) with respect to the scalar field $\phi$, which is now just an auxiliary field, we get
\begin{align}
\label{unimodular-FR-15}
R=\e^{\pm\kappa\sqrt{\frac{2}{3}}\varphi}\Bigg[ 4\kappa^2 \bigg(V(\varphi)+\lambda h(\varphi) \bigg)\pm2\kappa \sqrt{\frac{2}{3}}\bigg(V'(\varphi)+\lambda h'(\varphi)\bigg)\Bigg]\,.
\end{align}
In order to find the Jordan frame unimodular $F(R)$ gravity, one
needs to invert Eq.~(\ref{unimodular-FR-15}) to get the scalar
field $\phi (R)$, as a function of $R$ only, and so independent of
$\lambda$. This directs us to choose
\begin{align}
h(\varphi)=\e^{-2\kappa\sqrt{\frac{3}{2}}\varphi}\, .
\end{align}
Hence, in this case we can invert to find $\varphi=\varphi(R)$, and consequently, the resulting unimodular $F(R)$ gravity theory takes the form
\begin{align}
S = \int d^4 x \Bigg[\sqrt{-g} \bigg( F(R) - \lambda \bigg) + \lambda \Bigg],
\end{align}
where the expression of $F(R)$ is given by
\begin{align}
F(R)= \e^{\pm\kappa\sqrt{\frac{2}{3}}\varphi(R)}{2\kappa^2}R-\e^{\pm 2\kappa\sqrt{\frac{2}{3}}\varphi(R)}V(\varphi(R))\, .
\end{align}
Now, having the correspondence between the Jordan and Einstein
frames, in the following we discuss the correspondence of the
finite-time singularities in the two frames. We again note that
for all the physical quantities in the Einstein frame we use
tilde to differ it from the Jordan frame, and additionally, we
set the gravitational coupling constant to be unity throughout the
discussions, i.e., $\kappa=1$.

\begin{center}
 {\it (i) Power law evolution}
\end{center}

We consider the power law cosmology characterized by the following
scale factor expressed in terms of $\widetilde{\tau}$ (the time
coordinate of the unimodular Einstein frame FLRW
metric~(\ref{FLRW-unimodular-FR-8})) as follows
\begin{align}
\label{unimodular-power-law}
\widetilde{a}\left( \widetilde{\tau} \right) =\widetilde{a}_{c}\left(\frac{\widetilde{\tau}}{\widetilde{\tau}_{c}}\right)^{p}\, ,
\end{align}
where $\widetilde{\tau}_{c}$ is some fiducial time and $p$ is a
positive real number. For this scale factor, the unimodular Hubble
parameter becomes $\widetilde{K} (\tau)=p\widetilde{\tau}^{-1}$.
Thus, one can see that at $\widetilde{\tau}=0$, the unimodular
Hubble rate diverges, and hence, for the power law cosmology in the
Einstein frame, a Big Bang singularity is realized. The above
scale factor is a solution of the Einstein frame Friedmann and Raychaudhuri
equations~(\ref{EQ1-unimodular-FR-11})-(\ref{EQ2-unimodular-FR-12})
when the potential has an exponential form. In this case the
scalar field takes the form
\begin{align}
\label{unimodular-power-law-scalar-field}
\varphi\left( \widetilde{\tau} \right)=\pm\sqrt{2p(1-3p)}\,\log \left(\widetilde{\tau}/\widetilde{\tau}_{c}\right)\,.
\end{align}

Now, using this solution
(\ref{unimodular-power-law-scalar-field}), it is possible to
change all the variables from the Einstein frame to the Jordan
frame with the use of the following conformal transformation
\begin{align}
g_{\mu\nu}=\e^{\pm\sqrt{\frac{2}{3}}\varphi}\hat{g}_{\mu\nu} \,.
\label{unimodular-power-law-transform}
\end{align}
As discussed above, the time coordinate $\tau$ in the Jordan frame
is equivalent to the original time coordinate in the Einstein
frame. Thus, using the
transformation~(\ref{unimodular-power-law-transform}), one can
express scale factor in the Jordan frame as follows
\begin{align}
a(\tau)=\e^{\pm\frac{1}{2}\sqrt{\frac{2}{3}}\varphi}\widetilde{a}(\widetilde{\tau})\sim \tau ^{p\pm\sqrt{\frac{p(1-3 p)}{3}}} \,.
\label{unimodular-power-law2}
\end{align}

Now, when the minus sign is taken in the conformal transformation,
we see that for $0<p<1/6$ the quantity $p-\sqrt{\frac{p(1-3
p)}{3}}$ is negative, so at $\tau=0$ the scale factor diverges,
this means that in the Jordan frame we have a contracting
Universe. For $p=1/6$ the Universe becomes static. When $1/6<p\leq
1/3$, in the Jordan frame, the Universe conserves the Big Bang
singularity at $\tau=0$, and finally, for $p>1/3$ the exponent in
(\ref{unimodular-power-law2}) becomes complex, meaning that this
case does not make physical sense.


\begin{center}
 {\it (ii) The case $F(R)=R^{-n}$ in the Jordan Frame}
\end{center}

Here we investigate a model starting in the Jordan frame and
conformally transform it to the Einstein frame. Let us consider
the vacuum unimodular $F(R)$ gravity having a typical form
$F(R)\sim R^{-n}$. In this case, the scale factor evolves as
\begin{align}
a(\tau)\sim(\tau_s-\tau)^{\frac{1+3n+2n^2}{5+10n+6n^2}}\,,
\end{align}
where we have used the $\tau$ coordinate of the unimodular FLRW
metric. One can see that a Big Rip singularity occurs in the
$\tau$ coordinate, if the parameter $n$ lies in the range
$-1<n<-1/2$. Now, in terms of the original FLRW cosmic time $t$,
the scale factor takes the form
\begin{align}
a(t)\sim t^{\frac{(2n+1)(n+1)}{2+n}}\;,
\end{align}
which shows that a Big Rip singularity in the $\tau$ coordinate, appears in the $t$ coordinate if $n<-2$, or, $-1<n<-1/2$.

We shall now investigate what happens in the corresponding
Einstein frame scalar-tensor theory. The scalar field $\sigma$ of
the conformal factor is given by
\begin{align}
\sigma \sim (n+1)\ln R \sim -\left(\frac{2(n+1)(2+n)}{5+10n+6n^2}\right)\ln (\tau_s-\tau)\, ,
\end{align}
where the Ricci scalar in the unimodular time parameter is given by
\begin{align}
R\sim (\tau_s-\tau)^{-2\left(\frac{2+n}{5+10n+6n^2}\right)}\, .
\end{align}
This means that under a conformal transformation, the scale factor transforms as
\begin{align}
\widetilde{a}\left( \widetilde{\tau} \right)&=\e^{\sigma/2} a(\tau),
\end{align}
where the parameter $\widetilde{\tau}$ is the same as the original time coordinate $\tau $, that means, $\widetilde{\tau}=\tau$. Therefore, we get
\begin{align}
 \widetilde{a}\sim (\tau_{s}-\tau )^{\frac{n^2-1}{6 n^2+10n+5}}\,.
\end{align}
These conditions show that we could have more values of $n$ for
which the power of the scale factor is negative. As one can see
that for $-1<n<1$, the power of the scale factor becomes
negative. Thus, we conclude that a Type I singularity will appear
for more values of $n$.


\begin{center}
 {\it (iii) A singular cosmological model}
\end{center}

Here we consider a toy model where the unimodular Hubble parameter in the Einstein frame is given by
\begin{align}
\widetilde{K}(~\widetilde{\tau}~)&= f_s (\widetilde{\tau}-\widetilde{\tau}_{s})^{\alpha}\,,
\label{toy-unimodular-F(R)}
\end{align}
where $k_{s}$, $\alpha$ are real numbers and
$\widetilde{\tau}_{s}$ refers to a specific time instant of
$\widetilde{\tau}$. For this Hubble parameter, the scale factor
can be given by
\begin{align}
\widetilde a\left( \widetilde{\tau} \right)&=a_{s} \exp\left[\frac{f_s\left( \widetilde{\tau}-\widetilde{\tau}_{s} \right)^{\alpha+1}}{\alpha+1}\right]\,.
\end{align}
where $a_{s}$ is a constant.

Now, to convert this Einstein frame solution to the Jordan frame,
one needs to find the scalar field $\varphi$ giving rise to such a
solution~(\ref{toy-unimodular-F(R)}). To do this it is essential
to solve the differential equation obtained by subtracting the
unimodular Friedmann equations~(\ref{EQ1-unimodular-FR-11}) and (\ref{EQ2-unimodular-FR-12}) as follows
\begin{align}
0= \dot{\varphi}^2+6 f_s^2 \left( \widetilde{\tau} -\widetilde{\tau}_{s} \right)^{2 \alpha}+2 f_s \alpha \left( \widetilde{\tau}-\widetilde{\tau}_{s} \right)^{\alpha-1} \,.
\label{diff-eqn-unimodular-F(R)}
\end{align}
However, this equation cannot be solved in general for an
arbitrary power of $\alpha$, hence, one needs to find an
approximate solution to the above differential equation. Thus, in
order to proceed, one needs to approximate the solution around the
singularity $\widetilde{\tau}=\widetilde{\tau}_s$. This can be
done for two cases as follows: when $\alpha<-1$ and the Hubble
rate is that of a Type I singularity, and when $\alpha>-1$ and the
other types of singularities are present. For $\alpha<-1$, and
around the singularity, the second term
of~(\ref{diff-eqn-unimodular-F(R)}) dominates over the third term
and the differential equation~(\ref{diff-eqn-unimodular-F(R)}) is
approximated to be
\begin{align}
\dot{\varphi}^2+6 f_s^2 \left( \widetilde{\tau} -\widetilde{\tau}_{s} \right)^{2 \alpha}&\sim0\,.
\label{approx-solution1-unimodular-F(R)}
\end{align}
In this case, the solution for $\varphi$ becomes imaginary, and
hence, such a Hubble rate could only be described by a phantom
scalar field. For such a phantom field, the corresponding Jordan
frame $F(R)$ becomes complex and this is unphysical.

However, for $\alpha>-1$, and around the singularity, the third
term of (\ref{diff-eqn-unimodular-F(R)}) dominates over the
second term and therefore the scalar field behaves as
\begin{align}
\varphi\left( \widetilde{\tau} \right)&\sim \pm \frac{2 \sqrt{-2f_s\alpha} }{\alpha+1} \left( \widetilde{\tau} -\widetilde{\tau}_{s} \right)^{\frac{\alpha+1}{2}}\, ,
\end{align}
which is real if $-2 f_s \alpha \geq 0$ and nontrivial if $-2 f_s \alpha \neq 0$.
Thus, in this case one can continue further and conformally transform to the Jordan frame.
Applying the conformal transformation~(\ref{conf-unimodular-FR-14}) and using the fact
that the time coordinate is unchanged, i.e.,
$\tau=\widetilde{\tau}$, the scale factor in the Jordan frame reads
\begin{align}
\label{sf-singular-unimodular-F(R)}
a(\tau) \sim a_{s} \exp\left(\frac{3 f_s (\tau -\tau_{s})^{\alpha+1}\pm 2 \sqrt{-3 \alpha f_s}\; (\tau -\tau_{s})^{\frac{\alpha+1}{2}}}{3 (\alpha+1)}\right)\,.
\end{align}

Now from this scale factor (\ref{sf-singular-unimodular-F(R)}), one
can describe a variety of finite-time singularities as follows:
\begin{itemize}
 \item For $-1<\alpha<1$, a Type III singularity occurs.
 \item For $1<\alpha<3$, a Type II singularity occurs.
 \item For $3<\alpha$, a Type IV singularity occurs.
\end{itemize}
In TABLE~\ref{tab:correspondence-unimodular-F(R)} we show how the
singularities change their types from one frame to
another\footnote{We note that in
TABLE~\ref{tab:correspondence-unimodular-F(R)}, we do not keep
Type I singularity since it only appears when $\alpha<-1$, and hence, we have a phantom scalar field and consequently for such a
phantom scalar field, the corresponding Jordan frame $F(R)$
becomes complex and this is unphysical.}. One can observe that the
unimodular $F(R)$ case behaves similarly to the standard $F(R)$
case. From TABLE~\ref{tab:correspondence-unimodular-F(R)}, we
observe that a Type II singularity in the Einstein frame is
modified to the more severe Type III singularity in the Jordan
frame. The Type IV singularity in the Einstein frame may become a
more severe Type II singularity in the Jordan frame if the
parameter $\alpha$ lies in the range $1<\alpha<3$.
\begin{table}
\caption{We show the correspondence of finite-time singularities
in the Einstein and Jordan frames, for the cosmological evolution
(\ref{toy-unimodular-F(R)}) in the Einstein frame for the case
$\alpha>-1$. }
\label{tab:correspondence-unimodular-F(R)}
\begin{center}
\renewcommand{\arraystretch}{1.4}
\begin{tabular}{|c@{\hspace{1 cm}}|@{\hspace{1 cm}} c|}
\hline
\textbf{Singularity in the Einstein Frame} & \textbf{Singularity in the Jordan Frame}\\
\hline\hline
Type II & Type III \\
Type III & Type III \\
Type IV & Type IV or Type II \\
\hline
\end{tabular}
\end{center}
\end{table}

\begin{center}
{\it 2.1. Qualitative Analysis of the Phase structure of Unimodular $F(R)$ gravity near Finite-Time Singularities}
\end{center}

Here we discuss the qualitative behavior of the dynamical system
corresponding to the vacuum unimodular $F(R)$ gravity near the
finite-time singularities. To begin with, we introduce the
following variables~\cite{Bahamonde:2016wmz}
\begin{align}
\label{variables-unimodular-F(R)}
x_1=-\frac{1}{K F'(R)}\frac{dF'(R)}{d\tau}\, , \quad x_2=-\frac{F(R)}{6K^2 F'(R) a^6}\, , \quad x_3=\frac{R}{6K^2 a^6}\, , \quad x_4= \frac{\lambda}{6a^6 K^2 F'(R)}\,,
\end{align}
which allows one to write the Friedmann equation as
\begin{align}
x_1+x_2+x_3+x_4 = 1\, .
\label{constraint-friedmann1-unimodular-F(R)}
\end{align}
This constraint tells us that in order to describe the dynamics one needs only three variables as the fourth variable can be obtained using the remaining three variables.
Using the above variables, the Raychaudhuri equation becomes
\begin{align}
\label{friedmann2-unimodular-F(R)}
\frac{1}{F'(R) K^2}\frac{d^2 F'(R)}{d\tau^2}=1+5x_1+3x_2+x_3+3x_4\, .
\end{align}
Now, differentiating the variables as defined in
Eq.~(\ref{variables-unimodular-F(R)}) with respect to a
cosmological time $dN= K(\tau)d\tau$, one gets the following
dynamical system \cite{Bahamonde:2016wmz}
\begin{align}
\label{dyn-system-unimodular-F(R)}
x_1'=&\, -1+x_1^2-x_1 x_3-3x_2-x_3-3x_4 \, , \nonumber \\
x_2'=&\, -m+4x_2+x_1x_2-2x_2x_3+50-16x_3 \, , \nonumber \\
x_3'=&\, m+20x_3-2x_3^2-50 \, , \nonumber \\
x_4'=&\, x_4(x_1-2x_3+4) \, ,
\end{align}
where $m=\ddot{K}/K^3$. Let us note that from the constraint
Eq.~(\ref{constraint-friedmann1-unimodular-F(R)}), in reality, one
only has three dynamical equations in
(\ref{dyn-system-unimodular-F(R)}). In fact, one can disregard the
fourth equation which is a combination of the other three. The
dynamical system, as one can notice, is non-autonomous due to the
existence of the term $m$.
The critical points of the system~(\ref{dyn-system-unimodular-F(R)}) have been shown in
TABLE~\ref{tab-critical-unimodular-F(R)}. For an autonomous system
$m$ should vanish. Now if we consider that
$K(\tau)=f_s(\tau-\tau_s)^{\alpha}$ with $\alpha<-1$ which
corresponds to the Big Rip singularity as $\tau\to \tau_s$, then
the parameter $m$ is equal to, $m=\frac{(-1+\alpha ) \alpha (\tau -\tau_s)^{-2-2 \alpha }}{f_s^2}\sim 0$,
and consequently, the dynamical system of Eq.~(\ref{dyn-system-unimodular-F(R)}) becomes
autonomous near the Big Rip singularity. Thus, one may expect that
finding the fixed points of the dynamical system near the Big Rip
singularity could offer some insights on the non-autonomous
system, however, from TABLE~\ref{tab-critical-unimodular-F(R)},
one can clearly see that if one sets $m =0$, then $x_1$ and $x_2$
become complex, that means the critical points are not real. In
fact, if we assume that $m$ is a constant, then in order to have
real critical points one needs $m\gtrsim 17$. This indicates that
the dynamics near the Big Rip singularity behaves very strangely.

\begin{table}[h]
\caption{The table shows the critical points of the dynamical system (\ref{dyn-system-unimodular-F(R)}). }
\label{tab-critical-unimodular-F(R)}
\begin{center}
\begin{tabular}{|c|c|c|c|c|}
 \hline
 Critical Point & $x_1$ & $x_2$ & $x_3$ & $x_4$ \\
 \hline

 $P_1$ & $\frac{1}{2 \sqrt{2}}\left(-\sqrt{m}-\sqrt{m+4 \sqrt{2} \sqrt{m}-40}+2 \sqrt{2}\right)$ & $\frac{1}{4} \left(3 \sqrt{2} \sqrt{m}+\sqrt{2} \sqrt{m+4 \sqrt{2} \sqrt{m}-40}-20\right)$ &$5-\frac{\sqrt{m}}{\sqrt{2}}$ & $0$\\
 \hline
 $P_2$ &$\frac{1}{2 \sqrt{2}}\left(\sqrt{m}-\sqrt{m+4 \sqrt{2} \sqrt{m}-40}+2 \sqrt{2}\right)$ & $\frac{1}{4} \left(3 \sqrt{2} \sqrt{m}-\sqrt{2} \sqrt{m+4 \sqrt{2} \sqrt{m}-40}-20\right)$ & $5-\frac{\sqrt{m}}{\sqrt{2}}$ & $0$ \\
 \hline
 $P_3$ & $\frac{1}{2\sqrt{2}}\left(\sqrt{m}+\sqrt{m-4 \sqrt{2} \sqrt{m}-40}+2 \sqrt{2}\right)$ & $\frac{1}{4} \left(-3 \sqrt{2} \sqrt{m}-\sqrt{2} \sqrt{m-4 \sqrt{2} \sqrt{m}-40}-20\right)$ & $5+\frac{\sqrt{m}}{\sqrt{2}}$ & $0$ \\
 \hline
 $P_4$ & $\frac{1}{2\sqrt{2}}\left(\sqrt{m}-\sqrt{m-4 \sqrt{2} \sqrt{m}-40}+2 \sqrt{2}\right)$ & $\frac{1}{4} \left(-3 \sqrt{2} \sqrt{m}+\sqrt{2} \sqrt{m-4 \sqrt{2} \sqrt{m}-40}-20\right)$ & $5+\frac{\sqrt{m}}{\sqrt{2}}$ & $0$ \\
 \hline
\end{tabular}
\end{center}
\end{table}

\subsection{$F(G)$ Gravity}
\label{sec-F(G)-gravity}

Apart from the popular modified gravity models, namely, $F(R)$ and
$F(T)$, another very interesting class of modified gravity models
which can explain the late-time cosmic acceleration is the
string-inspired modified Gauss--Bonnet gravity $-$ the so called
$F(G)$-gravity
\cite{Nojiri:2005jg,Nojiri:2005am,Cognola:2006eg,Cognola:2006sp,Brevik:2006nh,Li:2007jm,Boehmer:2009fey,Sadeghi:2009hi,Elizalde:2010jx,MohseniSadjadi:2010pc,Garcia:2010xz,Banijamali:2012zz,Rodrigues:2012qu,Bamba:2014mya,Momeni:2014bua,Huang:2015kca,Abdolmaleki:2015iaa,Astashenok:2015haa,Kusakabe:2015ida,Oikonomou:2015qha,Carloni:2017ucm,Oikonomou:2017ppp,Terrucha:2019jpm,Inagaki:2019rhm,Bahamonde:2019swy}
where $F (G)$ is any arbitrary function of the Gauss-Bonnet
invariant defined in (\ref{GB}). Even though this class of
modified gravity models can explain the late-time accelerating
expansion of the Universe, however, it has been found that they
can lead to finite-time future singularities
\cite{Bamba:2008ut,Bamba:2010wfw}. In this section, we shall
describe the appearance of finite-time singularities in this class
of modified gravity models. Unfortunately this model includes
ghosts \cite{DeFelice:2009ak} although the ghost-free
modifications have been proposed \cite{Nojiri:2018ouv,
Nojiri:2019dwl}. In this section, however, we show the structure
of the singularities based on the original models for the purpose
of the illustration.

We start with the action of
$F(G)$-gravity which is given by~\cite{Nojiri:2005jg}
\begin{align}
S=\int d^{4}x \sqrt{-g} \left[ \frac{1}{2\kappa^2}
\bigg(R+F(G) \bigg) +{L}_{\mathrm{matter}}
\right]\,.\label{action-F(G)}
\end{align}
In the background of a spatially flat FLRW Universe in (\ref{FLRWk0}), one can write down the
equation of motion for this gravity as follows
\begin{align}
24H^{3}\dot{F}'(G)+6H^{2}+F(G)-G F'(G)=&\, 2\kappa^{2}\rho\,,
\label{eom-FG-1} \\
8H^{2}\ddot{F}'(G)+16H\dot{F}'(G)\left(\dot{H}+H^{2}\right)
+\left(4\dot{H}+6H^{2}\right)+F(G)-G F'(G)=&\, -2\kappa^{2}p\,,
\label{eom-FG-2}
\end{align}
where the overhead dot stands for the derivative with respect to the cosmic time and the prime denotes the differentiation with respect to $G$. The above two equations (\ref{eom-FG-1}) and (\ref{eom-FG-2}) can also be alternatively represented as
\begin{align}
\rho_{\mathrm{eff}}=\frac{3}{\kappa^{2}}H^{2}\,,
\quad
p_{\mathrm{eff}}=-\frac{1}{\kappa^{2}} \left( 2\dot H+3H^{2} \right)\,,
\label{MFE}
\end{align}
where $\rho_{\mathrm{eff}}$ and $p_{\mathrm{eff}}$ are
the effective energy density and pressure of the Universe, respectively.
We note that the energy density and pressure of the
matter sector described by ${L}_{\mathrm{matter}}$ in
Eq.~(\ref{action-F(G)}) are contained in $\rho_{\mathrm{eff}}$ and
$p_{\mathrm{eff}}$. Now, using the expressions of $R$ and $G$ in the FLRW Universe, given by
$R = 6 \left(2H^{2}+\dot H \right)$ and $G = 24H^{2} \left( H^{2}+\dot H \right)$,
one can visualize $\rho_{\mathrm{eff}}$ and $p_{\mathrm{eff}}$ of
Eq.~(\ref{MFE}) as follows:
\begin{align}
\rho_{\mathrm{eff}} =&\, \frac{1}{2\kappa^{2}} \left[ -F(G)+24H^{2}
\left(H^{2}+\dot H\right)F'(G)-24^{2}H^{4}
\left(2\dot H^{2}+H\ddot H+4H^{2}\dot H\right)F''(G)
\right]+\rho\,,
\label{eq:rho-eff-1}
\\
p_{\mathrm{eff}} =&\, \frac{1}{2\kappa^{2}}\Bigg[F(G)-24H^{2}
\left(H^{2}+\dot H\right)F'(G)
+ (8 \times 24)\; H^{2}\Bigl\{ 6\dot{H}^{3}+8 H \dot{H} \ddot{H}
+ 24\dot{H}^{2} H^2 + 6H^3\ddot{H}
\nonumber\\
&\, + 8H^4\dot{H}+H^{2} \dddot{H} \Bigr\}F''(G)+ (8 \times 24^{2})\;H^{4}
\left(2\dot{H}^2+H\ddot{H}+4H^{2}\dot{H}\right)^{2}F'''(G)\Bigg]+p\,,
\label{eq:p-eff-1}
\end{align}
Assuming the matter sector with a constant EoS parameter $w \equiv
p/\rho$, and then combining the Friedmann and Raychaudhuri
equations in Eq.~(\ref{MFE}), one leads to the following equation:
\begin{align}
{G}\left(H,\dot{H}, \ddot{H},\dddot{H},...\right)=
-\frac{1}{\kappa^2}\left[ 2\dot{H}+3(1+w)H^2 \right],
\label{eqn-derivatives}
\end{align}
where
\begin{align}\label{eqn-def}
{G}\left(H,\dot{H}, \ddot{H},\dddot{H},...\right)=p_{\mathrm{eff}}-w\rho_{\mathrm{eff}}.
\end{align}
The Eq.~(\ref{eqn-derivatives}) brings in a very crucial physical
insight in terms of the new function ${G} \left(H,\dot{H},\ddot{H},\dddot{H},...\right)$ which involves $H$, $\dot{H}$ and the
higher order derivatives of $H$. When a cosmological model is
prescribed in terms of the Hubble rate $H=H(t)$, the
right-hand side of Eq.~(\ref{eqn-derivatives}) then reduces to a
function, $f(t)$ of the cosmic time $t$. Now, if the function
${G}\left(H,\dot{H}, \ddot{H},\dddot{H},...\right)$ in
Eq.~(\ref{eqn-def}) is chosen in such a way so that
${G}\left(H,\dot{H}, \ddot{H},\dddot{H},...\right)$ reproduces the
above function $f(t)$, then the aforementioned cosmology given by
$H = H (t)$ can be realized. Therefore, the function
${G}\left(H,\dot{H}, \ddot{H},\dddot{H},...\right)$ plays a very
crucial role to judge the viability of a given cosmological
scenario characterized by the Hubble rate \cite{Bamba:2008ut}.

The mathematical form of ${G} \left(H,\dot{H},\ddot{H},\dddot{H},... \right)$ can be determined from the prescribed
gravitational theory. In the context of $F(G)$-gravity, inserting
Eqs.~(\ref{eq:rho-eff-1}) and (\ref{eq:p-eff-1}) into
Eq.~(\ref{eqn-def}), one obtains,
\begin{align}
{G}\left(H,\dot{H}, \ddot{H},\dddot{H},...\right) =
\frac{1}{2\kappa^{2}} &\, \left[ (1+w)F(G)-24(1+w)H^2\left(H^2+\dot{H}\right)
F'(G)+ (8 \times 24)H^{2} \left\{ 6\dot{H}^{3} +8 H \dot{H}\ddot{H} \right. \right.
\nonumber \\
&\, \left. + 6(4+w)\dot{H}^{2} H^2 + 3(2+w)H^3 \ddot{H}+4(2+3w)H^4 \dot{H}+H^{2}
\dddot{H} \right\} F''(G)
\nonumber\\
&\, \left. + (8 \times 24^{2})H^{4}
\left(2\dot{H}^2+H\ddot{H}+4H^{2}\dot{H}\right)^{2}F'''(G) \right]\,.
\label{form-F(G)}
\end{align}

\subsubsection{Finite-time Singularities}

In this section, we investigate the possibility of finite-time
future singularities in $F(G)$ gravity models. To start with we
consider the following expression of the Hubble parameter in (\ref{H3a-F(R)-sp}). From
Eq.~(\ref{H3a-F(R)-sp}) one can clearly notice that if
$\beta>0$, the $H (t)$ becomes singular in the limit $t\rightarrow t_{s}$.
Therefore, $t_{s}$ determines the time when a singularity
appears in the above model. While on the other hand, for
$\beta<0$, $H (t)$ does not exhibit any singular behavior, but for
non-integer negative values of $\beta$, some derivative of $H (t)$
may be singular and consequently the curvature could exhibit
singular behavior~\cite{Bamba:2008ut}. Therefore, we see that for
both positive and negative values of $\beta$, singularities in $H
(t)$ or in the curvature may appear. Therefore, we assume
$\beta\neq 0$ because the case $\beta=0$ corresponds to the de
Sitter space, which has no singularity.

Thus, for the above cosmology characterized by the Hubble rate
in Eq.~(\ref{H3a-F(R)-sp}), one can try to find the equivalent
$F(G)$ models following the reconstruction technique described in
Refs.~\cite{Nojiri:2006su,Nojiri:2006gh,Bamba:2008ut,Bamba:2010wfw}.
Now, with the choice of suitable functions $P(t)$ and $Q(t)$ of a
scalar field $t$ which in this case is identified with the cosmic
time, the action integral in Eq.~({\ref{action-F(G)}}) can be
expressed as
\begin{align}
S=\int d^{4}x \sqrt{-g} \left[
\frac{1}{2\kappa^2}\bigg( R+P(t)G+Q(t) \bigg)
+{{L}}_{\mathrm{matter}}\right]\,.
\label{action-F(G)-rewritten}
\end{align}
The variation of the action in Eq.~(\ref{action-F(G)-rewritten})
with respect to $t$ gives,
\begin{align}
\frac{d P(t)}{dt}G+\frac{d Q(t)}{dt}=0\,,
\label{PQ}
\end{align}
from which one can in principle express $t=t(G)$. Now,
substituting $t=t(G)$ into Eq.~(\ref{action-F(G)-rewritten}), one
can express the action in terms of $F(G)$ where
\begin{align}
F(G)=P\left(t (G)\right)G+Q \left(t (G)\right)\,.
\label{form-F(G)}
\end{align}

Now, subtracting the equations of motion, namely,
Eqs.~(\ref{eom-FG-1}) and (\ref{eom-FG-2}), we obtain a second
order differential equation
\begin{align}\label{second-order-diff-eqn}
8 \frac{d}{dt}\left(H^2 \frac{dP}{dt}\right) - 8 H^3 \frac{dP}{dt} + 4 \dot{H} + 2 \kappa^2 \rho_{0} (1+w) a^{-3 (1+w)} = 0\, ,
\end{align}
where $\rho_{0}$ refers to the present energy density of $\rho$.
We note that instead of considering only one fluid in Eqs.
(\ref{eq:rho-eff-1}) and (\ref{eq:p-eff-1}), if one would consider
several matter components with $w_i = p_i/\rho_i$ being the EoS
parameter of the $i$-th fluid, then in this case, the last term of
the equation (\ref{second-order-diff-eqn}) will be replaced by $
2\kappa^2 \sum_{i} (1+w_i) \rho_{i0} a^{-3 (1+w_i)}$. If one can
solve for $P (t)$ from Eq.~(\ref{second-order-diff-eqn}), then from
Eq. (\ref{eom-FG-1}),
 one can find $Q (t)$ as follows
\begin{align}\label{eqn-form-Q(t)}
Q (t) = - 24 H^3 \frac{dP}{dt} - 6 H^2 + 2 \kappa^2 \rho_{0}
a^{-3 (1+w)}\,.
\end{align}

Thus, one can see that any cosmology which is prescribed in terms of any Hubble rate,
then this cosmology can be realized by some specific $F(G)$
gravity model. Just for simplicity, we neglect the matter term
from the above equations and focus on the specific $F(G)$ models
described by the Hubble rate given in
Eq.~(\ref{H3a-F(R)-sp}). Therefore, under the absence of
matter term, Eqs.~(\ref{second-order-diff-eqn}) and
(\ref{eqn-form-Q(t)}) become,
\begin{align}
\label{second-order-diff-eqn-3}
8 \frac{d}{dt}\left(H \frac{dP}{dt}\right) - 8 H^3 \frac{dP}{dt} + 4 \dot{H} = 0\, ,
\end{align}
and
\begin{align}
\label{eqn-form-Q(t)-3}
Q (t) = - 24 H^3 \frac{dP}{dt} - 6 H^2\, .
\end{align}
In the following we show a correspondence between the finite-time
singularities and the associated $F(G)$ model.

\begin{center}
 (i) {\it Big Rip singularity:}
\end{center}

Let us focus on the Big Rip singularity which is realized by some
of the $F(G)$ gravity models. To begin with we consider $\beta=1$
and $H_{s}=0$ in Eq.~(\ref{H3a-F(R)-sp}), for which the
Hubble rate $H (t)$ takes the form of (\ref{BigRip}) \cite{Bamba:2010wfw}
and consequently one can derive $G$ as follows,
\begin{align}
G = \frac{24h_s^3}{(t_{s}-t)^4}(1+h_s)\,.
\label{eqn-G-big-rip}
\end{align}
Thus, with the above choice of $H$, the
most general solution of the second order differential equation~(\ref{second-order-diff-eqn-3}) is \cite{Bamba:2010wfw}
\begin{align}
\label{soln-P}
P(t)=\frac{1}{4h_s(h_s-1)}(2t_{s}-t)t+c_{1}\frac{(t_{s}-t)^{3-h_s}}{3-h_s}+c_{2}\,,
\end{align}
where $c_{1}$ and $c_{2}$ are arbitrary constants. Having the
solution for $P (t)$ as in Eq.~(\ref{soln-P}), from
Eq.~(\ref{eqn-form-Q(t)-3}), one can derive $Q (t)$ as
\cite{Bamba:2010wfw}
\begin{align}
Q(t)=-\frac{6h_s^{2}}{(t_{s}-t)^2}-\frac{24 h_s^{3}}{(t_{s}-t)^3} \left[
\frac{(t_{s}-t)}{2h_s(h_s-1)}-c_{1}(t_{s}-t)^{2-h_s} \right]\,.
\end{align}
Further, from Eq.~(\ref{PQ}) we obtain
\begin{align}
t - t_s = \left[ \dfrac{24(h_s^{3}+h_s^{4})}{G} \right]^{1/4}\, ,
\end{align}
which, as one can see, is consistent with
Eq.~(\ref{eqn-G-big-rip}). Thus, having all the above information,
using Eq.~(\ref{form-F(G)}), one can write down the most general
form of $F(G)$ realizing the Big Rip singularity as follows
\cite{Bamba:2010wfw}:
\begin{align}
F(G)= \left(\frac{\sqrt{6h_s^{3}(1+h_s)}}{h_s(1-h_s)} \right)\sqrt{G}
+c_{1}G^{\frac{1+h_s}{4}}+c_{2}G\,.
\label{solution-F(G)-bigrip}
\end{align}

For $h_s=1$, that means when the Hubble rate becomes $H(t) = (t_s- t)^{-1}$,
one can also find
another exact solution for $P(t)$ as follows,
\begin{align}
P(t)=\alpha(t_{s}-t)^{q}\ln \left[ \gamma(t_{s}-t)^{z} \right]\,,
\end{align}
where $\gamma$ is a positive real number,
From Eq.~(\ref{eqn-form-Q(t)-3}), one can derive,
\begin{align}
Q(t)=-\frac{12}{(t_{s}-t)^{2}}\ln \left[ \gamma (t_{s}-t) \right]\,,
\label{schiappa}
\end{align}
Thus, using the expressions for $P (t)$ and $Q (t)$,
the form
of $F(G)$ is given by \cite{Bamba:2010wfw}
\begin{align}
F(G)= \frac{\sqrt{3}}{2}\sqrt{G} \ln(\gamma G)\,.
\end{align}
This form of $F(G)$ corresponds to the Hubble rate (\ref{H3a-F(R)-sp}) with $h_s =1$ and $H_s = 0$ and it realizes the Big Rip singularity. In general, for large values of $G$, $F (G) \sim \alpha \sqrt{G} \ln(\gamma G)$ with $\alpha > 0$, $\gamma > 0$, and the Big Rip singularity may appear. Moreover, the Big Rip singularity may also appear for $F (G) \sim \alpha \sqrt{G} \ln(\gamma G^z + G_s)$ with $\alpha >0$, $\gamma >0$, $z >0$ where $G_s$ is a constant such that $\gamma G^z + G_s >0$.

\begin{center}
 (ii) {\it Other types of singularities:}
\end{center}

Let us now discuss the appearance of other types of singularities
in $F (G)$-gravity models. We consider the Hubble rate of
Eq.~(\ref{H3a-F(R)-sp}) with $H_{s}=0$ but $\beta \neq 1$. In
this case, the evolution of the scale factor $a(t)$ turns out to
be
\begin{align}
a(t) = a_0\exp \left[ \frac{h_s}{\beta-1}\left( (t_{s}-t)^{1-\beta}-(t_{s}-t_0)^{1-\beta}\right) \right]\,,
\end{align}
and we observe the emergence of following singularities:

\begin{itemize}

\item When $\beta>1$, then $H$ and $G$ are given by
\begin{align}
H = \frac{h_s}{(t_{s}-t)^{\beta}}\,,
\quad G \sim \frac{24h_s^4}{(t_{s}-t)^{4\beta}}\, .
\label{HII}
\end{align}

A solution of the Eq.~(\ref{second-order-diff-eqn-3}) in the limit
$t\rightarrow t_{s}$ can be found as
\begin{align}
P(t) \simeq \frac{\alpha}{(t_{s}-t)^{z}}\, ,
\end{align}
where $z=-2\beta$ and $\alpha=-1/4h_s^{2}$. The expression of $F(G)$ now follows
\begin{align}
F(G)=-12\sqrt{\frac{G}{24}}\,.
\label{primo}
\end{align}
Hence, if for large values of $G$, $F(G)\sim -\alpha\sqrt{G}$
with $\alpha>0$, then a Type I singularity could appear.

\item When $0<\beta<1$ for which
the forms of $H$ and $G$ are given by
\begin{align}
H = \frac{h_s}{(t_{s}-t)^{\beta}}\,,
\quad G \sim \frac{24h_s^3\beta}{(t_{s}-t)^{3\beta+1}}\, .
\label{Hsuper}
\end{align}

In this case, an asymptotic solution of
Eq.~(\ref{second-order-diff-eqn-3}) in the limit $t\rightarrow
t_{s}$ can be found as
\begin{align}
P(t) \simeq \frac{\alpha}{(t_{s}-t)^{z}}\;,
\end{align}
where $z=-(1+\beta)$ and $\alpha=1/2h_s(1+\beta)$. The form of $F(G)$
now becomes
\begin{align}
F(G)=\frac{6h_s^{2}}{(\beta+1)}(3\beta+1)\left(\frac{|G|}{24h_s^{3}|\beta|}
\right)^{2\beta/(3\beta+1)}\,.
\label{secondo}
\end{align}
Hence, if for large values of $G$, $F(G)$ has the following form
\begin{align}
F(G)\sim \alpha |G|^{\gamma}\,,
\quad
\gamma = \frac{2\beta}{3\beta+1}\,,
\label{solution-F(G)-other-types}
\end{align}
where $\alpha>0$ and $0<\gamma <1/2$, then
because we are assuming $0<\beta<1$ and in that case, a Type III singularity could emerge.

Finally, for a general value of $\beta$ we have the following:

\begin{itemize}

\item If for $G\rightarrow-\infty$, $F(G)$ has the form as in
Eq.~(\ref{solution-F(G)-other-types}) with $\alpha>0$ and
$-\infty<\gamma<0$, then we find $-1/3<\beta<0$ and a Type II
(sudden) singularity could appear.

\item If for $G\rightarrow 0^{-}$, $F(G)$ takes the form of
Eq.~(\ref{solution-F(G)-other-types}) with $\alpha<0$ and
$1<\gamma<\infty$, then we obtain $-1<\beta<-1/3$ and a Type II
singularity could occur.

\item If for $G\rightarrow 0^{-}$, $F(G)$ assumes the form as in
Eq.~(\ref{solution-F(G)-other-types}) with $\alpha>0$ and
$2/3<\gamma<1$, then we obtain $-\infty<\beta<-1$ and a Type IV
singularity could appear. We also require that
$\gamma\neq2n/(3n-1)$, i.e., $\beta\not=n$, where $n$ is a natural
number.

\end{itemize}

Let us note that we can generate all the possible Type II singularities
as shown above except for $\beta=-1/3$,
i.e., for $H=h_s (t_{s}-t)^{1/3}$ because $\gamma=0$. In this case, $G$ takes the form:
\begin{align}
G=24h_s^{3}\beta+24h_s^{4}(t_{s}-t)^{4/3}<0\,.
\end{align}
\end{itemize}

Thus, in order to express $t$ in terms of $G$, it is essential to consider the whole expression of $G$, and by considering the leading term involving $(t_{s}-t)$ in (\ref{second-order-diff-eqn}) and (\ref{eqn-form-Q(t)}) with $\rho_0=0$, we obtain
\begin{align}
F(G)\simeq \frac{1}{4 \sqrt{6}h_s^{3}}G(G+8h_s^{3})^{1/2}
+\frac{2}{\sqrt{6}}(G+8h_s^{3})^{1/2}\,,
\end{align}
which satisfies Eq.~(\ref{eqn-derivatives}) in the limit
$t\rightarrow t_{s}$. Consequently, the specific model $F(G)=
\sigma_1 G(G+c_{3})^{1/2}+ \sigma_2 (G+c_{3})^{1/2}$, where
$\sigma_1$, $\sigma_2$ and $c_{3}$ are positive constants, can
generate a Type II singularity.

\subsection{$F(R,G)$ Gravity}
\label{sec-f(R,G)-gravity}

The $F(R, G)$-gravity is a very generalized gravitational theory
where $F$ is a generalized function of the Ricci scalar $R$
and the Gauss-Bonnet invariant is defined in Eq. (\ref{GB}). One can
clearly see that the $F (R, G)$ gravity theory can recover
$F(R)$-gravity and $F(G)$-gravity as special cases. Moreover, with
the suitable choice for $F$ leading to $F (R, G) = R$, one
recovers Einstein's GR as well. Thus, one can see that this
modified gravity theory being the generalized version of both $F(R)$ and $F (G)$ gravity theories can offer some appealing consequences and very
soon of its introduction, $F (R, G)$ gravity theory received
significant attention in the community
\cite{Bamba:2010wfw,Elizalde:2010jx,DeFelice:2010hg,DeFelice:2011ka,delaCruz-Dombriz:2011oii,Makarenko:2012gm,Atazadeh:2013cz,SantosDaCosta:2018bbw,Wu:2018jve,Odintsov:2018nch,Camci:2018apx,Barros:2019pvc,KumarSanyal:2019rpa,Shekh:2019rib,Inagaki:2019rhm,Elizalde:2020zcb,Navo:2020eqt,Odintsov:2021nim,Lohakare:2021yuo,Nojiri:2021mxf}.
In the present section we shall investigate the finite-time future
singularities appearing in this generalized modified gravitational
theory.
This model includes ghosts \cite{DeFelice:2009ak} and the ghost-free modifications
have been proposed in Ref. \cite{Nojiri:2021mxf}.
In this section, however, again for the illustration, we show the structure of the singularities based on the original models.

The action of the ${F}(R,G)$-gravity is given by
\cite{Bamba:2010wfw,delaCruz-Dombriz:2011oii,Atazadeh:2013cz}:
\begin{align}
S = \int d^4 x \sqrt{-g} \left[ \frac{{F}(R,G)}{2\kappa^2}
+{{L}}_{\mathrm{matter}} \right]\,,
\label{action-f(R,G)}
\end{align}

For the action (\ref{action-f(R,G)}), one can derive the gravitational equations as follows
\begin{align}
{{F}}'_{R}\left( R_{\mu\nu}-\frac{1}{2}R g_{\mu\nu}\right)
=&\, \kappa^2 T^{(\mathrm{matter})}_{\mu \nu}
+\frac{1}{2}g_{\mu\nu} \left({F}-{{F}}'_{R}R\right)
+{\nabla}_{\mu}{\nabla}_{\nu}
{{F}}'_{R} -g_{\mu\nu} \Box {{F}}'_{R}
\nonumber \\
&\, + \left(-2RR_{\mu\nu} +4R_{\mu\rho}R_{\nu}{}^{\rho}
-2R_{\mu}{}^{\rho\sigma\tau}R_{\nu\rho\sigma\tau}
+4g^{\alpha\rho}g^{\beta\sigma}R_{\mu\alpha\nu\beta}R_{\rho\sigma}
\right){{F}}'_{G}
\nonumber \\
&\, +2\left({\nabla}_{\mu}{\nabla}_{\nu} {{F}}'_{G} \right)R
-2g_{\mu \nu}\left(\Box {{F}}'_{G} \right)R
+4\left(\Box {{F}}'_{G} \right)R_{\mu \nu}
-4\left({\nabla}_{\rho}{\nabla}_{\mu} {{F}}'_{G} \right)
R_{\nu}{}^{\rho}
\nonumber \\
&\, -4\left({\nabla}_{\rho}{\nabla}_{\nu} {{F}}'_{G} \right)
R_{\mu}{}^{\rho}
+4g_{\mu \nu}\left({\nabla}_{\rho}{\nabla}_{\sigma}
{{F}}'_{G} \right)R^{\rho\sigma}
-4\left({\nabla}_{\rho}{\nabla}_{\sigma} {{F}}'_{G} \right)
g^{\alpha\rho}g^{\beta\sigma}R_{\mu\alpha\nu\beta}\,,
\label{gravitational-eqns-f(R,G)}
\end{align}
where ${\nabla}_{\mu}$ is the covariant derivative
operator associated with the metric tensor $g_{\mu \nu}$;
$\Box \equiv g^{\mu \nu} {\nabla}_{\mu} {\nabla}_{\nu}$
denotes the covariant d'Alembertian for a scalar field;
$T^{(\mathrm{matter})}_{\mu \nu} = \mathrm{diag} \left(\rho, p, p, p \right)$ is
the energy-momentum tensor of the matter sector that includes all the ordinary matter fluids
with $\rho~(= \sum _{i} \rho_i)$ and $p~(= \sum_{i} p_i)$ are the total energy density
and pressure of all the fluids, respectively
 ($\rho_i$, $p_i$ being the representatives of the $i$-th fluid), and
\begin{align}
{{F}}'_{R} = \frac{\partial {F}(R,G)}{\partial R}\,, \quad
{{F}}'_{G} = \frac{\partial {F}(R,G)}{\partial G}\,.
\end{align}

In the background of a spatially flat FLRW Universe (\ref{FLRWk0}), the
gravitational field equations in
(\ref{gravitational-eqns-f(R,G)}) can be expressed as
\begin{align}
\rho_{\mathrm{eff}}=\frac{3}{\kappa^{2}}H^{2}\,, \quad
p_{\mathrm{eff}}=-\frac{1}{\kappa^{2}} \left( 2\dot H+3H^{2} \right)\,,
\end{align}
where $\rho_{\mathrm{eff}}$ and $p_{\mathrm{eff}}$, termed as the
effective energy density and pressure of the Universe,
respectively, are given by
\begin{align}
\rho_{\mathrm{eff}} =&\, \frac{1}{{{F}}'_{R}} \left[ \rho + \frac{1}{2\kappa^{2}}
\left\{ \left( {{F}}'_{R}R-{F} \right) -6H{\dot{{F}}}'_{R}
+G{{F}}'_{G}-24H^3{\dot{{F}}}'_{G} \right\} \right]\,,
\label{eq:eff-rho-f(R,G)} \\
p_{\mathrm{eff}} =&\,
\frac{1}{{{F}}'_{R}} \left[ p + \frac{1}{2\kappa^{2}} \left\{
-\left( {{F}}'_{R}R-{F} \right)
+4H{\dot{{F}}}'_{R}+2{\ddot{{F}}}'_{R}-G{{F}}'_{G}
+16H\left(\dot{H} +H^2 \right){\dot{{F}}}'_{G}
+8H^2 {\ddot{{F}}}'_{G}
\right\}
\right]\,.
\label{eq:eff-p-f(R,G)}
\end{align}

Let us now proceed towards the investigations of the finite-time
singularities in the ${F}(R,G)$-gravity. Similar to the earlier
section \ref{sec-F(G)-gravity} on $F(G)$, we reconstruct the
${F}(R,G)$-gravity models that can produce finite-time
singularities. In order to do so, we consider the pure
gravitation action of ${F}(R,G)$-gravity, that means action
(\ref{action-f(R,G)}) without matter sector exactly what we have
considered in the earlier section \ref{sec-F(G)-gravity}. In this
case, from Eqs.~(\ref{eq:eff-rho-f(R,G)}) and
(\ref{eq:eff-p-f(R,G)}), one can write down the gravitational
equations as
\begin{align}
0=&\, 24H^{3}{\dot{{F}}}'_{G}
+6H^{2}{{F}}'_{R}
+6H{\dot{{F}}}'_{R}+({F}-R{{F}}'_{R}
-G{{F}}'_{G})\,,
\label{gfe1-f(R,G)-no-matter} \\
0=&\, 8H^{2}{\ddot{{F}}}'_{G}+2{\ddot{{F}}}'_{R}
+4H{\dot{{F}}}'_{R}+16H{\dot{{F}}}'_{G}(\dot H+H^{2})
+{{F}}'_{R}(4\dot H+6H^{2})
+{F}-R {{F}}'_{R}
-G {{F}}'_{G}\,.
\label{gfe2-f(R,G)-no-matter}
\end{align}
We note that in the case of pure gravity (i.e., when no matter sector is present in the gravitational action),
the above equations (\ref{gfe1-f(R,G)-no-matter}) and (\ref{gfe2-f(R,G)-no-matter}) are linearly dependent.

Now, following the similar fashion as adopted in the earlier
section (\ref{sec-F(G)-gravity}), with the choice of the proper
functions $P(t)$, $Z(t)$ and $Q(t)$ of a scalar field (where we
identify the scalar field with time $t$), one can rewrite
Eq.~(\ref{action-f(R,G)}) without ${{L}}_{\mathrm{matter}}$ as
\begin{align}
S=\frac{1}{2\kappa^2}\int d^{4}x \sqrt{-g} \Bigg( P(t)R+Z(t)G+Q(t) \Bigg)\,.
\label{action-f(R,G)-rewritten}
\end{align}
By varying the action in Eq.~(\ref{action-F(G)-rewritten}) with
respect to $t$, we obtain
\begin{align}
P'(t)R+Z'(t)G+Q'(t)=0\,,
\label{eqn-t-diff-eqn-f(R,G)}
\end{align}
where the prime denotes differentiation with respect to $t$. From
Eq.~(\ref{eqn-t-diff-eqn-f(R,G)}), one can in principle express $t$
as a function of $R, G$, i.e., $t=t(R,G)$. Now, with the
substitution of $t=t(R,G)$ in Eq.~(\ref{action-F(G)-rewritten}),
we can express ${F}(R,G)$ as
\begin{align}
{F}(R,G)=P\left(t (R, G) \right)R+Z \left(t (R, G) \right) G+Q \left(t (R, G) \right)\, .
\label{expression-F(R,G)}
\end{align}

\subsubsection{Finite-time Singularities}

We consider, once again, the following expression of the Hubble parameter in (\ref{H3a-F(R)-sp}) \cite{Bamba:2010wfw}.
 From Eq.~(\ref{H3a-F(R)-sp}), one can express the scale factor as
\begin{align}
\label{scale-factor-f(R,G)}
a(t) = \bar{a} \exp\left(g(t)\right)\, ,
\end{align}
where $\bar{a}$ is a constant and $g(t)$ is some differentiable
function of $t$ satisfying $\dot{g} (t) = H(t)$. Using the
conservation law and Eq.~(\ref{gfe1-f(R,G)-no-matter}), we arrive
at the following second order differential equation
\begin{align}
P''(t)+4 \dot{g}^{2}(t)Z''(t)-\dot g(t)P'(t)+(8 \dot{g} \ddot{g}
-4 \dot{g}^{3}(t))Z'(t)+2 \ddot{g}(t)P(t)=0\,,
\label{eqn-ode-f(R,G)}
\end{align}
where we have implemented the expression of the scale factor as
given in Eq.~(\ref{scale-factor-f(R,G)}). Now, from
Eq.~(\ref{gfe1-f(R,G)-no-matter}), one can derive $Q(t)$ as
\begin{align}
Q(t)=-24\dot{g}^{3}(t)Z'(t)-6\dot{g}^{2}(t)P(t)-6\dot{g}(t)P'(t)\,.
\label{eqn-for-Q-F(R,G)}
\end{align}

If $P(t)\neq 0$, then ${F}(R,G)$ can be written as
\begin{align}
 {F}(R,G)=R\; \psi (R,G)+f(R,G)\,,
\label{zuzzurellone}
\end{align}
where $\psi (R,G)\neq 0$ and $f(R,G)$ is any generic function of $R$ and $G$.
Now, from Eqs.~(\ref{eq:eff-rho-f(R,G)}) and
(\ref{eq:eff-p-f(R,G)}), we obtain,
\begin{align}
\rho_{\mathrm{eff}} = -\frac{1}{2\kappa^{2}g(R,G)}\biggl[
24H^{3}{\dot{{F}}}'_{G}
+6H^{2}\left( R\frac{d \psi(R,G)}{d R}+\frac{d f(R,G)}{d R} \right)
+6H{\dot{{F}}}'_{R}
+({F}-R{{F}}'_{R}
-G{{F}}'_{G})\biggr],
\label{tic}
\end{align}
and
\begin{align}
p_{\mathrm{eff}} =&\, \frac{1}{2\kappa^{2}g(R,G)}\biggl[
8H^{2}{\ddot{{F}}}'_{G}+2{\ddot{{F}}}'_{R}
+4H{\dot{{F}}}'_{R}+16H{\dot{{F}}}'_{G}(\dot H+H^{2})\nonumber\\
&\, +\left(R\frac{d \psi (R,G)}{d R}+\frac{d f(R,G)}{d R}\right)(4\dot H+6H^{2}) + {F}-R {{F}}'_{R}-G {{F}}'_{G}\biggr]\,,
\label{tac}
\end{align}
and similar to the earlier Section~\ref{sec-F(G)-gravity}, the function ${G} \left(H, \dot{H}, \ddot{H},... \right)$ can be written as
\begin{align}
{G}\left(H,\dot{H}, \ddot{H},\dddot{H},...\right)=
-\frac{1}{\kappa^2}\left[ 2\dot{H}+3(1+w)H^2 \right],
\label{eqn-derivatives-f(R,G)}
\end{align}
where more explicitly,
\begin{align}
{G}(H,\dot{H},\ddot{H},...) =&\, p_{\mathrm{eff}}-w\rho_{\mathrm{eff}}
\nonumber \\
=&\, \frac{1}{2\kappa^{2} \psi(R,G)} \Bigg[ (1+w)({F}
-R{{F}}'_{R}-G{{F}}'_{G})
+\left(R\frac{d\psi(R,G)}{dR}+\frac{df(R,G)}{dR}\right)
\left(6H^2(1+w)+4\dot{H}\right)
\nonumber \\
&\, + H{\dot{{F}}}'_{R}(4+6w)
+8H{\dot{{F}}}'_{G}\left(2\dot{H}+ H^{2}(2+3w)\right)
+2{\ddot{{F}}}'_{R}+8 H^{2}{\ddot{{F}}}'_{G} \Bigg]\,,
\label{G(H,dotH)-f(R,G)}
\end{align}
where $w$ is the EoS parameter of the matter. It is important to note that the use of the above equation (\ref{G(H,dotH)-f(R,G)}) demands $\psi (R,G) \neq 0$ on the solution.
Let us now explicitly describe the finite-time singularities
allowed in various ${F}(R,G)$-gravity models.

\begin{center}
 {\it (i) Big Rip singularity}
\end{center}

Let us first discuss the appearance of the Big Rip singularity in
this modified gravity model. We consider the Hubble rate of
Eq.~(\ref{H3a-F(R)-sp}) with $\beta = 1$ and $H_s = 0$ which
gives the Hubble rate in (\ref{BigRip})
and consequently, one can derive $R$ and $G$ for
Eq.~(\ref{BigRip}) as
\begin{align}
\label{BigRipRG}
R = \frac{6 h_s}{(t_{s}-t)^{2}}(2h_s+1)\,, \quad
G = \frac{24 h_s^{3}}{(t_{s}-t)^{4}}(1+h_s)\,.
\end{align}

A very trivial solution of Eq. (\ref{eqn-ode-f(R,G)}) can be given by
\begin{align}
P(t) = \alpha(t_{s}-t)^{z}\,, \quad
Z(t) = \delta(t_{s}-t)^{x}\,,
\end{align}
where $\alpha$ and $\delta$ be any real numbers; $x=3-h_s$, and $z$ can be found from
\begin{align}
z_{\pm}
=\frac{1-h_s\pm \sqrt{h_s^{2}-10 h_s+1}}{2}\,.
\end{align}

Therefore, the most general solution of $P(t)$ can be given as
\begin{align}
P(t)=\alpha_{1}(t_{s}-t)^{z_{+}}+\alpha_{2}(t_{s}-t)^{z_{-}}\,,
\end{align}
where $\alpha_{1}$ and $\alpha_{2}$ are any real numbers. Now,
from Eq.~(\ref{eqn-for-Q-F(R,G)}), one can derive
\begin{align}
Q(t)=\dfrac{24h_s^{3}\delta(3-h_s)}{(t_{s}-t)^{h_s+1}}+\dfrac{6h_s\alpha_{1}(z_{+}-h_s
)}{(t_{s}-t)^{2-z_{+}}}+\dfrac{6h_s\alpha_{2}(z_{-}-h_s)}{(t_{s}-t)^{2-z_{-}}}\, .
\end{align}

Now for the condition $0<h_s<5-2\sqrt{6}$ or $h_s>2+\sqrt{6}$, the solution of
${F}(R,G)$ is given by
\begin{align}
{F}(R,G)=\bar{\alpha}_{1}R^{1-\frac{z_{+}}{2}}+\bar{\alpha}_{2}R^{1-\frac{z_{-}}{2}}
+\bar{\delta} G^{\frac{h_s+1}{4}} \,,
\label{trivial}
\end{align}
where some factors have been absorbed into the constants. Note also that $z_{\pm} \neq 0$, otherwise $h_s$ vanishes which is a contradiction.

Another solution of Eq.~(\ref{eqn-ode-f(R,G)}) can be given by
\begin{align}
P(t) = \frac{\alpha}{(t_{s}-t)^{z}}\,, \quad
Z(t) = \frac{\delta}{(t_{s}-t)^{x}}\,,
\end{align}
where $\delta$, $x$ are real numbers; $z =x+2$ and $\alpha$ is given by
\begin{align}
\alpha=\dfrac{4 h_s^{2}\delta x (h_s-x-3)}{x^{2}+(5-h_s)x+6}\,.
\end{align}
From Eq.~(\ref{eqn-for-Q-F(R,G)}), one finds
\begin{align}
Q(t)=-\frac{6h_s}{(t_{s}-t)^{x+4}}\left[4h_s^{2}\delta x+\alpha(x+2+h_s)
\right]\,.
\end{align}

The solution of the Eq.~(\ref{eqn-t-diff-eqn-f(R,G)}) is given by
\begin{align}
t = t (R, G) =t_s- \left[\;\;\dfrac{-\alpha(x+2)R\pm
\sqrt{\alpha^{2}(x+2)^{2}R^{2}+24h_s\bigl(4h_s^{2}\delta x+\alpha(x+2+h_s)\bigr)
(x+4)\delta x G}}{2\delta x G}\;\;\right]^{1/2}\,,
\label{conditions-for-real-soln}
\end{align}
where $x\neq 0$ and $\delta\neq 0$. In order to get real
solutions, one has to ensure that first of all, we should have
$\Delta = \alpha^{2}(x+2)^{2}R^{2}+24h_s\bigl(4h_s^{2}\delta
x+\alpha(x+2+h_s)\bigr)(x+4) \geq 0$ in
Eq.~(\ref{conditions-for-real-soln}) and secondly, the entire term
inside the third brace of Eq.~(\ref{conditions-for-real-soln}) has
to be positive. For $h_s>0$, the principal cases with the real
solutions of (\ref{conditions-for-real-soln}) are the following:

\begin{itemize}
\item Case I: For $x>0$, $\delta>0$, $1+x \leq h_s <
x+5+\frac{6}{x}$, we need to use the sign $+$ in the r.h.s. of Eq.~(\ref{conditions-for-real-soln}).

\item Case II: For $-\frac{3}{2} \leq x<0$, $\delta<0$, $h_s \geq
x+1$, we need to use the sign $+$ in the r.h.s. of
Eq.~(\ref{conditions-for-real-soln}).

\item Case III: For $-4<x<-\frac{3}{2}$, $\delta<0$,
$h_s>x+5+\frac{6}{x}$, we must use the sign $+$ in the r.h.s. of Eq.~(\ref{conditions-for-real-soln}).

\item Case IV: For $x>0$, $\delta<0$, $x+5+\frac{6}{x}>h_s \geq
1+x$, we need to use the sign $-$ in the r.h.s. of
Eq.~(\ref{conditions-for-real-soln}).

\item Case V: For $-\frac{3}{2} \leq x<0$, $\delta>0$, $h_s \geq
x+1$, we must use the sign $-$ in the r.h.s. of
Eq.~(\ref{conditions-for-real-soln}).

\item Case VI: For $-4<x<-\frac{3}{2}$, $\delta>0$,
$h_s>x+5+\frac{6}{x}$, we need to use the sign $-$ in the right
hand side of Eq.~(\ref{conditions-for-real-soln}).

\item Case VII: For $x=-4$, $\delta>0$, we must use the sign $-$
in the r.h.s. of Eq.~(\ref{conditions-for-real-soln}).

\item Case VIII: For $x=-4$, $\delta<0$, we must use the sign $+$
in the r.h.s. of Eq.~(\ref{conditions-for-real-soln}).

\end{itemize}
The solution of ${F}(R,G)$ is then given by

\begin{align}
{F}(R,G)= \left(\frac{\alpha}{(t_s-t(R,G))^{x+2}} \right)R+ \left(\frac{\delta}{(t_s-t(R,G))^{x}} \right)G
- \left(\frac{6h_s}{(t_s-t(R,G))^{x+4}} \right)\Bigg[4h^{2}\delta x+\alpha(x+2+h_s)\Bigg]\,,
\label{solution-f(R,G)}
\end{align}
where $t(R,G)$ is given by Eq.~(\ref{conditions-for-real-soln}).
The expression for ${F}(R,G)$ in Eq.~(\ref{solution-f(R,G)})
represents an exact solution of the equation of motion in
Eqs.~(\ref{eq:eff-rho-f(R,G)}) and (\ref{eq:eff-p-f(R,G)})
allowing the Big Rip singularity. Let us show some specific
examples of ${F}(R,G)$-gravity models allowing the Big Rip
singularity obtained for some specific values of $\alpha$ and $x$.

\begin{itemize}
\item For $\alpha=1$ and $x=-2$, we find
\begin{align}
{F}(R,G)=R+ \left(\dfrac{\sqrt{6}\sqrt{h_s(1+h_s)}}{(1-h_s)} \right)\sqrt{G}\,,
\quad
h\neq 1\,.
\end{align}

\item If $\alpha=0$ and $x=h_s-3$ (this case corresponds to Case (I) $-$ Case (VI), presented above), then one finds
\begin{align}
{F}(R,G)=\delta\; G^{\frac{h_s+1}{4}}\,,
\quad
\delta \neq 0\,,
\end{align}
which is actually equivalent to Eq.~(\ref{trivial}) for
$\alpha_{1}=\alpha_{2}=0$.

\item If $x=-4$, then the model reduces to
\begin{align}
{F}(R,G)=
\dfrac{16h_s^{4}\delta}{(1+2h_s^{2})^{2}}\left[(9+21h_s+6h_s^{2})-(1+h_s)^{2}
\frac{R^{2}}{G}\right]\,,
\quad
\delta \neq 0\,.
\label{Dante}
\end{align}
Thus, we see that for large values of $R$ and $G$,
${F}(R,G)\sim\pm\alpha\mp\delta(R^{2}/G)$ with
$\alpha>0$ and $\delta>0$, then the Big Rip singularity could appear.

\item For $x=h_s-1$, the model
becomes (by absorbing some constants)
\begin{align}
{F}(R,G)=
\delta \left( \dfrac{R}{G} \right)^{\frac{1-h_s}{2}} G\,,
\quad
\delta \neq 0\,,
\quad
h_s \neq 1\,.
\label{Sturmtruppen}
\end{align}
Thus, we see that for large values of $R$ and $G$,
${F}(R,G)\sim \delta G^{\gamma}/R^{\gamma-1}$
with $\delta \neq 0$ and $\frac{1}{2} < \gamma < 1$ or $1<\gamma<+\infty$,
and the Big Rip singularity could appear.

\end{itemize}

We close the discussion with a general model
that allows Big Rip singularity.
One can verify that the model
\begin{align}
{F}(R,G)=\gamma\frac{G^{m}}{R^{n}}\,,
\label{Kriegsmarine}
\end{align}
with $\gamma$ being any real number, is a solution of
Eqs.~(\ref{eq:eff-rho-f(R,G)}) and (\ref{eq:eff-p-f(R,G)}) in the
case of the Big Rip singularity for some value of $h_s$. In
general, it is possible to obtain solutions for $h_s>0$ if $m>0$,
$n>0$ and $m>n$. For example, the case $n=2$ and $m=3$ realizes
the Big Rip singularity in $h_s=5$; the case $n=1$ and $m=3$
realizes the big rip singularity in $h_s=4+\sqrt{19}$ and so on.
This is a generalization of Eq.~(\ref{Sturmtruppen}). It is
important here to mention that for $m=-1$ and $n=-2$, we do not
recover a physical scenario because in this case we obtain
$h_s=-3$. However, this case (i.e., $m=-1$ and $n=-2$), has a
similarity with Eq.~(\ref{Dante}) allowing the Big Rip
singularity. Finally, for the case with $m=0$ or $n=0$, we recover
the model in Eq.~(\ref{trivial}).

\begin{center}
 {\it (ii) Other types of singularities}
\end{center}

Here we discuss other types of singularities that may appear in various ${F} (R, G)$-gravity models.
To begin with we consider the following expansion of the Hubble rate in (\ref{Hsin}).
For the above choice of the Hubble rate, an exact solution of
the differential Eq.~(\ref{eqn-ode-f(R,G)}) can be found as
\begin{align}
P(t) = -\lambda(4h_s^{2})(t_{s}-t)\,, \quad
Z(t) = \lambda(t_{s}-t)^{2\beta+1}\,,
\end{align}
where $\lambda$ is a constant. The form of $Q(t)$ from
Eq.~(\ref{eqn-for-Q-F(R,G)}) can now be derived as
\begin{align}
Q(t)=\dfrac{24h_s^{4}\lambda}{(t_{s}-t)^{2\beta-1}}+\dfrac{48h_s^{3}\beta}{(t_{s
}-t)^{\beta}}\,.
\end{align}
Now we examine the cases when $\beta >1$ or $\beta < 1$.
\begin{itemize}
\item For $\beta>1$, one may obtain the asymptotic real solution
of Eq.~(\ref{eqn-t-diff-eqn-f(R,G)}):
\begin{align}
t =t(R,G)=t_s-2^{1/2\beta}\left[\dfrac{h_s^{2}R+\sqrt{h_s^{4}R^{2}+6h_s^{4}(4\beta^{2}-1)G}}{(1+2\beta)G}\right]^{1/2\beta}\,.
\end{align}
Thus, the mathematical form of ${F}(R,G)$ can now be expressed as
\begin{align}
{F}(R,G)=-4h_s^{2}\lambda (t_s-t(R,G))R+\lambda
(t_s-t(R,G))^{1+2\beta}G+24h_s^{4}\lambda (t_s-t(R,G))^{1-2\beta}\,,
\quad \beta>1\,.
\end{align}
This is an asymptotic solution of
Eq.~(\ref{eqn-derivatives-f(R,G)}) for $\beta>1$, when
\begin{align}
-\frac{1}{\kappa^{2}}\left[2\dot H +3(1+w)H^{2}\right]\sim
-\dfrac{3(1+w)h_s^{2}}{\kappa^{2}}(t_{s}-t)^{-2\beta}\,.
\label{auto}
\end{align}
Interestingly, for $\beta \gg 1$, the mathematical form of ${F}(R,G)$ is given by
\begin{align}
{F}(R,G) \simeq \lambda\left(\dfrac{\alpha G}{R+\sqrt{R^{2}+\gamma G}}-R \right)\,,
\quad \alpha>0\,,
\quad \gamma>0\,,
\quad \lambda \neq 0\,,
\end{align}
and this is the asymptotic behavior of a ${F}(R,G)$ model where
a ``strong'' Type I singularity ($\beta \gg 1$) may appear.

We consider some other explicit cases where Type I singularity appears as follows.
Taking
\begin{align}
{F}(R,G)=\gamma\frac{G^{m}}{R^{n}}\,,
\label{miseriaccia}
\end{align}
and using Eqs.~(\ref{eq:eff-rho-f(R,G)}) and
(\ref{eq:eff-p-f(R,G)}), one can verify that
the function ${G}\left(H, \dot{H}, \ddot{H},... \right)$ in
Eq.~(\ref{G(H,dotH)-f(R,G)}) becomes,
\begin{align}
{G}(H,\dot{H},..)\simeq
-\frac{3 h_s^{2}(2m-n-1)(1+w)}{\kappa^{2}(t_{s}-t)^{2\beta}}\,,
\end{align}
which under the restriction $2m-n-1>0$, represents an asymptotic
solution of Eq.~(\ref{eqn-derivatives-f(R,G)}) for $\beta>1$.
Thus, we see that for the model ${F}(R,G)\simeq \gamma
G^{m}/R^{n}$ with $m>(n+1)/2$, the Type I singularity may appear.
This observation has some important consequences in the context of
other modified gravity models because one can see that the
modified gravity models where either $F(R)=R^{n}$ with $n>1$ or
$F(G)=G^{m}$ with $m>1/2$ can allow singularities.

One can also find some other models allowing the Type I singularities and we can construct these models following section \ref{sec-F(G)-gravity} as follows.
The Type I singularities actually correspond to the asymptotic limits for $R$ and $G$
\begin{align}
R\sim 12 H^{2}\,, \quad G\sim 24 H^{4}\,.
\end{align}
Since $R$ and $H$ in the asymptotic limits are nothing but the functions of the Hubble parameter only, therefore, one can write that,
\begin{align}
\lim_{t\rightarrow t_{s}} 24\left( \frac{R}{12} \right)^{2}
= \lim_{t\rightarrow t_{s}} G\,.
\label{Limite}
\end{align}
Now, if we substitute $G$ for $R$ in Eq.~(\ref{primo}) considering
Eq.~(\ref{Limite}), then we obtain a zero function\footnote{The
reason for obtaining the zero function is that, in this case
Eq.~(\ref{primo}) becomes zero on the singularity solution. }.
Thus, if we instead substitute $G$ for $G/R$, we therefore obtain
the following model
\begin{align}
{F}(R,G)= R-\dfrac{6G}{R}\,,
\label{zap}
\end{align}
which is an asymptotic solution of
Eq.~(\ref{eqn-derivatives-f(R,G)}) such as Eq.~(\ref{auto}). Thus,
we see that Type I singularity appears in the model ${F}(R,G)\sim
R-\alpha(G/R)$ with $\alpha>0$.
\item For the Hubble rate in Eq.~(\ref{Hsin}) with $\beta<1$, it is not
possible to express $G$ and $R$ as the functions of the same
variable (i.e., $H$ or the same combination of $H$ and $\dot H$).
However, if we examine the asymptotic behavior of $G$ and $R$, we have
\begin{align}
R \simeq \frac{6h_s\beta}{(t_{s}-t)^{\beta+1}}\,, \quad
G \simeq \frac{24 h_s^{3}\beta}{(t_{s}-t)^{3\beta+1}}\,,
\end{align}
and
\begin{align}
\frac{G}{R}\sim (t_s-t)^{-2\beta}\sim G^{\frac{2\beta}{3\beta+1}}\label{Todt}
\end{align}
If we replace $G$ by $G/R$ (as given in Eq.~(\ref{Todt})) in
Eq.~(\ref{secondo}), then we find that the asymptotic time
dependence in Eq.~(\ref{eqn-derivatives-f(R,G)}) for $\beta<1$ is
exactly same as follows:
\begin{align}
 -\frac{1}{\kappa^{2}}\left[2\dot H +3(1+w)H^{2}\right]\sim
\frac{\alpha}{(t_{s}-t)^{\beta+1}}+\frac{\gamma}{(t_{s}-t)^{2\beta}}\,.
\end{align}
With this consideration, one can derive
a specific ${F}(R,G)$-gravity model (by setting some parameters)
from Eq.~(\ref{secondo}) as
\begin{align}
{F}(R,G)=R+\frac{3(3\beta+1)}{2(\beta+1)}\frac{G}{R}\,,
\end{align}
which may encounter with other types of singularities.
So, in the model ${F}(R,G)\sim R+\bar{\alpha}(G/R)$ with $\bar{\alpha}>0$,
Type II, Type III and Type IV singularities may appear.
Then, by substituting $G$ for $R$ we get
\begin{align}
{F}(R,G)
\simeq
R+\bar{\delta}
|R|^{\frac{2\beta}{1+\beta}}\,,
\end{align}
which is a well-know result in cosmology. Now we have the following observations:
\begin{itemize}
\item In the model ${F}(R\rightarrow\infty)\sim R+\bar\delta R^{\frac{2\beta}{1+\beta}}$,
for
$0<\beta<1$ and $\bar\delta>0$,
a Type III singularity may appear.
\item In the model ${F}(R\rightarrow -\infty)\sim R+\bar\delta|R|^{\frac{2\beta}{1+\beta}}$,
for
$-1<\beta<0$
and $\bar\delta>0$,
a Type II singularity may appear.
\item In the model ${F}(R\rightarrow 0^{-})\sim R+\bar\delta |R|^{\frac{2\beta}{1+\beta}}$,
for
$\beta<-1$ and $\beta\not=-n$
(being $n$ a natural number)
and $\bar\delta<0$, a Type IV singularity may appear.
\end{itemize}
\end{itemize}

\subsection{$F(T)$ Gravity}
\label{sec-singularities-F(T)}

One of the alternative gravitational theories beyond Einstein's GR
is the Teleparallel Equivalent of GR (TEGR) where instead of the
curvature scalar $R$ defined by the Levi-Civita connection one
uses the concept of torsion scalar $T$ defined by the
Weitzenb\"{o}ck connection \cite{Hehl:1976kj}. The idea of this
gravitational theory was originally introduced by Einstein in 1928
under the name ``Fern-Parallelismus'' or ``distant parallelism''
or ``teleparallelism'' \cite{Einstein,2005physics...3046U}. It
was found that the modified teleparallel gravity theory $F (T)$ in
which $F$ is an arbitrary function of the torsion scalar $T$, can
explain the late-time accelerating expansion of the Universe for
some suitable choices of $F (T)$
\cite{Bengochea:2008gz,Linder:2010py}. Moreover, it was also
recognized that the $F (T)$ models can also take an active role to
describe the early inflationary phase
\cite{Rezazadeh:2017edd,Awad:2017ign,Bamba:2016wjm,Rezazadeh:2015dza,Nashed:2014lva}.
With such appealing consequences, the modified teleparallel
gravitation models got massive attention in the astrophysical and
cosmological community
\cite{Paliathanasis:2021uvd,Paliathanasis:2017htk,Paliathanasis:2016vsw,Bamba:2016gbu,Haro:2014xtk,Paliathanasis:2014iva,Amoros:2013nxa,Basilakos:2013rua,Bamba:2012vg}.
We refer to two recent reviews on $F (T)$ teleparallel gravity
models \cite{Cai:2015emx,Bahamonde:2021gfp}.
In Ref. \cite{Izumi:2012qj,Ong:2013qja}, however, it has been proved that superluminal modes appear in the $F(T)$ gravity and therefore the $F(T)$ gravity model is physically inconsistent.
In this section, however, just for the illustration, we show the structure of the singularities in the
framework of the $F(T)$ gravity.

In order to discuss the teleparallelism, we use the orthonormal
tetrad components $e_A (x^{\mu})$. We assume that the index $A$
runs over $0$, $1$, $2$, $3$ for the tangent space at each point
$x^{\mu}$ of the manifold and $e_A^\mu$ forms the tangent vector
of the manifold. The relation between the metric $g^{\mu\nu}$ and
the tetrad components is given by $g_{\mu\nu}=\eta_{A B} e^A_{\mu}
\e^B_\nu$ where $\eta_{A B} = \mathrm{diag}(-1, 1, 1, 1)$. Compared to
the General theory of Relativity where the torsionless Levi-Civita
connection is used, in Teleparallelism the curvatureless
Weitzenb\"{o}ck connection is used \cite{Weitzenbock}, hence, the
gravitational field is described by the non-null torsion tensor

\begin{align}\label{eqn-F(T)-1}
T^\rho_{\ \mu\nu} \equiv
\e^\rho_A \left( \partial_\mu e^A_\nu - \partial_\nu e^A_\mu
\right)\, ,
\end{align}

The torsion scalar $T$, one of the main ingredients of the teleparallel gravity, is defined
by \cite{Hayashi:1979qx,Maluf:2013gaa}:
\begin{align}\label{eqn-F(T)-2}
T \equiv
S_\rho^{\ \mu\nu} T^\rho_{\ \mu\nu}\, ,
\end{align}
where
\begin{align}\label{eqn-F(T)-3}
S_\rho^{\ \mu\nu}
\equiv \frac{1}{2} \left(K^{\mu\nu}_{\ \ \rho} + \delta^\mu_\rho
T^{\alpha \nu}_{\ \ \alpha}
 - \delta^\nu_\rho T^{\alpha \mu}_{\ \ \alpha}\right)\, ,
\end{align}
and $K^{\mu\nu}_{\verb||\rho}$, the contorsion tensor is given by

\begin{align}\label{eqn-F(T)-4}
K^{\mu\nu}_{\ \ \rho} \equiv -\frac{1}{2}
\left(T^{\mu\nu}_{\ \ \rho} - T^{\nu \mu}_{\ \ \rho}
 - T_\rho^{\ \ \mu\nu}\right)\, .
\end{align}

The action of the modified teleparallel gravity as
\cite{Linder:2010py}
\begin{align}
\label{eqn-F(T)-5-action}
S = \int d^4x |e| \left[ \frac{F(T)}{2{\kappa}^2}
+{{L}}_{\mathrm{matter}} \right]\, ,
\end{align}
$|e|= \det \left(e^A_\mu \right)=\sqrt{-g}$.
Now, varying the action in Eq.~(\ref{eqn-F(T)-5-action}) with
respect to the vierbein vector field $e_A^\mu$ one can obtain the
gravitational field equations \cite{Bengochea:2008gz}
\begin{align}
\label{eqn-F(T)-6}
\frac{1}{e}
\partial_\mu \left( eS_A^{\ \mu\nu} \right) F'
 - e_A^\lambda T^\rho_{\ \mu \lambda} S_\rho^{\ \nu\mu} F'
+S_A^{\ \mu\nu} \partial_\mu T F'' +\frac{1}{4} e_A^\nu F
= \frac{{\kappa}^2}{2} e_A^\rho {T^{(\mathrm{M})}}_{\rho}^{\ \nu}\, ,
\end{align}
where prime denotes the differentiation with respect to the torsion scalar $T$ and ${T^{(\mathrm{M})}}_{\rho}^{\ \nu}$ is the energy-momentum tensor describing the entire matter sector.

Now, let us assume that the background manifold is described well
by the spatially flat FLRW Universe (\ref{FLRWk0}).
Then, vierbein takes the form$e^A_{\mu} = \left(1, a (t), a (t), a (t) \right)$ where $a (t)$ is the
scale factor of the FLRW Universe. One can quickly derive the line
element of the FLRW Universe (\ref{FLRWk0}).
For this tetrad, using Eqs.~(\ref{eqn-F(T)-1}), (\ref{eqn-F(T)-3}), and (\ref{eqn-F(T)-4}), one
can derive the torsion scalar $T= - 6 H^2$.
In this flat FLRW background, one can write down the gravitational equations
\begin{align}
H^2 =&\, \frac{\kappa^2}{3} (\rho + \rho_\mathrm{GDE}) = \frac{\kappa^2}{3} \rho_\mathrm{eff},\label{efe-F(T)-1}\\
\dot{H} =&\, - \frac{\kappa^2}{2} (\rho + p+ \rho_\mathrm{GDE} + p_\mathrm{GDE}) = - \frac{\kappa^2}{2} (\rho_\mathrm{eff} + p_\mathrm{eff})\, ,
\label{efe-F(T)-2}
\end{align}
where $\rho_\mathrm{eff}$ and $p_\mathrm{eff}$ are the
total energy density and total pressure of the Universe, respectively, $\rho$
and $p$ denote the energy density and pressure of all the perfect
fluids comprising the entire matter sector where they satisfy the
conservation law $\dot{\rho} + 3 H (p + \rho) = 0$. Further,
$\rho_\mathrm{GDE}$ and $p_\mathrm{GDE}$ are given by
\begin{align}
\label{rhoeff-peff-F(T)}
\rho_\mathrm{GDE} = \frac{1}{2{\kappa}^2}
\left( -T -F +2T F' \right) \, , \quad p_\mathrm{GDE}
= -\frac{1}{2{\kappa}^2} \left[ \left(4 - 4 F' - 8 T F''\right) \dot H
 -T -F +2T F' \right]\, .
\end{align}
Notice that the effective fluid characterized by $\rho_\mathrm{GDE}$ and $p_\mathrm{GDE}$ satisfies
the continuity equation $\dot{\rho}_\mathrm{GDE} + 3 H (p_\mathrm{GDE}+ \rho_\mathrm{GDE}) = 0$.
As one can clearly notice that the choice of $F (T)$ highly
influences the gravitational equations. In the following we shall
focus on the finite-time future singularities that may appear in
the context of modified teleparallel gravity.

We consider, as in the previous sections, the following Hubble rate in (\ref{BigRip}) and (\ref{H3a-F(R)-sp}) \cite{Bamba:2012vg}
\begin{align}
H \sim&\, \frac{h_{\mathrm{s}}}{ \left( t_{\mathrm{s}} - t \right)^{\beta}}\,, \quad
\mathrm{for} \,\,\, \beta> 0\,,
\label{eqn-F(T)-H-1} \\
H \sim&\, H_{\mathrm{s}} + \frac{h_{\mathrm{s}}}{
\left( t_{\mathrm{s}} - t \right)^{\beta}}\,,
\quad
\mathrm{for} \ \beta<-1\, ,\ \mbox{and}\ -1< \beta < 0\, .
\label{eqn-F(T)-H-2}
\end{align}
One can clearly notice that as $t \rightarrow t_s$, finite-time
singularities may appear depending on the nature of $\beta$.
This can be illustrated as follows: when $t\to t_{\mathrm{s}}$, for
$\beta > 0$, both $H \sim h_{\mathrm{s}} \left( t_{\mathrm{s}} - t \right)^{-\beta}$ and
$\dot{H} \sim \beta h_{\mathrm{s}} \left( t_{\mathrm{s}} - t \right)^{-\left(\beta+1\right)}$ diverge
to infinity; for $-1 < \beta < 0$, $H$ is finite, but $\dot{H}$
becomes infinity; for $\beta < -1$, but $\beta$ is not any
integer, both $H$ and $\dot{H}$ are finite, but the higher
derivatives of $H$ can diverge to infinity.

From Eq.~(\ref{eqn-F(T)-H-1}) one can derive the evolution of the
scale factor as
\begin{align}
a \sim&\, a_{\mathrm{s}} \exp \left[ \frac{h_{\mathrm{s}}}{\beta-1}
\left( t_{\mathrm{s}} - t
\right)^{-\left(\beta-1\right)}
\right]\, ,
\quad
\mathrm{for}\,\,\,
0<\beta<1,\; \mbox{and} \,\,\, \beta>1\,,
\label{eqn-F(T)-a1} \\
a \sim&\, a_{\mathrm{c}}
\left(\frac{t_{\mathrm{s}} - t_{\mathrm{c}}} {t_{\mathrm{s}} - t}\right)^{h_{\mathrm{s}}},
\quad
\mathrm{for}\,\,\, \beta = 1\,,
\label{eqn-F(T)-a2}
\end{align}
where $a_{\mathrm{s}}$ and $a_{\mathrm{c}}$ are some {positive} constants. Now, from Eq.~(\ref{eqn-F(T)-a1}) one can see that,
when $t\to t_{\mathrm{s}}$,
for $\beta \geq 1$,
$a \to \infty$, whilst for $\beta < 0$ and $0 < \beta < 1$,
$a \to a_{\mathrm{s}}$.
Additionally, one can further notice that from $\rho_\mathrm{eff} = 3H^2/\kappa^2$ (Eq.~(\ref{efe-F(T)-1})) and (\ref{eqn-F(T)-H-1}) that,
for $\beta > 0$,
$H \to \infty$ and as a result,
$\rho_{\mathrm{eff}} = 3 H^2/\kappa^2 \to \infty$,
while on the other hand,
for $\beta < 0$, $H$ asymptotically becomes finite
and $\rho_{\mathrm{eff}}$ asymptotically approaches to a finite
constant value $\rho_{\mathrm{s}}$.
Furthermore, from
$\dot{H} \sim \beta h_{\mathrm{s}} \left( t_{\mathrm{s}} - t
\right)^{-\left(\beta+1\right)}$ and Eq.~(\ref{eqn-F(T)-H-2}) one can see that
for $\beta>-1$, $\dot{H} \to \infty$ and as a consequence,
$P_\mathrm{eff} = -\left(2\dot H + 3H^2\right)/\kappa^2 \to \infty$.
While for $\beta < -1$, but $\beta$ is not any integer,
$a$, $\rho_{\mathrm{eff}}$ and $P_{\mathrm{eff}}$ are finite since both $H$ and $\dot{H}$ are finite, however, the higher derivatives of
$H$ diverges.

Thus, we see that for the Hubble parameter described in
Eqs.~(\ref{eqn-F(T)-H-1}) and (\ref{eqn-F(T)-H-2}) the $F(T)$
gravity models can encounter with the finite-time singularities
for the following values of $\beta$:
\begin{itemize}

\item For $\beta \geq 1$, Type I (``Big Rip'') singularity appears.

\item For $-1<\beta<0$, Type II (``sudden'') singularity appears.

\item For $0<\beta<1$, Type III singularity appears.

\item For $\beta< -1$, but $\beta$ is not any integer,
Type IV singularity appears.

\end{itemize}

\begin{table*}[tbp]
\caption{The table summarizes the emergence of the finite-time
future singularities for different values of $\beta$ in the
expressions of the Hubble parameter given in
Eqs.~(\ref{eqn-F(T)-H-1}) and (\ref{eqn-F(T)-H-2}) { along with the behavior of $H$ and} $\dot{H}$ in the limit of $t \to t_{\mathrm{s}}$. Table courtesy
Ref.~\cite{Bamba:2012vg}. }
\begin{center}
\begin{tabular}{cccccccc}
\hline
\hline
Region of $\beta~(\neq 0, \, -1)$ \quad 
& Type of the Singularity \quad & $H$ ($t \to t_{\mathrm{s}}$) \quad 
& $\dot{H}$ ($t \to t_{\mathrm{s}}$)\\
\hline
$\beta \geq 1$ & \quad Type I & $H \to \infty$
& $\dot{H} \to \infty$ \\

$-1 < \beta < 0$ & \quad Type II 
& $H \to H_{\mathrm{s}}$
& $\dot{H} \to \infty$ \\

$0 < \beta < 1$ & \quad Type III 
& $H \to \infty$
& $\dot{H} \to \infty$\\

$\beta < -1$, but $\beta$ is not any integer
& \quad Type IV & $H \to H_{\mathrm{s}}$
& $\dot{H} \to 0$ \\
&
&
& 
(and higher order
\\
&
&
&
derivatives
of $H$ diverge)\\

\hline
\hline
\end{tabular}
\end{center}
\label{table-F(T)-singularities}
\end{table*}
In TABLE~\ref{table-F(T)-singularities}, we summarize the
appearance of finite-time singularities out of the expressions of
the Hubble parameter as given in Eqs.~(\ref{eqn-F(T)-H-1}) and
(\ref{eqn-F(T)-H-2}). Thus, one can see that the above expressions
of the Hubble parameter can be used to reconstruct the $F (T)$
gravity models which may allow the finite-time singularities. We
focus on the finite-time singularities when the geometrical sector
dominates over the matter sector, that means when $\rho_\mathrm{eff}
\simeq \rho_\mathrm{GDE}$ and $p_\mathrm{eff} \simeq p_\mathrm{GDE}$, that
means $w_\mathrm{eff} = p_\mathrm{eff}/\rho_\mathrm{eff} \simeq p_{\rm
GDE}/\rho_\mathrm{GDE}$. Following this, one can have

\begin{align}
w_{\mathrm{eff}} \simeq w_\mathrm{GDE} = - \frac{\left(4 - 4 F' - 8 T F''\right) \dot H
 -T -F +2T F'}{-T -F +2T F'}\, .
\end{align}

On the other hand, from Eq.~(\ref{rhoeff-peff-F(T)}) one can write

\begin{align}
\label{relation-rhoGDE-pGDE}
p_\mathrm{GDE} = - \rho_\mathrm{GDE} + I (H, \dot{H})\, ,
\end{align}
where $I(H, \dot{H}) = -\frac{1}{\kappa^2} \left[2 - 2 F' - 4 T F''\right]\dot{H}$.
From Eq.
(\ref{efe-F(T)-2})
one can quickly write $p_\mathrm{GDE} = -\rho_\mathrm{GDE} -2 \dot{H}/\kappa^2$, where we have using that the geometrical sector dominates over the matter one. Now, comparing this with (\ref{relation-rhoGDE-pGDE}),
we obtain the following differential equation

\begin{align}
\label{FT-ode}
\dot{H} + \frac{\kappa^2}{2} I (H, \dot{H}) = 0 \quad \Longrightarrow \quad
\dot{H} \left[ F' + 2 T F'' \right] = 0\, .
\end{align}

In the following we shall illustrate the the appearance of the finite-time singularities in a specific $F(T)$ gravity model having the power law form: 

\begin{align}
\label{power-law-FT}
F(T) = A T^{\alpha}\,,
\end{align}
where $A~(\neq 0)$ and $\alpha~(\neq 0)$ are constants. With this
choice of $F (T)$, Eq.~(\ref{FT-ode}) becomes
\begin{align}
\label{FT-ode-powerlaw}
\left(2\alpha-1\right)
H^{2\left(\alpha-1\right)}
= 0\,.
\end{align}
Notice that for both Eqs.~(\ref{eqn-F(T)-H-1}) and
(\ref{eqn-F(T)-H-2}), $\dot{H} \neq 0$. In the limit of $t
\rightarrow t_s$, Eq.~(\ref{FT-ode-powerlaw}) needs to be
satisfied. Now, from Eqs.~(\ref{eqn-F(T)-H-1}) and
(\ref{eqn-F(T)-H-2}), one can see that for $\beta> 0$ [here two different cases may arise in this way: (i) $\beta \geq 1$
(i.e., Type I singularity) and $0<\beta<1$ (Type III
singularity)], $\alpha < 1$, so that
Eq.~(\ref{FT-ode-powerlaw}) is asymptotically satisfied. While on
the other hand, for $\beta < 0$ [here one can similarly explore two
different cases: $-1<\beta<0$ (Type II singularity and $\beta<-1$
(Type IV singularity)], $\alpha = 1/2$, in which
Eq.~(\ref{FT-ode-powerlaw}) is always satisfied. At this point it
is very crucial to mention that the conditions: (i) for $\beta >
0$ (Type I singularity corresponds to $\beta \geq 1$ and Type III
singularity corresponds to $0<\beta<1$), $\alpha < 0$, while (ii)
for $\beta < 0$ (Type II singularity corresponds to $-1<\beta<0$
and Type IV singularity corresponds to $\beta<-1$), $\alpha = 1/2$
are ``necessary conditions'' for the appearance of the finite-time
future singularities but they are not the sufficient conditions.
Because if $\alpha < 0$, then the Type I singularity with $\beta
\geq 1$ appears rather than the Type III singularity with
$0<\beta<1$, since in the limit of $t \to t_{\mathrm{s}}$, both $H$
and $\dot{H}$ with $\beta \geq 1$ diverge more rapidly than those
with $0<\beta<1$. This is because of the absolute value of the power $\beta$, namely, the absolute value of $\beta$ for the Type
I singularity ($\beta \geq 1$) is larger than the absolute value
of $\beta$ for the Type III singularity ($0<\beta<1$). As a
consequence, the Type I singularity is realized faster than the
Type III singularity, and this results in the appearance of the
Type I singularity. In a similar fashion, if $\alpha = 1/2$, then
Type IV singularity with $\beta<-1$ appears compared to the Type
II singularity with $-1<\beta<0$ because again in the limit of $t
\to t_{\mathrm{s}}$, $H \to H_{\mathrm{s}}$ and $\dot{H} \to 0$
with $\beta<-1$ are realized more quickly than $H \to
H_{\mathrm{s}}$ and $\dot{H} \to \infty$ with $-1<\beta<0$. This
is again because of the absolute value of the power $\beta$,
namely, the absolute value of $\beta$ for the Type IV singularity
($\beta<-1$) is larger than the absolute value of $\beta$ for the
Type II singularity ($-1<\beta<0$). Hence, the Type IV singularity
appears faster than the Type II singularity, and Type IV
singularity appears as a result.

Finally, we note that Type V (``$w$'') singularity can also appear in this model. In the Type V singularity,
a specific choice of the scale factor can be considered~\cite{Dabrowski:2009kg}
\begin{align}
a (t) =&\, a_{\mathrm{s}} \left( 1-\frac{3\sigma}{2}
\left\{ \frac{n-1}{n-\left[ 2/\left(3\sigma\right) \right]}
\right\}^{n-1}\right)^{-1}
+ \frac{1-2/\left(3\sigma \right)}{n-2/\left(3\sigma \right)}
\times
n a_{\mathrm{s}}
\left( 1-\frac{2}{3\sigma}
\left\{ \frac{n-\left[ 2/\left(3\sigma\right) \right]}{n-1}
\right\}^{n-1}\right)^{-1}
\left( \frac{t}{t_{\mathrm{s}}}
\right)^{2/\left(3\sigma \right)}
\nonumber \\
&\, +a_{\mathrm{s}} \left( \frac{3\sigma}{2}
\left\{ \frac{n-1}{n-\left[ 2/\left(3\sigma\right) \right]}
\right\}^{n-1} -1 \right)^{-1}
\left[ 1 - \frac{1-2/\left(3\sigma \right)}{n-2/\left(3\sigma \right)}
\frac{t}{t_{\mathrm{s}}} \right]^{n} \,,
\label{FT-type-V-singul-scale-factor}
\end{align}
where $n$ $\sigma$ are any arbitrary real numbers. In the limit of
$t \to t_{\mathrm{s}}$, we have $H(t \to t_{\mathrm{s}}) \to 0$
and $\dot{H} (t \to t_{\mathrm{s}}) \to 0$. The effective EoS
$w_{\mathrm{eff}} = \left(1/3\right) \left(2q -1\right) \to
\infty$ where $q \equiv -\ddot{a}a/\dot{a}^2$ is the deceleration
parameter of the Universe. Thus, for the power law
model~(\ref{power-law-FT}) with $\alpha >1$,
Eq.~(\ref{FT-ode-powerlaw}) can be satisfied asymptotically.
Hence, the Type V (``$w$'') singularity can appear in this model.

\subsection{Non-local Gravity}
\label{sec-singularities-nolocal}

Along with the known modified gravitational theories presented in
the earlier sections, an addition to the class of modified gravity
theories is the non-local gravitational theory, i.e., non-local
additions to GR \cite{Deser:2007jk}. These non-local corrections
naturally arise due to quantum effects and with such new
corrections, it is also possible to explain the accelerating
expansion of the Universe \cite{Deser:2007jk}. In this section we
shall describe how the finite-time future singularities may appear in this gravitational theory.

The action of the non-local gravity is given by \cite{Bamba:2012ky}
\begin{align}
\label{action-nonlocal}
S=\int d^4 x \sqrt{-g}\Bigg[
\frac{1}{2\kappa^2}\left\{ R\left(1 + f(\Box^{-1}R )\right) -2 \Lambda \right\}
+ {L}_\mathrm{matter} \left(Q; g\right)
\Bigg]\,,
\end{align}
where $R$ is the Ricci scalar, $g$ is the determinant of the metric tensor $g_{\mu\nu}$, $\Box \equiv g^{\mu \nu} {\nabla}_{\mu} {\nabla}_{\nu}$
with ${\nabla}_{\mu}$ being the covariant derivative
is the covariant d'Alembertian for a scalar field, and $f$ is any arbitrary function of $\Box^{-1}R$, $\Lambda$ is a cosmological constant,
and ${L}_{\mathrm{matter}}\left(Q; g\right)$ denotes the matter Lagrangian in which
$Q$ stands for the matter fields.

Introducing two scalar fields $\eta$
and $\xi$, we can rewrite the above action (\ref{action-nonlocal}) as
\begin{align}
\label{non-local-action2}
S =&\, \int d^4 x \sqrt{-g}\Bigg[
\frac{1}{2\kappa^2}\Bigg\{R\left(1 + f(\eta)\right)
+ \xi\left(\Box\eta - R\right) - 2 \Lambda \Bigg\}
+ {L}_\mathrm{matter} \Bigg] \nonumber \\
=&\, \int d^4 x \sqrt{-g}\Bigg[
\frac{1}{2\kappa^2}\Bigg\{R\left(1 + f(\eta)\right)
 - \partial_\mu \xi \partial^\mu \eta - \xi R - 2 \Lambda \Bigg\}
+ {L}_\mathrm{matter}
\Bigg]
\,. 
\end{align}
In the context of a spatially flat
FLRW metric (\ref{FLRWk0}), one can write down the
gravitational field equations as
\begin{align}
0=&\, - 3 H^2\left(1 + f(\eta) - \xi\right) + \frac{1}{2}\dot\xi \dot\eta
 - 3H\left(f'(\eta)\dot\eta - \dot\xi\right) + \Lambda
+ \kappa^2 \rho \, , \label{non-local-field-eq01}\\
0=&\, \left(2\dot H + 3H^2\right) \left(1 + f(\eta) - \xi\right)
+ \frac{1}{2}\dot\xi \dot\eta
+ \left(\frac{d^2}{dt^2} + 2H \frac{d}{dt} \right) \left( f(\eta) - \xi \right) - \Lambda + \kappa^2 p\, ,\label{non-local-field-eq02}
\end{align}
where we have considered that the scalar fields $\eta$ and $\xi$ depend only on time \cite{Bamba:2012ky}. { Note that here $\rho$ and $p$ are respectively the energy density and pressure of the matter sector.}
One can further write down
the equations of motion for the scalar fields $\eta$ and $\xi$ as
\begin{align}
0=&\, \ddot \eta + 3H \dot \eta + 6 \dot H + 12 H^2 \, ,
\label{nonlocal-eta} \\
0=&\, \ddot \xi + 3H \dot \xi - \left( 6 \dot H + 12 H^2\right)f'(\eta) \, ,
\label{nonlocal-xi}
\end{align}
where we have used $R = 6\dot{H} + 12H^2$. Let us now focus on
the emergence of the finite-time singularities in this
gravitational theory.

\subsubsection{Finite-time Singularities}

We start with the following expression of the
Hubble parameter in (\ref{BigRip}) or (\ref{Hsin}) (see \cite{Bamba:2012ky}).
The scale factor is also given in (\ref{BigRip}).
Now with the use of $\ddot{\eta} + 3H\dot{\eta} =a^{-3} d \left( a^3 \dot{\eta} \right)/dt$ and
Eq.~(\ref{nonlocal-eta}), one can express $\eta$ as
\begin{align}
\eta = -\int^{t} \frac{1}{a^3}
\left( \int^{\bar{t}}
Ra^3 d\bar{t} \right)dt\,.
\label{eqn-eta}
\end{align}
We note that in the limit $t\to t_s$: for $\beta>1$, $\dot{H} \ll H^2$, and hence, $R \sim 12 H^2$,
whilst for $-1 < \beta< 0$ and $0 < \beta < 1$, $\dot{H} \gg H^2$, and hence, $R \sim 6\dot{H}$. By
applying these relations to Eq.~(\ref{eqn-eta}) and then taking
the leading term in terms of $\left( t_{\mathrm{s}} - t \right)$,
one can express $\eta$ for different regions of $\beta$ as follows
\cite{Bamba:2012ky}:
\begin{align}
\eta \sim&\, -\frac{4h_{\mathrm{s}}}{\beta-1}
\left( t_{\mathrm{s}} - t \right)^{-\left(\beta-1\right)}
+ {\eta}_{\mathrm{c}}
\quad
(\beta > 1)\,,
\label{eq-eta2} \\
\eta \sim&\, -\frac{6h_{\mathrm{s}}}{\beta-1}
\left( t_{\mathrm{s}} - t \right)^{-\left(\beta-1\right)}
+ {\eta}_{\mathrm{c}}
\quad
(-1 < \beta < 0\,, \, 0 < \beta < 1)\,,
\label{eq-eta3}
\end{align}
where ${\eta}_{\mathrm{c}}$ is an integration constant. 
As discussed in \cite{Bamba:2012ky}, for 
${\eta}_{\mathrm{c}} =0$, the finite-time future singularities depicted by
the Hubble parameter in Eq. (\ref{BigRip}) or (\ref{Hsin}) do not appear, however, for ${\eta}_{\mathrm{c}} \neq 0$, finite time future singularities can occur.

For ${\eta}_{\mathrm{c}} \neq 0$, note that, if the power in terms of $\left( t_{\mathrm{s}} - t \right)$ is negative, i.e. $-\left( \beta -1\right) < 0$, then the first term proportional to $\left( t_{\mathrm{s}} - t \right)^{-\left(\beta -1\right)}$ is the leading one. On the other hand, if the power in terms of $\left( t_{\mathrm{s}} - t \right)$ is positive, i.e. $-\left( \beta -1\right) > 0$, then the second constant term becomes the leading one. Thus, 
for $\beta > 1$, the first term 
is the leading one, i.e., 
$\eta \propto \left( t_{\mathrm{s}} - t \right)^{-\left(\beta-1\right)}$, while for 
$-1 < \beta < 0$ and $0 < \beta < 1$, the second term 
is the leading one, i.e., $\eta \sim {\eta}_{\mathrm{c}}$. 
For $\beta =1$ in Eq. (\ref{BigRip}) or (\ref{Hsin}), it follows from Eq.~(\ref{nonlocal-eta}) that, 
\begin{eqnarray}\label{non-local-new-sp01}
 \eta \sim 6h_{\mathrm{s}} \left[ 
\left( 1+2h_{\mathrm{s}}
\right)/\left( 1+3h_{\mathrm{s}} \right) \right] 
\ln \left( t_{\mathrm{s}} - t \right) + {\eta}_{\mathrm{c}}.
\end{eqnarray}
We continue by taking a particular form of $f(\eta)$ as \cite{Bamba:2012ky}

\begin{equation}\label{non-local-new-sp02}
f(\eta) = f_{\mathrm{s}} \eta^{\sigma}\,, 
\end{equation}
where $f_{\mathrm{s}}$ and $\sigma$ are non-zero constants. 
Now, using
$\ddot{\xi} + 3H\dot{\xi} =a^{-3} d \left( a^3 \dot{\xi} \right)/dt$ 
and Eq.~(\ref{nonlocal-xi}), 
$\xi$ can be expressed as 
\begin{equation} 
\xi = \int^{t} \frac{1}{a^3} \left( \int^{\bar{t}} 
\frac{d f(\eta)}{d \eta} 
Ra^3 d\bar{t} \right)dt\,.
\label{non-local-new-sp03}
\end{equation}

Now we apply the conditions $R \sim 12 H^2$ (for $ \beta >1$) and $R \sim 6\dot{H}$ (for $ \beta <1$) 
to Eq.~(\ref{non-local-new-sp03}) and 
taking into account the leading term in terms of $\left( t_{\mathrm{s}} - t \right)$, we obtain \cite{Bamba:2012ky}
\begin{eqnarray} 
&& \xi \sim - f_{\mathrm{s}} \left(-\frac{4h_{\mathrm{s}}}{\beta -1} \right)^{\sigma} 
\left( t_{\mathrm{s}} - t \right)^{-\left(\beta-1\right) \sigma} 
+ {\xi}_{\mathrm{c}}~, 
\quad 
(\beta > 1)\,,
\label{non-local-new-sp04} \\ 
&& \xi \sim 
\frac{6 f_{\mathrm{s}} h_{\mathrm{s}} \sigma 
{\eta}_{\mathrm{c}}^{\sigma -1}}{\beta -1} 
\left( t_{\mathrm{s}} - t \right)^{-\left(\beta -1\right)} 
+ {\xi}_{\mathrm{c}}~, 
\quad 
(-1 < \beta < 0\,~, \, 0 < \beta < 1)\,,
\label{non-local-new-sp05}
\end{eqnarray}
where ${\xi}_{\mathrm{c}}$ is the constant of integration. 
Now, if the power in terms of $\left( t_{\mathrm{s}} - t \right)$ is negative, i.e. $-\left(\beta-1\right) \sigma < 0$, the first term proportional to 
$\left( t_{\mathrm{s}} - t \right)^{-\left(\beta-1\right) \sigma}$ is the leading one. On the other hand, if the power in terms of $\left( t_{\mathrm{s}} - t \right)$ is 
positive, i.e. $-\left(\beta - 1\right) \sigma > 0$, then the second constant term becomes the leading one. 
Therefore, we see that for $\beta > 1$, $\sigma > 0$, 
$\xi \propto \left( t_{\mathrm{s}} - t \right)^{-\left(\beta-1\right) \sigma}$, while 
for ($\beta > 1$, $\sigma < 0$) and 
($-1 < \beta < 0\,, \, 0 < \beta < 1$), 
$\xi \sim {\xi}_{\mathrm{c}}$. Therefore one arrives at the following three cases \cite{Bamba:2012ky}: 

\begin{itemize}
 
\item For $\beta > 1$, $\sigma > 0$: $\eta \propto \left( t_{\mathrm{s}} - t \right)^{-\left(\beta-1\right)}$ 
and 
$\xi \propto \left( t_{\mathrm{s}} - t \right)^{-\left(\beta-1\right) \sigma}$.

\item For $\beta > 1$, $\sigma < 0$: $\eta \propto \left( t_{\mathrm{s}} - t \right)^{-\left(\beta-1\right)}$ 
and 
$\xi \sim {\xi}_{\mathrm{c}}$. 

\item For $-1 < \beta < 0\,, \, 0 < \beta < 1$:\ 
$\eta \sim {\eta}_{\mathrm{c}}$ and $\xi \sim {\xi}_{\mathrm{c}}$. 

\end{itemize}
Now, having all of these, following \cite{Bamba:2012ky} one can further investigate the possibility of the finite time future singularities in this gravitational theory when the Hubble rate is given in Eq. (\ref{BigRip}) or (\ref{Hsin}) and the results are as follows 
\begin{itemize}
 
\item For $\sigma < 0$, with ${\eta}_{\mathrm{c}} \neq 0$ and ${\xi}_{\mathrm{c}} = 1$, Type I (Big Rip) singularity can occur for $\beta > 1$.

\item If ${\eta}_{\mathrm{c}} \neq 0$, and 
$f_{\mathrm{s}} {\eta}_{\mathrm{c}}^{\sigma-1} 
\left( 6\sigma - {\eta}_{\mathrm{c}} 
\right) + {\xi}_{\mathrm{c}} -1 = 0$, 
then Type II (sudden) singularity can occur for $-1 < \beta < 0$.

\item If ${\eta}_{\mathrm{c}} \neq 0$, and 
$f_{\mathrm{s}} {\eta}_{\mathrm{c}}^{\sigma-1} 
\left( 6\sigma - {\eta}_{\mathrm{c}} 
\right) + {\xi}_{\mathrm{c}} -1 = 0$, 
there Type III singularity can occur for $0 < \beta < 1$.

\end{itemize}

\subsection{Non-minimal Maxwell-Einstein Gravity}
\label{sec-Maxwell-Einstein-gravity}

In this section, we discuss the finite-time future singularities
appearing in the non-minimal Maxwell-Einstein gravity with general
gravitational coupling. To begin with, we consider the following
action~\cite{Bamba:2008ja,Bamba:2008ut}:
\begin{align}
S_{\mathrm{GR}} = \int d^{4}x \sqrt{-g}
\left[\frac{1}{2\kappa^2} R - \frac{1}{4} I(R)
F_{\mu\nu}F^{\mu\nu} \right]\,,\label{action-ME-action}
\end{align}
where $I(R)$ is any function of the Ricci scalar, $F_{\mu\nu} = {\partial}_{\mu}A_{\nu} - {\partial}_{\nu}A_{\mu}$
is the electromagnetic field-strength tensor in which $A_{\mu}$ denotes the $U(1)$
gauge field. It is well-known that the coupling between the scalar curvature and
the Lagrangian of the electromagnetic field, as shown in action (\ref{action-ME-action}), arises in curved
spacetime due to one-loop vacuum-polarization effects in
Quantum Electrodynamics~\cite{Drummond:1979pp}.

Now. taking the variations of the action in
Eq.~(\ref{action-ME-action}) with respect to the metric
$g_{\mu\nu}$ and the $U(1)$ gauge field $A_{\mu}$, one can find
the gravitational field equations and the equation of motion of
$A_{\mu}$, respectively, as~\cite{Bamba:2008ja,Bamba:2008ut}
\begin{align}
R_{\mu \nu} - \frac{1}{2}g_{\mu \nu}R
= \kappa^2 T^{(\mathrm{EM})}_{\mu \nu}\,,
\label{eqn-GFE-EM}
\end{align}
and
\begin{align}
 -\frac{1}{\sqrt{-g}}{\partial}_{\mu}
\left( \sqrt{-g} I(R) F^{\mu\nu}
\right) = 0\,,\label{eqn-Amu}
\end{align}
where $T^{(\mathrm{EM})}_{\mu \nu}$ in Eq.~(\ref{eqn-GFE-EM})
denotes the contribution to the energy-momentum tensor from the
electromagnetic field:
\begin{align}
T^{(\mathrm{EM})}_{\mu \nu}
=&\, I(R) \left( g^{\alpha\beta} F_{\mu\beta} F_{\nu\alpha}
 -\frac{1}{4} g_{\mu\nu} F_{\alpha\beta}F^{\alpha\beta} \right) \nonumber \\
&\, +\frac{1}{2} \left[ I^{\prime}(R)
F_{\alpha\beta}F^{\alpha\beta} R_{\mu \nu}
+ g_{\mu \nu} \Box \left( I^{\prime}(R)
F_{\alpha\beta}F^{\alpha\beta} \right)
 - {\nabla}_{\mu} {\nabla}_{\nu}
\left( I^{\prime}(R) F_{\alpha\beta}F^{\alpha\beta} \right) \right]\,,
\label{eq:Tmunu-ME}
\end{align}
in which the prime refers to the derivative with respect to $R$;
$\nabla_{\mu}$ denotes the covariant derivative operator
associated with $g_{\mu \nu}$, and $\box \equiv g_{\mu \nu}
\nabla_{\mu} \nabla_{\nu}$ is the covariant d'Alembertian for a
scalar field. To understand the dynamics of the Universe within
this gravitational theory, we consider the flat FLRW metric and following Ref.
\cite{Bamba:2008ut} we consider the case where only the effects of
magnetic fields are present that means the effects of electric
fields are negligible. Additionally, we further consider that only
one component of the magnetic field $\Vec{B}$ is non-zero, that
means other two components are zero. Thus, it follows from
$\mathrm{div} \Vec{B} = 0$ that the off-diagonal components of the
last term of the r.h.s. of Eq.\ (\ref{eq:Tmunu-ME}) for
$T^{(\mathrm{EM})}_{\mu \nu}$, i.e., ${\nabla}_{\mu}
{\nabla}_{\nu} \left[ I^{\prime}(R) F_{\alpha\beta}F^{\alpha\beta}
\right]$ are zero. Hence, all of the off-diagonal components of
$T^{(\mathrm{EM})}_{\mu \nu}$ are zero. Furthermore, since we have
assumed that we only have the magnetic fields as background
quantities at the $0^\mathrm{th}$ order, therefore, the magnetic
fields remain independent of the space components $\Vec{x}$.

In the FLRW spacetime, the equation of motion for the
$U(1)$ gauge field in the Coulomb gauge, ${\partial}^jA_j(t,\Vec{x}) =0$,
and the case of $A_{0}(t,\Vec{x}) = 0$, becomes
\begin{align}
{\ddot{A}}_i(t,\Vec{x})
+ \left( H + \frac{\dot{I}}{I}
\right) {\dot{A}} (t,\Vec{x})
 - \frac{1}{a^2}\mathop{\Delta}\limits^{(3)}\, A_i(t,\Vec{x}) = 0\,,
\label{eqn:ME-1}
\end{align}
where $\mathop{\Delta}\limits^{(3)} = {\partial}^i {\partial}_i$ is the flat 3-dimensional Laplacian.
From Eq.~(\ref{eqn:ME-1}) it follows that the Fourier mode $A_i(k,t)$
satisfies the equation
\begin{align}
{\ddot{A}}_i(k,t) + \left( H + \frac{\dot{I}}{I} \right)
 {\dot{A}}_i(k,t) + \frac{k^2}{a^2} A_i(k,t) = 0\,,
\label{eqn:ME-2}
\end{align}
which can alternatively be expressed (by replacing the cosmic time $t$ in terms of the conformal time
$\eta $) as
\begin{align}
\frac{\partial^2 A_i(k,\eta)}{\partial \eta^2} + \frac{1}{I(\eta)} \frac{d I(\eta)}{d \eta}
\frac{\partial A_i(k,\eta)}{\partial \eta} + k^2 A_i(k,\eta) = 0\,.
\label{eqn:ME-3}
\end{align}

We note that finding the exact solution of Eq.~(\ref{eqn:ME-3})
for any arbitrary coupling function $I (\eta)$ is not possible while
with the use of the Wentzel--Kramers--Brillouin (WKB) approximation on subhorizon scales and the
long-wavelength approximation on superhorizon scales, and finally matching these
solutions at the horizon crossing~\cite{Bamba:2006ga,Bamba:2007sx}, an
approximate solution can be found as
\begin{align}
\left|A_i(k,\eta)\right|^2
= |\bar{C}(k)|^2
= \frac{1}{2kI(\eta_k)}
\left|1- \left[ \frac{1}{2}\frac{1}{kI(\eta_k)}\frac{d I(\eta_k)}{d \eta}
+ i \right]k\int_{\eta_k}^{{\eta}_{\mathrm{f}}}
\frac{I(\eta_k)}{I \left(\widetilde{\widetilde{\eta}} \right)}
d\widetilde{\widetilde{\eta}}\,\right|^2\,,
\label{eqn:ME-4}
\end{align}
where $\eta_k$ and ${\eta}_{\mathrm{f}}$ denotes the conformal time
at the horizon-crossing and the conformal time at the end of inflation, respectively.
 From Eq.~(\ref{eqn:ME-4}),one may obtain the amplitude of the proper magnetic fields in the position space
\begin{align}
|{B_i}^{(\mathrm{proper})}(t)|^2 = \frac{k|\bar{C}(k)|^2}{\pi^2}\frac{k^4}{a^4}\,,
\label{eqn:ME-5}
\end{align}
on a comoving scale $L=2\pi/k$. From Eq.~(\ref{eqn:ME-5}) one can
see that the proper magnetic fields evolve as
$|{B_i}^{(\mathrm{proper})}(t)|^2 = |\bar{B}|^2/a^4$, where
$|\bar{B}|$ is a constant \cite{Bamba:2008ut}. This means that the
impact of the coupling function $I$ on the value of the proper
magnetic fields exists only during the inflationary stage. On the
other hand, the conductivity of the Universe,
${\sigma}_\mathrm{c}$, is extremely small during the period of
inflation since at that time only a few charged particles exist.
After the reheating stage, a number of charged particles are
produced, so that the conductivity immediately jumps to a large
value~${\sigma}_\mathrm{c} \gg H$. Consequently, for a large
enough conductivity at the reheating stage, the proper magnetic
fields evolve in proportion to $a^{-2}(t)$ in the
radiation-dominated stage and in the subsequent matter-dominated
stage~\cite{Ratra:1991bn}.

Now, from (\ref{eq:Tmunu-ME}), one can express the effective
energy density and effective pressure of the Universe as follows
\begin{align}
\rho_\mathrm{eff} =&\,
\left\{ \frac{I(R)}{2} + 3\left[
 -\left( 5 H^2 + \dot{H} \right) I^{\prime}(R) + 6 H \left( 4H\dot{H} + \ddot{H} \right) I^{\prime\prime}(R)
\right] \right\}
\frac{|\bar{B}|^2}{a^4}\,,
\label{rhoeff-ME} \\
p_\mathrm{eff} =&\, \biggl[ -\frac{I(R)}{6} +\left( -H^2 + 5\dot{H} \right) I^{\prime}(R) - 6 \left( -20H^2\dot{H} + 4\dot{H}^2 - H\ddot{H} + \dddot{H} \right)
I^{\prime\prime}(R)
\nonumber\\
&\, -36\left( 4H\dot{H} + \ddot{H} \right)^2 I^{\prime\prime\prime}(R)
\biggr]
\frac{|\bar{B}|^2}{a^4}\,,
\label{peff-ME}
\end{align}
where the following relations under the assumption of
negligible electric fields have been used:
$
g^{\alpha\beta} F_{0\beta} F_{0\alpha}
 -\left(1/4\right) g_{00} F_{\alpha\beta}F^{\alpha\beta}
= |{B_i}^{(\mathrm{proper})}(t)|^2/2,
$
and
$
F_{\alpha\beta}F^{\alpha\beta}
= 2|{B_i}^{(\mathrm{proper})}(t)|^2
$.

We recall that
in terms of the effective energy density and pressure, the Friedmann and Raychaudhuri
equations for this gravitational theory can be written as
\begin{align}
\frac{3}{\kappa^2}H^2\ =&\, \rho_\mathrm{eff}\,,
\label{efe1-ME} \\
-\frac{1}{\kappa^2}\left(2\dot H + 3H^2\right) =&\, p_\mathrm{eff}\,,
\label{efe2-ME}
\end{align}
where $\rho_\mathrm{eff}$ and $p_\mathrm{eff}$ are
given in Eqs.~(\ref{rhoeff-ME}) and (\ref{peff-ME}), respectively.
Since the Friedmann and Raychaudhuri equations (\ref{efe1-ME}) and (\ref{efe2-ME}) includes $I (R)$, therefore, depending on the functional forms of $I (R)$,
the Maxwell-Einstein gravity may allow finite-time future singularities. In the following we discuss the possible finite-time future singularities.

\begin{center}
 (i) {\it Big Rip singularity}
\end{center}

We consider the Hubble rate in (\ref{H-F(R)-big-rip}).
For this Hubble rate, one can derive the scale factor and the Ricci scalar as in (\ref{BigRipa}) and (\ref{BigRipRG}).
We assume that for the large curvature, $I(R)$ behaves as
\begin{align}
\label{choice-I(R)-ME}
I(R) \sim I_s R^\alpha,
\end{align}
where $I_s$ and $\alpha$ are constants. Thus, with the choice of
$I (R)$ in Eq.~(\ref{choice-I(R)-ME}), $\rho_\mathrm{eff}$ in
Eq.~(\ref{rhoeff-ME}) behaves as $\left(t_s - t\right)^{-2\alpha + 4h_s}$ while from the l.h.s. of the Friedmann equation, i.e.,
Eq.~(\ref{efe2-ME}) $\rho_\mathrm{eff}$ behaves as $\left(t_s - t\right)^{-2}$. From the consistency in $\rho_\mathrm{eff}$, we have,
\begin{align}
\label{eqn-alpha-h0}
 - 2 = -2 \alpha + 4h_s, \quad \mbox{i.e., }\quad \alpha = 1+ 2h_s\, .
\end{align}

Now, from Eq.~(\ref{efe2-ME}) one can write,
\begin{align}
\frac{3h_s^2}{\kappa^2} =&\, I_s \left( 12h_s^2 + 6h_s \right)^{\alpha -2} \nonumber \\
&\, \times \left\{ \frac{\left( 12h_s^2 + 6h_s \right)^2}{2} 
+ 3\left[ - \alpha \left( 12h_s^2 + 6h_s \right)\left( h_s + 5h_s^2 \right)
+ 6\alpha \left( \alpha -1 \right) h_s \left( 2h_s + 4 h_s^2 \right)\right]
\right\} \frac{|\bar{B}|^2}{a_s^4}~,
\end{align}
which can be simplified with the use of $\alpha$ from (\ref{eqn-alpha-h0}) as
\begin{align}
\frac{3h_s^2}{\kappa^2} = - \frac{I_s h_s\left( 12h_s^2 + 6h_s \right)^\alpha
|\bar{B}|^2}{2 a_s^4}\, ,
\end{align}
which clearly infers that $I_s$ must be negative.
This tells that $I(R)$ is also negative
and therefore the gauge field becomes ghost and the model becomes physically inconsistent.

Thus, it follows from Eqs.~(\ref{choice-I(R)-ME}) and
(\ref{eqn-alpha-h0}) that the Big rip singularity as described
through the Hubble rate in Eq.~(\ref{H-F(R)-big-rip}) may
emerge when for the large curvature, $I(R)$ behaves as
$R^{1+2h_s}$. However, for other choices of $I (R)$, we do not get
the Big rip singularity. We in fact note that
for $I(R) = I_s R^\alpha$, $H=h_s/\left(t_s - t\right)$ is an exact solution.

\newpage 
\begin{center}
 (ii) {\it Other types of singularities}
\end{center}

We now focus on other types of singularities that may arise for a more general choice of $I (R)$.
Thus, we consider the general case where the Hubble parameter takes the form of (\ref{H-F(R)-big-rip}).
For this choice of the Hubble rate, one can find the scale factor and the scalar curvature as follows
\begin{align}
\label{scale-factor-R-general-ME}
a = a_s \exp \left[ \frac{h_s}{\beta-1}
\left( t_s -t \right)^{-\left(\beta -1 \right)} \right]\, , \quad R = 6h_s \left[\beta + 2h_s \left( t_s -t \right)^{-\left(\beta-1 \right)}
\right] \left( t_s -t \right)^{-\left(\beta+1 \right)}\,,
\end{align}
where $a_s$ is a positive constant. Similar to the previous case,
we assume that in the large curvature, $I (R)$ behaves as in
Eq.~(\ref{choice-I(R)-ME}). We note that the case with $\beta >1$
is unphysical because, for $\beta >1$, $a \to \infty$ in the limit
$t \rightarrow t_s$, and as a consequence, $\rho_\mathrm{eff} \to 0$ and $p_\mathrm{eff} \to 0$ since
$\rho_\mathrm{eff} \propto a^{-4}$ and $p_\mathrm{eff} \propto a^{-4}$. On the other hand, $H\to \infty$ as $t \rightarrow t_s$. Thus, we see that for
$\beta >1$, Eqs.~(\ref{efe1-ME}) and (\ref{efe2-ME}) are not satisfied
leading to a unphysical case. Therefore, we restrict ourselves to
$\beta <1$ and explore the possible singularities as follows.

\begin{itemize}
\item If $0< \beta <1$ and $\alpha > 0$, then $\rho_\mathrm{eff}$ in
Eq.~(\ref{efe1-ME}) evolves as $ \rho_\mathrm{eff} \propto \left(t_s
- t\right)^{-\alpha \left( \beta + 1 \right)}$, however, looking
at Eq.~(\ref{efe2-ME}), we see that $\rho_\mathrm{eff}$ evolves as
$\rho_\mathrm{eff} \propto \left(t_s - t\right)^{-2\beta}$.
Therefore, from the consistency, we must have,
\begin{align}
\label{relation-alpha-beta-ME}
 - 2\beta = -\alpha \left( \beta + 1 \right) \ \Longrightarrow \
\beta = \frac{\alpha}{2-\alpha}, \quad \mbox{or} \quad
\alpha = \frac{2\beta}{\beta+1}\,.
\end{align}

Now, from Eq.~(\ref{efe2-ME}), one can find,
\begin{align}
\label{EM-sp1}
\frac{3h_s^2}{\kappa^2} = - \frac{I_s \left( 6h_s \beta \right)^\alpha \left( 1-\beta \right)
|\bar{B}|^2}{2 a_s^4 \left( \beta + 1 \right)}\,,
\end{align}
where we have used Eq.~(\ref{relation-alpha-beta-ME}). Notice that
Eq.~(\ref{EM-sp1}) demands $I_s$ to be negative. Thus, in this
case (i.e., $\alpha > 0$ and $0< \beta <1$), we see that, in the
limit $t \to t_s$, $a \to a_s$, $R \to \infty$, $\rho_\mathrm{eff}
\to \infty$, and $|p_\mathrm{eff}| \to \infty$. Hence, the Type
III singularity appears.
\item If $-1< \beta <0$ and $\left( \beta-1 \right)/\left( \beta+1 \right) < \alpha <0$, then in the limit $t \to t_s$, $a
\to a_s$, $R \to \infty$, $\rho_\mathrm{eff} \to 0$, and
$|p_\mathrm{eff}| \to \infty$. Although we can see that the final
value of $\rho_\mathrm{eff}$ is not a finite value but it vanishes
($\rho_\mathrm{eff} \rightarrow 0$ as $t \rightarrow t_s$),
however, this singularity can be considered to be the singularity
of Type II. Because, when $I$ and $H$ are given by $I = 1 + I_s R^\alpha$ and $H=H_s+h_s \left(t_s -t\right)^{-\beta}$
(where $H_s$ is a constant), respectively, then in the limit $t \rightarrow t_s$, $\rho_\mathrm{eff} \to \rho_s$, a constant, where
$\rho_{s}$ can be found from Eqs.~(\ref{rhoeff-ME}) and
(\ref{efe1-ME}) as $\rho_s = 3H_s^2/\kappa^2 = |\bar{B}|^2/\left(2 _s^4\right)$.
\item If $\beta <-1$, then
in the limit $t \to t_s$, $a \to a_s$ and $R \to 0$. Now,
for $\alpha \geq \left( \beta-1 \right)/\left( \beta+1 \right)$,
in the limit $t \to t_s$, $\rho_\mathrm{eff} \to 0$,
$|p_\mathrm{eff}| \to 0$, and higher derivatives of $H$
diverge. That means, Type IV singularity appears.
\end{itemize}

We further note here that for $-1< \beta <0$ and $\alpha > 0$,
$\rho_\mathrm{eff} \to \infty$, but $H \to 0$ in the limit $t \rightarrow t_s$, which shows that Eq.~(\ref{efe1-ME}) is not satisfied.
That means, the case with $\alpha >0$ and $-1< \beta <0$ is not physical in the sense that the Friedmann equation
is inconsistent.
We also find that If $\alpha < 0$ and $0< \beta <1$, then $\rho_\mathrm{eff} \to 0$, but $H \to \infty$.
Hence, Eq.~(\ref{efe1-ME}) is again not satisfied.
Further, if $-1< \beta <0$ and $\alpha \leq \left( \beta-1 \right)/\left( \beta+1 \right)$,
then we have unphysical scenario because in this case, in the limit $t \to t_s$, $a \to a_s$, $R \to \infty$,
$\rho_\mathrm{eff} \to 0$, and $|p_\mathrm{eff}| \to 0$, but
$\dot{H} \to \infty$. Thus, we see that Eq.~(\ref{efe2-ME}) is not
satisfied.

We mention that if $I(R)$ is a constant\footnote{Let us note
that the case $I(R)=1$ corresponds to the ordinary Maxwell theory.}, we do not find any finite-time future singularity.
We close this section with two examples of $I (R)$ as follows. Suppose
$I(R)$ takes the Hu--Sawicki
form given by~\cite{Hu:2007nk}
\begin{align}
I(R) = I_{\mathrm{HS}}(R) \equiv \frac{c_1 \left(R/m^2 \right)^n}
{c_2 \left(R/m^2 \right)^n + 1}\,,
\label{ME-HS-model}
\end{align}
where $c_1$ and $c_2$ are dimensionless constants, $n$ is a
positive constant, and $m$ denotes a mass scale. We see that the
model in Eq.~(\ref{ME-HS-model}) satisfies the following conditions: (i)
$\lim_{R\to\infty} I_{\mathrm{HS}}(R) = c_1/c_2 = \mbox{const}$ and (ii) $ \lim_{R\to 0} I_{\mathrm{HS}}(R) = 0$.
The second example of $I (R)$ having the same features as with the
Hu--Sawicki form of $I (R)$ in Eq.~(\ref{ME-HS-model}) takes the
form~\cite{Nojiri:2007as}
\begin{align}
I(R) = I_{\mathrm{NO}}(R) \equiv
\frac{\left[ \left(R/M^2\right) - \left(R_\mathrm{c}/M^2\right) \right]^{2q+1}
+ {\left(R_\mathrm{c}/M^2\right)}^{2q+1}} {c_3 + c_4 \left\{
\left[ \left(R/M^2\right) - \left(R_\mathrm{c}/M^2\right) \right]^{2q+1} + {\left(R_\mathrm{c}/M^2\right)}^{2q+1} \right\}}\,,
\label{ME-NO-model}
\end{align}
where $c_3$ and $c_4$ are dimensionless constants, $q$ is a
positive integer, $M$ denotes a mass scale, and $R_\mathrm{c}$ is
current curvature. One can check that $I (R)$ given in
Eq.~(\ref{ME-NO-model}) satisfies the conditions: (i)
$\lim_{R\to\infty} I_{\mathrm{NO}}(R) = 1/c_4 = \mbox{const}$ and
(ii) $\lim_{R\to 0} I_{\mathrm{NO}}(R) = 0.$ If $\beta < -1$ and
$I(R)$ is given either by $I_{\mathrm{HS}}(R)$ in
Eq.~(\ref{ME-HS-model}) or $I_{\mathrm{NO}}(R)$ in
Eq.~(\ref{ME-NO-model}), in the limit $t \to t_s$, $a \to a_s$, $R \to 0$, $\rho_\mathrm{eff} \to 0$, and $|p_\mathrm{eff}| \to 0$.
In addition, higher derivatives of $H$ diverge. Thus, the Type IV
singularity emerges. Thus, we see that the Maxwell theory which is
non-minimally coupled to Einstein gravity may produce finite-time
future singularities depending on the form of non-minimal
gravitational coupling.

We conclude that the general conditions on $I(R)$ for which the finite-time future
singularities characterized by Eq.~(\ref{H-F(R)-big-rip})
cannot emerge are that in the limit $t \to t_s$, $I(R) \to \bar{I}$
(where $\bar{I} (\neq 0)$ is a finite constant),
$I^{\prime}(R) \to 0$, $I^{\prime\prime}(R) \to 0$, and
$I^{\prime\prime\prime}(R) \to 0$.

\begin{center}
 {\bf J.1:} Influence of non-minimal gravitational coupling on the
finite-time future singularities in modified gravity
\end{center}

In this section, we discuss the influence of non-minimal
gravitational coupling of the electromagnetic field on the $F (R)$
gravity theory \cite{Bamba:2008ut}. This is a generalization of
the previous section \ref{sec-Maxwell-Einstein-gravity}. In this
case, the total energy density and pressure of the Universe are contributed as,
$\rho_{\mathrm{tot}} = \rho_\mathrm{eff} + \rho_{\mathrm{MG}}$ and
$p_{\mathrm{tot}} = p_\mathrm{eff} + p_{\mathrm{MG}}$, respectively, where $\rho_\mathrm{eff}$ and
$p_\mathrm{eff}$ are given by Eqs.~(\ref{rhoeff-ME}) and
(\ref{peff-ME}), respectively. And $\rho_{\mathrm{MG}}$ and
$p_{\mathrm{MG}}$ can be found from Eqs.~(\ref{rhoeff-F(R)}) and
(\ref{peff-F(R)}) as follows
\begin{align}
\rho_{\mathrm{MG}} =&\,
\frac{1}{\kappa^2}\left[-\frac{1}{2}f(R) + 3\left(H^2 + \dot H\right) f'(R)
 - 18 \left(4H^2 \dot H + H \ddot H\right)f''(R)\right]\,,
\label{sp-new-ME-rhoeff-1} \\
p_{\mathrm{MG}} =&\,
\frac{1}{\kappa^2}\left[\frac{1}{2}f(R) - \left(3H^2 + \dot H \right)f'(R)
+ 6 \left(8H^2 \dot H + 4{\dot H}^2
+ 6 H \ddot H + \dddot H \right)f''(R) 
+ 36\left(4H\dot H + \ddot H\right)^2 f'''(R) \right]\,.
\label{sp-new-ME-rhoeff-2}
\end{align}
In this case, it follows from Eqs.~(\ref{FLRWs}), (\ref{rhoeff-ME}), (\ref{peff-ME}),
(\ref{sp-new-ME-rhoeff-1}) and (\ref{sp-new-ME-rhoeff-2}) that the flat Friedmann
and Raychaudhuri equations can be given by \cite{Bamba:2008ut}
\begin{align}
\frac{3}{\kappa^2}H^2 = \rho_\mathrm{tot}
=&\, \left\{ \frac{I(R)}{2} + 3\left[
-\left( 5H^2 + \dot{H} \right) I^{\prime}(R) + 6 H \left( 4H\dot{H} + \ddot{H} \right) I^{\prime\prime}(R)
\right] \right\} \frac{|\bar{B}|^2}{a^4} \nonumber \\
&\, +\frac{1}{\kappa^2}\left[-\frac{1}{2}\left( F(R)-R \right) + 3\left(H^2 + \dot H\right) \left( F^{\prime}(R)-1 \right)
 -18 \left(4H^2 \dot H + H \ddot H\right) F^{\prime\prime}(R)\right]\,,
\label{sp-new-ME-flrw1} \\
 -\frac{1}{\kappa^2}\left(2\dot H + 3H^2\right) = p_\mathrm{tot} =&\,
\biggl[ -\frac{I(R)}{6} + \left( - H^2 + 5\dot{H} \right) I^{\prime}(R) - 6 \left( -20H^2\dot{H} + 4\dot{H}^2 - H\ddot{H} + \dddot{H}
\right) I^{\prime\prime}(R) \nonumber \\
&\, -36\left( 4H\dot{H} + \ddot{H} \right)^2 I^{\prime\prime\prime}(R) \biggr]
\frac{|\bar{B}|^2}{a^4}
+\frac{1}{\kappa^2} \biggl[ \frac{1}{2}\left( F(R)-R \right)
 -\left(3H^2 + \dot H \right) \left( F^{\prime}(R)-1 \right) \nonumber \\
&\, + 6 \left(8H^2 \dot H + 4{\dot H}^2
+ 6 H \ddot H + \dddot H \right)F^{\prime\prime}(R) + 36\left(4H\dot H + \ddot H\right)^2 F^{\prime\prime\prime}(R) \biggr]\,.
\label{sp-new-ME-flrw2}
\end{align}

Now, using Eqs.~(\ref{sp-new-ME-flrw1}) and (\ref{sp-new-ME-flrw2}), we find
\begin{align}
\Bigg[ \frac{I(R)}{3} + 2\left(-8H^2 + \dot{H} \right) I^{\prime}(R) + 6\left(32 H^2 \dot{H} -4\dot{H}^2 + 4H\ddot{H} -\dddot{H} \right)
I^{\prime\prime}(R)
 -36\left(4H\dot{H} + \ddot{H} \right)^2 I^{\prime\prime\prime}(R)
\Bigg] \frac{|\bar{B}|^2}{a^4} \nonumber \\
+\frac{1}{\kappa^2} \left[ 2\dot{H}F^{\prime}(R)
+6\left( -4H^2 \dot{H} +4\dot{H}^2 + 3H\ddot{H} + \dddot{H} \right)
F^{\prime\prime}(R)
+36\left(4H\dot H + \ddot H\right)^2 F^{\prime\prime\prime}(R)
\right] = 0\, .
\label{sp-new-ME-01}
\end{align}

Recalling that in modified gravity with the ordinary Maxwell
theory, if $F (R)$ behaves as in Eq.~(\ref{sec-F(R)-expression-F(R)-sp}), i.e.,
\begin{align}
\label{sp-new-ME-02}
F(R) \sim F_0 R + F_1 R^q\, , \quad \mbox{where $q$ is a constant}\, ,
\end{align}
a finite-time future singularity appears.
Thus, one may investigate the possibility of finite-time singularities when the
non-minimal gravitational electromagnetic theory is taken into
account in this $F(R)$-gravity model.

Now, when by putting $\beta=-u$ in (\ref{Hsin}), $H$ behaves as \cite{Bamba:2008ut}
\begin{align}
\label{sp-new-ME-03}
H \sim h_s\left( t_s -t \right)^u\,,
\end{align}
where $u$ is an positive integer, then there exists no finite-time
future singularity. In what follows, we consider the case when $u \geq 2$.
From Eq.~(\ref{sp-new-ME-03}), one finds that
\begin{align}
R \sim 6\dot{H}
\sim -6u h_s \left( t_s -t \right)^{u-1}\,,
\quad
a \sim a_s \exp \left[-\frac{h_s}{u+1}
\left( t_s -t \right)^{u+1} \right]\,,
\label{sp-new-ME-04}
\end{align}
where in the expression of $R$ we have considered only the leading term.

Now, we examine the non-minimal gravitational coupling of the
electromagnetic field characterized by $I(R)$ which gives the
solution in Eq.~(\ref{sp-new-ME-04}). We again consider that $I(R)$ behaves as in Eq.~(\ref{choice-I(R)-ME}).
Concerning Eq.~(\ref{sp-new-ME-01}), its first and second terms in the left
hand side are coming from the non-minimal
gravitational electromagnetic coupling and the modified-gravity
sectors, respectively. When $t$ is close to $t_s$, the leading term of the
non-minimal gravitational electromagnetic coupling sector evolves
as $\left( t_s -t \right)^{\left( u-1 \right) \left( \alpha-1 \right) -2}$.
On the other hand, if $q \leq 1$, or $q>1$ and $u < q/\left(q-2 \right)$, then the modified gravity part, i.e.,
the second term of the l.h.s. of Eq.~(\ref{sp-new-ME-01})
evolves as $\left( t_s -t \right)^{\left( u-1 \right) \left( q-1 \right) -2}$.
The consistency thus gives $\alpha = q$.
Additionally, if we consider that the leading term of the
non-minimal gravitational electromagnetic coupling sector should
not diverge in the limit $t \to t_s$, then $\alpha$ must satisfy
the relation $\alpha \geq \left(u+1\right)/\left( u-1 \right)$.
Taking only the leading terms in Eq.~(\ref{sp-new-ME-01}) and
using $\alpha = q$, one gets \cite{Bamba:2008ut}
\begin{align}
I_s = \frac{a_s^4 F_1}{|\bar{B}|^2 \kappa^2}\,.
\label{sp-new-ME-05}
\end{align}
When $q>1$ and $u \geq q/\left(q-2 \right)$, then the leading term of
the modified-gravity sector evolves as
$\left( t_s -t \right)^{u-1}$.
Thus, the consistency here gives $\alpha = 2u/\left( u-1 \right)$.
In this case, considering only the leading terms in Eq.~(\ref{sp-new-ME-01}) and
using $\alpha = 2u/\left( u-1 \right)$, one gets \cite{Bamba:2008ut}
\begin{align}
I_s = \frac{a_s^4 F_0}{|\bar{B}|^2 \kappa^2}
\frac{u-1}{6 u^2 \left( u+1 \right)}
\left(-6h_s u \right)^{-2/\left(u-1 \right)}\,.
\label{sp-new-ME-06}
\end{align}
Consequently, one may observe that
the non-minimal gravitational coupling of the electromagnetic field given by
$I(R) \sim I_s R^\alpha$ with the specific values of $I_s$ and $\alpha$
stated above can avoid the finite-time future singularities arising
in pure modified gravity.

However,, on the contrary, it may also happen that the non-minimal
gravitational coupling of the electromagnetic field is unable to
remove the finite-time singularity while such coupling could make
the singularity stronger (or weaker). We further assume that for
the large curvature, $I(R)$ behaves exactly as in
Eq.~(\ref{choice-I(R)-ME}). Now, using the result of
Eq.~(\ref{F(R)-BR-section-sp-new1}), we consider that for large
curvature, $F(R)$ behaves as $F(R) \propto R^{\bar{q}}$, where
$\bar{q} = 1 - \alpha_-/2 <1$. In this case, the Big Rip
singularity characterized as in eq.~(\ref{H-F(R)-big-rip})
appears. It follows from Eq.~(\ref{eqn-alpha-h0}) that if $\alpha =1 + 2h_s$,
then in the limit $t \to t_s$, $\rho_\mathrm{eff} \to \infty$ and $|p_\mathrm{eff}| \to \infty$
which means that the singularity becomes stronger when the non-minimal gravitational
coupling of the electromagnetic field is taken into account.

Let us now consider a more general case. We assume $H$ evolves as
in Eq.~(\ref{Hsin}) with $0<\beta<1$ which
corresponds to a Type III singularity and this singularity appears
when $F(R)$ takes the form of
Eq.~(\ref{F(R)-form-beta-between-0-1}), and $\alpha >0$, then in
the limit $t \to t_s$, $\rho_\mathrm{eff} \to \infty$ and
$|p_\mathrm{eff}| \to \infty$. Thus, we see that the non-minimal
gravitational coupling of the electromagnetic field makes the
singularity stronger. If $-1<\beta<0$, then there exists a Type II
singularity, which can appear when the $F(R)$ takes the form of
Eq.~(\ref{F(R)-form-beta-less-0}), and $\left( \beta-1 \right)/\left( \beta+1 \right) < \alpha <0$,
then $\rho_\mathrm{eff} \to 0$ and $|p_\mathrm{eff}| \to \infty$.
For $|p_{\mathrm{tot}}| > |p_{\mathrm{MG}}|$ (or $|p_{\mathrm{tot}}| < |p_{\mathrm{MG}}|$),
the non-minimal gravitational coupling of the
electromagnetic field can make the singularity stronger (weaker).
In conclusion, we observe that the non-minimal gravitational
coupling in the Maxwell theory could qualitatively influence the
future dynamics of the Universe.

\subsection{Semi-classical Gravity}
\label{sec-singularities-semi-classical}

In Einstein's GR where the gravitational equations are given by
$G_{\mu \nu} = 8 \pi G T_{\mu \nu}$, the spacetime geometry and
the matter distribution are classical in nature. However, in a
physical scenario where the matter evolution follows the
principles of quantum mechanics, there the energy-momentum tensor
should be an operator $\hat{T}_{\mu \nu}$ in the quantum world.
Thus, in order to realize a consistent picture, the spacetime
geometry $G_{\mu \nu}$ needs to be quantized
\cite{Belenchia:2018szb,Wald:2020jow}, however, this theory is
under construction. While on the other hand, an alternative
semi-classical approach can be furnish where the spacetime
geometry remains classical but it is sourced by the quantum
expectation of the energy-momentum tensor operator
\cite{Liu:2022wog}, i.e., $G_{\mu\nu}=8\pi\langle \psi|\hat T_{\mu\nu}|\psi\rangle$
($|\psi\rangle$ is the quantum state of
matter which evolves with the spacetime) proposed by M{\"o}ller
and Rosenfeld\,\cite{Mueller1962,Rosenfeld:1963hjy}.
In this section, we discuss the finite-time future singularities appearing
in semi-classical gravity. We follow the notation of
\cite{Davies:1977ti} where we consider the spatially flat FLRW
Universe characterized by the line element
$ds^2=dt^2-a^2(t)\delta_{ij} dx^i dx^j$, and thus, the
stress-tensor for a perfect fluid is given by $T_{\mu}^{\nu}= -p\delta_{\mu}^{\nu}+(\rho+p)u^{\mu}u_{\nu}$.
Then, if one considers some massless fields conformally coupled to gravity,
the vacuum stress-tensor acquire an anomalous trace given by
\cite{Davies:1977ti}
\begin{align}
T_\mathrm{vac}= \alpha \Box R-\frac{\beta}{2}G\, ,
\end{align}
where $R$ is the Ricci scalar and $G$ is the Gauss-Bonnet invariant defined in (\ref{GB}).
In terms of the Hubble rate one has
\begin{align}
T_\mathrm{vac}=6\alpha\left(\dddot{H}+12H^2\dot{H}+7H\ddot{H}+4\dot{H}^2 \right)
 -12\beta(H^4+H^2\dot{H})\, .
\end{align}

On the other hand, the coefficients $\alpha$ and $\beta$ are fixed by the regularization process.
For instance, if one uses adiabatic regularization one has \cite{Fischetti:1979ue}
\begin{align}
\alpha= &\, \frac{1}{2880\pi^2}(N_0+6N_{1/2}+12N_1)>0\, , \nonumber \\
\beta= &\, - \frac{1}{2880\pi^2}(N_0+\frac{11}{2}N_{1/2}+62N_1)<0\, ,
\end{align}
while point splitting gives \cite{Davies:1977ti}
\begin{align}
\alpha= &\, \frac{1}{2880\pi^2}(N_0+3N_{1/2}-18N_1)\, , \nonumber \\
\beta= &\, - \frac{1}{2880\pi^2}(N_0+\frac{11}{2}N_{1/2}+62N_1)\, ,
\end{align}
where $N_0$ being the number of scalar fields, $N_{1/2}$ the number of four-component neutrinos and $N_1$ the number of electromagnetic fields.

What is important, as pointed out in Ref.~\cite{Wald:1978pj}, is
that the coefficient $\alpha$ is arbitrary and it is influenced by
the regularization method and also by the fields present in the
Universe, but $\beta$ is independent of the regularization scheme
and it is always negative.

Now, we are interested in the value of the vacuum energy density, namely $\rho_\mathrm{vac}$.
Since the trace is given by $T_\mathrm{vac}=\rho_\mathrm{vac}-3p_\mathrm{vac}$, inserting this expression
in the conservation equation $\dot{\rho}_\mathrm{vac}+3H(\rho_\mathrm{vac}+p_\mathrm{vac}) = 0$ one gets
\begin{align}
 \dot{\rho}_\mathrm{vac}+4H\rho_\mathrm{vac}-HT_\mathrm{vac}=0\, ,
\end{align}
which is a first order linear differential equation which could integrated using the variation of constants method, leading to
\begin{align}
\rho_\mathrm{vac}=6\alpha\left(3H^2\dot{H}+H\ddot{H}- \frac{1}{2}\dot{H}^2 \right)-3\beta H^4+Ca^{-4}\, .
\end{align}
where $C$ is a constant of integration which for the flat FLRW spacetime vanishes.
This could be understanding as follows: for a static spacetime $\rho_\mathrm{vac}$ reduces to $Ca^{-4}$,
and the flat FLRW spacetime reduces to the Minkowski one, for which $\rho_\mathrm{vac}=0$.
Therefore, in semi-classical gravity the Friedmann equation becomes
\begin{align}
\label{semiclassical}
H^2=\frac{(\rho+\rho_\mathrm{vac})\kappa^2}{3}\, .
\end{align}

Now first of all, we consider the simplest case: $\alpha=0$.
The Friedmann equation will become
\begin{align}
\label{Friedmannsemi}
H^2=\frac{\rho\kappa^2}{3}-\beta\kappa^2H^4\, ,
\end{align}
which implies that $H^2\leq \frac{1}{-\beta\kappa^2}= \frac{1}{|\beta|\kappa^2}$, because the energy density $\rho$ must be positive.

Differentiating Eq. (\ref{Friedmannsemi}) with respect to the
cosmic time and using the conservation equation, one gets the
following Raychaudhuri equation
\begin{align}
\dot{H}=-\frac{1}{2}\frac{(\rho+p)\kappa^2}{(1-{2|\beta|}H^2\kappa^2)}\, ,
\end{align}
which for a fluid with linear EoS, $p=(\gamma-1)\rho$ could be
written as
\begin{align}
\label{Raychaudurysemi}
\dot{H}=-\frac{3\gamma}{2}H^2\frac{(1-H^2|\beta|\kappa^2)}
{(1-{2}H^2|\beta|\kappa^2)}\, .
\end{align}
Then, for a phantom fluid ($\gamma<0$), there are two different situations:
\begin{enumerate}
\item $0< H < \frac{1}{\sqrt{2|\beta|}\kappa}$. In this case $\dot{H}$ becomes positive, and thus, the Hubble rate always increases.
Since when $H$ approaches to $\frac{1}{\sqrt{2|\beta|}\kappa}$ the velocity of the Hubble rate increases,
one can deduce that the value $\frac{1}{\sqrt{2|\beta|}\kappa}$ is reached in a finite-time $t_s$, at with $\dot{H}$ diverges
(and also the Ricci scalar), meaning that we have a Type IV singularity.
\item $\frac{1}{\sqrt{2|\beta|}\kappa}<H< \frac{1}{\sqrt{|\beta|}\kappa}$. Now, $\dot{H}$ becomes negative,
and once again the value of the Hubble rate $\frac{1}{\sqrt{2|\beta|}\kappa}$ is reached in a finite-time, and thus,
$\dot{H}$ becomes singular at finite-time and a Type IV singularity is obtained.

In fact, the time $t_s$ could be analytically calculated because Eq. (\ref{Raychaudurysemi}) may be integrated and leads to \cite{Haro:2011zzb}
\begin{align}
 -\frac{1}{H(t)}+ \frac{2}{h}\ln\left( \frac{h-H(t)}{h-H_i} \frac{h+H(t)}{h+H_i} \right)
=-\frac{3\gamma}{2}(t-\bar{t})\, ,
\end{align}
where to simplify, we have introduced the notation $h\equiv \frac{1}{\sqrt{|\beta|}\kappa}$, and we have defined
$\bar{t}=-\frac{2}{3\gamma H_i}$ with $H_i=H(0)$. Then, when $H=\frac{h}{\sqrt{2}}$ one gets
\begin{align}
t_s=\bar{t}+\frac{2\sqrt{2}}{3\gamma h}\left[1-\sqrt{2}\ln\left(
(\sqrt{2}-1)^2\frac{h+H_i}{h-H_i} \right) \right]>0\, .
\end{align}
\end{enumerate}

Finally, we consider the case $\alpha\not=0$, and we look for future singular solutions whose leading term for the energy density is $\rho_s(t_s-t)^{\mu}$.
Inserting this expression in the conservation equation
$\dot{\rho}=-3H(\rho+p)$, one gets the leading term of the Hubble rate
\begin{align}
H(t)\sim \frac{\mu}{3\gamma (t_s-t)}\, .
\end{align}
Plugging this expression in the modified Friedmann equation and
picking up the leading term, one gets $\mu=-4$, meaning that
$\gamma$ must be negative (i.e. phantom fluid) in order to have a
positive Hubble rate, but in that case one has $\rho_s<0$ when
$\alpha$ is positive and $\beta$ is negative, obtaining a negative
value of the energy density of the form,
\begin{align}\label{sing}
\rho_s=-\frac{\mu^2}{\gamma^2}\left[\alpha\left(\frac{2\mu}{3\gamma}+1 \right) - \beta\frac{\mu^2}{27\gamma^2}\right].
\end{align}
Thus, we see that for positive values of the parameter $\alpha$,
this is unrealistic, and indicates that when $\alpha>0$ there is
no future singular solutions in the expanding phase.

However, if one allows all the values of the parameter $\alpha$,
as we can see from the equation (\ref{sing}), then there exists
finite-time future singularities. For example, for a phantom fluid
when $\alpha<0$ and $\beta<0$ with
$\alpha\left(\frac{2\mu}{3\gamma}+1 \right)
-\beta\frac{\mu^2}{27\gamma^2}<0$, one gets $\mu=-4$ but now with
$\rho_s>0$. In fact, a very deep study was done in
Ref.~\cite{Haro:2011zzb}, where it was shown that the Universe
bounces, and for $-1<\frac{\beta}{3\alpha}<0$, the Universe
develops a future singularity at finite-time in the contracting
phase. On the contrary, for $\frac{\beta}{3\alpha}<-1$ one gets a
Universe bouncing infinitely many times, and thus, without future
singularities.

Now, we consider the non-linear EoS, $p=-\rho-f(\rho)$, with
$f(\rho)=\frac{A}{\sqrt{3}\kappa}\rho^{\nu+\frac{1}{2}}$ as in
Section~\ref{sec-classification-singularities} (see also the
analysis done in \cite{Nojiri:2004ip}). As discussed in
Section~\ref{sec-classification-singularities}, for this
non-linear EoS, the Hubble rate and the energy density evolve
respectively as $H(t)=h_s(t_s-t)^{-\frac{1}{2\nu}}$ and
$\rho(t)={\rho}_s(t_s-t)^{-1/\nu}$ when $\nu$ does not vanish,
and one can quickly recall that:
\begin{enumerate}
\item For $\nu<-1/2$ and $A<0$, there is a Type II (Sudden singularity).
\item For $-1/2<\nu<0$ and $A<0$, there is a Generalized Sudden singularity.
 \item For $0<\nu<1/2$ and $A>0$ (phantom fluid), there is a Big Rip singularity.
 \item For $\nu>1/2$ and $A>0$ (phantom fluid), there is a Big Freeze singularity.
\end{enumerate}
Then, allowing all the values of the parameter $\alpha$ and taking
$\beta<0$, our goal is to investigate if quantum effects are able
to avoid or mitigate these singularities.

Once again we look for future solutions whose leading term of the energy density is $\rho(t)\sim \rho_s(t_s-t)^{\mu}$. Inserting it into the conservation equation
$\dot{\rho}= 3Hf(\rho)$, one gets
\begin{align}
H(t)\sim -\frac{\mu\kappa\rho_s^{\frac{1}{2}-\nu}}{\sqrt{3}A}(t_s-t)^{\frac{\mu}{2}-1-\mu\nu }\equiv
 \bar{h}_s(t_s-t)^a\, .
\end{align}

Now, plugging the leading terms of the energy density and the Hubble rate in the semi-classical Friedmann equation (\ref{semiclassical}), one realizes
a variety of possibilities:
\begin{enumerate}
\item $\mu<0 ~(\Longrightarrow A>0)$ and $a\geq 0$:
In this case, for a phantom fluid, the energy density is singular but the Hubble rate vanishes at the singularity or is constant ($a=0$).
The dominant terms of $\rho_\mathrm{vac}$ are
\begin{align}
 -3\alpha\dot{H}^2\sim -3\alpha{a^2}\bar{h}_s^2 (t_s-t)^{2a-2}\, ,
\end{align}
and
\begin{align}
6\alpha H\ddot{H}\sim 6\alpha a(a-1)\bar{h}_s^2 (t_s-t)^{2a-2}\, .
\end{align}

Then, equaling these terms to $-\rho$ one gets
\begin{align}
2a-2=\mu \Longleftrightarrow \mu\nu=-2\, ,
\end{align}
and $\rho_s=3\alpha a(a-2)\bar{h}_s^2$, meaning that this solution is only obtained for negative values of the parameter $\alpha$, because as we will see $0\leq a<1$.

On the other hand, the conditions $\mu<0$ and $a\geq 0$
are equivalent to
\begin{align}
 -2\leq \mu<0\ \Longleftrightarrow\ 1\leq \nu<\infty\, ,
\end{align}
where we have used the relation $\mu\nu=-2$. And since $a=\frac{\mu}{2}-1-\mu\nu=\frac{\mu}{2}+1$ one conclude that $0\leq a<1$ as we have already explained.

Finally, the leading terms of the Hubble rate and the energy density are given by
\begin{align}
\label{sing}
H(t)\sim -\frac{2\kappa\rho_s^{\frac{1}{2}-\nu}}{\sqrt{3}A\nu}(t_s-t)^{-\frac{1}{\nu}+1}, \quad
\mbox{and} \quad
\rho(t)\sim \rho_s(t_s-t)^{-\frac{2}{\nu}}\, .
\end{align}

\item $\mu<0~(\Longrightarrow A>0)$ and $-1<a<0$:
In this case the dominant terms of the energy density of the vacuum are, as in the case $a>0$,
$-3\alpha\dot{H}^2$ and $6\alpha H \ddot{H}$. In this case, we also have $\mu\nu=-2$ and
$\rho_s=3\alpha a(a-2)\bar{h}_s^2$, but now $\alpha$ must be positive because we are assuming that $a$ is negative.

On the other hand, the condition $-1<a<0$ leads to
\begin{align}
 -4<\mu<-2 \Longleftrightarrow \frac{1}{2}<\nu<1\, .
\end{align}

So, in this case for a phantom fluid, the Hubble rate and the energy density evolve as in Eq. (\ref{sing}), but both the quantities are singular at $t=t_s$.

\item $\mu<0~(\Longrightarrow A>0)$ and $a=-1$: All the terms of
$\rho_\mathrm{vac}$ scale as $(t_s-t)^{-4}$, which means that $\mu$
must be $-4$ and $\nu=1/2$, so the EoS is linear and this case
reduces to the one previously studied.

\item $\mu<0~(\Longrightarrow A>0)$ and $a<-1$:
Now the dominant term of $\rho_\mathrm{vac}$ is
$-3\beta H^4\sim -3\beta \bar{h}_s^4 (t_s-t)^{4a}$,
and the leading terms must satisfy
\begin{align}
\rho_s(t_s-t)^{\mu}-3\beta \bar{h}_s^4(t_s-t)^{4a}=0\, ,
\end{align}
which means that $\mu=4a$ and $\rho_s=3\beta \bar{h}_s^4<0$, showing that no singular solutions arise in this case.

Note that the condition $\mu=4a$ together with $a<-1$ leads to
\begin{align}
\mu<-4 \quad \Longleftrightarrow \quad \frac{1}{4}<\nu<\frac{1}{2}\, .
\end{align}

\end{enumerate}

Summing up, for a phantom fluid with EoS,
$p=-\rho-\frac{A}{\sqrt{3}\kappa}\rho^{\nu+\frac{1}{2}}$, taking
into account the vacuum effects due to conformally coupled
massless fields, we have the following observations:
\begin{enumerate}
\item For $1\leq \nu<\infty$, there are future singular solutions where the energy density diverges but not the Hubble rate, when the parameter $\alpha$ is negative.
\item For $\frac{1}{2}<\nu<1$, there are future singular solutions where both the energy density and the Hubble rate diverge, when the parameter $\alpha$ is positive.
\item For $\nu=\frac{1}{2}$, it is the linear case, and, as we have already seen, only there are singular solutions when the condition
$\alpha\left(\frac{2\mu}{3\gamma}+1 \right) -\beta\frac{\mu^2}{27\gamma^2}<0$ with $\mu=-4$ and $\gamma=-A$ is satisfied.
\item For $\nu<\frac{1}{2}$, no future singular solutions arise.
\end{enumerate}

\section{Singularities in Braneworld Models}
\label{sec-singularities-braneworld}

In this section, we shall discuss the emergence of finite-time
singularities of unusual form and nature admitted in the
braneworld theory, a fascinating theory where it is argued that
our observable Universe could be a $1+3$-dimensional surface
(named as ``brane'') immersed into a $1+3+d$-dimensional spacetime
(named as ``bulk'') with standard model particles and fields
confined on the brane while gravity can freely access the bulk
\cite{Randall:1999ee,Randall:1999vf}. The braneworld gravity
caught the attention of the scientific community and the models have been investigated widely from both theoretical and observational perspectives,
see for instance Refs.
\cite{Nojiri:2000gb,Sahni:2000ubn,Gregory:2001dn,Himemoto:2001hu,Deffayet:2001pu,Sahni:2002dx,Nojiri:2003jn,Koley:2005nv,Pal:2005pa,Pal:2006hg,Pal:2006yi,Banerjee:2007qi,Ge:2007yu,Bandyopadhyay:2006yk,Banerjee:2007sg,Pal:2007ap,Das:2007qn,Sheykhi:2008qs,Pal:2007gp,Pal:2008hr,Sheykhi:2009zza,Feng:2009hr,Lombriser:2009xg,Chakraborty:2009zz,Mukherji:2008hs,Guha:2010vyk,Dutta:2010zza,Dutta:2010mp,Bazeia:2013uva,Banerjee:2011wk,Ghaffari:2014pxa,Biswas:2015zka,Dutta:2015jaq,Heydarzade:2015dba,Bazeia:2015oqa,Ghaffari:2015foa,Wang:2018jsw}
(also see the review articles on the braneworld gravity and
cosmology \cite{Brax:2003fv,Brax:2004xh,Maartens:2010ar}). One of
the pioneering works in braneworld geometry is the Randall--Sundrum
scenario described by the following action 
\cite{Brax:2003fv,Brax:2004xh}
\begin{align}
 S=\int dx^5\sqrt{-g^{(5)}}(2R^{(5)}+\Lambda_5)+\int dx^4\sqrt{-g}\sigma + S_\mathrm{matter}\, ,
\end{align}
where $g_{\mu\nu}^{(5)}$ is the bulk metric, $R^{(5)}$ the Ricci scalar in the bulk, $\Lambda_5$ is the bulk cosmological constant,
$\sigma$ is the brane tension and $S_\mathrm{matter}$ stands for the action of the matter sector as in (\ref{correspondence-action-unimodular-FR-1}).
For this action, corresponding Friedmann equation is given by
\begin{align}
H^2=\frac{\kappa^2 \rho}{3}\left(1+\frac{\rho}{2\sigma}\right)\, .
\end{align}
As we will see and study in Section~\ref{sec-singularities-LQG}, when the brane tension is negative, defining $\rho_c=-2\sigma$,
one obtains the so-called {\it holonomy corrected Friedmann equation}.
Since, we will study the negative case in Section~\ref{sec-singularities-LQG},
we consider here $\sigma>0$. Then, the Friedmann equation in the flat FLRW spacetime can be written as follows
\begin{align}
\rho=\sigma \left(-1+\sqrt{1+\frac{6\kappa^2H^2}{\sigma}} ~\right)\,,
\end{align}
where one can see that when the brane tension goes to infinite, one recovers the usual Friedmann equation of GR.

It is not difficult to see that when the brane tension is
positive, one obtains the same kind of singularities as in GR.
Here, we only show that for a linear EoS of the form $p=w\rho$ with $w<-1$, one
has the Big Rip singularity. Effectively, the conservation
equation is given by
\begin{align}
\dot{\rho}=-\sqrt{3\rho}\kappa \sqrt{1+\frac{\rho}{2\sigma}}\rho (1+w)\, ,
\end{align}
which for large value of the energy density reduces to
\begin{align}
 \dot{\rho}=-\sqrt{\frac{3}{2\sigma}}\kappa \rho^2 (1+w)\, ,
\end{align}
whose solution is given by
\begin{align}
 \rho(t)=-\sqrt{\frac{2\sigma}{3}}\frac{1}{\kappa(1+w)}\frac{1}{t_s-t}\, ,
\end{align}
where we have introduced the notation
$t_s\equiv t_0-\sqrt{\frac{2\sigma}{3}}\frac{1}{\kappa\rho_0(1+w)}$, being $\rho_0$ the energy density at the present time $t_0$.
Then, since $w<-1$, we can see that $t_s>t_0$, and thus, the singularity appears as a Big Rip one.

Now we consider the more elaborated action \cite{Shtanov:2002ek}
\begin{align}
S=M^3\sum_i\left[ \int_{\mathrm{bulk}}(\mathcal{R}-2\Lambda_i)-2\int_{\mathrm{brane}} K\right]
+\int_{\mathrm{brane}}(m^2R-2\sigma)+\int_{\mathrm{brane}}
L(h_{\alpha\beta}, \phi)\, ,
\end{align}
where $M$ is the fundamental scale of the theory, $\Lambda_i$ is the cosmological constant in the $i$-th bulk component, $\sigma$ the brane tension,
$h_{\alpha\beta}$ the induced metric, $\mathcal{R}$ the Ricci scalar for the induced metric, $K$ the extrinsic curvature,
$L(h_{\alpha\beta}, \phi)$ corresponds to the presence of matter fields $\phi$ on the
brane interacting with the induced metric and describes their dynamics and $m^2$ arises when one considers quantum effects generated by matter fields residing in the brane.

Dealing with the $Z_2$ symmetry of reflection, which requires equal cosmological constants on the two sides of the brane, i.e.,
$\Lambda_1 = \Lambda_2 \equiv \Lambda$, the corresponding Friedmann equation is given by \cite{Collins:2000yb,Shtanov:2000vr,Shtanov:2002ek}
\begin{align}
\label{brane}
H^2+\frac{k^2}{a^2}=\frac{\rho+\sigma}{3m^2}+ \frac{2}{{ l}^2}\left[
1\pm \sqrt{1+{ l}^2\left( \frac{\rho+\sigma}{3m^2}-\frac{\Lambda}{6}-\frac{C}{a^4}
\right) }\right]\, ,
\end{align}
where $k$ denotes the spatial curvature, $l^2=m^2/M^3$ is the length scale
and the term $C/a^4$, sometimes referred to as the {\it dark radiation},
arising due to the projection of the bulk gravitational degrees of freedom onto the brane.

The two new singularities that we will discuss in this Section
are connected with the fact that the expression under the square
root of Eq.~(\ref{brane}) turns to zero at some point during the
evolution of the Universe, so that solutions of the cosmological
equations cannot be continued beyond this point.

These two types of {\it quiescent} singularities display the following behavior \cite{Shtanov:2002ek}:

\begin{enumerate}
\item The first type of singularity (labeled as `S1') which is
essentially induced by the presence of the dark radiation term in
the square root of Eq.~(\ref{brane}) arises one of the following
two cases:
\begin{enumerate}
\item $C > 0$ and the density of matter increases slower than
$a^{-4}$ in the limit $a \rightarrow 0$. Such singularities may
appear if the Universe is filled with the matter sector having EoS
$ \frac{p}{\rho} < \frac{1}{3}$. An example is the pressure-less
matter (dust) for which $\rho \propto a^{-3}$. In addition, a
special case, in which it also occurs, is an empty Universe (i.e.,
where $\rho=0$).

\item The energy density of the Universe is radiation-dominated,
that means $\rho=\frac{\rho_0}{a^4}$ and in addition, $C> \rho_0$.

\end{enumerate}

The above singularities may occur either in the past of an
expanding Universe or in the future of a collapsing one, i.e., when the spatial curvature is $k=1$. 

\item A second type of singularity (labeled as `S2') arises when
\begin{align}
l^2\left( \frac{\sigma}{3m^2}-\frac{\Lambda}{6}\right)<-1\, .
\end{align}
In this case, it is important to realize that the combination
$(\rho/3m^2-C/a^4)$ decreases monotonically with the expansion of
the Universe. As an effect, the expression under the square root
of Eq.~(\ref{brane}) can become zero at a suitable late-time
beyond that the cosmological solution cannot be extended. Let us
note that (S2) has some interesting features than (S1), because
\begin{enumerate}
\item It may appear during the late-time expansion of the Universe.
\item It may occur even if dark radiation is totally
absent, i.e., $C=0$.
\end{enumerate}
\end{enumerate}

Note that in both (S1) and (S2), the scale factor and its first
time derivatives remain finite, however, all the higher
derivatives of $a(t)$ with respect to the cosmic time tend to
infinity as the singularity is approached. This is due to the fact
that when one takes the temporal derivative of Eq.~(\ref{brane})
the square root appears in the denominator, and thus, since it
vanishes at the singularity, all the derivatives of the Hubble
rate tend to infinite at the singular time.

Now let us consider the general braneworld without $Z_2$ symmetry.
In this case, the Friedmann equation for the brane embedded into
the five-dimensional bulk is given by \cite{Shtanov:2001pk}
\begin{align}
\label{brane1}
m^2\left(H^2+\frac{k}{a^2}-\frac{\rho+\sigma}{3m^2} \right)^2=4M^6\left(H^2+\frac{k}{a^2}-\frac{\Lambda}{6}- \frac{C}{a^4}\right)-\frac{M^{12}}{36m^4}\left[
\frac{E/a^4}{H^2+k/a^2-(\rho+\sigma)/3m^2}
\right]^2\, ,
\end{align}
where $E$ is a constant of integration. In this case we can see that the singularity (S2) is always
present in the past of the expanding brane. The reason for this
rests in the negative character of the last term on the right-hand
side of Eq.~(\ref{brane1}), which rapidly grows by absolute value
as $a\rightarrow 0$, while the left-hand side of this equation is
constrained to remain positive. In the case of an expanding brane,
the last term on the right-hand side of Eq.~(\ref{brane1}) rapidly
decays and becomes unimportant. Therefore, provided
Eq.~(\ref{brane}) is satisfied, the expanding Universe will
encounter an (S2) singularity in the future.

Finally, we note that the singularities (S1) and (S2) do not
appear for $m=0$, the usual singularities in GR appear only.
Effectively, in this situation Eq.~(\ref{brane1}) turns out to be
\begin{align}
\label{brane2}
H^2+\frac{k}{a^2}=\frac{\Lambda}{6}+ \frac{C}{a^4} + \frac{(\rho+\sigma)^2}{36M^6}+\frac{M^6E^2}{16a^8(\rho+\sigma)^2}\, ,
\end{align}
which only admits cosmological singularities, when the scale factor vanishes, associated with an infinite density of matter and
dark radiation ($C/a^4$) or the last term in Eq.~(\ref{brane2}).

\section{Singularities in Matter Creation Models}
\label{sec-singularities-matter-creation}

The theory of matter creation or particle creation, plays a
crucial role in the understanding of the dynamical evolution of
our Universe. A consistent framework of the continuous matter
creation was initiated by Parker and his collaborators
\cite{Parker:1968mv,Parker:1969au,Parker:1971pt,Parker:1972kp,Fulling:1974pu,Ford:1977in,Papastamatiou:1979rv},
and Zeldovich and others
\cite{Zeldovich:1970si,Zeldovich:1971mw,Grib:1974ym,Grib:1976pw,Zeldovich:1977vgo,Grib:1980au}
through the investigations of the material content of the
Universe. Within this framework, the existing material content of
the Universe is a result of the continuous creation of radiation
and matter particles due to the gravitational field of the
expanding Universe acting on the quantum vacuum. These produced
particles have their mass, energy and momentum. The matter
creation theory gained significant attention after the pioneering
work by Prigogine et al \cite{Prigogine:1989zz} who showed how to
insert the creation of matter into Einstein's gravitational
equations. This was achieved through the modification of the usual
conservation equation as follows \cite{Prigogine:1989zz}

\begin{align}
 (n u^{\mu})_{; \mu} = n \Gamma\;,
\end{align}
where $n$ is the particle number density, $u^{\mu}$ is the usual
particle four velocity and $\Gamma$ denotes the particle creation
or matter creation rate. The particle creation rate $\Gamma$ is
the heart of this theory since this quantity controls the dynamics
of the Universe by modifying its expansion history. From the
thermodynamical point of view, as the entropy flux vector of the
matter field which allows the particle creation, $s^{\mu} = n
\sigma u^{\mu}$, where $\sigma$ is the entropy per particle
(specific entropy) must satisfy the second law of thermodynamics,
i.e., $s^{\mu}_{; \mu} \geq 0$, therefore, one can derive that
$\Gamma \geq 0$ \cite{Calvao:1991wg}. According to the Parker's
theorem \cite{parker_toms_2009}, in the radiation dominated era,
the production of particles, is heavily suppressed, and hence,
this quantity is considered to have no effect during the radiation
dominated era. After Prigogine et al \cite{Prigogine:1989zz}, the
thermodynamics of particle creation was discussed in detail
through the covariant formalism \cite{Calvao:1991wg,Lima:1992np}.
A special attention was given to the `adiabatic' or `isentropic'
particle creation where the entropy per particle remains constant.
The cosmological scenarios driven by such adiabatic particle
creation have been investigated widely over the years with many
interesting results
\cite{Lima:1995xz,Abramo:1996ip,Lima:1996mp,Gunzig:1997tk,Lima:1999rt,Alcaniz:1999hu,Zimdahl:1999tn,
Lima:2008qy,Steigman:2008bc,Lima:2009ic,Basilakos:2010yp,Lima:2011hq,Lima:2012cm,Lima:2014qpa,
Chakraborty:2014ora,Lima:2014hda,Chakraborty:2014fia,Nunes:2015rea,Baranov:2015eha,deHaro:2015hdp,
Pan:2016jli,Nunes:2016aup,Pan:2016bug,Bhattacharya:2017lvr,Ivanov:2019zvm}.
In particular, particle creation mechanism can explain the
late-time accelerating expansion without any need of DE or
modified gravity theory
\cite{Lima:2008qy,Steigman:2008bc,Lima:2009ic,Basilakos:2010yp,Lima:2012cm,Lima:2014qpa}.
Moreover, such adiabatic particle creation process can also
explain the early time inflationary phase
\cite{Lima:1995xz,Abramo:1996ip,Gunzig:1997tk,Zimdahl:1999tn}. It
is essential to note that a non vanishing particle creation rate
$\Gamma$ is dynamically equivalent to a bulk pressure scenario
\cite{HU1982375,Barrow:1986yf,Barrow:1987kw,Barrow:1988yc,Sussman:1994uy,Gariel:1995kh,Desikan:1996da,
Belinchon:1998ij,Zimdahl:2000zx,Zimdahl:2000zm,Fabris:2005ts,Colistete:2007xi,Hipolito-Ricaldi:2009xbk,
Avelino:2008ph,Li:2009mf,Avelino:2010pb,Hipolito-Ricaldi:2010wrq,Gagnon:2011id,Piattella:2011bs,Meng:2012mb,
Velten:2013qna,Li:2014bsa,Acquaviva:2014vga,Disconzi:2014oda,Brevik:2017msy,Yang:2019qza,Yang:2023qqz},
however, as pointed out in Ref.~\cite{Lima:1992np}, such bulk
viscosity and the particle creation processes are not
thermodynamically equivalent. In the following we describe the
most basic equations of a cosmological scenario where adiabatic
matter creation is allowed.

In a spatially flat FLRW Universe (\ref{FLRWk0}), the gravitational equations
endowed with particle creation can be written as
\cite{Lima:2008qy,Steigman:2008bc,deHaro:2015hdp}
\begin{align}
H^2= \frac{\kappa^2 \rho}{3}\, , \label{MC-friedmann1} \quad
\dot{H}= -\frac{\kappa^2}{2} \left(1-\frac{\Gamma}{3H}\right) (p+ \rho)\, , 
\end{align}
where $H$ is the Hubble rate of the FLRW Universe, $p$, $\rho$ are
the pressure and energy density of the 
Universe, respectively, and $\Gamma$ is the creation rate of particles.
Using the gravitational equations of (\ref{MC-friedmann1}), one can see that
the conservation equation of the matter sector endowed with the
particle creation is modified as
\begin{align}
 \dot{\rho} + 3 H \left(1 - \frac{\Gamma}{3H}\right) (p+\rho) = 0\, ,
\end{align}
from which one can see that under the condition $\Gamma/3H \ll 1$,
the standard conservation equation is recovered. If the rate of
the particle production, $\Gamma$, is prescribed, then the
dynamics of the Universe can in principle be determined. In
general there is no such guiding principle available yet to derive
the particle creation rate, thus, one usually proposes some
phenomenological choices for $\Gamma$. This approach is similar to
the phenomenological model buildings of DE.

\subsection{Constant Matter Creation Rate}

In this section, we consider the simplest case when the particle
creation rate $\Gamma$ is a constant, i.e., $\Gamma = \Gamma_c
>0$ \cite{deHaro:2015hdp}. Assuming that the fluid obeys the
linear EoS $p=(\gamma-1)\rho$, where $\gamma$ is a constant, using
Eq.~(\ref{MC-friedmann1}), for the constant matter creation rate,
one arrives at
\begin{align}
\dot{H}= -\frac{3\gamma}{2} \left(1-\frac{\Gamma_c}{3H}\right)H^2\, ,
\end{align}
which can be integrated into
\begin{align}
H (t) = \frac{\Gamma_c }{3} \left[\frac{\frac{H_0}{H_0-\Gamma_c/3}
\exp\left(\frac{\Gamma_c \gamma}{2} (t-t_0)\right)}{\frac{H_0}{H_0-\Gamma_c/3}\exp\left(\frac{\Gamma_c\gamma}{2} (t-t_0)\right)-1}\right]\, ,
\label{MC-Hubble-solution}
\end{align}
where $t_0$, and $H_0$ denote, once again, the
present values of the cosmic time, respectively, and the Hubble parameter.
One can clearly see from Eq.~(\ref{MC-Hubble-solution}) that if for
some finite-time $t = t_s$, the denominator of the right hand
side of Eq.~(\ref{MC-Hubble-solution}) vanishes, then the model
allows a finite-time singularity. The condition for finite-time singularity is,
\begin{align}
\frac{H_0}{H_0-\Gamma_c/3}\exp\left(\frac{\Gamma_c \gamma}{2} (t_s-t_0)\right)-1 = 0\, ,
\end{align}
which actually provides with the following condition
\begin{align}
t_s = t_0 + \frac{2}{\Gamma_c \gamma} \ln \left(\frac{H_0 - \frac{\Gamma_c}3}{H_0}\right) \, ,
\end{align}
where the condition $H_0 - \Gamma_c/3 >0$ must be satisfied. Naturally, one can see that $t_s < t_0$ as $H_0-\Gamma_c/3 < H_0$
provided that $\gamma>0 \rightarrow p/\rho > -1$ (non-phantom fluid), that is one has a Big Bang singularity.
On the contrary, when $\gamma<0 \rightarrow p/\rho <-1$ (phantom fluid), we have $t_s>t_0$ obtaining a Big Rip singularity.
In fact, one could check that
\begin{align}
H (t)= \frac{\Gamma_c}{3} \left[\frac{\exp\left({\frac{\Gamma_c \gamma}{2}(t-t_s)}\right)}{\exp\left({\frac{\Gamma_c \gamma}{2}(t-t_s)}\right)-1}
\right] \, ,
\label{MC-Hubble-singularity-time}
\end{align}
with $H (t) \rightarrow \infty$ as $t \rightarrow t_s$, which confirms our conclusion.

\subsection{Variable Matter Creation Rate}

For the time dependent matter creation rate, we realize a more generalized cosmological scenario with new possibilities.
We consider the following time dependent matter creation rate \cite{Pan:2016jli}
\begin{align}
\label{MC-Gamma-general}
\Gamma(H)=-\Gamma_c + mH + \frac{n\Gamma_c^2}{H}\, ,
\end{align}
where $m$ and $n$ are dimensionless parameters and $\Gamma_c$ is a
constant. Notice that for specific values of $m$, $n$, one can
recover a number of matter creation rates, for example, $\Gamma
(H) = -\Gamma_c$ for $m = n =0$; $\Gamma (H) \propto H$ by setting
$\Gamma_c =0$ and $n =0$; $\Gamma \propto H^{-1}$ under the
choices of $\Gamma_c =0$ and $m =0$; and several more. Here we
restrict to the matter creation scenario when the matter creation
rate takes the full expression as in Eq.~(\ref{MC-Gamma-general}).
For this general matter creation rate, assuming a linear EoS for
the perfect fluid given by $p=(\gamma-1)\rho$, the Raychaudhuri
equation is given by
\begin{align}
\dot{H}=-\frac{\gamma}{2} \Bigg[(3-m)H^2+\Gamma_cH-\Gamma_c^2n \Bigg]\, .
\end{align}

Since the dynamical system is depicted by a first order autonomous differential equation
\begin{align}
\dot{H}=F(H) = -\frac{\gamma}{2}\Bigg[(3-m)H^2+\Gamma_cH-\Gamma_c^2n \Bigg]\, ,
\end{align}
thus, to understand the dynamics we only need to find the fixed
points of the system, i.e., the points satisfying $F(H)=0$. And
then, given a fixed point, namely $H_*$, when
$\frac{dF(H_*)}{dH}<0$ the fixed point is asymptotically stable
(an attractor) and when $\frac{dF(H_*)}{dH}>0$ the fixed point is unstable (a repeller).
For the present matter creation rate, i.e.,
Eq.~(\ref{MC-Gamma-general}), the fixed points are given by
\begin{align}
H_{\pm}=\frac{\Gamma_c}{2(m-3)}\left( 1\pm \sqrt{1+4(3-m)n} \right)\, .
\end{align}
It is also important to note that the nature of the fluid depends on the parameter $\gamma$. For $\gamma<0$,
it behaves like a phantom fluid while $\gamma>0$, it behaves like a quintessence fluid.
Here, we will consider the case of a phantom fluid (the non-phantom case has been studied in detail in Ref.~\cite{Pan:2016jli}).
A simple calculation shows that for our model
\begin{align}
\frac{dF(H_{\pm})}{dH}=\pm \frac{\gamma\Gamma_c}{2} \sqrt{1+4(3-m)n}~.
\end{align}

Then, for a phantom fluid and for a positive $\Gamma_c$,
$H_+$ is always an attractor and $H_-$ is always a repeller. In the following we show that there are six different regions in the plane of parameters $(m,n)$:
\begin{enumerate}

\item $\Omega_1 =\{(m,n): m-3>0, n<0\}$.
One has $H_{+}>0$ and $H_{-}<0$. In addition, since
$H_+$ is an attractor, then for any initial condition $H_\mathrm{ini}$ greater than $H_+$, the Universe converges at late times asymptotically to a de Sitter phase
with $H=H_+$, and going back in time, the Universe has a Big Bang singularity, because for large values of the Hubble rate,
the Raychaudhuri equation will become $\dot{H}\sim -\frac{\gamma (3-m)}{2}H^2$.

\item $\Omega_2 =\{(m,n): m-3<0, n\geq 0\}$.
One has $H_+<0$ and $H_->0$. Now, since $H_-$ is a repeller, for any initial condition $H_\mathrm{ini}$ greater than $H_-$,
the Hubble rate diverges in the future in a finite-time because, once again, for large values of $H$ the Raychaudhuri equation becomes
$\dot{H}\sim -\frac{\gamma (3-m)}{2}H^2$, and thus, a Big Rip singularity is obtained.

\item $\Omega_3 =\{(m,n): m-3>0, n> 0, 4(3-m)n>-1\}$.
In this situation one has $H_+>H_->0$, and for any initial condition $H_\mathrm{ini}$ greater than $H_+$, there is a Big Bang singularity
and the Universe ends in a de Sitter phase with $H=H_+$. For an initial condition between $H_-$ and $H_+$,
the Universe is no-singular starting at $H_-$ and ending in an infinite-time to $H_+$.
Finally, for an initial condition less than $H_-$, the Universe enters into the contracting phase forever.

\item $\Omega_4 =\{(m,n): m-3<0, n\leq 0, 4(3-m)n>-1\}$.
Now $H_+<H_-<0$, and thus, for any initial condition greater than $H_-$, there is a Big Rip singularity.

\item $\Omega_5 =\{(m,n): m-3>0, n> 0, 4(3-m)n<-1\}$. There are no fixed points and $\dot{H}$ is always negative, meaning that for any initial condition there is a Big Bang singularity.

\item $\Omega_6 =\{(m,n): m-3<0, n< 0, 4(3-m)n<-1\}$. Once again there are no fixed points, but now $\dot{H}$ is positive, and thus, a Big Rip singularity always occurs.

\end{enumerate}

On the other hand, considering once again a phantom fluid but taking $\Gamma_c<0$,
one has the following observations:

\begin{enumerate}

\item $\Omega_1=\{(m,n): m-3>0, n<0\}$.
One has $H_->0$ and $H_+<0$. In addition, since
$H_-$ is an attractor, then for any initial condition $H_\mathrm{ini}$ greater than $H_-$, the Universe at late times converges asymptotically
to a de Sitter phase with $H=H_-$, and going back in time, the Universe has a Big Bang singularity, because for large values of the Hubble rate,
the Raychaudhuri equation becomes $\dot{H}\sim -\frac{\gamma (3-m)}{2}H^2$.

\item $\Omega_2=\{(m,n): m-3<0, n\geq 0\}$.
One has $H_-<0$ and $H_+>0$. Now, since $H_+$ is a repeller, then for any initial condition $H_\mathrm{ini}$ greater than $H_+$,
the Hubble rate diverges in the future in a finite-time because, once again, for large values of $H$,
the Raychaudhuri equation is $\dot{H}\sim -\frac{\gamma (3-m)}{2}H^2$, and thus, a Big Rip singularity occurs.

\item $\Omega_3=\{(m,n): m-3>0, n> 0, 4(3-m)n>-1\}$.
In this situation one has $H_+<H_-<0$,
and for any initial condition $H_\mathrm{ini}$ greater than $H_-$, there is a Big Bang singularity, and the Universe enters
in the contracting phase and it ends in a de Sitter phase with $H=H_-$. For an initial condition between $H_-$ and $H_+$,
the Universe is non-singular starting at $H_+$ and ending in an infinite-time to $H_-$.
Finally, for an initial condition less than $H_+$, the Universe enters into the contracting phase forever.

\item $\Omega_4=\{(m,n): m-3<0, n\leq 0, 4(3-m)n>-1\}$.
Now $H_+>H_->0$, and thus, for any initial condition greater than $H_+$, there is a Big Rip singularity.

\item $\Omega_5=\{(m,n): m-3>0, n> 0, 4(3-m)n<-1\}$.
There are no fixed points and $\dot{H}$ is always negative, meaning that for any initial condition there is a Big Bang singularity.

\item $\Omega_6=\{(m,n): m-3<0, n< 0, 4(3-m)n<-1\}$.
Once again there are no fixed points, but now $\dot{H}$ is positive, and thus, a Big Rip singularity always occurs.

\end{enumerate}

Finally, we mention that as the matter creation rate is not
properly known, thus, one can consider a different matter creation
rate to investigate the possible occurrence of finite-time
singularities.

\section{Singularities in Loop Quantum Cosmology}
\label{sec-singularities-LQG}

An approach to quantum cosmology that could avoid finite time
singularities is Loop Quantum Cosmology (LQC). For a general overview on LQC, we refer to
the following
works~\cite{Bojowald:1999tr,Bojowald:2001xe,Bojowald:2002gz,
Ashtekar:2003hd,Singh:2003au,Date:2004fj,Bojowald:2005epg,
Ashtekar:2006es,Sami:2006wj,Bojowald:2006qu,Szulc:2006ep,Vandersloot:2006ws,Ashtekar:2007tv,Samart:2007xz,
Wei:2007rp,Zhang:2007bi,Parisi:2007kv,Szulc:2007uk,Ashtekar:2008zu,Mielczarek:2008zv,Rovelli:2008aa,
Fu:2008gh,Corichi:2009pp,Ashtekar:2011ni,Corichi:2011pg,Singh:2011gp,Pawlowski:2011zf,
Wilson-Ewing:2012lmx,Bamba:2012ka,Barrau:2013ula,Rovelli:2013zaa,Amoros:2014tha,deHaro:2014kxa,
Odintsov:2014gea,Ashtekar:2015dja,deBlas:2016puz,Langlois:2017hdf,Zhu:2017jew}.
The main idea is that LQC assumes a discrete nature of space which
leads, at quantum level, to consider a Hilbert space where quantum
states are represented by almost periodic functions of the
dynamical part of the connection \cite{Ashtekar:2011ni}.
Unfortunately, the connection variable does not correspond to a
well defined quantum operator in this Hilbert space and therefore
we re-express the gravitational part of the Hamiltonian in terms
of almost periodic function. It could be executed from a process
of regularization~\cite{Ashtekar:2003hd}. This new regularized
Hamiltonian introduces a quadratic modification ($\rho^2$) in the
Friedmann equation at high energies, which give rise to a bounce
when the energy density becomes equal to a critical value of the
order of the Planck energy density. Thus, the {\it holonomy
corrected} Friedmann equation is given by
\cite{Singh:2006im,Singh:2009wm,Haro:2010wmu,Dzierzak:2009ip}
\begin{align}
\label{Friedmannlqc}
H^2=\frac{\rho\kappa^2}{3}\left(1-\frac{\rho}{\rho_c} \right)\, ,
\end{align}
where $\rho_c$ is the so-called critical density.
A way to obtain this Friedmann equation goes as follows:
The gravitational part of the Hamiltonian in GR is given by
\begin{align}
\mathcal{H}=- \frac{3}{\kappa^2}H^2 a^3=-\frac{3}{\gamma^2\kappa^2}\beta^2V\, ,
\end{align}
where we have introduced the canonically conjugate variable
\begin{align}
\beta=\gamma H\qquad \mbox{and}\qquad V=a^3\, ,
\end{align}
whose Poisson braked is $\{\beta, V\}=\frac{\gamma}{2}$, where $\gamma=0.2375$ is the Immirzi parameter.
Then, to capture the holonomy effects one makes the replacement
$\beta\rightarrow \frac{\sin(\lambda \beta)}{\lambda}$, where $\lambda$ is a parameter with the dimension of
length, whose numerical value is determined by invoking the quantum nature of the geometry, that is,
identifying its square with the minimum eigenvalue of the area operator in Loop Quantum Gravity (LQG) \cite{Singh:2008uxj}.
Thus, the full Hamiltonian in LQC is given by
\begin{align}
\mathcal{H}_\mathrm{LQC}= -3V\frac{\sin^2(\lambda \beta)}{\gamma^2\lambda^2\kappa^2}+\rho V\, ,
\end{align}
and the Hamiltonian constraint $H_\mathrm{LQC}=0$ leads to
$\frac{\sin^2(\lambda \beta)}{\gamma^2\lambda^2}=\frac{\kappa^2\rho}{3}$.
On the other hand, the Hamiltonian equation
\begin{align}\label{Hamiltonian-LQC-SP}
\dot{V}=\{V,\mathcal{H}_\mathrm{LQC}\}=-\frac{\gamma}{2}\frac{\partial \mathcal{H}_\mathrm{LQC}}{\partial \beta}\Longrightarrow
H=\frac{\sin(2\lambda\beta)}{2\gamma\lambda}\, ,
\end{align}
and by taking the square of this last equation (\ref{Hamiltonian-LQC-SP})
\begin{align}
H^2=\frac{\sin^2(2\lambda\beta)}{4\gamma^2\lambda^2}
= \frac{\sin^2(\lambda\beta)}{\gamma^2\lambda^2}\Bigg[1-\sin^2(\lambda\beta)\Bigg]\,,
\end{align}
and using the Hamiltonian constraint $\frac{\sin^2(\lambda \beta)}{\gamma^2\lambda^2}=\frac{\kappa^2\rho}{3}$, together with 
the critical density defined as, $\rho_c=\frac{3}{\gamma^2\lambda^2\kappa^2}$, one arrives at the holonomy corrected Friedmann equation.

Next, the Raychaudhuri equation in LQC could be obtained taking the temporal derivative of the holonomy corrected Friedmann equation
and using the conservation one.
The result is
\begin{align}
\dot{H}=-\frac{\kappa^2(\rho+p)}{2}\left(1-\frac{2\rho}{\rho_c} \right) \, .
\end{align}

As we will see when $\rho\ll \rho_c$ the equation
(\ref{Friedmannlqc}) coincides with the standard Friedmann
equation of GR. From the holonomy corrected Friedmann and
Raychaudhuri equation, the future singularities were studied in
several works such as
\cite{Sami:2006wj,Singh:2009mz,Bamba:2012ka}, where the authors
concluded that only the singularities of type I and III are
removed in LQC, because the difference appears at high energy
densities, because the standard Friedmann equation
$H^2=\frac{\kappa^2\rho}{3}$ depicts in the plane $(H,\rho)$ a
parabola, which is an unbounded curve, and thus, singularities
with a divergent energy density are allowed. On the contrary, in
LQC the holonomy corrected Friedmann equation depicts, in the
plane $(H,\rho)$, an ellipse, where the energy density is always
bounded by $0\leq \rho\leq \rho_c$, meaning that the singularities
of Type I and III are forbidden in LQC. As a consequence, the
Big Rip and Big Bang singularities do not exist in this theory,
and it is argued that the Big Bang singularity is replaced by a
Big Bounce which is produced when the energy density reaches the
value $\rho_c$. In fact, from the conservation equation one can
see that for a non-phantom fluid, the movement along the ellipse
depicted by the holonomy corrected Friedmann equation is clockwise
and anti-clockwise for a phantom fluid.

On the other hand, the singularity of Type II appears in LQC
provided that the value of the energy density at the singularity,
namely $\rho_s$ is less than the critical one. This is simple to
understand: the Type II singularity appears when the pressure
diverges for a finite value of the energy density which happens
for the nonlinear EoS $p=-\rho-f(\rho)$ considered in 
section \ref{sec-future-singularities},
with $\nu<-1/2$ and $A>0$. In the same way, the Type IV
singularity appears near $0\cong \rho\ll \rho_c$, that is, when
the holonomy corrections could be disregarded and GR is recovered. This kind of singularity has been discussed in section \ref{sec-future-singularities}.

Another approach to LQC is the Dapor-Lieger model which depicts an
emergent Universe from a de Sitter regime in the contracting phase
\cite{Dapor:2017rwv,Assanioussi:2018hee}. For this model the
corresponding Friedmann equation, which is more complicated than
in standard LQC, is given by \cite{Li:2018opr,deHaro:2018khb}
\begin{align}
H^2_{\pm}=\frac{\kappa^2\rho}{3(\gamma^2+1)}\left( 1-\frac{\rho}{\rho_{max}}\right) \left[
1+\frac{2\gamma^2}{1\pm\sqrt{1-\frac{\rho}{\rho_\mathrm{max}}}}
\right]\, ,
\end{align}
where $\gamma\cong 0.2375$ is, once again, the well-known Immirzi parameter and $\rho_\mathrm{max}=\frac{\rho_c}{4(\gamma^2+1)}$
is the maximum value reached by the energy density in this theory.
Then, since the curves depicted by this Friedmann equation are also bounded, one gets the same kind of singularities as in standard LQC,
that is, only singularities of Type II and IV could appear in LQC.


\section{Cosmological Finite-time Singularities in Modified
Gravity and in Interacting Multifluid Cosmology: Dynamical System versus Finite-time Cosmological Singularities}
\label{sec-dyn-system-vs-singularities}

Modified gravity in its various forms makes possible the
realization of finite-time future cosmological singularities
without the need of phantom scalars, as in the case of
ordinary GR \cite{Caldwell:2003vq}. In the literature, the GR
perspective of DE can marginally accommodate a phantom fluid
evolution and cosmic singularities without invoking phantom
scalars, however, modified gravity can perfectly generate a
phantom DE era and even cosmic singularities. Another interesting
perspective along with modified gravity is to take into account
the interacting multifluids that can be viscous or not
\cite{Barrow:1994nx,Zimdahl:1998rx,Tsagas:1998jm,Hipolito-Ricaldi:2009xbk,Gorini:2005nw,Kremer:2003vs,Carturan:2002si,
Buchert:2001sa,Hwang:2001fb,Cruz:2011zza,Oikonomou:2017mlk,Brevik:2017msy,Nojiri:2005sr,Capozziello:2006dj,
Nojiri:2006zh,Elizalde:2009gx,Elizalde:2017dmu,Brevik:2016kuy,Balakin:2012ee}.

Most of the multifluid approaches in cosmology invoke an
interaction between the DE and DM, which is supported
theoretically by the fact that DE dominates over DM after the
formation of the galactic structures. Also, the interaction between
the dark sectors is further supported by the degeneracy of the DE
models, since the DM density parameter $\Omega_m$ cannot
accurately be measured \cite{Kunz:2007rk}. As already argued, cosmology with
interacting DM-DE sectors is vastly considered in the literature
\cite{Amendola:1999er,Gondolo:2002fh,Farrar:2003uw,Huey:2004qv,Cai:2004dk,Guo:2004xx,Wang:2005jx,Pavon:2005yx,Das:2005yg,Barrow:2006hia,Amendola:2006dg,delCampo:2006vv,Wang:2006qw,Bertolami:2007zm,delCampo:2008sr,Valiviita:2008iv,delCampo:2008jx,He:2008tn,Boehmer:2008av,Gavela:2009cy,Jackson:2009mz,Koyama:2009gd,Majerotto:2009np,Boehmer:2009tk,Jamil:2009eb,He:2010im,Li:2010ju,Chimento:2011pk,Chimento:2012zz,Pettorino:2013oxa,Costa:2013sva,Chimento:2013rya,Chakraborty:2012gr,Yang:2014gza,yang:2014vza,Yang:2014hea,Pan:2012ki,Nunes:2016dlj,vandeBruck:2016jgg,Pourtsidou:2016ico,vandeBruck:2016hpz,Mukherjee:2016shl,An:2017crg,Sharov:2017iue,DiValentino:2017iww,Yang:2017yme,vandeBruck:2017idm,Cai:2017yww,Kumar:2017dnp,Yang:2017ccc,Yang:2017zjs,VanDeBruck:2017mua,Mifsud:2017fsy,Pan:2017ent,Yang:2018euj,Yang:2018xlt,Gonzalez:2018rop,Yang:2018qec,Pan:2019jqh,Paliathanasis:2019hbi,Yang:2019uog,Li:2019loh,Yang:2019vni,vonMarttens:2019ixw,Li:2019ajo,Pan:2019gop,Mifsud:2019fut,Yang:2020uga,Pan:2020bur,DiValentino:2019ffd,DiValentino:2019jae,Lucca:2020zjb,Yang:2020tax,Pan:2020mst,DiValentino:2020kpf,Anchordoqui:2021gji,Gao:2021xnk,Yang:2021hxg,Johnson:2021wou,Yang:2021oxc,Paliathanasis:2021egx,Lucca:2021dxo,Gariazzo:2021qtg,Bonilla:2021dql,Mukherjee:2021ggf,Nunes:2022bhn,Yao:2022kub,Yang:2022csz,Pan:2022qrr,Zhai:2023yny}
and it is a quite popular research line. It is notable though that
some specific interacting DM-DE models are plagued with
instabilities, due to the fact that the growth of matter
perturbations can be affected by an existing non-trivial
interaction between the components of the dark sectors. On the
other hand, baryonic fluids cannot be coupled to any of the
components of the dark sector since such a coupling would
eventually result in a fifth force, which is not a physically
acceptable feature in contemporary physics. Since the discovery of
the late-time accelerating epoch of the Universe in the late 90's,
many proposals try to model the DE fluid, which is a negative
pressure fluid, with a quintessence, de Sitter or slightly phantom
EoS parameter. Also, it is possible that finite-time singularities
may occur in the Universe, and it is quite hard to model these in
standard GR since a phantom scalar would be required, as we
already mentioned. Finite-time cosmological singularities are not
perfectly understood, since they share the problems of all
cosmological singularities, the main problem being the geodesics
incompleteness, at least for crushing types of singularities. In this
section, we shall study finite-time singularities using the
dynamical system approach, in the context of $F(R)$ gravity and
coupled DE-DM models. The study of the cosmological phase space
for the aforementioned systems offers many insights for the
complete understanding of finite-time singularities, among which
the understanding of the fixed points and of the stability of the
solutions, and studying cosmological systems in terms of their
dynamical systems are quite popular in the literature, see for
example
\cite{Bahamonde:2017ize,Boehmer:2014vea,Boehmer:2010jqg,Goheer:2007wu,Leon:2014yua,Leon:2010pu,deSouza:2007zpn,Giacomini:2017yuk,Kofinas:2014aka,Leon:2012mt,Gonzalez:2006cj,Alho:2016gzi,Biswas:2015cva,Muller:2014qja,Mirza:2014nfa,Rippl:1995bg,Ivanov:2011vy,Khurshudyan:2016qox,Boko:2016mwr,Odintsov:2017icc,Odintsov:2017tbc,Oikonomou:2017ppp,Odintsov:2015wwp}.
For our analysis in this section, we shall form the $F(R)$ gravity
field equations for a spatially flat FLRW spacetime in terms of an
autonomous dynamical system, by using specific choices of
variables. The resulting dynamical system will be studied in the
vicinity of finite-time future singularities, emphasizing mainly on the Big Rip singularity. An important outcome of our analysis is
that a Big Rip singularity in the context of $F(R)$ gravity always
occurs while the Universe is accelerating. This is due to the fact
that for the Big Rip case, the dynamical system is attracted to a
stable accelerating cosmological attractor, which is also the
final attractor of the asymptotically autonomous $F(R)$ gravity
dynamical system. For the Big Rip case, we also reveal which is
the leading order $F(R)$ gravity term which can realize a Big Rip
finite-time singularity, and we also consider the occurrence of
Type II, III and IV singularities in $F(R)$ gravity cosmological
systems. We also stress the fact that dynamical systems
singularities may not necessarily indicate the presence of a
physical finite-time singularity, which is quite important to
note. Apart from $F(R)$ gravity finite-time singularities, we
shall also study multifluid cosmology using again the dynamical
systems approach. Specifically, we shall consider a three-fluid
system consisting of interacting DE-DM fluids in the presence of a
non-interacting baryonic fluid. By using the cosmological
equations, we shall construct an autonomous dynamical system, and
we shall study the phase space of this system. We shall focus on
analyzing the dynamical system singularities, using the dominant
balance technique developed in Ref.~\cite{GORIELY2000422}, and we
shall examine when a dynamical system singularity is actually a
true finite-time singularity of the cosmological system. As we
shall show, the cosmological dynamical system has no global
attractors which may drive the cosmological system to a
finite-time singularity. Finally, the analysis we shall perform
indicates that the dynamical system possesses de Sitter fixed
points, the occurrence of which depends on the interaction between
DE and DM.

\subsection{Analysis of Finite-time Singularities in $F(R)$ Gravity via the Autonomous $F(R)$ Gravity Dynamical System}

In order to study the phase space of $F(R)$ gravity near cosmic
finite-time singularities, let us first form the dynamical system
of $F(R)$ gravity in an autonomous form \cite{Odintsov:2017tbc}.
Consider the vacuum $F(R)$ gravity action (\ref{action-F(R)}) with ${L}_\mathrm{matter}=0$.
Now, by varying the gravitational action (\ref{action-F(R)}) 
with respect to the metric tensor $g_{\mu \nu}$ we obtain the
field equations for $F(R)$ gravity in vacuum,
\begin{align}
\label{eqnmotion}
F'(R)R_{\mu \nu}(g)-\frac{1}{2}F(R)g_{\mu \nu}-\nabla_{\mu}\nabla_{\nu}F(R)+g_{\mu \nu}\square F'(R)=0\, ,
\end{align}
which takes the following form,
\begin{align}
\label{modifiedeinsteineqns}
R_{\mu \nu}-\frac{1}{2}Rg_{\mu \nu}=\frac{\kappa^2}{F(R)}\Bigg[T_{\mu \nu}+\frac{1}{\kappa^2}
\left(\frac{F(R)-RF'(R)}{2}g_{\mu\nu}+\nabla_{\mu}\nabla_{\nu}F'(R)-g_{\mu \nu}\square F'(R)\right)\Bigg]\, .
\end{align}
For a flat FLRW metric,
the $F(R)$ gravity field equations take the forms in (\ref{different-frames-eq-2}) and (\ref{different-frames-eq-3}).

The autonomous
dynamical system of $F(R)$ gravity can be formed by using the
dimensionless variables $x_1$, $x_2$, and $x_3$, which are defined
as follows,
\begin{align}
\label{variablesslowdown}
x_1=-\frac{\dot{F'}(R)}{F'(R)H}\, ,\quad x_2=-\frac{F(R)}{6F'(R)H^2}\, , \quad
x_3= \frac{R}{6H^2}\, .
\end{align}
We shall choose the $e$-foldings number $N$ to quantify the
dynamical evolution of the variables $x_1$, $x_2$, and $x_3$,
instead of the cosmic time.
Using the transformation,
\begin{align}
\label{specialderivative}
\frac{d }{d N}=\frac{1}{H}\frac{d }{d t}\, ,
\end{align}
and also by expressing the field equations with respect to the
variables $x_1$, $x_2$, and $x_3$ defined in Eq.~(\ref{variablesslowdown}),
the vacuum $F(R)$ gravity dynamical system takes the form,
\begin{align}
\label{dynamicalsystemmain}
\frac{d x_1}{d N}=&\, -4-3x_1+2x_3-x_1x_3+x_1^2\, , \nonumber \\
\frac{d x_2}{d N}=&\, 8+m-4x_3+x_2x_1-2x_2x_3+4x_2 \, , \nonumber \\
\frac{d x_3}{d N}=&\, -8-m+8x_3-2x_3^2 \, ,
\end{align}
with the parameter $m$ being defined as follows,
\begin{align}
\label{parameterm}
m=-\frac{\ddot{H}}{H^3}\, .
\end{align}
This parameter contains cosmic time-dependent quantities, or
equivalently, variables that depend explicitly on the $e$-foldings
number $N$. Obviously, the presence of this parameter renders the
dynamical system non-autonomous. The only case for which the
$F(R)$ gravity dynamical system is autonomous is when $m$ takes
constant values. This narrows the options for studying the full
phase space thus by choosing $m$ to be constant, we examine
subspaces of the full $F(R)$ gravity phase space which correspond
to specific cosmological evolutions. For example the case $m=0$
among other things, describes a quasi-de Sitter evolution or the
evolution near a quasi-de Sitter fixed point, thus the study of
the dynamical system will reveal the behavior of the dynamical
system for a quasi-de Sitter evolution. The full analysis of the
$F(R)$ gravity dynamical system was performed in Ref.
\cite{Odintsov:2017tbc}. We shall focus on the $m=0$ case which is
relevant for the studies of some finite time cosmological singularities. The total EoS
of the system in terms of the variables $x_i$
$\left(i=1,2,3\right)$ takes the form \cite{Nojiri:2017ncd},
\begin{align}
\label{equation of state1}
w_\mathrm{eff}=-\frac{1}{3} (2 x_3-1)\, .
\end{align}
Furthermore, the dimensionless variables $x_i$ $\left(i=1,2,3\right)$ satisfy
the following constraint,
\begin{align}
\label{friedmanconstraint1}
x_1+x_2+x_3=1\, ,
\end{align}
which stems from the Friedmann equation.
Let us present the phase space structure of the $F(R)$ gravity dynamical system for the
case $m=0$ which is of particular importance.
Following \cite{Odintsov:2017tbc}, the fixed points of the dynamical system
in terms of a general constant value of the parameter $m$ read,
\begin{align}
\label{fixedpointsgeneral}
\phi_*^1=&\, \left( - \frac{\sqrt{-2m} + \sqrt{-2 m+20 \sqrt{-2m}+4} + 2}{4}\, ,\ \frac{3 \sqrt{-2m}+\sqrt{-2 m+20 \sqrt{-2m}+4}-2}{4}\, ,\
\frac{4-\sqrt{-2m}}{2} \right)\, , \nonumber \\
\phi_*^2=&\, \left( - \frac{\sqrt{-2m} - \sqrt{-2 m+20 \sqrt{-2m}+4} + 2}{4}\, ,\ \frac{3 \sqrt{-2m}-\sqrt{-2 m+20 \sqrt{-2m}+4}-2}{4}\, ,\
\frac{4-\sqrt{-2m}}{2}\right)\, , \nonumber \\
\phi_*^3=&\, \left(\frac{\sqrt{-2m}-\sqrt{-2 m-20 \sqrt{-2m}+4}-2}{4}\, ,\ - \frac{3 \sqrt{-2m} - \sqrt{-2 m-20 \sqrt{-2m}+4} + 2}{4}\, ,\
\frac{\sqrt{-2m}+4}{2} \right)\, , \nonumber \\
\phi_*^4=&\, \left(\frac{\sqrt{-2m}+\sqrt{-2 m-20 \sqrt{-2m}+4}-2}{2}\, ,\ - \frac{\sqrt{-2m-20 \sqrt{-2m}+4} + 3 \sqrt{-2m} + 2}{4}\, ,\
\frac{ \sqrt{-2m}+4}{2} \right)\, .
\end{align}
Hence, for $m\simeq 0$, the fixed points are,
{\begin{align}\label{fixedpointdesitter}
\phi_*^1=\phi_*^3=(-1,0,2)\, ,\quad \phi_*^2=\phi_*^4=(0,-1,2)\, .
\end{align}}
Neither of the above two fixed points are hyperbolic for $m=0$,
therefore the stability analysis must be performed numerically.
This analysis was performed in Ref.~\cite{Odintsov:2017tbc} and as
it was shown that the fixed point $\phi_*^1$ is stable, while
$\phi_*^2$ is unstable. These results are important for the
analysis of the dynamical system near finite-time singularity
using an analytical approach. These results are important both for
physical and mathematical reasons, especially the ones related to
the fixed point $\phi_*^1=\phi_*^3=(-1,0,2)$, which is
essential for the Big Rip singularity analysis. For this fixed
point, since $x_3=2$, the EoS becomes $w_\mathrm{eff}=-1$. Thus, we
shall show the extremely important result that as the Universe
approaches a Big Rip singularity, it accelerates prior to approaching the singularity.

\subsubsection{Finite-time Singularities of $F(R)$ Cosmology and its Dynamical System}

Since we are interested in studying the phase space of vacuum
$F(R)$ gravity near finite-time singularities, let us consider the
cases for which the Hubble rate can be approximated as follows,
\begin{align}
\label{hubblerate}
H(t)\simeq h_s(t_s-t)^{-\beta}\, ,
\end{align}
with $t_s$ signifying the time instance at which the singularity
occurs, $\beta$ is a real parameter which determines the type of
the finite-time singularity and $h_s$ is some free parameter with
dimensions (GeV)$^{^{-\beta+1}}$. Since the finite-time singularity
at $t=t_s$ is a future singularity, we have $t_s>t$. Now,
according to the following values of $\beta$, the following types
of singularities may be developed,
\begin{itemize}
\item For $\beta > 1$, a Type I (Big Rip) singularity occurs.

\item For $0<\beta<1$, a Type III singularity occurs.

\item For $-1<\beta<0$, a Type II (pressure) singularity occurs.

\item For $\beta < -1$, a Type IV singularity occurs.
\end{itemize}
As we will show later on in this section, the parameter $\beta$
crucially affects the behavior of $F(R)$ gravity near a cosmic singularity.
Now let us investigate which is the form of the
dynamical system of vacuum $F(R)$ gravity near a finite-time
singularity of the form (\ref{hubblerate}). By expressing the
cosmic time as a function of the $e$-foldings number, using the
definition of the latter,
\begin{align}
\label{efoldingsdefinition}
N=\int^tH(t)d t\, ,
\end{align}
we get,
\begin{align}
\label{tsasfunctionofefoldings}
t_s-t=\left(\frac{(\beta-1) (N-N_c)}{h_s}\right)^{\frac{1}{1-\beta }}\, ,
\end{align}
with $N_c$ being an integration constant which basically depends on the chosen initial conditions.
For the form of the Hubble rate (\ref{hubblerate}), the parameter $m$ can be calculated as a
function of the $e$-foldings number $N$, and it has the following form,
\begin{align}
\label{parametermasfunctionofN}
m=-\frac{\beta (\beta +1)}{(\beta -1)^2 (N_c-N)^2}\, .
\end{align}
The above relation is a universal relation which covers all the
cases of cosmic singularities that may occur for various values of
the parameter $\beta$, which stems from the Hubble rate (\ref{hubblerate}).
However, as $N$ takes different values for
each type of singularity as $t$ approaches $t_s$, the parameter
$m$ will take distinct values too for the various types of
singularities which correspond to different limiting values of the $e$-foldings number $N$.
Upon substituting $m$ from Eq.~(\ref{parametermasfunctionofN}) in Eq.~(\ref{dynamicalsystemmain}),
the dynamical system of vacuum $F(R)$ gravity takes the form,
\begin{align}
\label{dynamicalsystemnonautonomousgeneral}
\frac{d x_1}{d N}=&\, -4+3x_1+2x_3-x_1x_3+x_1^2\, , \nonumber \\
\frac{d x_2}{d N}=&\, -\frac{\beta (\beta +1)}{(\beta -1)^2 (N_c-N)^2}+8-4x_3+x_2x_1-2x_2x_3+4x_2 \, ,\nonumber \\
\frac{d x_3}{d N}=&\, \frac{\beta (\beta +1)}{(\beta -1)^2 (N_c-N)^2}-8+8x_3-2x_3^2 \, .
\end{align}
The dynamical system of Eq.~(\ref{dynamicalsystemnonautonomousgeneral}) is non-autonomous, so
it is quite hard to tackle it since most theorems governing the
autonomous dynamical systems do not apply.
However, as we shall see, in some limiting cases, the dynamical system is rendered
autonomous, in the cases of interest, the behavior of the phase
space can be revealed near cosmological evolutions developing finite-time singularities.
In the next subsections, we shall
analyze the behavior of the $F(R)$ gravity phase space near
crushing and non-crushing types of singularities. In the same
research line we shall investigate which approximate $F(R)$
gravity can generate such singularities, based on the fixed points
of the phase space near singularities.
The major outcome is the behavior of the phase space near Big Rip singularities, in which
case in the context of $F(R)$ gravity the Big Rip singularity
always occurs while the Universe is accelerating.

\subsubsection{The Case of the Big Rip Singularity}

We shall begin our analysis with the most severe type of a
finite-time singularity, namely, the Big Rip, which is a
crushing type singularity, meaning that geodesics incompleteness
occurs. From a practical and quantitative point of view, this case
is the easiest to tackle since the dynamical system near the singularity is rendered autonomous.
For a full analysis of this study, see Ref.~\cite{Odintsov:2018uaw}.
For the Big Rip singularity, the parameter $\beta$ in the Hubble rate
(\ref{hubblerate}) takes the values $\beta>1$, so as the cosmic
time approaches the singularity time $t_s$, as we can see from Eq.~(\ref{tsasfunctionofefoldings}),
this corresponds to $N\to \infty$.
Therefore, in this case the parameter $m$ of Eq.~(\ref{parametermasfunctionofN}) approaches zero as the singularity
time instance is approached, therefore, remarkably the dynamical
system near the Big Rip singularity is rendered asymptotically autonomous.
From a mathematical point of view, this is a great simplification, since it will enable us to study the structure of
the phase space in a semi-analytic way.
This simplification only occurs for the Big Rip singularity which is the most interesting
case, thus this is a remarkable feature of the theory.

\begin{figure}[h]
\centering
\includegraphics[width=0.55\textwidth]{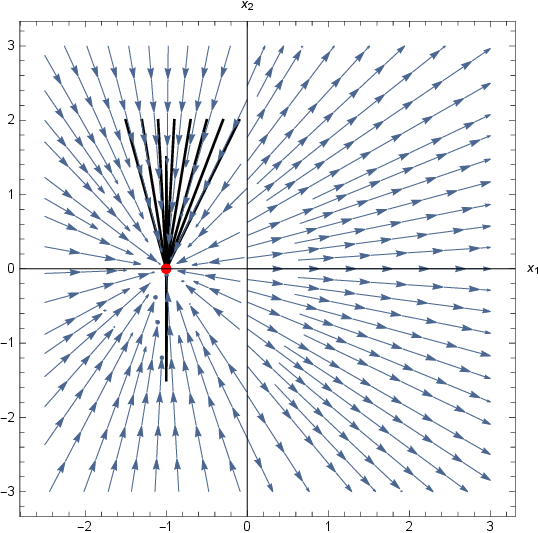}
\caption{{\it{The vector flow and trajectories in the $x_1-x_2$
plane for the $F(R)$ gravity dynamical system near a Big Rip
singularity.}}} \label{plot1}
\end{figure}

Let us now analyze in some detail the phase space behavior of
$F(R)$ gravity near a Big Rip singularity.
Since for a Big Rip singularity, the parameter $\beta$ takes values $\beta>1$, we
consider first the case that $\beta$ takes large values, that is,
$\beta\gg 1$. This case offers great simplifications since in
this case, the parameter $m$ becomes $m\simeq -\frac{1}{(N-N_c)^2}$, therefore the dynamical system can be
integrated analytically.
The solutions for $x_1(N)$, $x_2(N)$, and $x_3(N)$ in this case are,
\begin{align}
\label{generalsolutionsforx1x2x3}
x_1(N)=&\, \frac{-\frac{3 \sqrt{\pi } N_c \e^N \mathrm{Erf}\left(\sqrt{N-N_c}\right)}{\sqrt{N-N_c}}+2 \e^{N_c} N-2 \e^{N_c} N_c
+2 \e^{N_c}}{\e^{N_c} (2 N_c-2 N+1)-2 \sqrt{\pi } \e^N (N-N_c)^{3/2} \mathrm{Erf}\left(\sqrt{N-N_c}\right)} \nonumber \\
&\, +\frac{\frac{2 \sqrt{\pi } N_c^2 \e^N \mathrm{Erf}\left(\sqrt{N-N_c}\right)}{\sqrt{N-N_c}}+\frac{2 \sqrt{\pi }
\e^N N \mathrm{Erf}\left(\sqrt{N-N_c}\right)}{\sqrt{N-N_c}}}{\e^{N_c} (2 N_c-2 N+1)-2 \sqrt{\pi } \e^N (N-N_c)^{3/2} \mathrm{Erf}\left(\sqrt{N-N_c}\right)} \nonumber \\
&\, -\frac{\frac{2 \sqrt{\pi } \e^N N^2 \mathrm{Erf}\left(\sqrt{N-N_c}\right)}{\sqrt{N-N_c}}+\frac{4 \sqrt{\pi } N_c
\e^N N \mathrm{Erf}\left(\sqrt{N-N_c}\right)}{\sqrt{N-N_c}}}{\e^{N_c} (2 N_c-2 N+1)-2 \sqrt{\pi } \e^N (N-N_c)^{3/2} \mathrm{Erf}\left(\sqrt{N-N_c}\right)} \nonumber \\
&\, +\frac{\sqrt{\pi } \e^N N \mathrm{Erf}\left(\sqrt{N-N_c}\right)}{\sqrt{N-N_c} \left(\e^{N_c} (2 N_c-2 N+1)-2 \sqrt{\pi }
\e^N (N-N_c)^{3/2} \mathrm{Erf}\left(\sqrt{N-N_c}\right)\right)} \, ,\nonumber \\
x_2(N)= &\, \frac{\mathcal{C}_2 \e^{N_c-N} (N_c-N)}{2 \sqrt{\pi } (N-N_c)^{3/2} \mathrm{Erf}\left(\sqrt{N-N_c}\right)+\e^{N_c-N} (2 N_c-2 N+1)} \nonumber \\
&\, -\frac{\e^{-N} (N_c-N) \left(\frac{\e^{N_c} (8 N_c-8 N+1)}{2 (N_c-N)^2}-\frac{4 \sqrt{\pi }
\e^N \mathrm{Erf}\left(\sqrt{N-N_c}\right)}{\sqrt{N-N_c}}\right)}{2 \sqrt{\pi } (N-N_c)^{3/2} \mathrm{Erf}\left(\sqrt{N-N_c}\right)+\e^{N_c-N} (2 N_c-2 N+1)} \, , \nonumber \\
x_3(N)= &\, \frac{1}{2 N_c-2 N}+2\, ,
\end{align}
with $\mathrm{Erf}(x)$ being the error function, and furthermore,
the parameter $\mathcal{C}_2$ is a freely chosen integration
constant. Asymptotically, in the case $N\to \infty$, the solutions
$x_1(N)$, $x_2(N)$, and $x_3(N)$ are,
\begin{align}
\label{asymptoticformsofthevariables}
x_1(N)\simeq -\frac{(N_c-N)^2}{N^2}\simeq -1\, ,\quad
x_2(N)\simeq -\frac{2}{N}\simeq 0\, , \quad x_3(N)\simeq 2\, .
\end{align}
Remarkably, as the Big Rip singularity is approached
asymptotically for $N\to \infty$, the functions approach the phase
space point $(x_1,x_2,x_3)=(-1,0,2)$. Recall that from
Eq.~(\ref{fixedpointdesitter}) the trajectories in the phase space
of $F(R)$ gravity approaching a Big Rip singularity are attracted
to the stable de Sitter point. Hence, in a nutshell the
asymptotically autonomous dynamical system of $F(R)$ gravity near
a Big Rip singularity approaches the unstable de Sitter point
$\phi_*^1$ of the de Sitter subspace of the $F(R)$ gravity phase
space. In general, this is not the general rule in dynamical
systems. Proving that a solution of an autonomous dynamical system
is also a solution of the non-autonomous dynamical system is quite
difficult formally, but in our case we proved it through the
analytic solutions we obtained. Now apart from the major feature
apart from the fact that the dynamical system trajectories near a
Big Rip singularity are attracted to a stable de Sitter attractor,
it is important to highlight another major feature, and
specifically that the total EoS parameter near the singularity is
$w_\mathrm{eff}\simeq -1$. This stems from the fact that the fixed
point $\phi_*^1$ has $x_3=2$, hence, the total EoS parameter which
is given in Eq.~(\ref{equation of state1}) tends to the value
$w_\mathrm{eff}\simeq -1$ as the Big Rip singularity is
approached. Thus this shows the important feature of our analysis
that as the Big Rip singularity is approached, the Universe
accelerates, thus the singularity is approached in an accelerating
way. In order to further highlight the behavior of the $F(R)$
gravity phase space near a Big Rip singularity, we shall analyze
the dynamical system in a numerical way. In FIG.~\ref{plot1} upper
left, we showcase the vector-flow and the trajectories of the
phase space in the $x_1-x_2$ for $x_3=2$, using various initial
conditions near the fixed point $(x_1,x_2)=(-1,0)$. It is worth to
further study the behavior of the phase space, so in
FIG.~\ref{plot1}, we present the vector flow and trajectories in
the plane for the dynamical system that is obtained for $x_3=2$,
for various initial conditions near the point, with the red dot
indicating the fixed point $(x_1,x_2)=(-1,0)$.
In all cases, the trajectories with initial conditions near the
Big Rip de Sitter fixed point values, tend asymptotically to the
fixed point, which is stable. This means that once the
trajectories are attracted to the fixed point, they remain there
permanently.

Now let us address another important issue, namely that of the
behavior of the $F(R)$ gravity function near the Big Rip singularity.
As it is conceivable, only approximate forms of $F(R)$ gravity can be obtained due to the complexity of the field equations.
To this end, we shall utilize the functional forms of
the fixed point variables, namely $(x_1,x_2,x_3)=(-1,0,2)$ and
their definition in terms of the $F(R)$ gravity function and the
Hubble rate. Since asymptotically we have $x_1\simeq -1$ we get,
\begin{align}
\label{diffeqx1}
-\frac{\dot{F'}}{F'H}=-1\, ,
\end{align}
hence, by using Eq.~(\ref{hubblerate}) and also the fact that at
leading order near the Big Rip singularity the Ricci scalar at
leading order is,
\begin{align}
\label{leadingorderricciscalar}
R(t)\simeq 12 h_s^2 (t_s-t)^{-2 \beta }\, ,
\end{align}
we get the following solution for the derivative of the $F(R)$
gravity function,
\begin{align}
\label{frbigripcase}
F'(R)\simeq \exp \left(\gamma R^{\frac{\beta -1}{2 \beta }}\right)+\Lambda_I\, ,
\end{align}
with $\Lambda_I$ being an arbitrary integration constant, and the
parameter $\gamma$ is equal to,
\begin{align}
\label{gamma}
\gamma=\frac{h_s}{(\beta-1)(12h_s^2)^{\frac{\beta-1}{2\beta}}}\, .
\end{align}
Upon integration of Eq.~(\ref{frbigripcase}) with respect to $R$,
we get the functional form of the $F(R)$ gravity as the Big Rip
singularity is approached, which is,
\begin{align}
\label{frfinalbigrip1}
F(R)\simeq \Lambda_\mathrm{I}\,R+\frac{2 \beta \gamma ^{-\frac{2 \beta }{\beta -1}} \Gamma \left(\frac{2 \beta }{\beta -1},-R^{\frac{\beta -1}{2 \beta }} \gamma \right)}{\beta -1}+\Lambda_\mathrm{II}\, ,
\end{align}
with $\Lambda_\mathrm{II}$ being an arbitrary integration constant.
We can further simplify the $F(R)$ gravity utilizing the fact that as
$t\to t_s$, the Ricci scalar blows up $R\to \infty$, and also due
to the fact that $\beta>1$, the $F(R)$ gravity of Eq.~(\ref{frfinalbigrip1}) can be further approximated as,
\begin{align}
\label{frfinalbigrip2}
F(R)\simeq \Lambda_\mathrm{I}\,R-\frac{\left(2 \beta \gamma
^{\frac{\beta +1}{\beta -1}-\frac{2 \beta }{\beta -1}}\right)
R^{\frac{\beta +1}{2 \beta }} \e^{\gamma R^{\frac{\beta -1}{2\beta }}}}{\beta -1}+\Lambda_\mathrm{II}\, .
\end{align}
We need to note that for consistency we need to require that the
parameter $\beta$ has the form $\beta=2n/(2m+1)$, with $n$ and
$m$ being positive integers. The resulting expression for $F(R)$
gravity in Eq.~(\ref{frfinalbigrip2}) is functionally similar to
the ones found in Refs.~\cite{Bamba:2008ut,Nojiri:2008fk}, however,
the general analytic treatment of finding the $F(R)$ gravity near
general forms of singularities could be quite demanding.

To recapitulate the findings of this section, let us highlight the most important findings.
As a Big Rip singularity is approached, the $F(R)$ gravity dynamical system becomes asymptotically
autonomous, and its solutions tend asymptotically to the stable de Sitter fixed point of the autonomous $F(R)$ gravity dynamical system for a de Sitter evolution.
This is quite remarkable and it indicates two things: firstly and more importantly that the
Universe as it approaches a Big Rip singularity driven by an
$F(R)$ gravity, it approaches in an accelerating way.
Simply state that in $F(R)$ gravity, a Big Rip singularity is approached during the DE era. Secondly the fact that the autonomous de Sitter $F(R)$ gravity dynamical system and the $F(R)$ gravity dynamical system near a Big Rip singularity, which is render autonomous near
the Big Rip singularity, share the same fixed point solution, or
stated differently, the same final attractors in the theory.
Let us note that in general, the addition of an $R^2$ term in the
$F(R)$ gravity Lagrangian may significantly affect the development
of the singularity since $R^2$ terms are known that they remedy
finite-time singularities
\cite{Bamba:2008ut,Nojiri:2008fk,Capozziello:2009hc,Nojiri:2009pf,Elizalde:2010ts}.
In addition, the combined presence of an $R^2$ term may provide a
unified description of inflation with DE era
\cite{Bamba:2008ut,Nojiri:2008fk,Capozziello:2009hc,Nojiri:2009pf,Elizalde:2010ts,Oikonomou:2022tux,Oikonomou:2020qah,Odintsov:2020iui,Odintsov:2020nwm}.
However, the addition of an $R^2$ term greatly obscures the
mathematical appearance of the dynamical system and thus makes it
impossible to reveal the behavior of the trajectories in this case.

Finally, let us note that the correspondence between the Einstein
and Jordan frames, may reveal important relationships between the
two frames, for example a Big Rip singularity in the Jordan frame
may correspond to a Type IV singularity in the Einstein frame
\cite{Bamba:2008hq}.

\subsubsection{The Cases of Type III, Type II and Type IV
Singularities}

We shall now consider the cases of Type III, Type II, and Type IV
singularities, in which cases, as the cosmic time tends to the
singularity occurring value $t_s$, we have that $N\to N_c$, see
Eq.~(\ref{tsasfunctionofefoldings}). Hence, in this case we have,
\begin{align}
\label{alpjarequirements}
\beta=\frac{2m}{2n+1}\, ,
\end{align}
with $n$ and $m$ being positive integers. Let us analyze first the
case of the Type III singularity, and since as $N\to N_c$, the
parameter $m$ defined in Eq.~(\ref{parametermasfunctionofN}) diverges.
The analytic treatment of both the Type III and Type II
singularities is difficult to tackle, however, for the case of the
Type IV singularity, things are easier, in the case that
$\beta\ll -1$. In this case, $m$ becomes $m\simeq -\frac{1}{(N-N_c)^2}$ and thus the dynamical
system~(\ref{dynamicalsystemnonautonomousgeneral}) has the solutions
$x_1(N)$, $x_2(N)$, and $x_3(N)$ which are identical with the ones
in Eq.~(\ref{generalsolutionsforx1x2x3}), for a general $N$, but
for the case at hand, at $N=N_c$, the parameters $x_1(N)$,
$x_2(N)$, and $x_3(N)$ become,
\begin{align}
\label{asymptoticformsofthevariables1}
x_1(N)\simeq&\, \left(\frac{4 N_c}{3}+12\right) (N-N_c)+2\, , \nonumber \\
x_2(N)\simeq&\, -\frac{1}{2(N-N_c)}-3-(\mathcal{C}_2-12)(N-N_c)\, , \nonumber \\
x_3(N)=&\, \frac{1}{2 N_c-2 N}+2\, ,
\end{align}
and therefore, as $N\to N_c$, the trajectories of the dynamical
system approaches $(x_1,x_2,x_3)=(2,-\infty,\infty)$ in the phase space.
It is noticeable that although $x_2$ and $x_3$ diverge as
the singularity is approached, the Friedmann constraint is still
satisfied because the singularities in the variables $x_2$ and $x_3$ cancel.
We can reveal the functional form of the $f(R)$
gravity near the Type IV singularity, simply by solving the
differential equation which stems form the condition $x_1=2$.
Following the same procedure as in the Big Rip case, as $N\to N_c$
we have at leading order,
\begin{align}
\label{dominanttypeivcurvature}
R\simeq 6h_s\beta(t_s-t)^{-\beta-1}\, ,
\end{align}
hence, the derivative of the $F(R)$ gravity reads,
\begin{align}
\label{frtypeiiibig}
F'(R)\simeq \Lambda_\mathrm{III}+\exp \left(-\gamma_I R^{-\frac{1-\beta }{\beta +1}}\right)\, ,
\end{align}
with $\Lambda_\mathrm{III}$ being an integration constant and the
parameter $\gamma_I$ is,
\begin{align}
\label{gammaI}
\gamma_I=\frac{2h_s}{(1-\beta)(6h_s |\beta|)^{-\frac{1-\beta }{\beta +1}}}\, .
\end{align}
The integration of Eq.~(\ref{frtypeiiibig}) yields the following
functional form of the $F(R)$ gravity,
\begin{align}
\label{frfinaltypeiii}
F(R)\simeq \Lambda_\mathrm{III}\,R+\frac{(\beta +1) \gamma_I^{\frac{\beta +1}{1-\beta }}
\Gamma \left(\frac{\beta +1}{\beta -1},R^{\frac{\beta -1}{\beta +1}} \gamma_I\right)}{1-\beta }+\Lambda_\mathrm{IV}\, ,
\end{align}
with $\Lambda_\mathrm{IV}$ being an integration constant.
We can further simplify the above functional form of the $F(R)$ gravity by
exploiting the fact that as $N\to N_c$ the Ricci scalar tends to
zero because $\beta\ll -1$, hence, we have at leading order,
\begin{align}
\label{leadingorderfrtypeiii}
F(R)\simeq R+\Lambda_\mathrm{III}\,R+\frac{(\beta +1) \gamma_I^{\frac{\beta +1}{1-\beta }}
\Gamma \left(\frac{\beta +1}{\beta -1}\right)}{1-\beta }-\frac{(\beta +1) \gamma_I R^{\frac{2 \beta }{\beta +1}}}{2 \beta }+\Lambda_\mathrm{IV}\, .
\end{align}
Furthermore for the extremely soft Type IV singularity $\beta\ll -1$, we have,
\begin{align}\label{typeivfinalfr}
F(R)\simeq R+\Lambda_\mathrm{III}\,R-\frac{1}{\gamma_I}-\frac{\gamma_I R^2}{2}+\Lambda_\mathrm{IV}\, .
\end{align}

\subsubsection{The Case of non-vacuum $F(R)$ Gravity}

In the previous subsections we considered the dynamical system of
$F(R)$ gravity near finite-time singularities in the absence of matter perfect fluids.
In this subsection, we shall consider the
inclusion of perfect matter fluids in the dynamical system of $F(R)$ gravity.
As we shall evince, this will perplex things in
the dynamical system when finite-time singularities are approached.
With regard to perfect fluids we shall consider
non-relativistic matter and radiation perfect fluids.
In the presence of matter perfect fluids, the field equations of $F(R)$
gravity for a FLRW metric read,
\begin{align}
\label{JGRG15new}
0 =&\, -\frac{F(R)}{2} + 3\left(H^2 + \dot H\right)
F'(R) - 18 \left( 4H^2 \dot H + H \ddot H\right) F''(R)+\kappa^2 \rho_{\text{matter}}\, ,\\
\label{Cr4bb}
0 =&\, \frac{F(R)}{2} - \left(\dot H + 3H^2\right)F'(R) + 6 \left( 8H^2 \dot H + 4 {\dot H}^2 + 6 H \ddot H + \dddot H\right)
F''(R) + 36\left( 4H\dot H + \ddot H\right)^2 F''(R) \nonumber \\
&\, + \kappa^2p_{\text{matter}}\, ,
\end{align}
with $\rho_{\text{matter}}$ and $p_{\text{matter}}$ being the total effective
energy density and the total effective pressure of all the matter
fluids present. In this case we introduce the following
dimensionless variables,
\begin{align}
\label{variablesslowdownnew}
x_1=-\frac{\dot{F'}(R)}{F'(R)H}\, ,\quad x_2=-\frac{F(R)}{6F'(R)H^2}\, ,\quad x_3=\frac{R}{6H^2}\, ,\quad
x_4=\frac{\kappa^2\rho_r}{3FH^2}\, ,\quad x_5=\frac{\kappa^2\rho_M}{3FH^2}\, ,
\end{align}
where $\rho_r$ and $\rho_M$ respectively denotes the energy density of radiation and the matter fluid. In the presence of
matter fluids, using Eqs.~(\ref{JGRG15new}) and the variables
(\ref{variablesslowdownnew}), the dynamical system
(\ref{dynamicalsystemmain}) of the vacuum $F(R)$ gravity takes the
following form,
\begin{align}
\label{dynamicalsystemmain2}
\frac{d x_1}{d N}=&\, -4+3x_1+2x_3-x_1x_3+x_1^2+3x_5+4x_4\, , \nonumber \\
\frac{d x_2}{d N}=&\, 8+m-4x_3+x_2x_1-2x_2x_3+4x_2 \, , \nonumber \\
\frac{d x_3}{d N}=&\, -8-m+8x_3-2x_3^2 \, , \nonumber \\
\frac{d x_4}{d N}=& x_4x_1-2x_4x_3 \, ,\nonumber \\
\frac{d x_5}{d N}=&\, x_5+x_5x_1-2x_5x_3 \, ,
\end{align}
where the parameter $m$ is defined in Eq.~(\ref{parameterm}).
In the case the Hubble rate is given in Eq.~(\ref{hubblerate}), the
parameter $m$ is given in Eq.~(\ref{parametermasfunctionofN}),
but in this case, the dynamical system cannot be analytically
solved even in the limiting cases of the parameter $m$ considered
in the previous subsections.
In addition it is impossible to reveal the behavior of the dynamical system even numerically,
except for the variable $x_3$ which behaves in an identical way as
in the autonomous dynamical system case.
Now let us stress for the first time the difference between a finite-time cosmological
singularity and a singularity of the dynamical system variables.
If some of the variables $x_i$ of the dynamical system blow up at
some finite-time, this singular behavior does not necessarily
indicate a finite-time cosmological singularity.
Take for example the variable $x_1$, for the vacuum $F(R)$ gravity case.
As the Big Rip cosmological singularity is approached, the variable $x_1$
tends asymptotically to the value $x_1\to -1$ while when the rest
of the finite-time singularities are considered, the variable
$x_1\to 2$ as the finite-time singularities are approached.
Thus, the singularities in the field variables do not
necessarily indicate a cosmological singularity, except for the variables $x_4$ and $x_5$ in Eq.~(\ref{variablesslowdownnew})
which depend explicitly on the energy density.
Regarding finite-time dynamical system singularities, these can offer
insights toward the complete understanding of the behavior of the
trajectories in the phase space.
The most formal way to study finite-time dynamical system singularities is by studying the
dominant balances of the dynamical system based on a $\psi$-series
approach near a finite-time dynamical system singularity
\cite{GORIELY2000422}, see also Refs.~\cite{Odintsov:2018uaw,Odintsov:2018awm}
for recent cosmological applications.
This method only applies for autonomous dynamical
systems and thus cannot be applied for the $F(R)$ gravity case or
some other modified gravity. Interacting multifluid cosmologies
though are relatively simple theories and their dynamical systems
is in most cases autonomous, thus can be studied using the
dominant balances method. This is the subject of the next
subsection.

\subsection{Finite-time Cosmological and Dynamical Systems
Singularities in Interacting Multifluids Cosmology }

The dark sector is composed by the DM and DE fluids, which may or may not interact.
The perfect fluids can be different in nature, for example the DE fluid may be generated by some non-trivial
underlying modified gravity.
Without explicitly using the modified gravity we may model the DE fluid in an agnostic way by a dark
fluid which interacts with the other dark sector fluids, such as
the one of DM, adding an interaction among them.
The interaction between the dark sectors though must carefully be chosen because
it may have considerable effects on the primordial matter density
perturbations \cite{Eingorn:2015rma,Koshelev:2010umw}. Both the
dark sector fluids though cannot interact with the baryonic
perfect fluid in order to avoid unobservable fifth force effects.
We shall write the field equations corresponding to the three
fluids and we shall form them in an autonomous dynamical system
way in order to study its trajectories, its singularities and the
connection of finite-time dynamical system singularities with the
finite-time cosmological singularities.
For the dynamical systems singularities, we shall use the dominant balances method developed
in \cite{GORIELY2000422} which we shall briefly review.

For a flat FLRW metric, the three fluid cosmological field
equations in a modified gravity cosmology context reads,
\begin{align}
\label{flateinstein-IMF}
H^2=\frac{\kappa^2}{3}\rho_\mathrm{tot}\, ,
\end{align}
with $\rho_\mathrm{tot} =\rho_{\rm DM}+\rho_{\rm DE}+\rho_b$ denoting the total energy density of the
cosmological fluids where $\rho_{\rm DM}$, $\rho_{\rm DE}$, $\rho_b$ are respectively the energy density of the pressureless DM, DE and baryons. 
Now, differentiating Eq.~(\ref{flateinstein-IMF}) with respect to the cosmic time and using the conservation equation for the total fluid we get, 
\begin{align}
\label{derivativeofh-IMF}
\dot{H}=-\frac{\kappa^2}{2}\left( \rho_\mathrm{tot}+p_\mathrm{tot} \right)\, ,
\end{align}
with $p_\mathrm{tot}$ being the total pressure of the fluids,
which basically consists of the pressure of the DE fluid since DM and baryons are pressureless. For the
DE sector, we shall consider a generalized EoS of the form
\cite{Nojiri:2005sr},
\begin{align}
\label{darkenergyequation-of-state-IMF}
p_{\rm DE}=-\rho_{\rm DE}-A\kappa^4\rho_{\rm DE}^2\, ,
\end{align}
with $A$ being a real dimensionless parameter. From the
energy-momentum conservation equations, we have,
\begin{align}
\label{continutiyequations-IMF}
\dot{\rho}_b+3H\rho_b=&\, 0\, , \nonumber \\
\dot{\rho}_{\rm DM}+3H\rho_{\rm DM}=&\, Q\, , \nonumber \\
\dot{\rho}_{\rm DE}+3H(\rho_{\rm DE}+p_{\rm DE})=&\,-Q\, ,
\end{align}
where $Q$ denotes the interaction term among the DE and DM
fluids and the sign of $Q$ determines which fluid gains energy at
the expense of the other fluid which loses energy. Here, $Q > 0$ implies that energy flow takes place from DE to DM (i.e. DE fluid loses energy and DM fluid gains energy) and $Q < 0$ indicates that energy flows from DM to DE (i.e. DM fluid loses energy and DE fluid gains energy).
We shall assume that $Q$ has the following form,
\begin{align}
\label{qtermform-IMF}
Q=3H(c_1\rho_{\rm DM}+c_2\rho_{\rm DE})\, ,
\end{align}
which has many phenomenological supports
\cite{Caldera-Cabral:2008yyo,Pavon:2005yx,Quartin:2008px,Sadjadi:2006qp,Zimdahl:2005bk}.
Note that $c_1$, $c_2$, are the coupling parameters of the interaction model (\ref{qtermform-IMF}) and they are real constants.
We can construct an autonomous dynamical system for the cosmological
system at hand based on the equations (\ref{flateinstein-IMF}),
(\ref{derivativeofh-IMF}), and (\ref{continutiyequations-IMF}), so we 
choose the dimensionless variables of the dynamical system as follows,
\begin{align}
\label{variablesofdynamicalsystem-IMF}
x_1=\frac{\kappa^2\rho_{\rm DE}}{3H^2}\, ,\quad x_2=\frac{\kappa^2\rho_{\rm DM}}{3H^2}\, ,\quad x_3=\frac{\kappa^2\rho_b}{3H^2}\, ,\quad z=\kappa^2H^2\, .
\end{align}
The variables of the dynamical system $x_i$ $\left(i=1,2,3\right)$ satisfy
the Friedmann equation, which is the case at hand is,
\begin{align}
\label{friedmannconstraint-IMF}
x_1+x_2+x_3=1\, .
\end{align}
Also the total EoS, $w_\mathrm{eff}=\frac{p_{\rm DE}}{\rho_\mathrm{tot}}$,
of the cosmological system, is also expressed in terms of the
variables of the dynamical system in the following way,
\begin{align}
\label{equationofstatetotal-IMF}
w_\mathrm{eff}=-x_1-3Ax_1^2z\, .
\end{align}
In view of the cosmological equations (\ref{flateinstein-IMF}),
(\ref{derivativeofh-IMF}), and (\ref{continutiyequations-IMF}) considered
together with the dynamical system variables~(\ref{variablesofdynamicalsystem-IMF}), we get
\begin{align}
\label{dynamicalsystemmultifluid-IMF}
\frac{d x_1}{d N}=&\, -\frac{\kappa^2Q}{3H^3}+9Ax_1^2z^2+3x_1x_2+3x_1x_3-9Azx_1^3\, , \nonumber \\
\frac{d x_2}{d N}=&\, \frac{\kappa^2Q}{3H^3}-3x_2+3x_2^2+3x_2x_3-9Ax_1^2x_2z\, , \nonumber \\
\frac{d x_3}{d N}=&\, -3x_3+3x_3^2+3x_3x_2-9Ax_1^2x_3z\, , \nonumber \\
\frac{d z}{d N}=&\, -3x_2z-3x_3z+9Ax_1^2z^2\, ,
\end{align}
which hold true for a general interaction term $Q$.
Also we used the $e$-foldings number as a dynamical variable instead of the cosmic time.
If we use the form of the interaction term $Q$, we
further have two terms in the dynamical system~(\ref{dynamicalsystemmultifluid-IMF}), which are
\begin{align}
\label{additionalterms}
\frac{\kappa^2Q}{3H^3}=3c_1x_2+3c_2x_1\, ,
\end{align}
hence, the dynamical system~(\ref{dynamicalsystemmultifluid-IMF}) has
two additional linear contributions for the dynamical system
variables $x_1$ and $x_2$.


\subsubsection{Singularity Structure of Autonomous Dynamical
Systems Using the Dominant Balances Technique}

In this subsection we shall introduce the dominant balances
technique for studying the finite-time singularities of polynomial
autonomous dynamical systems of an arbitrary dimension.
This method was introduced by \cite{GORIELY2000422}, and the theorems that
apply may specify the nature of the finite-time dynamical system singularities.
Extra work is required in order to reveal whether a
finite-time dynamical system singularity is a physical finite-time
cosmological singularity though.
For brevity we shall refer to the dominant balances framework as ``dominant balance analysis''.
Let a general a $n$-dimensional dynamical system having the following form,
\begin{align}
\label{dynamicalsystemdombalanceintro-IMF}
\dot{x}=f(x)\, ,
\end{align}
with $x$ being some real vector of $R^n$, and also
$f(x)=\left(f_1(x),f_2(x),...,f_n(x)\right )$ is some vector
containing polynomials of $x$.
A finite-time singularity developed by this dynamical system is some sort of a
moving singularity, which is always related to and determined by
the initial conditions of the vector $x$.
We can quantify the terminology moving singularity, which is a singularity of the form
$(t-t_c)^{-p}$, with $t_c$ being simply an integration constant.
An example is given by the differential equation
$\frac{d y}{d x}=\frac{1}{x^2y^2}$, which has the
solution $y=(\frac{1}{x}-c)^{-1}$, with $c$ being an integration constant.
This solution develops a singularity which mainly
depends on the chosen initial conditions, at $\frac{1}{x}=c$, so
this justifies why it is called a moving singularity.
Now in order to find whether the autonomous dynamical system~(\ref{dynamicalsystemdombalanceintro-IMF}) develops finite-time
singularities is predominantly based on the decomposition, or
simply truncation, of the function $f$ in several possible
dominant and subdominant terms, which we will denote as
$\hat{f}(x)$ and $\breve{f}(x)$, so in effect the dominant terms
of the dynamical system yield,
\begin{align}
\label{dominantdynamicalsystem-IMF}
\dot{x}=\hat{f}(x)\, .
\end{align}
The dominant terms of the function $f(x)$ can be chosen in several
ways, and also recall that one polynomial term is allowed to be
inserted in the dominant vector $\hat{f}(x)$ each time.
Each component $x_i$, $i=1,2,...,n$ of the vector $x$ is written as follows,
\begin{align}
\label{decompositionofxi-IMF}
x_1(\tau)=a_1\tau^{p_1},\,\,\,x_2(t)=a_2\tau^{p_2},\,\,\,....,x_n(t)=a_n\tau^{p_n}\, ,
\end{align}
and also notice that we require that the full dynamical system
solution $x$ may be written in a form of $\psi$-series in terms of
$\tau=t-t_c$, with $t_c$ being the time instance that the singularity occurs.
Upon substituting the forms of the solutions
$x_i$'s of Eq.~(\ref{decompositionofxi-IMF}), in Eq.~(\ref{dominantdynamicalsystem-IMF}),
we equate the powers at each order of the resulting polynomials, for each distinct choice of $\hat{f}$.
By using this procedure, we will be able to determine
the parameters $p_i$ $\left( i=1,2,...,n \right)$, by having in mind that the
sole solutions accepted are fractional numbers or integers.
Accordingly, we form out of the $p_i$ $\left(i=1,2,...,n\right)$ the new
vector $\vec{p}=(p_1,p_2,...,p_n)$, which is an important
ingredient of the method we apply.
Accordingly we find the parameters $a_i$ $\left(i=1,2,...,n\right)$, which can be determined in a
unique way by simply equating the coefficients of the polynomials
that result from the dominant part $\hat{f}$.
Using these coefficients we form the vector $\vec{a}=(a_1,a_2,a_3,....,a_n)$,
which is called dominant balance.
For the dominant balance analysis, only non-zero values of the balances are allowed, real
or even complex numbers.
The two vectors $\vec{a}$ and $\vec{p}$ form the balance $(\vec{a},\vec{p})$.
Now following the theorem developed by Goriely and Hyde in \cite{GORIELY2000422}, we have that if the
dominant balance contains complex entries, then the autonomous
dynamical system under consideration
(\ref{dynamicalsystemdombalanceintro-IMF}) develops no finite-time
singularities. In the case that the dominant balance entries are
real, then in principle finite-time singularities occur in the
dynamical system, meaning that some trajectories in the phase
space will certainly blow-up for some initial conditions.
But the question is whether these initial conditions that drive the
singular trajectories, are generic or correspond to a limited set of initial conditions.
In order to further enlighten this situation, one needs extra criteria that need to be fulfilled.
To this end, we construct the Kovalevskaya matrix $K$, defined in the following way,
\begin{align}
\label{kovaleskaya-IMF}
K=\left(%
\begin{array}{ccccc}
\frac{\partial \hat{f}_1}{\partial x_1} & \frac{\partial \hat{f}_1}{\partial x_2} & \frac{\partial \hat{f}_1}{\partial x_3} & \cdots & \frac{\partial \hat{f}_1}{\partial x_n} \\
 \frac{\partial \hat{f}_2}{\partial x_1} & \frac{\partial \hat{f}_2}{\partial x_2} & \frac{\partial \hat{f}_2}{\partial x_3} & \cdots & \frac{\partial \hat{f}_2}{\partial x_n} \\
 \frac{\partial \hat{f}_3}{\partial x_1} & \frac{\partial \hat{f}_3}{\partial x_2} & \frac{\partial \hat{f}_3}{\partial x_3} & \cdots & \frac{\partial \hat{f}_3}{\partial x_n} \\
 \vdots & \vdots & \vdots & \ddots & \vdots \\
 \frac{\partial \hat{f}_n}{\partial x_1} & \frac{\partial \hat{f}_n}{\partial x_2} & \frac{\partial \hat{f}_n}{\partial x_3} & \cdots & \frac{\partial \hat{f}_n}{\partial x_n} \\
\end{array}%
\right)-\left(%
\begin{array}{ccccc}
 p_1 & 0 & 0 & \cdots & 0 \\
 0 & p_2 & 0 & \cdots & 0 \\
 0 & 0 & p_3 & \cdots & 0 \\
 \vdots & \vdots & \vdots & \ddots & 0 \\
 0 & 0 & 0 & \cdots & p_n \\
\end{array}%
\right)\, ,
\end{align}
and it should be evaluated at each non-zero dominant balance
$\vec{a}$ found earlier in the method.
The eigenvalues of the Kovalevskaya matrix then, should be of the form $(-1,r_2,r_3,...,r_{n})$.
Now in the case that the $r_2,r_3,...,r_{n}$ are positive, the dynamical system of
Eq.~(\ref{dynamicalsystemdombalanceintro-IMF}) has general
trajectories-solutions which will definitely develop a finite-time
dynamical system singularity. In this case, all the initial
conditions used in this case, drive the dynamical system to a
finite-time dynamical system singularity.
One the other hand if one of the eigenvalues $r_2,r_3,...,r_{n}$ turns out to be
negative, in this case only a limited set of initial conditions
will lead the trajectories to a finite-time singularity, since the
solution found is not general.
It is noticeable that if a singularity is developed by the dynamical system, then the
singularity occurs in the same orthant of $\vec{a}$ in the
$n$-dimensional phase space spanned by the variables $x_i$.
Practically, this means that if the dominant balance value $a_2$
is negative, then the singularity occurs at the orthant $x_2<0$ in the $R^n$ space of the dynamical system variables.
Let us here provide a simple example of how the dominant balance method works,
and to this end consider the dynamical system,
\begin{align}
\label{example1-IMF}
\dot{x}_1=x_1(\alpha+bx_2)\, ,\quad \dot{x}_2=cx_1^2+dx_2\, ,
\end{align}
with $b,c>0$. From the r.h.s. of the above dynamical
system, we can form the following 2-dimensional vector field $f(x_i)$,
\begin{align}
\label{totalf-IMF}
f(x_i)=\left(%
\begin{array}{c}
x_1(\alpha+b\,x_2) \\
c\,x_1^2+dx_2 \\
\end{array}%
\right)\, .
\end{align}
By applying the dominant balances method, we find easily that the
only truncation which is dominant, denoted as $\hat{f}(x_i)$, is the following,
\begin{align}
\label{truncation-IMF}
\hat{f}(x_i)=\left(%
\begin{array}{c}
 b\,x_1\,x_2 \\
 c\,x_1^2 \\
\end{array}%
\right)\, ,
\end{align}
and note that this dominant balance results in an acceptable
dominant balance $(\vec{a},\vec{p})$.
Specifically, the only balances found, denoted as $(\vec{a}_1,\vec{p}_1)$ and
$(\vec{a}_2,\vec{p}_1)$, are the following,
\begin{align}
\label{balancesexample1-IMF}
\vec{a}_1=\left(\frac{1}{\sqrt{b\,c}},-\frac{1}{b} \right)\, ,\quad
\vec{a}_2= \left(-\frac{1}{\sqrt{b\,c}},-\frac{1}{b} \right)\, ,\quad \vec{p}_1=(-1,-1)\, .
\end{align}
Accordingly, the evaluation of the Kovalevskaya matrix $K$ leads
to the eigenvalues $r_1=-1$ and $r_2=2$, hence, due to the main
theorem related to the Kovalevskaya matrix, we find that the
autonomous dynamical system~(\ref{balancesexample1-IMF}) contains
solutions that for general initial conditions are attracted to a
finite-time singularity.

\subsubsection{Dominant Balance Analysis of Multifluid Cosmology, Dynamical System Finite-time Singularities versus Physical Finite-time Singularities}

The analysis that follows was firstly considered in Ref.~\cite{Odintsov:2018uaw}.
In general, the functional form of a cosmological dynamical system variable may reveal whether a
dynamical system finite-time singularity may or may not be related
to a physical finite-time singularity.
Let us consider the variables~(\ref{variablesofdynamicalsystem-IMF}) and apparently, if
the variable $z$ for example is finite in terms of $N$, then the quantity $H^2$ is finite.
In such a case, if one of the variables $x_i$ $\left(i=1,2,3\right)$ diverges, then this would simply imply that one
of the following energy densities $\rho_{\rm DE}$, $\rho_{\rm DM}$, or $\rho_b$ diverges.

In turn, such a situation would imply the occurrence of a Big Rip
or a Type II singularity in general, however, in reality things
are more complicated that it seems.
Considering the Friedmann constraint~(\ref{friedmannconstraint-IMF}) which is satisfied by all
the variables $x_i$~($i=1,2,3$), for all the $e$-foldings numbers,
if finite-time singularities occur in the variables $x_i$~($i=1,2,3$),
then these must occur in such a way so that the
infinities cancel and the Friedman constraint is formally satisfied.
 From a physics standpoint, we first consider the case
that the baryon density is finite, thus we shall seek singular
behaviors in the dark sector variables $x_1$ and $x_2$, in which
case these would cancel in the Friedmann constraint.
Thus the singular behavior would occur in such a way so that the
singularities of the dark sector variables $x_1$ and $x_2$ cancel
when their sum is considered.
For the corresponding dominant balances, this would simply imply that $a_1=-a_2$ and furthermore $p_1=p_2$.
If a finite-time singularity is found in $z$, then the singularities should also occur in the dark sector variables, so
that the Friedmann constraint is finite.
Also note that if $z$ is singular, this indicates a physical singularity for sure, most likely a Type II, a Big Rip, or Type III.
Furthermore, let us not that in all cases the variable $x_3$ must be positive.
Apparently, the most easy case to handle mathematically is the occurrence of a
finite-time singularity in the variable $z$.
This would signify a pressure singularity probably.

In conclusion, the only case that a singularity is acceptable that
can be verified for sure, is if the variable $z$ diverges at
finite-time, and in turn this would indicate the possible presence
of a finite-time singularity.
The type of physical cosmological singularity that occurs in $z$ depends strongly on the dependence
of the Hubble rate on the $e$-foldings number $N$.
Let us see this by using the Hubble rate~(\ref{hubblerate}), which expressed in
terms of the $e$-foldings number $N$, takes the form,
\begin{align}
\label{hubbleefoldingsfirstapprox-IMF}
H(N)\sim (N-N_c)^{\frac{\beta}{\beta-1}}\, .
\end{align}
In the case of a Big Rip singularity ($\beta>1$) and also for the
case of a Type III singularity ($0<\beta<1$), as $t\to t_s$, we
have that $N\to \infty$, hence, the Hubble rate diverges as the
singularity is approached $t\to t_s$. For both the Type II and
Type IV cases, the Hubble rate $H(N)$ tends to zero.
Thus a singular behavior of the dynamical system variable $z$, indicates
either a Big Rip or a Type III singularity physical cosmological
singularity. Hence, with certainty, the only case that a dynamical
system singularity indicates a physical cosmological finite-time
singularity is when $z$ diverges.
It is noticeable that when the Hubble rate diverges, in which case a singularity in $z$ is
developed, the term $1/H(N)^2$ tends to zero, when $N\to \infty$,
in the Big Rip case and when $N\to N_c$, for the case of the Type III singularity.
Therefore, the dark sector dynamical system
variables $x_1$ and $x_2$ are finite when a Big Rip or a Type III
singularity is developed. We used this characteristic example just
to make the point stronger, although the Hubble rate~(\ref{hubblerate}) is not a solution of the cosmological system.
The most difficult dynamical system singularities to interpret are
singularities occurring in one of the dark sector dynamical system
variables $x_1$ and $x_2$, since it is hard to know if a physical
singularity occurs when these two blow-up.
The only certain about the latter case is that the singular behavior must definitely
cancel when the sum $x_1+x_2$ is considered.
Below we summarize the main outcomes of the above discussion when the dynamical system~(\ref{dynamicalsystemmultifluid-IMF}) is considered: 

\begin{itemize}

\item {\it Singularity in $z$:} This indicates a physical singularity, either a Big Rip or Type III.

\item {\it Singularity in $x_1$ and $x_2$:} Dynamical system singularities, we must have, $a_1=-a_2$ and $p_1=p_2$.

\item {\it Singularity in $x_3$:} This is not possible.

\item {\it Constraints on $x_3$:} $x_3>0$ always and non-singular.

\item {\it Singularities in $x_1$ and $x_2$:} If $z$ develops a singularity, then $\rho_{\rm DE}$ and $\rho_{\rm DM}$ diverge 
in such a way so that $\rho_{\rm DM}/H^2$ and $\rho_{\rm DE}/H^2$ are singular, and these two must satisfy $\rho_{\rm DE}=-\rho_{\rm DM}$. 
If the variable $z$ is regular, and furthermore $x_1$ and $x_2$ are singular, then $\rho_{\rm DE}$ and $\rho_{\rm DM}$ are not necessarily singular,
but they must satisfy $\rho_{\rm DE} =-\rho_{\rm DM}$.
\end{itemize}
We note that for all the above cases, the Friedmann constraint $x_1+x_2+x_3=1$ must hold true. 
Now, in the following, we shall analyze the dominant balances and the
corresponding singularities of the multifluid system we considered previously.
Before going to that, we will make some clarifications.
Firstly we shall assume that $x_1(N)$, $x_2(N)$,
$x_3(N)$, and $z(N)$, near the possible finite-time dynamical
system singularities, behave as follows at leading order,
\begin{align}
\label{decompositionofxiactualexample-IMF}
x_1(N)=a_1(N-N_c)^{p_1}\, ,\quad x_2(N)=a_2(N-N_c)^{p_2}\, ,\quad
x_3(N)=a_3(N-N_c)^{p_3}\, ,\quad z(N)=a_4(N-N_c)^{p_4}\, ,
\end{align}
hence, we shall seek for dominant balances $(\vec{a},\vec{p})$,
with the vectors $\vec{a}$ and $\vec{p}$ having the form,
\begin{align}
\label{balancesactualcase-IMF}
\vec{a}=(a_1,a_2,a_3,a_4)\, ,\quad \vec{p}=(p_1,p_2,p_3,p_4)\, .
\end{align}
We can rewrite the dynamical system~(\ref{dynamicalsystemmultifluid-IMF}) in the form
$\frac{d \vec{x}}{d N}=f(\vec{x})$, with
$\vec{x}=(x_1,x_2,x_3,z)$, and also the function
$f(x_1,x_2,x_3,z)$ is defined as,
\begin{align}
\label{functionfmultifluid-IMF}
f(x_1,x_2,x_3,z)=\left(%
\begin{array}{c}
 -c_1x_2-c_2x_1+9Ax_1^2z^2+3x_1x_2+3x_1x_3-9Azx_1^3 \\
 c_1x_2+c_2x_1-3x_2+3x_2^2+3x_2x_3-9Ax_1^2x_2z \\
 -3x_3+3x_3^2+3x_3x_2-9Ax_1^2x_3z \\
 -3x_2z-3x_3z+9Ax_1^2z^2 \\
\end{array}%
\right) \, .
\end{align}
Therefore we shall seek for all the consistent truncations of the
vector $f(x_1,x_2,x_3,z)$ defined in Eq.~(\ref{functionfmultifluid-IMF}), and using the method of dominant
balances we shall examine the behavior of the dynamical system
near finite-time singularities.

\subsubsection{Consistent Truncation I}

One truncation of (\ref{functionfmultifluid-IMF}) has the following form,
\begin{align}
\label{truncation1-IMF}
\hat{f}(x_1,x_2,x_3,z)=\left(
\begin{array}{c}
 9 A x^2_1(N) z^2(N) \\
 3 x^2_2(N) \\
 3 x_3^2(N) \\
 9 A x_1^2(N) z^2(N) \\
\end{array}
\right)\, ,
\end{align}
hence, by following the procedure we developed previously, we find
the following solution for the vector $\vec{p}$,
\begin{align}
\label{vecp1-IMF}
\vec{p}= \left( -\frac{1}{3}, -1, -1, -\frac{1}{3}\right )\, .
\end{align}
In the same way, for the above $\vec{p}$, we find the following
solutions $\vec{a}_1$ and $\vec{a}_2$

\begin{eqnarray}\label{balancesdetails1-IMF} \vec{a}_1=\left( -\frac {1} {3 A^{1/3}}, -\frac {1} {3}, -\frac {1} {3}, -\frac {1} {3 A^{1/3}} \right)\qquad \mbox{and}
\qquad \vec{a}_2=\left( \frac {1} {3 A^{1/3}}, -\frac {1} {3}, -\frac {1} {3}, \frac {1} {3 A^{1/3}} \right), 
 \end{eqnarray}

and the corresponding Kovalevskaya matrix for the truncation~(\ref{truncation1-IMF}) has the following form,
\begin{align}
\label{kobvalev1-IMF}
K=\left(
\begin{array}{cccc}
 18 A x_1 z^2+\frac{1}{3} & 0 & 0 & 18 A x_1^2 z \\
 0 & 6 x_2+1 & 0 & 0 \\
 0 & 0 & 6 x_3+\frac{2}{3} & 0 \\
 18 A x_1 z^2 & 0 & 0 & 18 A z x_1^2+\frac{1}{3} \\
\end{array}
\right)\, .
\end{align}
We can find the Kovalevskaya matrix for each of the vectors
$\vec{a}_i$ $\left(i=1,2\right)$ appearing in Eq.~(\ref{balancesdetails1-IMF}),
hence, for $\vec{a}_1$, the Kovalevskaya matrix has the following form,
\begin{align}
\label{kovalevskayacase1-IMF}
K(\vec{a}_1)=\left(
\begin{array}{cccc}
 -\frac{1}{3} & 0 & 0 & -\frac{2}{3} \\
 0 & -1 & 0 & 0 \\
 0 & 0 & -\frac{4}{3} & 0 \\
 -\frac{2}{3} & 0 & 0 & -\frac{1}{3} \\
\end{array}
\right)\, ,
\end{align}
and also the corresponding eigenvalues for the vector $\vec{a}_1$ are,
\begin{align}
\label{eigenvalues1-IMF}
\left( r_1,r_2,r_3,r_4 \right)=\left(-1, -\frac{4}{3},-1,\frac{1}{3}\right)\, .
\end{align}

The Kovalevskaya matrix for the vector
$\vec{a}_2$, has the same form as for the vector $\vec{a}_1$, and
also the same eigenvalues. In effect, the truncation
(\ref{truncation1-IMF}) shows us that the Kovalevskaya matrix contains, apart from $-1$, other negative eigenvalues, meaning that the singularities do not occur for the general set of initial conditions.

\subsubsection{Consistent Truncation II}

We can form the second consistent truncation of
(\ref{functionfmultifluid-IMF}) as follows,
\begin{align}
\label{truncation12-IMF}
\hat{f}(x_1,x_2,x_3,z)=\left(
\begin{array}{c}
 9 A x_1^2(N) z^2(N) \\
 3 x_2(N) x_3(N) \\
 3 x_2(N) x_3(N) \\
 9 A x_1^2(N) z^2(N) \\
\end{array}
\right)\, ,
\end{align}
which results in the following vector $\vec{p}$,
\begin{align}
\label{vecp12-IMF}
\vec{p}= \left( -\frac{1}{3}, -1, -1, -\frac{1}{3} \right) \, ,
\end{align}
and now there is only one vector
\begin{eqnarray}\label{balancesdetails12-IMF} \vec{a}_1=\left( -\frac {1} {3 A^{1/3}}, -\frac {1} {3}, -\frac {1} {3}, -\frac {1} {3 A^{1/3}}\right).
 \end{eqnarray}

In the same way, for the consistent truncation~(\ref{truncation12-IMF}),
the Kovalevskaya matrix takes the following form,
\begin{align}
\label{kobvalev12-IMF}
K=\left(
\begin{array}{cccc}
 18 A x_1 z^2+\frac{1}{3} & 0 & 0 & 18 A x_1^2 z \\
 0 & 3 x_3+1 & 3 x_2 & 0 \\
 0 & 3 x_3 & 3 x_2+1 & 0 \\
 18 A x_1 z^2 & 0 & 0 & 18 A z x_1^2+\frac{1}{3} \\
\end{array}
\right)\, ,
\end{align}
which for $\vec{a}_{1}$ leads to
\begin{align}
\label{kovalevskayacase12-IMF}
K(\vec{a}_1)=\left(
\begin{array}{cccc}
 -\frac{1}{3} & 0 & 0 & -\frac{2}{3} \\
 0 & 0 & -1 & 0 \\
 0 & -1 & 0 & 0 \\
 -\frac{2}{3} & 0 & 0 & -\frac{1}{3} \\
\end{array}
\right)\, ,
\end{align}
and its eigenvalues are,
\begin{align}
\label{eigenvalues12-IMF}
(r_1,r_2,r_3,r_4)= \left(-1,-1,1,\frac{1}{3} \right) \, .
\end{align}

For the dominant balances, we shall
now analyze the development of phase space singularities for the
dynamical system (\ref{dynamicalsystemmultifluid-IMF}).
It develops
finite-time singularities and further investigation is required in
order to see whether these occur for general initial conditions or for limited ones.
This will be revealed by the form of the
eigenvalues (\ref{eigenvalues12-IMF}), hence, since $r_2<0$, this
indicates that the singularities do not occur for a general set of
initial conditions, but for a limited set of initial conditions.
Furthermore, the singularities for the three phase space variables
$x_1$, $x_2$, and $x_3$, is not the wanted one, as it can be seen
from $\vec{p}$ in Eq.~(\ref{vecp12-IMF}).
This is due to the fact that at leading order, the exponents of $x_1$, $x_2$, and $x_3$, are not of 
the same order, hence, the Friedmann constraint of Eq.~(\ref{friedmannconstraint-IMF})
is not satisfied in this case.
Hence, in this case, the limited set which yields singular solutions,
does not yield physically acceptable solutions.
Therefore, the mathematically consistent truncation~(\ref{truncation12-IMF}) does not
yield singular solutions which are generated by a general set of
initial conditions.

We can form a third consistent truncation of
(\ref{functionfmultifluid-IMF}), which has the following form,
\begin{align}
\label{truncation13-IMF}
\hat{f}(x_1,x_2,x_3,z)=\left(
\begin{array}{c}
 9 A x_1^2(N) z^2(N) \\
 3 c_2 x_1(N) \\
 3 x_3^2(N) \\
 9 A x_1^2(N) z^2(N) \\
\end{array}
\right)\, ,
\end{align}
which results in $\vec{p}$ given by,
\begin{align}
\label{vecp13-IMF}
\vec{p}= \left( -\frac{1}{3}, \frac{2}{3}, -1, -\frac{1}{3}\right) \, ,
\end{align}
while the vectors $\vec{a}_i$ $\left(i=1,2\right)$ are,

\begin{eqnarray}\label{balancesdetails13-IMF} \vec{a}_1=\left( -\frac {1} {3 A^{1/3}}, -\frac{3 c_2}{2 A^{1/3}}, -\frac {1} {3}, -\frac {1} {3 A^{1/3}}\right)
 \qquad \mbox{and} \qquad
\vec{a}_2=\left( \frac {1} {3 A^{1/3}}, -\frac{3 c_2}{2 A^{1/3}}, -\frac {1} {3}, \frac {1} {3 A^{1/3}}\right).
 \end{eqnarray}
and the corresponding Kovalevskaya matrix has the following form,
\begin{align}
\label{kobvalev13-IMF}
K=\left(
\begin{array}{cccc}
18 A x_1 z^2+\frac{1}{3} & 0 & 0 & 18 A x_1^2 z \\
3 c_2 & -\frac{2}{3} & 0 & 0 \\
0 & 0 & 6 x_3+1 & 0 \\
18 A x_1 z^2 & 0 & 0 & 18 A z x_1^2+\frac{1}{3} \\
\end{array}
\right)\, .
\end{align}
For the vector $\vec{a}_1$, the Kovalevskaya matrix takes the following form,
\begin{align}
\label{kovalevskayacase13-IMF}
K(\vec{a}_1)=\left(
\begin{array}{cccc}
 -\frac{1}{3} & 0 & 0 & -\frac{2}{3} \\
 3 c_2 & -\frac{2}{3} & 0 & 0 \\
 0 & 0 & -1 & 0 \\
 -\frac{2}{3} & 0 & 0 & -\frac{1}{3} \\
\end{array}
\right)\, ,
\end{align}
and its eigenvalues are,
\begin{align}
\label{eigenvalues13-IMF}
(r_1,r_2,r_3,r_4)= \left( -1,-1,-\frac{2}{3},\frac{1}{3} \right)\, .
\end{align}

Hence, this truncation leads to a situation with degeneracies, thus, let us
further analyze the case $\vec{a}_1$.
As it occurs, the singularity analysis is identical with the truncation~(\ref{truncation12-IMF}).
Therefore, no general initial conditions can lead to
singularities in the dynamical system in this case too, and
furthermore, the singular solutions are non-physical since the
Friedmann constraint (\ref{friedmannconstraint-IMF}) is not satisfied.
The conclusion of this section is that the dominant balance analysis does not yield any
singular solutions which are generated by general initial conditions.
Thus, it is rather compelling to present in more detail
the phase space of the dynamical system, which we do in the next subsection.


\subsubsection{Phase Space Analysis of Multifluid Cosmological Model}

Now we shall analytically investigate the phase space structure of
the dynamical system in order to further understand the various
phase space structures that emerge in it.
We will focus in finding the fixed points of the dynamical system and we shall examine the
stability of the fixed points against linear perturbations, by
using the well-known Hartman--Grobman theorem, when it applies. The
standard way to analyze autonomous non-linear dynamical system is
based on the linearization techniques essentially quantified by
the Hartman--Grobman linearization theorem.
The latter reveals the stability of the various fixed points and further indicates
whether non-trivial topological structures exist in the phase
space of the dynamical system, only in the case when the fixed points are hyperbolic.
Let us describe in brief the theoretical approach we shall use.
Consider the vector field $\Phi (t)$ $\epsilon$ $R^n$, which satisfies the differential flow,
\begin{align}
\label{ds1}
\frac{d \Phi}{d t}=g(\Phi (t))\, ,
\end{align}
with $g(\Phi (t))$ being a locally Lipschitz continuous map
having the form $g:R^n\rightarrow R^n$. Let the fixed points of
the dynamical system (\ref{ds1}) be denoted as $\phi_*$ and also
let $\mathcal{J}(g)$ denote the Jacobian matrix which corresponds
to the linearized version of the dynamical system (\ref{ds1})
near some fixed point, with the Jacobian matrix being equal to,
\begin{align}
\label{jaconiab}
\mathcal{J}=\sum_i\sum_j\left[\frac{\mathrm{\partial f_i}}{\partial x_j}\right]\, .
\end{align}
In the case of a fixed point being hyperbolic, the Jacobian matrix
evaluated at the fixed point reveals the stability of the fixed
point. Let us remind here that a fixed point is called hyperbolic
only if the spectrum $\sigma (\mathcal{J})$ of the eigenvalues of
the Jacobian matrix is comprised by elements $e_i$ satisfying the
condition $\mathrm{Re}(e_i)\neq 0$.
In this case, following the Hartman--Grobman theorem, the linearized version of the dynamical system,
\begin{align}
\label{loveisalie}
\frac{d \Phi}{d t}=\left. \mathcal{J}(g)(\Phi)\right|_{\Phi=\phi_*} (\Phi-\phi_*)\, ,
\end{align}
is equivalent topologically, to the dynamical system of Eq.~(\ref{ds1}), at the vicinity of the hyperbolic fixed points $\phi_*$.
Specifically, the Hartman--Grobman theorem guarantees the
existence of a homeomorphism $\mathcal{F}:U\rightarrow R^n$, in a neighborhood $U$ of the hyperbolic fixed point $\phi_*$, with $U$ being an open set.
This homeomorphism generates the flow $\frac{d h(u)}{d t}$, satisfying,
\begin{align}
\label{fklow}
\frac{d h(u)}{d t}=\mathcal{J}h(u)\, .
\end{align}
In view of the Hartman--Grobman theorem, the flows in Eqs.~(\ref{fklow}) and (\ref{ds1}) are homeomorphic. Now regarding the
stability of the fixed point $\phi_*$, the Hartman--Grobman
predicts that in the case that the spectrum of the eigenvalues of
the Jacobian matrix are negative, that is
$\mathrm{Re}\left(\sigma (\mathcal{J}(g))\right)<0$, then the
fixed point is stable asymptotically. In all the other cases, the
fixed point leads to instabilities in the phase space.
We shall now apply the Hartman--Grobman theorem on the dynamical system of
Eq.~(\ref{dynamicalsystemmultifluid-IMF}) with the interaction term
being chosen as in Eq.~(\ref{qtermform-IMF}). We shall calculate the
fixed points and we examine their stability against linear perturbations.
Our analysis indicates that for general non-zero
values of $c_2$, the analytic study of the fixed points is very
complicated and somewhat leads to extended and complicated.
So we shall focus on the case $c_2=0$.
Interactions like the one in (\ref{qtermform-IMF}), with $c_2=0$ are frequently used in the
literature~\cite{Boehmer:2008av}.
The case with $c_2\neq 0$ can be
dealt numerically, and in TABLE~\ref{numericalanalysisofsystem}, we
gather our results, for various signs of the free parameters
$c_1$ and $c_2$ of the interaction function (\ref{qtermform-IMF}).
For all the cases appearing in the TABLE~\ref{numericalanalysisofsystem}, the
distinct fixed points and physically interesting fixed points are
unstable, without the result to be dependent on the values of the free parameter $A$.

\begin{table}[h]
\caption{\label{numericalanalysisofsystem}Stability of the Fixed
Points of the Multifluid Dynamical System
(\ref{dynamicalsystemmultifluid-IMF}) for general values of $c_1$ and $c_2$.}
\begin{center}
\renewcommand{\arraystretch}{1.4}
\begin{tabular}{|c@{\hspace{1 cm}}@{\hspace{1 cm}}c@{\hspace{1 cm}} c|}
\hline
\textbf{Case No.} & \textbf{Region of $c_1$ and $c_2$} & \textbf{Nature of Fixed Points}\\
\hline\hline
Case I  &\quad $c_1>0$, $c_2>0$, for every $A$ & Unstable \\

Case II  & \quad $c_1>0$, $c_2<0$, for every $A$ & Unstable \\

Case III & $c_1<0$, $c_2>0$, for every $A$ & Unstable \\

Case IV  &\quad $c_1<0$, $c_2<0$, for every $A$ & Unstable \\

\hline
\end{tabular}
\end{center}
\end{table}

Now let us consider the analytic form of the fixed points for
$c_2=0$, in which case the fixed points are,
\begin{align}
\label{fixedpointsc20}
\phi_1=&\, \left\{x_1\to 0,x_2\to 0,x_3\to 0 \right\}\, , \nonumber \\
\phi_2=&\, \left\{x_2\to 0,x_3\to 0,z\to 0\right\}\, , \nonumber \\
\phi_3=&\, \left\{x_1\to 0,x_2\to 0,x_3\to 1,z\to 0 \right\}\, , \nonumber \\
\phi_4=&\, \left\{x_1\to c_1,x_2\to 1-c_1,x_3\to 0,z\to 0 \right\}\, ,\nonumber \\
\phi_5=&\, \left\{x_1\to \frac{3 A c_1-\sqrt{9 A^2 c_1^2-12 A (c_1-1) c_1}}{6 A c_1}, \right. \nonumber \\
&\qquad \left. x_2\to \frac{1}{2} \left(3 A c_1-\sqrt{3} \sqrt{A c_1 (3 A c_1-4 c_1+4)}-2 c_1+2\right), x_3\to 0,z\to c_1\right\}\, , \nonumber \\
\phi_6=&\, \left\{ x_1\to \frac{\sqrt{9 A^2 c_1^2-12 A (c_1-1) c_1}+3 A c_1}{6 A c_1}, \right. \nonumber \\
& \qquad \left. x_2\to \frac{1}{2} \left(\sqrt{9 A^2 c_1^2-12 A (c_1-1) c_1}+3 A c_1-2 c_1+2\right),x_3\to 0,z\to c_1\right\}\, .
\end{align}
The Jacobian matrix $\mathcal{J}$ for the current scenario, has
the simple closed form,
\begin{align}
\label{jacobianmatrix}
\mathcal{J}={\footnotesize \left(
\begin{array}{cccc}
 3 x_2+3 x_3+9 A x_1 z (2 z-3 x_1) & 3 x_1-3 c_1 & 3 x_1 & -9 A x_1^2 (x_1-2 z) \\
 -9 A x_2 z & 3 (c_1+2 x_2+x_3-3 A x_1 z-1) & 3 x_2 & -9 A x_1 x_2 \\
 -18 A x_1 x_3 z & 3 x_3 & 3 \left(-3 A z x_1^2+x_2+2 x_3-1\right) & -9 A x_1^2 x_3 \\
 18 A x_1 z^2 & -3 z & -3 z & -3 \left(-6 A z x_1^2+x_2+x_3\right) \\
\end{array}
\right) }\, .
\end{align}
For the last two cases, the fixed points correspond to de Sitter
vacuum, since we have $z=c_1$ at the two fixed points.
Regarding the other cases, the variable $z$ is equal to zero at the fixed
point, hence, the most interesting cases from a phenomenological
point of view correspond to $\phi_5$ and $\phi_6$, for which $x_3$
tends to zero. Hence, in this case, asymptotically near the fixed
point, the baryonic fluid does not contribute to the dynamical
evolution of the Universe, asymptotically near the two fixed points.
Furthermore, in order to be consistent physically, the
interaction constant $c_1$ must be $c_1>0$.
The eigenvalues of the Jacobian matrix evaluated at the fixed points, for the first four
fixed points, have the following form,
\begin{align}
\label{firsfoureigenvalues}
\phi^*_1\to&\, \left\{-3,0,0,3 c_1-3 \right\}\, , \nonumber \\
\phi^*_2\to& \left\{-3,0,0,3 c_1-3\right\}\, , \nonumber \\
\phi^*_3 \to&\, \left\{-3,3,3,3 c_1\right\}\, , \nonumber \\
\phi^*_4 \to&\, \left\{3 (1-c_1)-3,-3 (1-c_1),3 (1-c_1),3 (1-c_1) \right\}\, .
\end{align}
Regarding the first two fixed points, these are not hyperbolic,
hence, we cannot use the Hartman--Grobman theorem for analyzing these two.
In contrast, the fixed points $\phi^*_3$ and $\phi^*_4$
are hyperbolic, but not stable because some eigenvalues are positive.
Hence, in this case the fixed points are unstable but
physically appealing because these lead to a Hubble rate that
tends to zero, so this might describe the late-time era.

Now we turn our focus on the fixed points $\phi^*_5$ and
$\phi^*_6$ in which case, the calculation of the eigenvalues
results in a complicated algebraic equation, which is too extended
to present here. By doing a numerical analysis though, for $c_1>0$
and for $A$ being positive and negative, we end up with unstable de
Sitter fixed points. The instability can be easily explained due
to the fact that $\rho$ enters the EoS as $-\rho$ and not as
$w_{\rm DE}\rho$, hence, instability occurs.

Also, a detailed analysis of the trajectories for the dynamical
system at hand verifies the above findings. Indeed, for a wide
range of values of the free parameters $A$ and $c_1$ indicates
that there exist limited initial conditions which lead to
dynamical system blow-ups at finite-time.
Also, there are regions in which the dynamical system has regular trajectories, a fact
which is also supported by the stability study of the fixed points.
Concluding, the dynamical system~(\ref{dynamicalsystemmultifluid-IMF}) which describes a multifluid
cosmology with non-trivial interactions between the dark sector
fluids, is unstable in general and has no global attractors
leading to finite-time singularities, but has limited trajectories
leading to finite-time blow-ups of the dynamical system variables.

\subsection{Dynamical System Analysis of Exponential Quintessence DE Models}

In this section, we shall analyze the dynamical system of
exponential quintessence models, which are relevant in swampland models.
For a full analysis of this model, see~\cite{Odintsov:2018zai}, on which the presentation of this
subsection will be based.
Consider the action of a quintessence scalar field model in vacuum,
\begin{align}
\label{generalactionforquintessence}
\mathcal{S}=\int
d^{4}x\sqrt{-g}\left(\frac{1}{\kappa^2}R-\frac{1}{2}\partial_{\mu}\phi\partial^{\mu}\phi-V(\phi)
\right)\, .
\end{align}

The field equations for this quintessence scalar
theory in a flat FLRW background are,
\begin{align}
\label{scalarequationsofmotion}
0=&\, \ddot{\phi}+3H\dot{\phi}+V'(\phi) \, ,\nonumber \\
\frac{3H^2}{\kappa^2}=&\, \frac{\dot{\phi}^2}{2}+V(\phi)\, ,
\end{align}
with the prime denoting differentiation with respect to $\phi$.
We shall be focusing on exponential quintessence models, hence, the
scalar potential $V(\phi)$ takes the form $V(\phi)=\e^{- \kappa \lambda \phi}$.
For this choice of the potential, the field equations~(\ref{scalarequationsofmotion}) can
form an autonomous dynamical system by introducing the
dimensionless variables,
\begin{align}
\label{variablesdynamicalsystem}
x=\frac{\kappa \dot{\phi}}{\sqrt{6}H}\, ,\quad y=\frac{\kappa \sqrt{V}}{\sqrt{3}H}\, ,
\end{align}
with the ``dot'' indicating the differentiation with respect to $t$.
As in the previous sections, we shall use the $e$-foldings number
$N$ to quantify the dynamical evolution variable, so by combining
Eqs.~(\ref{scalarequationsofmotion}) and
(\ref{variablesdynamicalsystem}), we can form the following
autonomous dynamical system,
\begin{align}
\label{dynamicalsystem}
\frac{d x}{d N}=&\, -3x+\frac{\sqrt{6}}{2}\lambda y^2+\frac{3}{2}x\left(x^2-y^2\right) \, , \nonumber \\
\frac{d y}{d N}=&\, -\frac{\sqrt{6}}{2}\lambda x y+\frac{3}{2}y\left(x^2-y^2\right)\, .
\end{align}
Furthermore, the dimensionless variables $x$ and $y$ satisfy the
Friedmann constraint,
\begin{align}
\label{friedmanconstraint}
x^2+y^2=1\, ,
\end{align}
which means that in the variables $x$ and $y$ the dynamical system does not develop singularities. The phase space structure of this simple autonomous dynamical system 
has been performed in detail in Ref.~\cite{Agrawal:2018own}.

\subsubsection{Singularity Structure of the Dynamical System
Describing Swampland DE Models}

In this subsection, we shall use the dominant balance analysis
described in a previous section for the dynamical system
of Eq.~(\ref{dynamicalsystem}), with the Friedmann constraint~(\ref{friedmanconstraint}) holding always true.
For doing so, we recast the dynamical system of Eq.~(\ref{dynamicalsystem}), in the following form,
\begin{align}
\label{dynsysnewform}
\frac{d \vec{x}}{d N}=f(\vec{x})\, ,
\end{align}
with $\vec{x}=(x,y)$, and furthermore, we defined the
vector-valued function $f(\vec{x})$ as follows,
\begin{align}
\label{functionfmultifluidclassicalcase}
f(x,y)=\left(%
\begin{array}{c}
 f_1(x,y) \\
 f_2(x,y) \\
\end{array}%
\right)\, ,
\end{align}
with the $f_i$'s appearing in Eq.~(\ref{functionfmultifluidclassicalcase}) being equal to,
\begin{align}
\label{functionsficlassicalcase}
f_1(x,y)=&\, -3x+\frac{\sqrt{6}}{2}\lambda y^2+\frac{3}{2}x\left(x^2-y^2\right)\, , \nonumber \\
f_2(x,y)=&\, -\frac{\sqrt{6}}{2}\lambda x y+\frac{3}{2}y\left(x^2-y^2\right) \, .
\end{align}
We can form several truncations of this vector function
$f(\vec{x})$, one of which is the following,
\begin{align}
\label{truncation1classicalcase}
\hat{f}(x,y)=\left(
\begin{array}{c}
 -\frac{3 x y^2}{2} \\
\frac{3 x^2 y}{2} \\
\end{array}
\right)\, .
\end{align}
By using the method of dominant balances, we find for this
truncation that the vector $\vec{p}$ takes the form,
\begin{align}
\label{vecp1classicalcase}
\vec{p}= \left( -\frac{1}{2},-\frac{1}{2} \right)\, ,
\end{align}
and accordingly, the following non-trivial vector $\vec{a}$ is
obtained,
\begin{align}
\label{balancesdetails1classicalcase}
\vec{a}_1= & \, \left( -\frac{i}{\sqrt{3}}, -\frac{1}{\sqrt{3}}\right),\; \vec{a}_2=\left( -\frac{i}{\sqrt{3}}, \frac{1}{\sqrt{3}}\right),\; \vec{a}_3=\left( \frac{i}{\sqrt{3}},\; -\frac{1}{\sqrt{3}}\right),\; 
\vec{a}_4=\left( \frac{i}{\sqrt{3}}, \frac{1}{\sqrt{3}}\right)\, .
\end{align}
All the vectors found above have complex entries, and also the
Friedmann constraint~(\ref{friedmanconstraint}) is satisfied for all the vectors $\vec{a}_i$.
Indeed, at leading order, the expression $x^2+y^2$ forming the Friedmann constraint at leading
order reads,
\begin{align}
\label{friedmannconstraintatasingularity}
\left(\pm \frac{i}{\sqrt{3}}\right)^2\tau^{-\frac{1}{2}}+\left(\pm \frac{1}{\sqrt{3}}\right)^2\tau^{-\frac{1}{2}}\, ,
\end{align}
which vanishes, and we should note that $\tau=N-N_c$.
Now let us investigate whether singular solutions can be found.
Due to the fact that the vectors $\vec{a}_i$ have complex entries, this indicates that no
singular solutions exist, as we have already shown from the Friedmann constraint (\ref{friedmanconstraint}).
What now remains to check is whether these solutions are general or not, meaning whether these
correspond to general initial conditions or not.
The answer to this will be given by the evaluation of the Kovalevskaya matrix
$K$, for the solutions $\vec{a}_i$, and for the truncation~(\ref{truncation1classicalcase}),
\begin{align}
\label{kovalev1classicalcase}
K=\left(
\begin{array}{cc}
 \frac{1}{2}-\frac{3 y^2}{2} & -3 x y \\
 3 x y & \frac{3 x^2}{2}+\frac{1}{2} \\
\end{array}
\right)\, .
\end{align}
The evaluation of the Kovalevskaya matrix $K$ for
$(x,y)=\vec{a}_1$, yields,
\begin{align}
\label{ra}
K(\vec{a})=\left(
\begin{array}{cc}
 0 & -i \\
 i & 0 \\
\end{array}
\right)\, ,
\end{align}
the eigenvalues of which are,
\begin{align}\label{eigenvalues1classicalcase}
(r_1,r_2)=(-1,1)\, .
\end{align}
The same set of eigenvalues are found for all the rest of solutions $\vec{a}_i$,$\left(i=2,3,4\right)$.
The form of the eigenvalues indicates that the dynamical system (\ref{dynamicalsystem}) has
not finite-time singularities, and also these non-singular
solutions correspond to a general set of initial conditions.
However, these non-singular solutions found do not exclude the
existence of an actual physical cosmological finite-time
singularity. What we proved is that the dynamical system variables
$x$ and $y$ do not become singular for a general set of initial
conditions, which does not exclude the case that an actual
physical singularity may be developed. In order to investigate the
occurrence of physical singularities, let us assume that the
Hubble rate has the general form,
\begin{align}
\label{hubblerategeneral}
H(t)=H_s(t)+h_s(t)(t-t_s)^{-\beta}\, ,
\end{align}
with the parameter $\alpha$ taking the general form
$\beta=\frac{2m}{2n+1}$, and $m,n$ are positive integers, and
also $H_s(t)$ and $h_s(t)$ are assumed to be regular at $t=t_s$
and furthermore, for consistency $H_s(t_s)\neq 0$, $h_s(t_s)\neq 0$.
Moreover, the first and second order derivatives of $H_s(t)$
and $h_s(t)$ are assumed to satisfy the same constraints.
Although that the Hubble rate~(\ref{hubblerategeneral}) may not be a
solution to the field equations, we will examine the case that the
dominant part of the solution is of the form~(\ref{hubblerategeneral}) near a singularity.
The results of the singularity structure of the dynamical system indicate that
$x\sim \kappa\frac{\dot{\phi}}{H}$ and $y\sim \kappa\frac{\sqrt{V(\phi)}}{H}$, never
become singular for all the cosmic times.
Now, the singularity structure of the Hubble rate~(\ref{hubblerategeneral}), strongly
depends on the values of $\beta$ and specifically,
\begin{itemize}
\item If $\beta>1$, a physical Big Rip singularity is developed.
\item If $0<\beta < 1$, a physical Type III singularity is developed.
\item If $-1<\beta < 0$, a physical Type II singularity is developed.
\item If $\beta <-1$, a physical Type IV singularity is developed.
\end{itemize}
By combining Eqs.~(\ref{scalarequationsofmotion}) and
(\ref{hubblerategeneral}), we can express the terms $\dot{\phi}$
and $V(\phi)$ as functions of the Hubble rate.
Indeed, the two equations needed are,
\begin{align}
\label{neweqnmotion}
\frac{2}{\kappa^2}\dot{H}=&\, -\dot{\phi}^2\, , \\
\label{potentialasfunctionof}
V(\phi)=&\, -\frac{\ddot{H}+6\dot{H}}{\kappa^2\lambda \dot{\phi}}\, ,
\end{align}
but the resulting expressions are too lengthy to quote here.
Having though these expressions at hand, we shall investigate the
occurrence of physical singularities, having in mind that the
variables $x$ and $y$ never become singular for the dynamical
system at hand.
Considering the case of a Type IV singularity,
this can never occur when $-2<\beta <-1$, since the term $\sim (t-t_s)^{-\beta-2}$ is present in the variable $y$.
On the contrary, when $\beta<-2$, the two variables $x$, $y$ never become
singular, so in the case of a Type IV singularity, this can be
developed by the physical system for the values of $\beta$ in the
range $\beta<-2$. Regarding the Type II case, which recall that it
occurs for $-1<\beta<0$, the dynamical system variable $x$ is
singular in this case since it depends on the term $\sim (t-t_s)^{-\beta-1}$, therefore a Type II singularity can never
occur for the model at hand. Regarding the Type III case, it
cannot also occur due to the term $\sim (t-t_s)^{\frac{3\beta}{2}-\frac{3}{2}}$, which is singular for $0<\beta<1$.
Regarding, the Big Rip singularity, it can always
occur, because for $\beta>1$, the dynamical system variables $x$
and $y$ never develop singular behavior.
The results of our analysis are presented in TABLE~\ref{table1b}.

\begin{table}[h]
\caption{\label{table1b}Type of Physical Singularity for
$H(t)=H_s(t)+h_s(t)(t-t_s)^{-\beta}$}
\begin{center}
\renewcommand{\arraystretch}{1.4}
\begin{tabular}{|c@{\hspace{1 cm}}|@{\hspace{1 cm}} c|}
\hline
\textbf{Type of the Singularity} & \textbf{Occurrence/Non-occurrence}\\
\hline\hline
 Type I &  It occurs for $\beta>1$\\
\hline
Type II &  Does not occur \\
\hline
Type III &  Does not occur \\
\hline
Type IV &  Does not occur for $-2<\beta<-1$, but occurs for $\beta<-2$\\

\hline
\end{tabular}
\end{center}
\end{table}
Therefore, in conclusion, if the dominant behavior of the Hubble rate
solution for the model at hand is given in Eq.~(\ref{hubblerategeneral}), the Type III, and Type II never
occur in the physical system. However, the Type IV and Type I
singularities occur when $\beta<-2$ and for any $\beta>1$, respectively.

\subsection{Dynamical System of Interacting Multifluids in LQC}

In this section, we shall investigate the occurrence of finite-time
singularities in the context of a LQC interacting multifluid
system. The full analysis of the subject of this section was
performed in Ref.~\cite{Odintsov:2018awm} on which the
presentation will be based.
The three fluids that are considered
are a non-interacting baryonic fluid and the interacting DE-DM
fluids. Also the DE fluid will contain a bulk viscosity term.
The Friedmann equation in the context of LQC in the presence of
the three fluids and for a FLRW spacetime, has the following form,
\begin{align}
\label{flateinstein-LQC}
H^2=\frac{\kappa^2\rho_\mathrm{tot}}{3}\left(
1-\frac{\rho_\mathrm{tot}}{\rho_c}\right)\, ,
\end{align}
where, once again, $\rho_c$ denotes the critical density of LQC and $\rho_\mathrm{tot} =\rho_{\rm DM}+\rho_{\rm DE}+\rho_b$ is the total energy density of all the three
fluids present in which $\rho_{\rm DM}$, $\rho_{\rm DE}$ and $\rho_b$ denote the energy density of DM, DE and baryons, respectively. 
Upon differentiation with respect to the cosmic time, Eq.~(\ref{flateinstein-LQC}) yields,
\begin{align}
\label{derivativeofh-LQC}
\dot{H}=-\frac{\kappa^2}{2}\left(\rho_{\rm tot}+p_\mathrm{tot}\right)\left[
1-2\left(\frac{\rho_{\rm tot}+p_\mathrm{tot}}{\rho_c} \right)\right]\, ,
\end{align}
where $p_\mathrm{tot}$ denotes the total pressure of the three
fluids, which is basically equal to the pressure of the DE fluid,
since DM and the baryonic fluids are pressureless. For the DE
fluid, we assume the following EoS \cite{Nojiri:2005sr},
\begin{align}
\label{darkenergyequation-of-state-LQC}
p_{\rm DE}=-\rho_{\rm DE}-A\kappa^4\rho_{\rm DE}^2\, ,
\end{align}
with $A$ being a real dimensionless parameter.
The energy-momentum conservation yields the following three continuity equations,
\begin{align}
\label{continutiyequations-LQC}
\dot{\rho}_b+3H\rho_b=0\, , \quad
\dot{\rho}_{\rm DM}+3H\rho_{\rm DM}=Q\, , \quad
\dot{\rho}_{\rm DE}+3H(\rho_{\rm DE}+p_{\rm DE})=-Q\, ,
\end{align}
where $Q$ denotes the interaction between the dark fluids, and once again $Q>0$ indicates that the DE fluid loses energy and $Q<0$ indicates
that the DE fluid gains energy at the expense of the DM fluid.
We shall consider the following interaction between the dark sector fluids,
\begin{align}
\label{qtermform-LQC}
Q=3H(c_1\rho_{\rm DM}+c_2\rho_{\rm DE})\, ,
\end{align}
which is phenomenologically interesting
\cite{Caldera-Cabral:2008yyo,Pavon:2005yx,Quartin:2008px,Sadjadi:2006qp,Zimdahl:2005bk} and here $c_1$, $c_2$ are the coupling parameters. 
For the three-fluid LQC system with
equations~(\ref{flateinstein-LQC}) and (\ref{derivativeofh-LQC}), we can
construct an autonomous dynamical system by appropriately
introducing some dimensionless variables.
After this, we shall investigate whether the resulting polynomial dynamical system
develops singularities at some finite-time, using the dominant
balance technique we presented in previous sections of this chapter.
We shall also discriminate the dynamical system
singularities from the actual physical finite-time singularities.
The major outcome of this section is the fact that
we shall analytically prove that the resulting LQC 
dynamical system has no finite-time singularities, which verifies
that LQC actually leads to non-singular behavior.
We now construct the autonomous dynamical system from Eqs.~(\ref{flateinstein-LQC}),
(\ref{derivativeofh-LQC}) and (\ref{continutiyequations-LQC}).
We choose the dimensionless dynamical system variables as follows,
\begin{align}
\label{variablesofdynamicalsystem-LQC}
x_1=\frac{\kappa^2\rho_{\rm DE}}{3H^2}\, ,\quad x_2=\frac{\kappa^2\rho_{\rm DM}}{3H^2}\, ,\quad
x_3=\frac{\kappa^2\rho_b}{3H^2}\, ,\quad z=\frac{H^2}{\kappa^2\rho_c}\, ,
\end{align}
which are constrained to satisfy the Friedmann constraint,
\begin{align}
\label{friedmannconstraint-LQC}
x_1+x_2+x_3-z\left(x_1+x_2+x_3\right)^2=1\, ,
\end{align}
for all cosmic times. Furthermore, the total EoS parameter
$w_\mathrm{eff}$ corresponding to the multifluid system, expressed
in terms of the dimensionless dynamical system variables
(\ref{variablesofdynamicalsystem-LQC}), takes the following form,
\begin{align}
\label{equationofstatetotal-LQC}
w_\mathrm{eff}=-x_1-3A\kappa^4 \rho_c x_1^2z\, .
\end{align}
Now we can form the polynomial autonomous dynamical system for the
three multi-fluids system, by combining Eqs.~(\ref{flateinstein-LQC}),
(\ref{derivativeofh-LQC}), (\ref{continutiyequations-LQC}), and
(\ref{variablesofdynamicalsystem-LQC}), so after some algebra, we get,
\begin{align}
\label{dynamicalsystemmultifluid-LQC}
\frac{d x_1}{d N}=&\, -\frac{\kappa^2Q}{3H^3}+9 A
x_1^3 z-27 A x_1^2 z+3 w_{\rm DE} x_1^2-3 w_{\rm DE} x_1-18 x_1^3 z+3 x_1^2+3
x_1 x_2+3 x_1 x_3-3 x_1 \nonumber \\
&\, -18 w_{\rm DE} x_1^3 z-18 w_{\rm DE} x_1^2
x_2 z-36 x_1^2 x_2 z-36 x_1^2 x_3 z-18 x_1 x_2^2 z \nonumber \\
&\, -54 A x_1^4 z^2-54 A x_1^3 x_2 z^2-54 A x_1^3 x_3 z^2-18 w_{\rm DE} x_1^2 x_3
z-36 x_1 x_2 x_3 z-18 x_1 x_3^2 z\, , \nonumber \\
\frac{d x_2}{d N}=&\, \frac{\kappa^2Q}{3H^3}+9 A x_1^2
x_2 z+3 w_{\rm DE} x_1 x_2-18 x_1^2 x_2 z+3 x_1 x_2+3 x_2^2+3 x_2 x_3-3 x_2 \nonumber \\
&\, -18 w_{\rm DE} x_1^2 x_2 z-18 w_{\rm DE} x_1 x_2^2 z-36 x_1
x_2^2 z-36 x_1 x_2 x_3 z-18 x_2^3 z \nonumber \\
&\, -54 A x_1^3 x_2 z^2-54 A x_1^2 x_2^2 z^2-54 A x_1^2 x_2 x_3 z^2-18 w_{\rm DE} x_1 x_2 x_3
z-36 x_2^2 x_3 z-18 x_2 x_3^2 z\, , \nonumber \\
\frac{d x_3}{d N}=&\, 9 A x_1^2 x_3 z-18 w_{\rm DE} x_1^2 x_3
z+3 w_{\rm DE} x_1 x_3-18 x_1^2 x_3 z+3 x_1 x_3+3 x_2 x_3+3 x_3^2-3 x_3 \nonumber \\
&\, -18 w_{\rm DE} x_1 x_2 x_3 z-18 w_{\rm DE} x_1 x_3^2 z-36 x_1 x_2 x_3
z-36 x_1 x_3^2 z-18 x_2^2 x_3 z-36 x_2 x_3^2 z \nonumber \\
&\, -54 A x_1^3 x_3 z^2-54 A x_1^2 x_2 x_3 z^2-54 A x_1^2 x_3^2 z^2-18 x_3^3 z \, , \nonumber \\
\frac{d z}{d N}=&\, -9 A x_1^2 z^2+18 w_{\rm DE} x_1^2 z^2-3 w_{\rm DE} x_1 z+18 x_1^2 z^2-3 x_1 z-3 x_2 z-3 x_3 z \nonumber \\
&\, + 18 w_{\rm DE} x_1 x_2 z^2+18 w_{\rm DE} x_1 x_3 z^2+36 x_1 x_2 z^2+36 x_1 x_3 z^2+18 x_2^2 z^2 \nonumber \\
&\, + 54 A x_1^3 z^3+54 A x_1^2 x_2 z^3+54 A x_1^2 x_3 z^3+36 x_2 x_3 z^2+18 x_3^2 z^2\, ,
\end{align}
and note that, as in the previous sections in this chapter, we
used the $e$-foldings number to quantify the dynamical evolution
of the variables, instead of the cosmic time.
Furthermore, by choosing the interaction term $Q$ as in Eq.~(\ref{qtermform-LQC}), the
$Q$-containing terms in the dynamical system~(\ref{dynamicalsystemmultifluid-LQC}), can be expressed in terms of
the variables $x_1$ and $x_2$ in the following way,
\begin{align}
\label{additionalterms-LQC}
\frac{\kappa^2Q}{3H^3}=3c_1x_2+3c_2x_1\, .
\end{align}
Now having the dynamical system~(\ref{dynamicalsystemmultifluid-LQC})
at hand, we can apply the dominant balance analysis in order to
reveal the dynamical system singularities and we shall also
investigate whether physical singularities occur too.
This is the subject of the next subsection.

\subsubsection{Dominant Balance Analysis of the three-fluid
Cosmological Dynamical System}

We shall use the method of dominant balances in order to
investigate whether the dynamical system~(\ref{dynamicalsystemmultifluid-LQC}) develops finite-time singularities.
As we shall show, the dynamical system is
singularity-free and in fact this result holds true for a general set of initial conditions.
One should have in mind of course that
there exists a limited set of initial conditions that may lead in
principle to finite-time dynamical system singularities, but this
limited set is not of interest.
Near a singularity, the dimensionless variables $x_1(N)$, $x_2(N)$, $x_3(N)$ and $z(N)$ at
leading order behave as follows,
\begin{align}
\label{decompositionofxiactualexample-LQC}
x_1(N)=a_1(N-N_c)^{p_1}\, ,\quad \,x_2(N)=a_2(N-N_c)^{p_2}\, ,\quad
x_3(N)=a_3(N-N_c)^{p_3}\, ,\quad z(N)=a_4(N-N_c)^{p_4}\, .
\end{align}
The dynamical system~(\ref{dynamicalsystemmultifluid-LQC}) can be
rewritten as follows
$\frac{d \vec{x}}{d N}=f(\vec{x})$, with $\vec{x}$
being $\vec{x}=(x_1,x_2,x_3,z)$, and also $f(x_1,x_2,x_3,z)$ is
defined to be equal to,
\begin{align}
\label{functionfmultifluid-LQC}
f(x_1,x_2,x_3,z)=\left(%
\begin{array}{c}
 f_1(x_1,x_2,x_3,z) \\
 f_2(x_1,x_2,x_3,z) \\
 f_3(x_1,x_2,x_3,z) \\
 f_4(x_1,x_2,x_3,z) \\
\end{array}%
\right)\, ,
\end{align}
and the corresponding functions $f_i(x_1,x_2,x_3)$, $i=1,2,3,4$ are,
\begin{align}
\label{functionsfi-LQC}
f_1(x_1,x_2,x_3,z)=&\, 3c_1x_2+3c_2x_1+9 A x_1^3 z-27 A x_1^2 z+3
w_{\rm DE} x_1^2-3 w_{\rm DE} x_1-18 x_1^3 z+3 x_1^2+3 x_1 x_2+3 x_1 x_3-3 x_1 \nonumber \\
&\, -18 w_{\rm DE} x_1^3 z-18 w_{\rm DE} x_1^2 x_2 z-36 x_1^2 x_2 z-36 x_1^2 x_3 z-18 x_1 x_2^2 z \nonumber \\
&\, -54 A x_1^4 z^2-54 A x_1^3 x_2 z^2-54 A x_1^3 x_3 z^2-18 w_{\rm DE} x_1^2 x_3 z-36 x_1 x_2 x_3 z-18
x_1 x_3^2 z\, , \nonumber \\
f_2(x_1,x_2,x_3,z)=&\, 3c_1x_2+3c_2x_1+9 A x_1^2 x_2 z+3 w_{\rm DE} x_1 x_2-18 x_1^2 x_2 z+3 x_1 x_2+3 x_2^2+3 x_2
x_3-3 x_2 \nonumber \\
&\, -18 w_{\rm DE} x_1^2 x_2 z-18 w_{\rm DE} x_1 x_2^2 z-36 x_1 x_2^2 z-36 x_1 x_2 x_3 z-18 x_2^3 z\nonumber \\
&\, -54 A x_1^3 x_2 z^2-54 A x_1^2 x_2^2 z^2-54 A x_1^2 x_2 x_3 z^2-18 w_{\rm DE} x_1 x_2 x_3
z-36 x_2^2 x_3 z-18 x_2 x_3^2 z\, , \nonumber \\
f_3(x_1,x_2,x_3,z)=&\, 9 A x_1^2 x_3 z-18 w_{\rm DE} x_1^2 x_3 z+3 w_{\rm DE} x_1
x_3-18 x_1^2 x_3 z+3 x_1 x_3+3 x_2 x_3+3 x_3^2-3 x_3 \nonumber \\
&\, -18 w_{\rm DE} x_1 x_2 x_3 z-18 w_{\rm DE} x_1 x_3^2 z-36 x_1 x_2 x_3 z-36 x_1
x_3^2 z-18 x_2^2 x_3 z-36 x_2 x_3^2 z \nonumber \\
&\, -54 A x_1^3 x_3 z^2-54 A x_1^2 x_2 x_3 z^2-54 A x_1^2 x_3^2 z^2-18 x_3^3 z \, , \nonumber \\
f_4(x_1,x_2,x_3,z)=&\, -9 A x_1^2 z^2+18 w_{\rm DE} x_1^2 z^2-3 w_{\rm DE}
x_1 z+18 x_1^2 z^2-3 x_1 z-3 x_2 z-3 x_3 z \nonumber \\
&\, + 18 w_{\rm DE} x_1 x_2 z^2+18 w_{\rm DE} x_1 x_3 z^2+36 x_1 x_2 z^2+36 x_1 x_3 z^2+18 x_2^2 z^2 \nonumber \\
&\, + 54 A x_1^3 z^3+54 A x_1^2 x_2 z^3+54 A x_1^2 x_3
z^3+36 x_2 x_3 z^2+18 x_3^2 z^2\, .
\end{align}
Now we shall seek for consistent truncations of the vector-valued
function $f(x_1,x_2,x_3,z)$ defined in Eq.~(\ref{functionfmultifluid-LQC}), and investigate whether singular
solutions exist or not.
Among many distinct truncations of the
vector function $f(x_1,x_2,x_3,z)$, one is the following,
\begin{align}
\label{truncation1-LQC}
\hat{f}(x_1,x_2,x_3,z)=\left(
\begin{array}{c}
 3 x_1(N) x_2(N) \\
 3 (w_{\rm DE}+1) x_1(N) x_2(N) \\
 -54 A x_1^2(N) x_3^2(N) z^2(N) \\
 54 A x_1^2(N) z^3(N) x_2(N) \\
\end{array}
\right)\, .
\end{align}
The corresponding vector $\vec{p}$,  which obviously corresponds to the solution $\rho_\mathrm{tot}=\rho_c$, is easily found,
\begin{align}\label{vecp1-LQC}
\vec{p}=( -1, -1, -1, 1 )\, ,
\end{align}
and the corresponding vectors $\vec{a}$ are,
\begin{eqnarray}\label{balancesdetails1-LQC} 
\vec{a}_1=\left( -\frac{1}{3 (w_{\rm DE}+1)}, -\frac{1}{3},
-\frac{1}{3}, \frac{w_{\rm DE}+1}{\sqrt{-2A}}\right) \qquad \mbox{and} \qquad
\vec{a}_2=\left( -\frac{1}{3 (w_{\rm DE}+1)}, -\frac{1}{3},
-\frac{1}{3}, -\frac{w_{\rm DE}+1}{\sqrt{-2A}}\right).\end{eqnarray}

The corresponding Kovalevskaya matrix is given by
\begin{align}
K(\vec{a}_1)=\left(
\begin{array}{cccc}
0 & -\frac{1}{w_{\rm DE}+1} & 0 & 0 \\
 -(w_{\rm DE}+1) & 0 & 0 & 0 \\
-2(w_{\rm DE}+1)& 0 & -1 & \frac{2\sqrt{-2A}}{3(w_{\rm DE}+1)} \\
-\frac{6(w_{\rm DE}+1)^2}{\sqrt{-2A}} & 
-\frac{3(w_{\rm DE}+1)}{\sqrt{-2A}} & 0 & 2 \\
\end{array}
\right)\, ,
\end{align}
with eigenvalues 
\begin{align}
(r_1,r_2,r_3,r_4)=(-1,2,1,-1)\, .
\end{align}

In conclusion, the major outcome of this section is the
fact that LQC utterly erases any finite-time singularities that
may have occurred in the classical interacting three fluid system.
We have to take into account that the variables $x_i$ $(i=1,2,3)$
diverge when the Hubble rate vanishes, and this happens when $\rho_{\mathrm{tot}}=\rho_c$ or $\rho_{\mathrm{tot}}=0$. That means, the singularity, in the $x_i$ variables, exhibits that the total energy density coincides with the critical one, and thus, this singularity in the variables $x_i$ is not a physical singularity in the sense that neither the Hubble rate diverges nor the corresponding energy densities diverge for our interacting three fluid system in the context of LQC.

We shall conclude the analysis by finding analytically the fixed
points of the dynamical system. Also, we shall reveal the behavior
of the total EoS parameter of Eq.~(\ref{equationofstatetotal-LQC}), in
terms of the coefficients $c_1$ and $c_2$ appearing in the interaction term $Q$. We can find the fixed points of the
dynamical system which well-known linearization techniques for
dynamical systems. In the case that the fixed point is hyperbolic,
in which case the eigenvalues of the matrix that results after the
linearization of the dynamical system contain non-zero real part,
the stability of the fixed point can be determined. In the case of
a hyperbolic fixed point, when the eigenvalue of the linearization
matrix has a negative real part, then the fixed stable, and if the
real part is negative, then the fixed point is unstable. We shall
denote the fixed points of the dynamical
system~(\ref{dynamicalsystemmultifluid-LQC}) as $\phi_*$, and the
Jacobian of the linearized dynamical system near the fixed points
as, $\mathcal{J}(g)$, with the latter being equal,
\begin{align}
\label{jaconiab-LQC}
\mathcal{J}=\sum_i\sum_j\left[ \frac{\partial f_i}{\partial x_j}\right]\, .
\end{align}
Now we shall solve the equation $f(x_1,x_2,x_3,z)=0$, with
$f(x_1,x_2,x_3,z)$ appearing in Eq.~(\ref{functionfmultifluid-LQC}),
and this process will reveal the fixed points of the dynamical
system~(\ref{dynamicalsystemmultifluid-LQC}), which are,
\begin{align}
\label{fixedpointsc20-LQC}
\phi_1^*=&\, \left\{x_1\to 0,x_2\to 0,x_3\to 0 \right\}\, , \nonumber \\
\phi_2^*=&\, \left\{x_1\to 0,x_2\to 0,x_3\to 0,z\to 0 \right\}\, , \nonumber \\
\phi_3^*=&\, \left\{x_1\to 0,x_2\to 0,x_3\to 1,z\to 0 \right\}\, , \nonumber \\
\phi_4^*=&\, \left\{x_1\to \frac{-\sqrt{(c_1-c_2-w_{\rm DE})^2+4 c_1 w_{\rm DE}}-c_1+c_2+w_{\rm DE}}{2 w_{\rm DE}}, \right. \nonumber \\
& \qquad \left. x_2\to \frac{\frac{c_1^2}{w_{\rm DE}}+\frac{c_1 \sqrt{(c_1-c_2-w_{\rm DE})^2+4 c_1 w_{\rm DE}}}{w_{\rm DE}}
 -\frac{c_1 c_2}{w_{\rm DE}}+c_1}{2 c_1},x_3\to 0,z\to 0\right\} \, , \nonumber \\
\phi_5^*=&\, \left\{x_1\to \frac{\sqrt{(c_1-c_2-w_{\rm DE})^2+4 c_1 w_{\rm DE}}-c_1+c_2+w_{\rm DE}}{2 w_{\rm DE}}, \right. \nonumber \\
& \qquad \left. x_2\to \frac{\frac{c_1^2}{w_{\rm DE}}-\frac{c_1 c_2}{w_{\rm DE}}-\frac{c_1 \sqrt{(c_1-c_2-w_{\rm DE})^2+4 c_1 w_{\rm DE}}}{w_{\rm DE}}+c_1}{2 c_1},
x_3\to 0,z\to 0\right\}\, .
\end{align}
Apparently, the fixed points $\phi_1^*$ and $\phi_2^*$ are not
hyperbolic, and the fixed points $\phi_3^*$, $\phi_4^*$ and
$\phi_5^*$ are hyperbolic. Omitting the exact form of the
eigenvalues for brevity, a thorough analysis of the phase space
indicates that the fixed points are unstable, regardless the
values of the parameters $w_{\rm DE}$, $c_1$, $c_2$. Also the total EoS
in Eq.~(\ref{equationofstatetotal-LQC}) must be studied for various
values of the free parameters. We shall consider the case that we
fix the DE EoS parameter $w_{\rm DE}$ to have various values, varying
from quintessential values ($w_{\rm DE}=-0.5$) to phantom values
($w_{\rm DE}=-1.5$). A numerical analysis of the total EoS
parameter~(\ref{equationofstatetotal-LQC}) indicates that in the
context of the three fluid interacting cosmology, it is possible
to realize various cosmological evolutions, for example
quintessential, phantom and also matter dominated eras.

\subsubsection{The Choice of the DE EoS and its Physical Implications}

For the analysis performed in the previous subsections, the choice
of the DE EoS was chosen to be that of
Eq.~(\ref{darkenergyequation-of-state-LQC}). However, a more general
EoS can be used, which may include higher powers of the Hubble
rate and also further effects of viscosity might be included. In
this subsection we report the implications of the chosen
EoS~(\ref{darkenergyequation-of-state-LQC}) for DE phenomenology. 
Using the formalism of
\cite{Astashenok:2012iy} which we extend in the context of LQC, 
a comparison of the classical theory with the LQC one can be
done for the chosen EoS~(\ref{darkenergyequation-of-state-LQC}). We note that when the LQC effects are not considered, the EoS of
Eq.~(\ref{darkenergyequation-of-state-LQC}) drives the cosmological
system to finite-time singularities of the Type III
form~\cite{Nojiri:2005sx}. On the other hand, in the case of LQC, with one fluid, the singularities are eliminated \cite{Sami:2006wj}, and as we
demonstrated previously, the same applies in the three fluid
system in which DM and DE are interacting with each other. Thus, LQC eliminates the finite-time singularities. Our
findings and comparisons are presented in TABLE~\ref{table1}.
\begin{table}[h]
\caption{\label{table1}Classical versus LQC Singularities for the DE
EoS $p_{\rm DE}=-\rho_{\rm DE}-A\kappa^4\rho_{\rm DE}^2$~. }
\begin{center}
\renewcommand{\arraystretch}{1.4}
\begin{tabular}{|c@{\hspace{1 cm}}@{\hspace{1 cm}}c@{\hspace{1 cm}} c|}
\hline
\textbf{Case Study} & \textbf{Cosmological Scenario} & \textbf{Occurrence of Singularities}\\
\hline\hline

Classical Case & Interacting Multi-fluids & Type III singularity occurs \\
\hline 

LQC  & One DE Fluid & No singularity occurs\\
\hline 

LQC   & Interacting Multi-fluids & No singularity occurs\\
\hline

\end{tabular}
\end{center}
\end{table}

\section{The Avoidance of Finite-time Future Singularities}
\label{sec-avoid-singularities}

Near the singularity, especially in the case of Type I and Type
III, the Hubble rate becomes very large, which means that the
curvature and the temperature become very large and therefore the
quantum effect and the thermal effect become important.
One may naturally wonder whether it is possible to avoid the finite-time
singularities from the cosmological picture so that we realize a non-singular Universe.
In this section, we shall examine the
possibility of avoiding the finite-time singularities in the cosmological scenarios.
More specifically, we shall discuss what
could happen with the future singular dark Universe when the
effects of quantum or thermal radiation are included.
This section is based on Refs.~\cite{Frampton:2011sp,Brevik:2011mm,Nojiri:2020sti,Carroll:2003st}. 

\subsection{Little Rip Scenario}
\label{subsection-little-rip}

The possibility of a phantom DE EoS (i.e., $w_\mathrm{DE} < -1$) is
very hard to exclude from the cosmological picture~\cite{DiValentino:2020vnx}.
The phantom DE can solve the Hubble
constant tension~\cite{DiValentino:2021izs} quite excellently.
However, one of the serious features that the phantom DE carries is the following.
For $w_\mathrm{DE} <-1$, the DE density increases
with the increasing scale factor and both of them can blow up at
some finite-time in the future leading to a Big Rip singularity~\cite{Caldwell:2003vq}.
However, this future singularity can be
avoided even if $w_\mathrm{DE} < -1$ and this heuristic scenario,
dubbed as ``Little Rip'' was introduced in Ref.~\cite{Frampton:2011sp}.
With such fantastic proposal, the Little Rip scenario received significant attention to the scientific  community~\cite{Frampton:2011sp,Brevik:2011mm,
Frampton:2011rh,Nojiri:2011kd,Granda:2011kx,Ito:2011ae,Xi:2011uz,Liu:2012iba,Saez-Gomez:2012uwp,Brevik:2012nt,Brevik:2012ka,
Balakin:2012ee,Frampton:2013fxa,Makarenko:2014nca,Bouhmadi-Lopez:2014cca,Albarran:2015cda,Morais:2016bev}.
In this section, we shall describe how the Big Rip singularity
could be avoided even if $w_\mathrm{DE}$ is less than $-1$.

In order to realize a non-singular scenario, one can proceed with a non-singular scale factor $a(t)$ having the following form
\begin{align}
\label{scale-factor-LR}
a = \e^{f(t)}\, ,
\end{align}
where $f(t)$ is any arbitrary non-singular function satisfying $\dot{f}=H$.
 From the Friedmann equation $H^2 = (\kappa^2/3) \rho$, one can derive that $\dot{f}^2 (t) = (\kappa^2/3) \rho$.
And from the condition that $\rho$ is an increasing
function of $a$ which implies $ \frac{d \rho}{da} = \frac{6}{\kappa^2\dot{a}} \dot{f} \ddot{f} > 0$, gives the restriction
\begin{align}
\label{LR-condition}
\ddot{f} > 0\, ,
\end{align}
because we are dealing with an expanding Universe, i.e., with $\dot{f}=H>0$.
Thus, one may note that all Little Rip scenarios
which are described by the evolution of the scale factor given by
Eq.~(\ref{scale-factor-LR}), with non-singular $f$ should satisfy
the Eq.~(\ref{LR-condition}).

To proceed further in this direction, let us now consider the
well-known non-linear EoS~\cite{Nojiri:2005sr}
\begin{align}
\label{equation of state-LR}
p = - \rho - f(\rho) \quad \Longrightarrow \quad w = -1 - \frac{f(\rho)}{\rho}\, ,
\end{align}
where $f(\rho) > 0$ ensures that the $\rho$ increases with the scale factor, because in that case $w<-1$.
 From the conservation equation, $\dot{\rho} + 3 H (p+\rho) = 0$, one can find the scale factor as
\begin{align}
a=a_0\exp \left(\int_{\rho_0}^\rho \frac{d\rho}{3f(\rho)} \right)\, ,
\label{scale-factor-LP2}
\end{align}
where we set $a_0$, the present value of the scale factor to be unity and $\rho_0$ denotes the present value of the energy density.
Now, using the conservation equation $\dot{\rho}=3Hf(\rho)$ and the Friedmann equation $H^2 = \frac{\kappa^2 \rho}{3}$, one finds
\begin{align}
\label{time-LR}
t-t_0=\frac{1}{\sqrt{3}\kappa}\int_{\rho_0}^\rho \frac{d\rho}{\sqrt{\rho}f(\rho)}\, .
\end{align}
We assume that $f(\rho)$ is given by the power law $f(\rho) = \frac{A}{\sqrt{3}\kappa} \rho^{\nu+\frac{1}{2}}$, where $A >0$ and $\nu$ are constants.
We select $\nu=0$ in the above form (i.e., we consider $f(\rho) = \frac{A}{\sqrt{3}\kappa} \sqrt{\rho}$) and for which one can find the time from (\ref{time-LR}) as
\begin{align}
\label{time-explicit-LR}
t-t_0=\frac{1}{A}\,\ln \frac{\rho}{\rho_0}\, ,
\end{align}
which clearly shows that $\rho \rightarrow \infty$ is not reached
in some finite-time. That means even if we have $w< -1$ but we can
avoid the future singularity and it is known as the Little Rip scenario.

Now, solving (\ref{scale-factor-LP2}) for $f(\rho) = \frac{A}{\sqrt{3}\kappa}\sqrt{\rho}$, one can
express $\rho$ as a function of $a$ as \cite{Brevik:2011mm}:
\begin{align}
\rho(a)=\rho_0 \left( 1+\frac{\sqrt{3}A}{2\kappa\sqrt{\rho_0}}\ln a \right)^2\, ,
\label{rho-LR}
\end{align}
and consequently, using the first Friedmann equation, one can explicitly find the scale factor in terms of the time as \cite{Brevik:2011mm}:
\begin{align}
a(t)=\exp \left[ \frac{2\kappa\sqrt{\rho_0}}{\sqrt{3}A}\left\{\exp \left(\frac{A}{2}(t-t_0)\right)-1 \right\} \right]\, ,
\label{scale-factor-explicit0LR}
\end{align}
which does not exhibit any finite-time singularity in the future.

\subsubsection{Little Rip in Viscous Universe}

The Little Rip scenario can also be realized in the viscous
Universe as well. We refer to
section~\ref{sec-singularities-viscous-cosmology} for the basic
dynamical equations for a viscous Universe. In the viscous
Universe, the effective pressure term as in Eq.~(\ref{p-VC}) or
(\ref{p-explicit-VC}) which includes the bulk viscosity component
plays the crucial role. We generalize (\ref{equation of state-LR})
in presence of the bulk viscosity as
\begin{align}
\label{peff-viscous-LR}
p_\mathrm{eff} = -\rho - f(\rho) - \eta (H)\, ,
\end{align}
where notice that we have generalized the last term in the right
hand side of Eq.~(\ref{p-explicit-VC}) as $3 \xi (t) H \rightarrow \eta (H)$ in which $\xi (t)$ refers to the coefficient of the bulk viscosity.
We investigate the cosmological scenario with
\begin{align}\label{constant-xi}
\eta(H)=\bar{\eta}= \mathrm{constant}\, .
\end{align}
For the constant $\eta (H)$, using the conservation equation $\dot{\rho} + 3 H (p_\mathrm{eff} + \rho) = 0$, one can find the scale factor as
\begin{align}
a=\exp \left(\frac{1}{3}\int_{\rho_0}^\rho \frac{d\rho}{\bar{\eta}+f(\rho)}\right)\, ,
\label{scale-factor-LR-viscous-const-xi}
\end{align}
where we have explicitly set $a_0=1$. 
Further, using the conservation and Friedmann equations, one can express the time as
\begin{align}
t-t_0=\frac{1}{\sqrt{3}\kappa}\int_{\rho_0}^\rho \frac{d\rho}{\sqrt{\rho}\,[\bar{\eta}+f(\rho)]}\, .
\label{time-LR-viscous-const-xi}
\end{align}

Now, for the model $f(\rho)=\frac{A}{\sqrt{3} \kappa}\sqrt{\rho}$ (in this case, the dimension of $A$, in natural units, is of the Planck's mass),
we can solve
\begin{align}
t-t_0=\frac{2}{A} \ln \left(\frac{\sqrt{3}\kappa\bar{\eta}+A\sqrt{\rho}}{\sqrt{3}\kappa\bar{\eta}+A\sqrt{\rho_0}}\right)\, .
\end{align}
and consequently,
\begin{align}
\rho(t)=\frac{1}{A^2}\left[ \left(\sqrt{3}\kappa{\bar{\eta}}+A\sqrt{\rho_0}\right) \exp (\frac{A}{2}(t-t_0))-\sqrt{3}\kappa{\bar{\eta}}\right]^2\, .
\end{align}
Thus, we can see that in order to realize $\rho \rightarrow \infty$, $t \rightarrow \infty$ is essential.
This is the Little Rip scenario in the viscous Universe.
However, it is important to mention that the above Little Rip scenario is dependent on the
above EoS (\ref{peff-viscous-LR}) with the constant $\eta (H)$.

Interestingly, one can also realize the Little Rip phenomenon in a more generalized scenario where in addition $\eta (H)$ is not strictly a constant.
To illustrate this, we consider the following example:
\begin{align}
f(\rho)=\frac{A}{\sqrt{3}\kappa} \, \rho^{\nu+1/2}, \qquad \xi (t) = \frac{b}{3\kappa^2} \, \rho^\gamma\, ,
\label{LR-general}
\end{align}
where $\nu$, $b$ and $\gamma$ are constants.\footnote{One can quickly note that for $\nu=0$ and $\gamma = - 1/2$,
we realize the previous scenario $f(\rho) = A \sqrt{\rho}$ and $\eta (H) = \mathrm{constant}$. } 
Thus, for the above choice, one can solve for the cosmic time as
\begin{align}
t-t_0=
\int_{\rho_0}^\rho \frac{d\rho}{\rho \,(A \, \rho^\nu + b \, \rho^\gamma)}\, .
\label{time-LR-general-bulk-viscous}
\end{align}
which can be expressed after integration in terms of the hypergeometric function as
\begin{align}
t-t_0 =
\Bigg[ \frac{\nu \, \rho^{-\gamma}}{\gamma b (\gamma-\nu)}
+ \frac{\rho^{-\gamma}}{\gamma b (\gamma-\nu)} \left\{ -\nu
+ (\nu -\gamma)\, {}_2F_1 \left( 1, \frac{\gamma}{\gamma-\nu}, 1 + \frac{\gamma}{\gamma-\nu}, -\frac{A}{\beta}\rho^{\nu -\gamma}\right)\right\}\Bigg]\, .
\label{time-hypergeom-LR}
\end{align}

For constant bulk viscous coefficient, i.e., $\xi(t) = \mathrm{constant}$ (equivalently, $\gamma=0$ in Eq.~(\ref{LR-general}))
together with $\nu =0$ in Eq.~(\ref{LR-general}), we obtain (note
that in that case $\nu$ and $\gamma$ have units of the Planck mass):
\begin{align}
t-t_0=\frac{1}{ (A+b)}\ln \frac{\rho}{\rho_0}\, ,
\label{time-LR-new-case}
\end{align}
and this presents the Little Rip phenomenon because $\rho \rightarrow \infty$ requires $t \rightarrow \infty$, that means
the energy density does not diverge at some finite-time~\cite{Brevik:2011mm}.

\subsection{Geometrical invariants to remove the finite-time future singularities}
\label{sec-geo-invariants}

In Section \ref{sec-singularities-MG} we have seen that different modified gravity theories may lead to a variety of finite time future singularities. However, with the proper choice of the geometrical invariants describing the underlying gravitational theory, 
it is possible to remove such finite-time singularities. In this section we shall describe briefly how one can avoid such finite time future singularities from various modified gravity theories.

In the context of $F(R)$ gravity theory, one can reconstruct the $F (R)$ gravity models in
such a way so that the finite-time future singularities may disappear from the picture. However, such
reconstructions should be consistent with the existing theories of the Universe and the reconstructed $F (R)$ model should pass the essential tests of gravity. This point was first raised in Ref.~ \cite{Abdalla:2004sw} and subsequently discussed in Refs.~\cite{Nojiri:2008fk,Bamba:2008ut}. We recall that the general action of $F (R)$ gravity in the Jordan frame that depends solely on the curvature [i.e. Eq. (\ref{action-F(R)}) without matter] having the form 

\begin{align}
\label{action-F(R)-avoidance-section-01}
S =\int d^4 x \sqrt{-g}\; \frac{F(R)}{2\kappa^2}\,,
\end{align}

Introducing the auxiliary fields $A$, $B$, one can rewrite the action (\ref{action-F(R)-avoidance-section-01}) as Eq. (\ref{different-frames-eq-4}) and consequently as Eq. (\ref{different-frames-eq-5}). Now, using the conformal transformation $g_{\mu \nu} = e^{\sigma} g_{\mu \nu}$ where $\sigma = -\ln F'(A)$, one can obtain the action in the Einstein frame in Eq. (\ref{different-frames-eq-8}) and the corresponding potential $V (A)$ in Eq. (\ref{different-frames-eq-9}) [see section \ref{sec-correspondence-singularities-FR} for more details]. 

Dealing with the model

\begin{eqnarray}\label{action-F(R)-avoidance-section-03}
 F(R)=A -\gamma R^{-n} + \eta R^2. 
\end{eqnarray}
we consider the following case $-1<n<-\frac{1}{2}$, $\gamma<0$, and $\eta>0$ and with these choices, the field $\sigma$ turns out to be $\sigma = - \ln \left(1 + n\gamma A^{-n-1} + 2\eta A\right)$ where $1 + n\gamma A^{-n-1} + 2\eta A>0$, and the potential has the following form

\begin{eqnarray}\label{action-F(R)-avoidance-section-04}
 V(A)=\frac{(n+1)\gamma A^{-n} + \eta A^2}{\left(1 + n\gamma A^{-n-1} + 2\eta A\right)^2}\ . 
\end{eqnarray}
Now, since 
\begin{eqnarray}
\frac{d\sigma}{dA}=-\frac{F''(A)}{F'(A)}=-\frac{-n(n+1)\gamma A^{-n-2} + 2\eta}{1 + n\gamma A^{-n-1} + 2\eta A},
\end{eqnarray}
then 
there is a branch point, at
$A=A_0\equiv \left\{\frac{n(n+1)\gamma}{2\eta}\right\}^{\frac{1}{n+2}}$ or
$\sigma=\sigma_0 \equiv - \ln \left( 1 + (n+2)\left(\frac{2\eta}{n+1}\right)^\frac{n+1}{n+2}
(n\gamma)^\frac{1}{n+2} \right)$ where $\frac{d\sigma}{dA}=0$ or $F''(A)=0$.
For small values of $A$, the behavior of the potential $V (A)$ 
approaches to $\frac{A^{n+2}}{(n+1)\gamma}$.
However, when $A$ is very large, then $V(A)$ approaches 
to a constant, $V(A)\to \frac{1}{4\eta}$. Also, the potential $V(A)$ vanishes at $A=A_1 \equiv \left\{-\frac{(n+1)\gamma}{\eta}\right\}^{1/(n+2)}$ (since $0<\frac{n\gamma}{2}<-\gamma$, hence, $A_0<A_1$). Further, looking at the expression for $V'(A)$ given by

\begin{eqnarray}
V'(A)=\frac{\left\{-n(n+1)\gamma A^{-n-2} + 2\eta\right\}
A\left\{1-(n+2)\gamma A^{-n-1}\right\}}{\left(1 + n\gamma A^{-n-1} + 2\eta A\right)^3},
\end{eqnarray}
one can notice that $V'(A)$ has an extremum at $A=A_0$. Now, as there is a branch point at $\sigma=\sigma_0$, then if we start from the small curvature,
the growth of the curvature stops at $R=A_0$, where $\sigma=\sigma_0$. 
In fact, at the branch point,
where $h''(A)=0$, the mass $m_\sigma ~(\propto \frac{d^2 V}{d\sigma^2})$ of $\sigma$ becomes infinite since
\begin{eqnarray}
\frac{d^2 V}{d\sigma^2}
= \frac{F'(A)}{F''(A)}\frac{d}{dA}\left(\frac{F'(A)}{F''(A)}\frac{dV(A)}{dA}\right)
= -\frac{3}{F''(A)} + \frac{A}{F'(A)} + \frac{2F(A)}{F'(A)^2} \to + \infty.
\end{eqnarray}
Note also that $F''(0)<0$ when $A<A_0$. 
Then the growth of $\sigma$ is finished at $\sigma=\sigma_0$.
Hence, with the addition of the $R^2$ term, cosmic doomsday does not occur but the Universe ends up in a 
de Sitter phase. 

The same situation arises for the Ho\v{r}ava--Lifshitz $F (R)$ gravity too. As explored in Ref. \cite{Carloni:2010nx}, the Ho\v{r}ava--Lifshitz $F (R)$ gravity has a quite rich cosmological structure and the inclusion of $R^2$ in $F (R)$ could avoid the finite time future singularities from the picture.

On the other hand, recalling the $F (T)$ gravitational theory, it is also possible to construct some models of $F (T)$ gravity where the finite-time future singularities can be avoided. We consider the correction of the power law model $F (T) = A T^{\alpha}$ of Eq.(\ref{power-law-FT}) as follows 

\begin{align}
F (T) = A T^{\alpha} + B T^{\delta},
\end{align}
where $B T^{\delta}$ ($\delta \neq 0$, $B \neq 0$) is the correction term. This model has some interesting features. For $\delta > 1$, all four types finite time future singularities can be avoided in this model \cite{Bamba:2012vg}. We emphasize that the model $F (T) = A T^{\alpha} + B T^{2}$ where $T^2$ is treated as the correction term, which is obtained for $\delta = 2$, the minimum integer satisfying the condition $\delta > 1$, therefore removes all four types of finite time future singularities. This has similarity with the $F (R)$ gravity theory where it has been observed that the inclusion of $R^2$ term could cure the finite time future singularities \cite{Abdalla:2004sw,Nojiri:2008fk,Bamba:2008ut}. 
Apart from the above models, there are other type of models which can avoid the finite-time future singularities: 

\begin{itemize}

\item {\bf Exponential model:} In the exponential model of type
\begin{align}
F (T) = C \exp (\lambda T),
\end{align}
where $C$ and $\lambda$ are nonzero constants, we do not realize any finite-time future
singularities \cite{Bamba:2012vg}.

\item {\bf Logarithmic model:} In the logarithmic model of the form
\begin{align}
F (T) = D \ln \left(\gamma T \right),
\end{align}
where $D~ (\neq 0)$ and $\lambda~(>0)$ are constants, we do not realize any finite-time
future singularities \cite{Bamba:2012vg}.

\end{itemize}

In the context of non-local gravity (section \ref{sec-singularities-nolocal}), one can also avoid the finite time future singularities. We consider the scenario where a correction term of the form 
$u R^2/\left(2\kappa^2\right)$ where $u~(\neq 0)$ is any real number, is added to the action (\ref{action-nonlocal}) as follows~\cite{Bamba:2012ky}: 
\begin{eqnarray} 
S=\int d^4 x \sqrt{-g}\Bigg[
\frac{1}{2\kappa^2}\bigg\{ R\left(1 + f(\Box^{-1}R )\right) 
+ u R^2 
-2 \Lambda \bigg\}
+ \mathcal{L}_\mathrm{matter} \left(Q; g\right) \Bigg]
\,.
\label{eqn-non-local-last-section-01} 
\end{eqnarray}
Along with the two fields $\eta$, $\xi$ as in section \ref{sec-singularities-nolocal}, 
introducing a scalar field $\zeta$. 
the action in Eq.~(\ref{eqn-non-local-last-section-01}) can be expressed as 
\begin{equation}
S = \int d^4 x \sqrt{-g}\Bigg[
\frac{1}{2\kappa^2}\bigg\{R\left(1 + f(\eta)\right) 
 - \partial_\mu \xi \partial^\mu \eta - \xi R 
+ u\left( 2 \zeta R - \zeta^2 \right) 
- 2 \Lambda \bigg\}
+ \mathcal{L}_\mathrm{matter} 
\Bigg]\,.
\label{eqn-non-local-last-section-02} 
\end{equation}
Note that by varying action (\ref{eqn-non-local-last-section-02}) with respect to $\zeta$, one obtains $\zeta = R$ and hence by substituting $\zeta = R$ in (\ref{eqn-non-local-last-section-02}), one gets back (\ref{eqn-non-local-last-section-01}).

Now, in the context of a flat FLRW universe, the 
gravitational field equations for the action (\ref{eqn-non-local-last-section-01}) can be written as \cite{Bamba:2012ky}
\begin{eqnarray}
- 3 H^2\left(1 + f(\eta) - \xi\right) + \frac{1}{2}\dot\xi \dot\eta 
 - 3H\left(f'(\eta)\dot\eta - \dot\xi\right) 
+ \Theta 
+ \Lambda 
+ \kappa^2 \rho_{\mathrm{m}} = 0\,,
\label{eqn-non-local-last-section-03} \\ 
\left(2\dot H + 3H^2\right) \left(1 + f(\eta) - \xi\right) 
+ \frac{1}{2}\dot\xi \dot\eta 
+ \left(\frac{d^2}{dt^2} + 2H \frac{d}{dt} \right) \left( f(\eta) - \xi \right) 
+ \Xi 
- \Lambda + \kappa^2 p_{\mathrm{m}} =0\,, 
\label{eqn-non-local-last-section-04} 
\end{eqnarray}
where $\rho_{\mathrm{m}}$, $p_{\mathrm{m}}$ are respectively the energy density and pressure of the matter sector; 
$\Theta$ and $\Xi$ are respectively the contributions from the additional 
term $u R^2/\left(2\kappa^2\right)$ given by 
\begin{eqnarray} 
\Theta \equiv u \left( -6H^2 R + \frac{1}{2}R^2 -6H\dot{R} \right) = 18u\left( -6H^2 \dot{H} + \dot{H}^2 -2H \ddot{H} \right)\,, 
\label{eqn-non-local-last-section-05} \\
\Xi \equiv u \left[ 
2 \left(2\dot{H} + 3H^2\right) R 
-\frac{1}{2}R^2 +2\ddot{R} +4H\dot{R} 
\right] =
6u \left( 9\dot{H}^2 +18H^2 \dot{H} + 2\dddot{H} + 12H\ddot{H} \right)\,. 
\label{eqn-non-local-last-section-06} 
\end{eqnarray} 

Now, for the Hubble parameter as in (\ref{BigRip}) or (\ref{Hsin}), in the limit $t\to t_{\mathrm{s}}$, 
$\Theta$ of Eq.~(\ref{eqn-non-local-last-section-05}) and $\Xi$ of Eq.~(\ref{eqn-non-local-last-section-06}) 
are approximately given by 
\begin{align} 
\Theta 
\sim 
18u\left[ 
-6 h_{\mathrm{s}}^2 \beta 
\left( t_{\mathrm{s}} - t \right)^{-\left(3\beta+1\right)} 
+ h_{\mathrm{s}}^2 \beta ^2 
\left( t_{\mathrm{s}} - t \right)^{-2\left(\beta+1\right)} 
- 2h_{\mathrm{s}}^2 \beta \left(\beta+1\right) 
\left( t_{\mathrm{s}} - t \right)^{-2\left(\beta+1\right)} 
\right]\,, 
\label{eqn-non-local-last-section-07}
\end{align}
\begin{align}
\Xi \sim 6u \left[ 
9 h_{\mathrm{s}}^2 \beta^2
\left( t_{\mathrm{s}} - t \right)^{-2\left(\beta+1\right)} 
+ 18 h_{\mathrm{s}}^3 \beta 
\left( t_{\mathrm{s}} - t \right)^{-\left(3\beta+1\right)} 
+ 2h_{\mathrm{s}} \beta \left(\beta+1\right) \left(\beta+2\right) 
\left( t_{\mathrm{s}} - t \right)^{-\left(\beta +3\right)} 
+ 12h_{\mathrm{s}}^2 \beta \left(\beta+1\right) 
\left( t_{\mathrm{s}} - t \right)^{-2\left(\beta +1\right)} 
\right]\,. 
\label{eqn-non-local-last-section-08} 
\end{align} 
We now examine the r.h.s. of Eq.~(\ref{non-local-field-eq01}) with $\Theta$ 
in Eq.~(\ref{eqn-non-local-last-section-05}). 
For $\beta > 1$, $\sigma < 0$, 
the first term of Eq.~(\ref{eqn-non-local-last-section-07}), i.e. $-108 u h_{\mathrm{s}}^2 \beta 
\left( t_{\mathrm{s}} - t \right)^{-\left(3\beta +1\right)}$, 
becomes the leading term. 
As $u \neq 0$, $h_{\mathrm{s}} \neq 0$ and $\beta \neq 0$, therefore, this leading term does not vanish, which means that the additional $R^2$ term can remove the finite-time future singularities. 
On the other hand, for $-1 < \beta < 0\,, \, 0 < \beta < 1$, 
the second and third terms of Eq.~(\ref{eqn-non-local-last-section-07}) become the leading terms. 
Again since $u \neq 0$, $h_{\mathrm{s}} \neq 0$, $\beta \neq 0$ and $\beta \neq -2$, hence, 
these leading terms do not vanish. 
This means that the additional $R^2$ term can remove the finite-time future singularities. 

We remark that similar to the $F (R)$ gravity, where the inclusion of a correction term $R^2$ could cure the singularities \cite{Bamba:2012ky}, in the non-local gravity too, this situation happens.

\subsection{Scalar field models avoiding finite-time future singularities}

According to~\cite{Carroll:2003st}, it is possible to construct scalar field phantom DE models where the big rip singularity can be avoided. Considering a phantom scalar field $\phi$ with potential $V (\phi)$ as in \cite{Carroll:2003st}, in the background of a FLRW universe, the energy density and pressure of the phantom field, are respectively given by \cite{Copeland:2006wr}
\begin{eqnarray}
\rho_{\phi} = - \dot{\phi}^2/2 + V (\phi), \quad p_{\phi} = - \dot{\phi}^2/2 - V (\phi), 
\end{eqnarray}
and the equation of state of this phantom field, $w = p_{\phi}/\rho_{\phi}$ takes the form 
\begin{eqnarray}
 w_{\phi} = \frac{\dot{\phi}^2/2 + V (\phi)}{\dot{\phi}^2/2 - V (\phi)}, 
\end{eqnarray}
which satisfies $w_{\phi} < -1$ if $\dot{\phi}^2/2 < V (\phi)$. 
The equation of motion of the phantom field is given by 
\begin{eqnarray}
\ddot\phi + 3H\dot\phi - a^{-2}\nabla^2\phi - V'(\phi) = 0, 
\end{eqnarray}
where prime denotes the derivative with respect to $\phi$ and $\nabla^2$ denotes the spatial Laplacian, i.e. $\nabla^2\phi =
\partial_x^2\phi + \partial_y^2\phi +\partial_x^2\phi$. 
In this model scenario, the dynamics of the universe heavily depends on the potential function $V (\phi)$. As argued in \cite{Carroll:2003st}, the following gaussian potential has some interesting consequences in the context of finite time singularities: 
 \begin{eqnarray}
 V(\phi) = V_0 e^{-(\phi^2/\sigma^2)}\ ,
\end{eqnarray}
where $V_0$ and $\sigma$ are constants. For this particular choice of the potential, the evolution of the scalar field, the density parameter of phantom scalar field and its equation of state have been numerically investigated in \cite{Carroll:2003st}. From the evolution of the equation of state $w_{\phi}$, one finds that \cite{Carroll:2003st}: during the initial stages of evolution, the phantom scalar field is
frozen by the expansion and it behaves like the cosmological constant (i.e. $w_{\phi}\simeq -1$). After that the field starts to evolve quite rapidly towards the maximum of its potential and the energy density of the phantom field becomes
dominant where $w_{\phi}$ crosses the phantom divide line $w_{\phi} = -1$ and becomes more negative. Finally, in the very late phase of the universe, the field
comes to rest at the maximum of the potential and the accelerating expansion begins with $w_{\phi} = -1$. In the late phase, as $w_{\phi}$ does not go beyond $-1$, hence no future singularity appears. In fact, the universe
approaches to a de~Sitter phase.

On the other hand, it is interesting to note that for $F (R) = R+ \alpha R^2$ (i.e. the Starobinksy model \cite{Starobinsky:1980te}) where $\alpha$ is a dimensionless constant, one can derive the corresponding quintessence scalar field potential of exponential type \cite{Chaichian:2022apa}. As due to the presence of $R^2$ in $F (R)$, no finite-time future singularities appear \cite{Abdalla:2004sw}, therefore, we conclude that quintessence scalar field model with exponential  potential could avoid the finite-time future singularities.

\subsection{Inhomogeneous equation of state}

In the context of various modified gravity theories, using the effective energy density $\rho_{\rm eff}$ and pressure $p_{\rm eff}$, one can derive an effective equation of state $p_{\rm eff} - w \rho_{\rm eff} = G (H, \dot{H}, \ddot{H})$, see Eq. (\ref{effective-eos-F(R)-MG}) for the $F (R)$ gravity case [while this is true for other gravity theories as well] in which $G (H, \dot{H}, \ddot{H}) = - \frac{1}{\kappa^2}\left(2\dot H + 3(1+w)H^2 \right)$, see Eq. (\ref{Geff-F(R)}). The explicit form of $G (H, \dot{H}, \ddot{H})$ that involves the geometric invariants and their time derivatives is given in Eq. (\ref{Geff-explicit-F(R)}). Therefore, the modified gravity theories can lead to various inhomogeneous equation of state
parameters depending on the choice of the underlying gravitational theory. In this section we shall investigate the conditions on $G (H, \dot{H}, \ddot{H})$ that may prevent the appearance of finite time future singularities. 
If Eq. (\ref{Geff-F(R)}) is found to be inconsistent for kind of finite time future singularities, then one can conclude that no finite time future singularities are realized in this case.

We consider the case when $H$ evolves as (\ref{HsinR}). In that case, the r.h.s. of (\ref{Geff-F(R)}) behaves as~\cite{Bamba:2008ut}

\begin{eqnarray}\label{sec-avoidance-inhomogeneous-01}
G(H, \dot{H}, \ddot{H}) = - \frac{1}{\kappa^2}\left(2\dot H + 3(1+w)H^2 \right) \sim
\left\{
\begin{array}{lcl}
 -\frac{3(1+w)h_s^2}{\kappa^2}\left(t_0 - t\right)^{-2\beta} & \mbox{when} & \beta>1 \\
 -\frac{2\beta h_s + 3(1+w)h_s^2}{\kappa^2}\left( t_0 - t \right)^{-2} && 0<\beta<1 \\
 -\frac{2\beta h_s}{\kappa^2} \left(t_0 - t\right)^{-\beta - 1} && 0>\beta>-1
\end{array} \right. \ .
\end{eqnarray}
Now for $\beta>-1$, $\beta \neq 0$, which correspond to Type I, II and III singularities, notice that the l.h.s. of (\ref{sec-avoidance-inhomogeneous-01}), i.e.,
$G\left(H, \dot H, \cdots\right)$ of (\ref{Geff-F(R)}) diverges. Thus, in order 
to avoid the appearance of such finite time singularities, $G$ must be bounded and in this case, Eq.~(\ref{Geff-F(R)}) becomes inconsistent with the behavior of
the r.h.s. of (\ref{sec-avoidance-inhomogeneous-01}).

An example in this direction is the following \cite{Bamba:2008ut}
\begin{eqnarray}
\label{sec-avoidance-inhomogeneous-02}
G\left(H, \dot H, \cdots\right) = G_0 \left(\frac{1 + a H^2}{1+ b H^2} \right)\,
\end{eqnarray}
where $a$ and $b$ are positive real numbers and $G_0$ is a constant.

We further note that for $\beta>0$ corresponding to Type I or III singularity, the r.h.s. of (\ref{sec-avoidance-inhomogeneous-01}) becomes negative. Hence, if $G\left(H, \dot H, \cdots\right)$ is
positive for large $H$, we do not realize any finite time future singularity. 

Another chance is that if $G\left(H, \dot H, \cdots\right)$ contains the term like
$\sqrt{ 1 - a^2 H^2}$, which becomes imaginary for large $H$, then Eq. (\ref{sec-avoidance-inhomogeneous-01}) 
becomes inconsistent.
Thus, singularities where the curvature blows up (Type I, II, III) could be avoided. 

We remark that such mechanism could be applied even if we have the phantom EoS, i.e. $w<-1$. In that case
one can add an extra term
$G_1 (H) = G_0 \left(\sqrt{1 - H^2/H_0^2} - 1\right)$ to $G\left(H, \dot H, \cdots\right)$. 
Here, $G_0$ and $H_0$ are real numbers. If $H_0$ is considered to be large enough, $G_1 (H)$ is not relevant for the small curvature but relevant for large scale and hence
the possibility of the curvature singularity is avoided.

\subsection{Future Singularities with the Account of Quantum Effects}
\label{sec-quantum-effects}

In this subsection we will extend our analysis of singularities
performed when we have studied semi-classical gravity.
It is well known that, when the Universe approaches the future singularity,
its curvature and other geometrical invariants grow up.
As a result, the quantum effects may change the behavior of the future spacetime singularity.
For example, one can show that quantum effects may change the structure of future singularity, see~\cite{Nojiri:2004ip,Nojiri:2004pf}
(see also~\cite{Kamenshchik:2013naa,Bates:2010nv,Tretyakov:2005en,Calderon:2004bi,Carlson:2016iuw}).
We will use simple qualitative arguments of Ref.~\cite{Nojiri:2010wj} to show the role of quantum effects in
conformally-invariant theories to future singularity.

It is well-known that the generalized conformal anomaly $T_A$ has the following form:
\begin{align}
\label{OVII}
T_A=b\left(\mathcal{F} + \frac{2}{3}\Box R\right) + b' {G} + b''\Box R\,,
\end{align}
where ${G}$ is the Gauss-Bonnet invariant in (\ref{GB}) and
$\mathcal{F}$ denotes the square of the 4D Weyl tensor, given by
\begin{align}
\label{GF}
\mathcal{F} = \frac{1}{3}R^2 -2 R_{\mu\nu}R^{\mu\nu} + R_{\mu\nu\rho\sigma}R^{\mu\nu\rho\sigma}\, .
\end{align}
In case when matter is conformally-invariant and there appear $N_0$ scalars,
$N_{1/2}$ spinors, $N_1$ vector fields, $N_2$
($= 1$) gravitons, and $N_\mathrm{HD}$ higher-derivative conformal
scalars, $b$ and $b'$, which are obtained using adiabatic regularization, take the following forms,
\begin{align}
\label{bs}
b= \frac{N_0 +6N_{1/2}+12N_1 + 611 N_2 - 8N_\mathrm{HD}}{120(4\pi)^2} \, ,\quad
b'=- \frac{N_0+11N_{1/2}+62N_1 + 1411 N_2 -28 N_\mathrm{HD}}{360(4\pi)^2}\, .
\end{align}
As given in (\ref{bs}), for the usual matter, $b$ is positive and $b'$ is negative. 
An exception is the higher-derivative conformal scalar.
Note that the value of $b''$ can always be shifted with the addition of $R^2$ to the classical action.

If one writes the energy density $\rho_A$ and pressure $p_A$
corresponding to the trace anomaly $T_A$, then one has $T_A=- \rho_A + 3p_A$.
Now with the use of the energy conservation law in the FLRW Universe, one gets
\begin{align}
\label{CB1}
0=\frac{d\rho_A}{dt} + 3 H\left(\rho_A + p_A\right)\, ,
\end{align}
which can be expressed by eliminating $p_A$ as
\begin{align}
\label{CB2}
T_A=-4\rho_A - \frac{1}{H}\frac{d\rho_A}{dt}\, ,
\end{align}
and with the help of integration, $\rho_A$ can be found as \cite{Nojiri:2005sx}:
\begin{align}
\label{CB3}
\rho_A = -\frac{1}{a^4} \int dt a^4 H T_A \, .
\end{align}
Now, using the above expression and identifying $\rho_\mathrm{eff}=\rho_A$,
we will consider the FLRW equation, later.
Before considering the FLRW equation, as in \cite{Nojiri:2010wj}, however,
we first consider the trace of the Einstein equation, for simplicity,
by including the trace anomaly, as follows,
\begin{align}
\label{CA1}
R = - \frac{\kappa^2}{2} \left(T_\mathrm{matter} + T_A \right)\, .
\end{align}
Here, $T_\mathrm{matter}$ refers to the trace of the matter
energy-momentum tensor. Now, for the FLRW Universe, $\mathcal{F}$ and
${G}$ are as follows
\begin{align}
\label{CA2}
\mathcal{F}=0\, ,\quad {G}=24\left(\dot H H^2 + H^4\right)\, .
\end{align}
What we would like to show is that, if there is a singularity, then the trace
equation (\ref{CA2}) cannot be consistent.
In particular, we will show that the contribution from the conformal anomaly in the r.h.s. of
Eq.~(\ref{CA1}) is more singular than the scalar curvature in the l.h.s. of Eq.~(\ref{CA1}). 
Note that a rigorous study on the quantum effects may be done following~\cite{Carlson:2016iuw}, however, this needs an extensive numerical investigations 
depending on the particles content of the Universe as well as effective dark fluid.

Now we assume that $H$ behaves as in (\ref{Hsin}) or (\ref{HsinR})
and neglect the contribution from matter and put $T_\mathrm{matter}=0$.
In the case of Type I singularity, the scalar curvature $R$, which is given by $R=12H^2 + 6\dot H$, $R$ behaves as
$R\sim \left(t_s - t\right)^{-2\beta}$ and in the case of Type III singularity, $R$
behaves as $R\sim \left(t_s - t\right)^{-\beta-1}$.
On the other hand, in the case of Type II singularity, $R$ behaves as $R\sim \left(t_s - t\right)^{-\beta-1}$.
When Type IV singularity appears, if $H_s(t) \neq 0$ in (\ref{HsinR}), $R$ is finite but if $H_s(t) =0$,
$R\sim \left(t_s - t\right)^{-\beta-1}$.

In the case of Type I singularity, near the singularity, $t\sim t_s$, as seen from Eq. (\ref{CA2}), the Gauss-Bonnet invariant
${G}$ behaves as ${G} \sim 24 H^4 \sim \left(t_s - t\right)^{-4\beta}$
and therefore ${G}$ becomes very large and the contribution
from the matter $T_\mathrm{matter}$ in (\ref{CA1}) can be neglected.
On the other hand, one finds $\Box R \sim \left(t_s - t\right)^{-2\beta -2}$.
Then since $R\sim \left(t_s - t\right)^{-2\beta}$, $T_A$ becomes much larger
than $R$ and therefore Eq.~(\ref{CA1}) cannot be satisfied.
This shows that the quantum effects coming from the conformal anomaly remove Type I singularity.

In the case of Type II singularity, we find that ${G}$ behaves as
${G}\sim 24 \dot H H^2 \sim \left(t_s - t \right)^{-3\beta -1}$.
Since $R \sim \left(t_s - t \right)^{-\beta - 1}$,
the Gauss-Bonnet term in $T_A$ is less singular and therefore negligible
compared with $R$ and the contribution from the matter. 
Therefore, the Gauss-Bonnet term in $T_A$ does not help to prevent Type II singularity.
Note, however, $\Box R$ behaves as $\Box R \sim \left(t_s - t\right)^{-\beta - 3}$, which is more singular than
the scalar curvature. Then if $2b/3 + b''\neq 0$,
the contribution from $T_A$ becomes much larger than $R$ near the
singularity $t\sim t_s$ and Eq.~(\ref{CA1}) cannot be satisfied.
Therefore, if $2b/3 + b''\neq 0$, even Type II singularity can also be 
prevented when the quantum effects due to conformal anomaly are included.

In the case of Type III singularity, the Gauss-Bonnet invariant behaves as
${G}\sim 24 \dot H H^2 \sim \left(t_s - t \right)^{-3\beta -1}$
and $\Box R$ behaves as $\Box R \sim \left(t_s - t\right)^{-\beta - 3}$.
Because the scalar curvature behaves as $R \sim \left(t_s - t \right)^{-\beta - 1}$, both of the terms, $\Box R$ and
${G}$, are more singular than the scalar curvature $R$ and Type III singularity is also prevented.
Thus, we demonstrated that quantum effects may remove finite-time future singularities.
Note that account of quantum gravity effects in specific models is also known to remove the Big Rip singularity~\cite{Elizalde:2004mq}.

\subsection{FLRW equation including Trace Anomaly}
\label{sec-trace-anomaly}

Here we consider the FLRW equation including the effective energy density $\rho_A$ induced by the trace anomaly in (\ref{CB3}).
By using (\ref{OVII}), (\ref{CA2}), and (\ref{CB3}), we find the explicit form of $\rho_A$ as follows
\begin{align}
\label{CB3B1}
\rho_A = -\frac{1}{a^4} \int dt a^4 H
\left\{ 24 b' \left(\dot H H^2 + H^4\right) - 6 \left(b + b''\right) \left(\frac{d^2}{dt^2} + 3 H \frac{d}{dt} \right) \left( \dot H + 2 H^2 \right) \right\}\, .
\end{align}
Then the FLRW equation is given by
\begin{align}
\label{BRHR2A1}
\frac{3}{\kappa^2} H^2 = \rho + \rho_A \, .
\end{align}
Here $\rho$ is the energy density corresponding to the phantom DE,
which behaves as $\rho \sim \rho_0 a^{-3\left(1 + w \right)}$ with
$w<-1$. By combining (\ref{CB3B1}) and (\ref{BRHR2A1}) and
assuming $\rho = \rho_0 a^{-3\left(1 + w \right)}$ (we take $a_0=1$), we obtain
\begin{align}
\label{AFLRW1}
\frac{3}{\kappa^2} H^2 a^4 = \rho_0 a^{1 -3 w} - \int dt a^4 H
\left\{ 24 b' \left(\dot H H^2 + H^4\right) - 6 \left(b + b''\right) \left(\frac{d^2}{dt^2} + 3 H \frac{d}{dt} \right) \left( \dot H + 2 H^2 \right) \right\}\, .
\end{align}
Differentiating the above expression with respect to $t$, we obtain
\begin{align}
\label{AFLRW2}
\frac{3}{\kappa^2} \left( \dot H + 4 H^2 \right) =&\, \left( 1 -3 w \right) \rho_0 a^{ -3 \left( 1 + w\right)} 
- \left\{ 24 b' \left(\dot H H^2 + H^4\right) - 6 \left(b + b''\right) \left(\frac{d^2}{dt^2} + 3 H \frac{d}{dt} \right) \left( \dot H + 2 H^2 \right) \right\}\, .
\end{align}
If we assume
\begin{align}
\label{Hq}
H(t) \sim {\widetilde h}_s \left( t_s - t \right)^{\widetilde\beta}\, ,
\end{align}
with a positive constant ${\widetilde h}_s$, we find
\begin{align}
\label{Hbehave}
& \dot H \sim \left( t_s - t \right)^{\widetilde\beta-1}\, , \quad H^2 \sim \left( t_s - t \right)^{2\widetilde\beta}\, , \quad
a^{ -3 \left( 1 + w\right)} \sim \left\{ \begin{array}{ll}
\e^{\frac{3 \left( 1 + w\right)}{\widetilde\beta +1}{\widetilde h}_s \left( t_s - t \right)^{\widetilde\beta+1}} & \mbox{if $\widetilde{\beta}\neq -1$} \\
\left( t_s - t \right)^{3 \left( 1 + w\right){\widetilde h}_s } & \mbox{if $\widetilde{\beta}= -1$}
\end{array} \right. \, , \nonumber \\
& \dot H H^2 \sim \left( t_s - t \right)^{3 \widetilde\beta-1} \, , \quad
H^4 \sim \left( t_s - t \right)^{4\widetilde\beta}\, , \quad
\frac{d^2 \dot H}{dt^2} \sim \left( t_s - t \right)^{\widetilde\beta-3}\, , \quad
H \frac{d \dot H}{dt} \sim \left( t_s - t \right)^{2\widetilde\beta-2}\, , \nonumber \\
& \frac{d^2 H^2}{dt^2} \sim \left( t_s - t \right)^{2\widetilde\beta-2}\, , \quad
H \frac{d H^2}{dt} \sim \left( t_s - t \right)^{3\widetilde\beta-1}\, .
\end{align}
By using the above behaviors, we investigate what kind of singularity can be realized or prohibited.

First, we consider the case of Type I singularity, where $\widetilde\beta\leq -1$.
If $\widetilde\beta<-1$, the energy density $\rho=\rho_0 a^{ -3 \left( 1 + w\right)}$ of matter grows up very rapidly when $t\to t_s$,
other terms in (\ref{AFLRW2}) cannot cancel the term and therefore Eq.~(\ref{AFLRW2}) is not satisfied.
If $\widetilde\beta=-1$, the l.h.s. of (\ref{AFLRW2}) behaves as
\begin{align}
\label{l1}
\frac{3}{\kappa^2} \left( \dot H + 4 H^2 \right) \sim \frac{3}{\kappa^2} \left( {\widetilde h}_s + 4 \widetilde {h}_s^2 \right) \left( t_s - t \right)^{-2}>0 \, .
\end{align}
On the other hand, the first term in the r.h.s. of (\ref{AFLRW2}), which comes from the matter, as found in (\ref{Hbehave}),
\begin{align}
\label{r1}
\left( 1 -3 w \right) \rho_0 a^{ -3 \left( 1 + w\right)} \sim \left( 1 -3 w \right) \rho_0 \left( t_s - t \right)^{3 \left( 1 + w\right){\widetilde h}_s } > 0 \, ,
\end{align}
and the second term in the r.h.s. of (\ref{AFLRW2}), which comes from the trace anomaly, behaves as
\begin{align}
\label{a1}
& - \left\{ 24 b' \left(\dot H H^2 + H^4\right) - 6 \left(b + b''\right) \left(\frac{d^2}{dt^2} + 3 H \frac{d}{dt} \right) \left( \dot H + 2 H^2 \right) \right\} \nonumber \\
=&\, - 24 b' \left(\widetilde {h}_s^3 + \widetilde{h}_s^4 \right)
+ 6 \left(b + b''\right) \left( 6 {\widetilde h_s} + 18 \widetilde {h}_s^2 + 12 \widetilde {h}_s^3 \right)
\left( t_s - t \right)^{-4} \, .
\end{align}
Because $b>0$ and $b'<0$ as in (\ref{bs}) and $b''$ is arbitrary, as long as $b + b''\geq 0$,
the second term in the r.h.s. of (\ref{AFLRW2}) is positive.
Compared with (\ref{l1}) and (\ref{a1}), we find that the l.h.s. of (\ref{AFLRW2}) cannot balance with the second term in the r.h.s. of (\ref{AFLRW2}).
Furthermore, because both of the first and second terms of (\ref{AFLRW2}) are positive as $b + b'' \geq 0$, they cannot cancel each other.
This tells that the behavior with $\widetilde\beta=-1$ does not satisfy Eq.~(\ref{AFLRW2}).
By combining the results with $\widetilde\beta<-1$ and $\widetilde\beta=-1$, we find that the Type I singularity is prohibited.

We now consider Type III singularity, where $-1 < \widetilde\beta <0$.
We should note that in Eq. (\ref{AFLRW2}), in the limit of $t\to t_s$, the first term in the r.h.s. is finite and the l.h.s. and the second term in the r.h.s. diverge. 
Therefore we may neglect the first term in the r.h.s. of (\ref{AFLRW2}).
Because now $2\widetilde\beta>\widetilde\beta - 1> 2\widetilde\beta - 1$, the l.h.s. cannot balance with the second term in the r.h.s.
and therefore (\ref{AFLRW2}) cannot be satisfied and the Type III singularity is prohibited.

In case of Type II, where $0< \widetilde\beta <-1$, n the limit of $t\to t_s$, the first term in the r.h.s. is finite, again.
On the other hand, in the limit, the l.h.s. (\ref{AFLRW2}) diverges as
$\left( t_s - t \right)^{\widetilde\beta-1}$ and the second term in the r.h.s. behaves as
$\left( t_s - t \right)^{\widetilde\beta-3}$ if $b + b'' \geq 0$, or $\left( t_s - t \right)^{3 \widetilde\beta-1}$ if $b + b'' = 0$,
the l.h.s. cannot balance with the second term in the r.h.s. and therefore (\ref{AFLRW2}) cannot be satisfied, again and the Type II singularity is prohibited.

By combining the above results, we find the Type I, II, and III singularities are prohibited by the
quantum effects coming from the trace anomaly.
The Type IV singularity can be allowed in general when there is any matter.

When the matter can be neglected, to see what could happen, we solve (\ref{AFLRW2}) by putting $\rho_0=0$ and by adjusting $b''$ so that $b+b''=0$.
Then we can rewrite (\ref{AFLRW2}) as follows,
\begin{align}
\label{AFLRW2B1}
1=& - \frac{\dot H \left( 1 + 8 \kappa^2 b' H^2 \right)}{4 H^2 \left( 1 + 2 \kappa^2 b' H^2 \right)} \, ,
\end{align}
which can be integrated to be
\begin{align}
\label{AFLRW2B1}
t - t_s = \frac{1}{4H} + \frac{3\kappa\sqrt{- 2 b'}}{8} \ln \left| \frac{1 + \kappa \sqrt{ -2 b'} H}{1 - \kappa \sqrt{ -2 b'} H} \right| \, .
\end{align}
The solution has two branches for positive $H$, that is, the
region $0<H<\frac{1}{\kappa \sqrt{ -2 b'}}$ and $\kappa \sqrt{ -2 b'}<H<+\infty$.
When $H\to 0$, $t$ goes to the plus infinity and $H\to \sqrt{ -2 b'}$, $t$ goes to the plus infinity, again.
That is, there is a local minimum in the region $0<H<\frac{1}{\kappa \sqrt{ -2 b'}}$.
In the region, $\kappa \sqrt{ -2 b'}<H<+\infty$, when $H\to \sqrt{ -2 b'}$, $t$ goes to the plus infinity and
after that $t$ monotonically decreases as a function of $H$ and
when $H\to +\infty$, $t-t_s$ vanishes as $t-t_s \sim \frac{1}{H}$.
Then there could be three scenarios for the cosmology.
In one scenario, the Universe starts from the finite $t$ corresponding to
the local minimum in the region $0<H<\frac{1}{\kappa \sqrt{ -2 b'}}$.
After that, there are two possibilities.
In one possibility, when $t$ increases, $H\to 0$ and in another
possibility, when $t$ increases, $H\to \frac{1}{\kappa \sqrt{ -2 b'}}$.
In another scenario, the Universe starts at $t\to t_s$ with $H\to +\infty$.
After that, $H$ monotonically decreases and goes
to $H\to \frac{1}{\kappa \sqrt{ -2 b'}}$ when $t\to +\infty$.

\subsection{Future Singularities with the Account of Thermal Effects}
\label{sec-thermal-effects}

We may remind that the large Hubble rate $H$ means the large temperature of the Universe.
The Hawking radiation effectively should be generated at the apparent horizon of the FLRW Universe~\cite{Gibbons:1977mu, Cai:2008gw}.
Eventually, it should give an important contribution to the energy density of the late-time
Universe, especially right before the Rip time. In other words, at
a large temperature that may even diverge at the Rip time, there
should appear thermal radiation.

\subsubsection{Type I Singularity with Thermal Effects: Transition to Type II Singularity}

Near the Type I (Big Rip) singularity, the temperature of the
Universe becomes large and we may expect the generation of thermal
radiation as in the case of the Hawking radiation.
The Hawking temperature $T$ is proportional to the inverse of the radius
$r_\mathrm{H}$ of the apparent horizon and the radius
$r_\mathrm{H}$ is proportional to the inverse of the Hubble rate $H$.
Therefore the temperature $T$ is proportional to the Hubble rate $H$.
As well-known in statistical physics, the energy-density
$\rho_\mathrm{t\_rad}$ of the thermal radiation is proportional to
the fourth power of the temperature.
Then when $H$ is large enough, we may assume that the energy-density of the thermal
radiation is given by
\begin{align}
\label{BRHR1}
\rho_\mathrm{t\_ rad} = \alpha H^4 \, ,
\end{align}
with a positive constant $\alpha$.
At the late time, the FLRW equation should be modified
by the account of thermal radiation,
\begin{align}
\label{BRHR2}
\frac{3}{\kappa^2} H^2 = \rho + \alpha H^4 \, .
\end{align}
Here $\rho$ is the energy density corresponding to the phantom DE,
which behaves as $\rho \sim \rho_0 a^{-3\left(1 + w \right)}$ with $w<-1$.
At the late-time but much before the Big Rip time, the
first term in the equation (\ref{BRHR2}) dominates and therefore
the Universe expands to the Big Rip singularity, where the Hubble
rate $H$ behaves as in (\ref{Hsin}) with $\beta=1$.
Then near the Big Rip time $t_s$, the second term in (\ref{BRHR2}) should
dominate and we obtain
\begin{align}
\label{BRHR3}
\frac{3}{\kappa^2} H^2 \sim \alpha H^4 \, ,
\end{align}
whose non-trivial solution is given by
\begin{align}
\label{BRHR4}
H^2 = H_\mathrm{crit}^2 \equiv \frac{3}{\kappa^2 \alpha} \, .
\end{align}
As $H$ goes to a constant, we might expect that the spacetime
goes to the asymptotically de Sitter spacetime but it is not true.
Even in the de Sitter spacetime, the scale factor $a$
becomes larger and larger as an exponential function of $t$, then
the first term in the equation (\ref{BRHR2}) should dominate finally.
The Hubble rate $H$ is, however, already larger than
$H_\mathrm{crit}$, there is no solution of (\ref{BRHR2}).
Then the Universe should end up at the finite-time with some kind of the singularity.

For more quantitative analysis, we solve (\ref{BRHR2}), with respect to $H^2$
as follows,
\begin{align}
\label{BRHR5}
H^2 = \frac{\frac{3}{\kappa^2} \pm \sqrt{\frac{9}{\kappa^4} - 4\alpha \rho_0
a^{-3\left( 1 + w \right)}}}
{2\alpha}\, .
\end{align}
As $H^2$ is a real number, we find that there is a maximum for the scale
factor $a$,
\begin{align}
\label{BRHR6}
a \leq a_\mathrm{max} \equiv \left( \frac{9}{4 \kappa^4 \alpha \rho_0}
\right)^{- \frac{1}{3\left( 1 + w \right)}} \, .
\end{align}
Then we consider the behavior of $a$ or $H$ around the maximal
$a=a_\mathrm{max}$ by writing the scale factor $a$ as
\begin{align}
\label{BRHR7}
a = a_\mathrm{max} \e^N \, .
\end{align}
Here, $N$ corresponds to the $e$-folding number but $N$ should be negative
because $a<a_\mathrm{max}$.
Furthermore, as we are interested in the region $a\sim a_\mathrm{max}$, we
assume $\left| N \right| \ll 1$.
Then by using $H=\frac{dN}{dt}$, Eq.~(\ref{BRHR5}) can be rewritten as
\begin{align}
\label{BRHR8}
\left( 1 \mp \frac{1}{2} \sqrt{3\left( 1 + w \right)N} \right) dN \sim dt
\sqrt{ \frac{3}{2\alpha \kappa^2}} \, ,
\end{align}
which can be integrated as
\begin{align}
\label{BRHR9}
N \mp \frac{1}{3} \left( -N\right)^\frac{3}{2} \sqrt{-3\left( 1 + w \right)}
\sim - \left( t_\mathrm{max} - t \right) \sqrt{ \frac{3}{2\alpha \kappa^2}} \, .
\end{align}
Here $a=a_\mathrm{max}$ when $t=t_\mathrm{max}$.
Because we are assuming $\left| N \right| \ll 1$, Eq.~(\ref{BRHR9}) can be
rewritten as
\begin{align}
\label{BRHR10}
N \sim - \left( t_\mathrm{max} - t \right) \sqrt{ \frac{3}{2\alpha \kappa^2}}
\mp \frac{\sqrt{-3\left( 1 + w \right)}}{3} \left( \left( t_\mathrm{max} - t \right)
\sqrt{ \frac{3}{2\alpha \kappa^2}} ~\right)^\frac{3}{2} \, .
\end{align}
Because $H=\frac{dN}{dt}$, we find
\begin{align}
\label{BRHR11}
H \sim& \sqrt{ \frac{3}{2\alpha \kappa^2}}
\mp \frac{\sqrt{-3\left( 1 + w \right)}}{2} \left(
\sqrt{ \frac{3}{2\alpha \kappa^2}} \right)^\frac{3}{2}
\left( t_\mathrm{max} - t \right)^{\frac{1}{2}} \, , \nonumber \\
\dot H \sim& \mp \frac{\sqrt{-3\left( 1 + w \right)}}{4} \left(
\sqrt{ \frac{3}{2\alpha \kappa^2}} \right)^\frac{3}{2}
\left( t_\mathrm{max} - t \right)^{-\frac{1}{2}} \, .
\end{align}
Then in the limit $t\to t_\mathrm{max}$, although $H$ is finite
but $\dot H$ diverges. Therefore, the Universe ends up with Type II
singularity at $t=t_\mathrm{max}$. Thus, we demonstrated that the
account of thermal effects near the Big Rip singularity changes
the Universe evolution to the finite-time Type II singularity.

\subsubsection{Type III Singularity with the Account of Thermal Effects: Transition to Type II Singularity}

The scale factor which generates Type III singularity
can be expressed as
\begin{align}
\label{III1}
a(t) = a_s \e^{ - \frac{h_s}{1-\beta} \left( t_s - t \right)^{1-\beta}} \, ,
\end{align}
with $a_s$, $t_s$, $\beta$, and $h_s$ some constants.
In order to generate Type III singularity we restrict the value of $\beta$ as
\begin{align}
\label{III2}
0<\beta<1 \, .
\end{align}
Then the Hubble rate $H$ is given by
\begin{align}
\label{III3}
H = h_s\left( t_s - t \right)^{-\beta} \, .
\end{align}
Hence, in the limit $t\to t_s$, $H$ diverges but the scale factor $a$ is finite.
 From Eq.~(\ref{FLRWs2}) it follows
\begin{align}
\label{III4}
\rho_\mathrm{eff} = \frac{3h_s^2}{\kappa^2}
 \left( t_s - t \right)^{-2\beta} \, , \quad
p_\mathrm{eff} = - \frac{1}{\kappa^2}
\left( - 2\beta h_s \left( t_s - t \right)^{-\beta -1}
+ 3 h_s^2 \left( t_s - t \right)^{-2\beta } \right)\, .
\end{align}
By deleting $\left( t_s - t \right)$, we find the following EoS,
\begin{align}
\label{III5}
p_\mathrm{eff} = - \rho_\mathrm{eff} - \frac{2 h_s \beta}{\kappa^2}
\left( \frac{\kappa^2 \rho_\mathrm{eff}}{3 h_s^2}
\right)^{\frac{\beta+1}{2\beta}}\, .
\end{align}
Using (\ref{III1}) and (\ref{III4}), one gets
\begin{align}
\label{III6}
\rho_\mathrm{eff} = \frac{3}{\kappa^2}
h_s^2 \Bigg( \frac{1-\beta}{h_s} \ln \left(
\frac{a_s}{a(t)}\right)
\Bigg)^{-\frac{2\beta }{1-\beta}}\, .
\end{align}
With the account of the thermal radiation, instead of
(\ref{BRHR2}), we have
\begin{align}
\label{III7}
\frac{3}{\kappa^2} H^2 = A \left( \ln \left( \frac{a_s}{a(t)}\right)
\right)^{-B} + \alpha H^4 \, , \quad
A \equiv \frac{3}{\kappa^2}
\left( \frac{h_s^2}{(1-\beta)^{\beta}} \right)^{\frac{1}{1-\beta}} \, ,
\quad B \equiv \frac{2 \beta}{1-\beta} > 0 \, .
\end{align}
Then instead of (\ref{BRHR5}), we obtain
\begin{align}
\label{III8}
H^2 = \frac{\frac{3}{\kappa^2} \pm \sqrt{\frac{9}{\kappa^4} - 4\alpha
A \left( \ln \left( \frac{a_s}{a(t)}\right) \right)^{-B}}}
{2\alpha} \, .
\end{align}
Then in order that $H^2$ to be real, we find that there is
a maximum $a_\mathrm{max}$ for $a(t)$,
\begin{align}
\label{III9}
a(t) \leq a_\mathrm{max} \equiv a_s
\e^{-\left(\frac{9}{4A \alpha \kappa^2}\right)^{- \frac{1}{B}}}
< a_s \, .
\end{align}
Because $a_\mathrm{max}$ is smaller than $a_s$, we find that dark
Universe with the future Type III singularity is transited to the
one with Type II singularity due to the account of thermal
effects.

\subsubsection{Thermal Radiation for Type II and Type IV Singularities}

When we consider Type II and Type IV singularities, $H_s(t)$ in (\ref{HsinR}) is finite and positive,
$ 0 < H= H_s(t_s) < \infty$
When one considers general matter,
the first FLRW equation where
usual matter and the thermal radiation as in (\ref{BRHR2}) are included,
is given by
\begin{align}
\label{CVT01B}
\frac{3}{\kappa^2} H^2 = \rho + \alpha H^4 \, .
\end{align}
Here, $\rho$ is the matter energy-density.
In case of Type II or Type IV singularity, if $H_s(t_s) \neq 0$,
near the singularity, the l.h.s. goes to a finite value
$\frac{3}{\kappa^2} H^2 \to \frac{3}{\kappa^2} H_s(t_s)^2$
and the contribution from the thermal radiation in the r.h.s. also
becomes finite, $\alpha H^4 \to \alpha H_s(t_s)^4$.
Therefore, the thermal radiation does not change the structure of the
singularity.
Even if $H_s(t_s)=0$, the r.h.s. behaves as $\left(t_s - t\right)^{-4\beta}$ and the
contribution from the thermal radiation behaves as
$\left(t_s - t\right)^{-2\beta}$.
Because $\beta<0$, the contribution from the
thermal radiation is less dominant and therefore the thermal radiation does not
change the structure of the singularity.

\subsection{Combination of Quantum Effect and Thermal Effect}
\label{sec-quantum-plus-thermal}

We now combine the quantum effect and the thermal effect.
In Subsection~\ref{sec-quantum-effects} the trace part of the Einstein equation is used.
As the radiation is usually conformal, the trace part of the energy-momentum tensor of the radiation
should vanish and the thermal radiation does not contribute to the trace equation.
We should be, however, more careful in the present situation.
The energy density of the thermal radiation is only determined by the temperature.
Therefore, the Universe expands and its volume with the thermal radiation increases, the total energy
should also be increased if the temperature is not changed or
increases as in the case of Type I (Big Rip) or Type III singularity.
In other words, say, in the phantom Universe, there should exist effectively negative pressure.
The energy of the thermal radiation is not conserved because the expansion produces
the new thermal radiation.
We should note, however, in order that the effective pressure, which includes the effect of the
expansion, is consistent with the FLRW equations, the
energy-density of the thermal radiation and the effective pressure
must satisfy the conservation law
\begin{align}
\label{CVT1}
0=\frac{d\rho_\mathrm{t\_rad}}{dt} + 3 H\left(\rho_\mathrm{t\_rad}
+ p_\mathrm{t\_rad}\right)\, .
\end{align}
To show the conservation law, we may start from the first FLRW equation where
usual matter and the thermal radiation as in (\ref{BRHR2}) are included,
\begin{align}
\label{CVT01}
\frac{3}{\kappa^2} H^2 = \rho + \rho_\mathrm{t\_rad} \, , \quad
\rho_\mathrm{t\_rad} = \alpha H^4 \, .
\end{align}
Here $\rho$ is matter energy-density.
By considering the derivative of Eq.~(\ref{CVT01}) with respect to time $t$,
we obtain,
\begin{align}
\label{CVT02}
\frac{6}{\kappa^2} H \dot H = \dot\rho + 4 \alpha H^3 \dot H\, .
\end{align}
Then by using the standard conservation law for matter,
\begin{align}
\label{CVT03}
0=\dot\rho + 3 H \left( \rho + p \right) \, ,
\end{align}
with the matter pressure, and combining (\ref{CVT01}) and
(\ref{CVT02}), we obtain
\begin{align}
\label{CVT04}
 - \frac{1}{\kappa^2} \left( 2\dot H + 3 H^2 \right) = p
 - \alpha \left( H^4 + \frac{4}{3}H^2 \dot H \right) \, ,
\end{align}
which is nothing but the second FLRW equation and we can identify
the effective pressure of the thermal radiation as follows,
\begin{align}
\label{CVT2}
p_\mathrm{t\_rad} = - \alpha \left( H^4 + \frac{4}{3}H^2 \dot H \right)\, .
\end{align}
Thus, effectively, the energy density and the
effective pressure of the thermal radiation satisfy the conservation law
(\ref{CVT1}) or we can find the exact and unique form of the effective pressure
in (\ref{CVT2}) directly by using the conservation law (\ref{CVT1}) and assuming
the form of the energy density of the radiation in (\ref{BRHR1}).

Then the trace part $T_\mathrm{t\_rad} = - \rho_\mathrm{t\_rad} + 3 p_\mathrm{t\_rad}$
of the energy-momentum tensor for the
radiation including the effect of the expansion of the Universe is given by
\begin{align}
\label{CVT2BB}
T_\mathrm{t\_rad} = -4 \alpha \left( H^4 + H^2 \dot H \right) \, .
\end{align}
Let us assume the behavior of the Hubble rate $H$ as in (\ref{Hsin}).
Then in the case of Type I (Big Rip) case ($\beta\geq 1$), near the singularity,
we find $T_\mathrm{t\_rad} \sim \left(t_s - t\right)^{-4\beta}$,
whose behavior is not so changed from that of $T_A$ although we need to
compare $b'$ with $\alpha$ to see which term is the dominant one.

In case of Type II singularity ($-1<\beta<0$),
we find $T_\mathrm{t\_rad} \sim \left(t_s - t \right)^{-3\beta -1}$.
As $R \sim \left(t_s - t \right)^{-\beta - 1}$,
the contribution from $T_\mathrm{t\_rad}$ is negligible.
In case $b''\neq 0$, which is arbitrary and can be put to vanish if we do not add
$R^2$ term, the contribution from $\Box R$ in $T_A$ dominates and the
Type II singularity does not occur.

In case of Type III singularity ($0<\beta<1$), we find
$T_\mathrm{t\_rad} \sim \left(t_s - t \right)^{-3\beta -1}$, whose
behavior is not changed from that of the Gauss-Bonnet invariant
${G}$ in $T_A$ but weaker than the behavior of $\Box R$.
Then if $b''\neq 0$, the contribution from the thermal radiation
is less dominant than that of the conformal anomaly $T_A$.
If $b''=0$, the contribution from $T_\mathrm{t\_rad}$ is not changed
from that from $T_A$ and we need to compare $b'$ with $\alpha$ to
see which could be dominant, again.
Thus, we demonstrated that when quantum effects dominate over thermal effects then future
singularities are removed.
However, in some cases which depend on the specific features of the theory under consideration, the
dominant contribution is due to the thermal effects. In this case, the
most possible 
future of the Universe is the Type II singularity.

\subsection{Quantum Effects May Change the Occurrence of a
Finite-time Future Singularity}

In this section, for the scalar-tensor theory, which often
generates the Big Rip singularity, we consider the quantum
correction found in Ref.~\cite{Barvinsky:1993zg}. The action of
the scalar-tensor theory with a single scalar is given by,
\begin{align}
\label{S1} L=\frac{1}{2\kappa^2}\left(R +
\frac{\tilde{\gamma}}{2}g^{\mu\nu}\partial_\mu\phi
\partial_\nu\phi - V(\phi)\right)\, ,
\end{align}
where $\tilde{\gamma}=\pm 1$. When $\tilde\gamma=1$, the scalar
field is the phantom universe and therefore often generates the
Big Rip singularity It would be interesting to investigate the
quantum properties of such scalar-tensor gravity.

The calculation of the one-loop effective action in the model
(\ref{S1}) has been performed as follows,
\begin{align}
\label{SS2} W_\mathrm{1-loop} =&\,
-\frac{1}{2}\ln\frac{L^2}{\mu^2}\int d^4x \sqrt{-g}\left\{
\frac{5}{2}V^2 - \tilde{\gamma}\left(V'\right)^2 +
\frac{1}{2}\left(V''\right)^2
+ \left[ \frac{\tilde{\gamma}}{2}V - 2V''\right]\phi_{,\mu}\phi^{,\mu} - \left[ \frac{13}{3}V + \frac{\tilde{\gamma}}{12}V''\right]R \right. \nonumber \\
& \left. + \frac{43}{60}R_{\alpha\beta}^2 + \frac{1}{40}R^2 -
\frac{\tilde{\gamma}}{6}R\phi_{,\mu}\phi^{,\mu} +
\frac{5}{4}\left(\phi_{,\mu}\phi^{,\mu}\right)^2 \right\} \, .
\end{align}
The above one-loop action is found in
Ref.~\cite{Barvinsky:1993zg}. We may regard this effective action
(\ref{SS2}) as a finite quantum correction to the classical one.
In (\ref{SS2}), the cut-off $L$ should be identified with the
corresponding physical quantity like the GUT scale or the Planck
scale. When the universe is approximated by the de Sitter
spacetime, the natural choice is $L^2=\left|R\right|$ because the curvature is large enough and constant~\cite{Buchbinder:1984giy,
Buchbinder:1992rb}. On the other hand, in the case that
$\left|V\right|\gg \left|R\right|$, $L^2$ should be identified with $|V|$.

The phantom terms may be induced even if $V=0$ and
$\tilde\gamma=-1$, which corresponds to the canonical scalar. This
happens if the universe goes through a region with negative
curvature. By the fine-tuning of $V$, there may appear the quantum
gravity-induced phantom theory, which subsequently may change the
universe's evolution.

Here, we consider the action where $L^2$ is replaced with
$\left|R\right|$ as a simple example,
\begin{align}
\label{SS3} W_\mathrm{1-loop}=&\, -\frac{1}{2}\int d^4x
\sqrt{-g}\ln \frac{\left|R\right|}{\mu^2}\left\{ \frac{5}{2}V^2 -
\tilde{\gamma}\left(V'\right)^2 + \frac{1}{2}\left(V''\right)^2
+ \left[ \frac{\tilde{\gamma}}{2}V - 2V''\right]\phi_{,\mu}\phi^{,\mu} \right. \nonumber \\
&\, \left. - \left[ \frac{13}{3}V +
\frac{\tilde{\gamma}}{12}V''\right]R +
\frac{43}{60}R_{\alpha\beta}^2 + \frac{1}{40}R^2 -
\frac{\tilde{\gamma}}{6}R\phi_{,\mu}\phi^{,\mu} +
\frac{5}{4}\left(\phi_{,\mu}\phi^{,\mu}\right)^2 \right\} \, .
\end{align}
The variations of this action with respect the scalar field $\phi$
and the metric $g_{\mu\nu}$ are given by
\begin{align}
\label{SS4} \frac{1}{\sqrt{-g}}\frac{\delta
W_\mathrm{1-loop}}{\delta\phi} =&\, -\frac{1}{2}\ln
\frac{\left|R\right|}{\mu^2}\left\{\left\{ \frac{5}{2}V^2 -
\tilde{\gamma}\left(V'\right)^2 + \frac{1}{2}\left(V''\right)^2
\right\}'
+ \left[\frac{\tilde{\gamma}}{2}V - 2V''\right]'\phi_{,\mu}\phi^{,\mu} \right. \nonumber \\
&\, \left. - 2 \nabla_\mu\left\{\left[ \frac{\tilde{\gamma}}{2}V -
2V''\right]\phi^{,\mu}\right\}
 - \left[ \frac{13}{3}V + \frac{\tilde{\gamma}}{12}V''\right]'R + \frac{\tilde{\gamma}}{3}\nabla_\mu\left\{R\phi^{,\mu}\right\}
 - 5 \nabla_\mu\left\{\left(\phi_{,\rho}\phi^{,\rho}\right)\phi^{,\mu}\right\}
\right\} \, , \\
\label{SS5} \frac{1}{\sqrt{-g}}\frac{\delta
W_\mathrm{1-loop}}{\delta g_{\mu\nu}} =&\, -\frac{1}{2}\ln
\frac{\left|R\right|}{\mu^2}\left[\frac{1}{2}g^{\mu\nu}\left\{\frac{5}{2}V^2
- \tilde{\gamma}\left(V'\right)^2
+ \frac{1}{2}\left(V''\right)^2 + \left[\frac{\tilde{\gamma}}{2}V - 2V''\right]\phi_{,\mu}\phi^{,\mu} \right.\right. \nonumber \\
&\, \left. - \left[ \frac{13}{3}V +
\frac{\tilde{\gamma}}{12}V''\right]R +
\frac{43}{60}R_{\alpha\beta}^2 + \frac{1}{40}R^2
 - \frac{\tilde{\gamma}}{6}R\phi_{,\mu}\phi^{,\mu} + \frac{5}{4}\left(\phi_{,\mu}\phi^{,\mu}\right)^2 \right\} \nonumber \\
&\, - \left[\frac{\tilde{\gamma}}{2}V -
2V''\right]\phi^{,\mu}\phi^{,\nu} + \left[ \frac{13}{3}V +
\frac{\tilde{\gamma}}{12}V''\right]R^{\mu\nu}
 - \left(\nabla^\mu \nabla^\nu - g^{\mu\nu}\nabla^2\right) \left[ \frac{13}{3}V + \frac{\tilde{\gamma}}{12}V''\right] \nonumber \\
&\, - \frac{43}{30}R^\mu_{\ \rho}R^{\nu\rho} +
\frac{43}{60}\left\{\left(\nabla_\alpha\nabla^\nu R^{\alpha\mu} +
\nabla_\alpha \nabla^\mu R^{\alpha\nu}\right) - \nabla^2
R^{\mu\nu}
 - g^{\mu\nu}\nabla_\rho \nabla_\sigma R^{\rho\sigma}\right\} + \frac{1}{20}R R^{\mu\nu} \nonumber \\
&\, + \frac{1}{20}\left(\nabla^\mu \nabla^\nu - g^{\mu\nu}
\nabla^2 \right)R +
\frac{\tilde{\gamma}}{6}R^{\mu\nu}\phi_{,\rho}\phi^{,\rho}
- \frac{\tilde{\gamma}}{6}\left(\nabla^\mu \nabla^\nu - g^{\mu\nu}\nabla^2\right) \left(\phi_{,\rho}\phi^{,\rho}\right) \nonumber \\
&\, + \frac{\tilde{\gamma}}{6}R \partial^\mu\phi \partial^\nu\phi
- \frac{5}{2} \phi_{,\rho}\phi^{,\rho} \phi^{,\mu} \phi^{,\nu} +
\left( - R^{\mu\nu} + \nabla^\mu \nabla^\nu -
g^{\mu\nu}\nabla^2\right)\left[ - \frac{1}{2R} \left\{
\frac{5}{2}V^2 - \tilde{\gamma}\left(V'\right)^2 + \frac{1}{2}\left(V''\right)^2 \right. \right. \nonumber \\
&\, \left. \left. + \left[\frac{\tilde{\gamma}}{2}V -
2V''\right]\phi_{,\rho}\phi^{,\rho}
 - \left[ \frac{13}{3}V + \frac{\tilde{\gamma}}{12}V''\right]R + \frac{43}{60}R_{\alpha\beta}^2
+ \frac{1}{40}R^2 -
\frac{\tilde{\gamma}}{6}R\phi_{,\rho}\phi^{,\rho} +
\frac{5}{4}\left(\phi_{,\rho}\phi^{,\rho}\right)^2 \right\}
\right] \, .
\end{align}
In the case when the Big Rip singularity occurs, the curvature
becomes quickly very large. This means that quantum effects (e.g.,
quantum gravity effects) become important not only for the early
universe but also for the future universe. These quantum effects
may even become dominant when the universe approaches the Big Rip
singularity. In fact, the quantum correction becomes dominant
because $W_\mathrm{1-loop}$ contains higher derivative terms, when
we may neglect the classical terms. In order to simplify the
situation more, we assume that the curvature and the scalar field
$\phi$ are constant $R_{\mu\nu}=\frac{3}{l^2}g_{\mu\nu}$,
$R=\frac{12}{l^2}$, and $\phi=c$. We also choose the potential
$V(\phi)$as the exponential function of $\phi$, $V(\phi)=V_0
\e^{-2\frac{\phi}{\phi_0}}$. Then by using (\ref{SS4}) and
(\ref{SS5}), we obtain
\begin{align}
\label{SS8}
0=&\, \frac{1}{\sqrt{-g}}\frac{\delta W_\mathrm{1-loop}}{\delta\phi} \nonumber \\
=&\, - \frac{1}{2}\ln \frac{\left|R\right|}{\mu^2}\left[ -
\frac{4}{\phi_0} \left(\frac{5}{2} -
\frac{4\tilde{\gamma}}{\phi_0^2} + \frac{8}{\phi_0^4}\right) V_0^2
\e^{-\frac{4c}{\phi_0}}
+ \frac{2}{\phi_0}\left(\frac{13}{3} + \frac{\tilde{\gamma}}{3\phi_0^2}\right)V_0 \e^{-\frac{2c}{\phi_0}}\frac{12}{l^2}\right]\, . \\
\label{SS9}
0=&\, \frac{1}{\sqrt{-g}} \frac{\delta W_\mathrm{1-loop}}{\delta g_{\mu\nu}} \nonumber \\
=&\, g^{\mu\nu}\left[ -
\frac{1}{4}\ln\left(\frac{12}{l^2\mu^2}\right)\left\{
\left(\frac{5}{2} - \frac{4\tilde{\gamma}}{\phi_0^2} +
\frac{8}{\phi_0^4}\right) V_0^2 \e^{-\frac{4c}{\phi_0}} -
\left(\frac{13}{3} + \frac{\tilde{\gamma}}{3\phi_0^2}\right)V_0
\e^{- \frac{2c}{\phi_0}}\frac{12}{l^2} + \frac{147}{5l^4}\right\}
\right. \nonumber \\
&\, \left. + \frac{1}{8}\left(\frac{5}{2} -
\frac{4\tilde{\gamma}}{\phi_0^2} + \frac{8}{\phi_0^4}\right)V_0^2
\e^{-\frac{4c}{\phi_0}} + \frac{3}{2l^2}\left(\frac{13}{3} +
\frac{\tilde{\gamma}}{3\phi_0^2}\right)V_0 \e^{-\frac{2c}{\phi_0}}
- \frac{441}{40l^4}\right]\, .
\end{align}
Eq.~(\ref{SS8}) can be solved with respect to $l^2$ as follows,
\begin{align}
\label{SS10} R=\frac{12}{l^2}=2 \left(\frac{5}{2} -
\frac{4\tilde{\gamma}}{\phi_0^2} + \frac{8}{\phi_0^4}\right)
\left(\frac{13}{3} +
\frac{\tilde{\gamma}}{3\phi_0^2}\right)^{-1}V_0
\e^{-\frac{2c}{\phi_0}}\, .
\end{align}
We should note, however, that Eq.~(\ref{SS9}) is not consistent
with the expression in Eq. (\ref{SS10}) in general. Then,
Eq.~(\ref{SS9}) might be regarded as an equation determining
$\mu$. We should also note that the r.h.s. in (\ref{SS10}) is not
always positive. In the case $\tilde{\gamma}>0$, when
$\tilde{\gamma}^2<5$, the r.h.s. in Eq. (\ref{SS10}) is positive, but
when $\tilde{\gamma}^2>5$, it is positive if $\phi_0^2 >
\frac{4}{5}\left(\tilde{\gamma} + \sqrt{\tilde{\gamma}^2 -
5}\right)$ or $\phi_0^2 < \frac{4}{5}\left(\tilde{\gamma} -
\sqrt{\tilde{\gamma}^2 - 5}\right)$. On the other hand, in a
phantom case $\tilde{\gamma}<0$, the r.h.s. in (\ref{SS10}) is
positive if $\phi_0^2 > - \frac{\tilde{\gamma}}{13}$. Anyway,
there may occur an (asymptotically) de Sitter solution. Thus, the
universe becomes a quantum de Sitter space before entering the Big
Rip singularity. This qualitative discussion indicates that the
finite time future singularity may never occur (or, at least may
become milder) under the conjecture that quantum effects become
dominant just before the Big Rip. Due to the sharp increase of the
curvature invariants near the Big Rip, such a conjecture looks
quite natural.

\section{Summary and Conclusions}
\label{sec-summary}

In this review we aimed to present an overview of how finite-time
cosmological singularities may arise in various cosmological
contexts and to show the nature of singularities.
There are various types of singularities that may be developed from
cosmological theories, and we focused to demonstrate how soft and
crushing types singularities may be developed.
In all the cases, it is vital to understand that these cosmic singularities cannot
be developed from standard GR approaches, unless a phantom
ingredient is added in the theory.
On the other hand, standard and non-standard modifications of GR give rise naturally to cosmic
singularities without the need of a phantom ingredient.
Thus, we presented all the distinct cases that may give rise to finite-time cosmological singularities.
Our aim was to provide the theoretical frameworks that allow this singularity occurrence to happen.
But our inherent reasoning for this work is to further motivate the study of finite-time cosmological singularities.
The point is that cosmic singularities have their own physical
significance, apart from the mathematical structures they imply.
The fabric of spacetime that can accommodate such spacelike future
singularities may have its own mysteries to reveal.
A singularity, especially if it is a crushing type singularity, indicates our
inability to describe physics adequately.
The reason might be unknown for the moment, but this should be the aim of future scientists.
The fact that GR cannot produce such singularities,
while some modifications of it can, indicates the fact that GR may
be an effective theory active in less strong gravity regimes,
while the modifications of it might be the more fundamental
theories that correctly describe to some extent nature in strong gravity limits.
The fact that the modifications of GR predict
cosmic singularities on the other hand, might be an indication
that these theories link the classical physics to the unknown
quantum nature of the Universe. Indeed, as gravity effects become
stronger, the modifications of GR come into play, so they might
act as a direct link between the classical GR physics and the
unknown quantum nature of our Universe.
The links are the singularities developed by the modifications of GR.
The singularities are always mysterious, but our experience from
electrodynamics indicates that they are the links between the
classical and quantum physics.
Their appearance indicates the need for a fundamental quantum theory.
GR itself does not lead to finite-time singularities, unless a phantom ingredient is added to the description.
On the other hand, GR modifications lead to
singularities, that in some sense touch the quantum theory, they
are in between the classical GR theory and the quantum theory of the Universe.
The singularities themselves might be a direct
indication that these theories are the correct description of
nature, and they are closer to the physical reality.
These theories are a step closer to the underlying fundamental quantum
theory that governs our Universe at high energy and at strong gravity regimes.
There are three things that are remarkable to
point out, firstly spacelike singularities in GR occur only at the
center of the Schwarzschild black holes, the nature of which is
unknown, regarding the interior of the black hole, thus the
singularity itself points to an underlying fundamental theory of
quantum nature beyond GR.
Secondly, string corrections to the GR
Lagrangian always introduce higher curvature terms in the
Lagrangian, and these corrections may lead to finite-time singularities.
Thirdly the Big Bang singularity, present in some
theories, indeed points out a quantum era for our Universe.
Thus singularities always point out the urge for an underlying
fundamental theory governing the Universe at strong gravity and
high energy regimes, that is yet to be found.

There are many fundamental questions that future scientists should
address at some point. Was spacetime itself created along with
matter? What is the relation of spacetime itself with matter, from
a fundamental point of view? Is spacetime evolving with matter or
it evolves independently? What is spacetime in the end, is it
another manifestation of matter or are these independent?
Obviously, the Big Bang singularity relates all these questions,
and of course a future crushing singularity brings all these
questions to the mainstream of theoretical physics. A singularity
in spacetime means geodesics incompleteness and the latter
indicates our inability to describe physics, to reach 
that era physically. This might indicate the change in the topology of the
Universe occurring at the time instance of the finite-time future
singularity. These questions are not easy to address. Thus, with this review we provided a comprehensive overview of all the distinct theories associated with the finite-time singularities, which may potentially yield indications of an underlying fundamental quantum theory. 

\section{Acknowledgments}

The work of JdH has been supported by grant
PID2021-123903NB-I00 funded by MCIN/AEI/10.13039/501100011033 and
by ``ERDF A way of making Europe''. This work was supported by MINECO (Spain), project
PID2019-104397GB-I00 and also partially supported by the program Unidad de Excelencia Maria de Maeztu CEX2020-001058-M, Spain (S.D.O). 
SP acknowledges the financial support
from the Department of Science and Technology (DST), Govt. of
India under the Scheme ``Fund for Improvement of S\&T
Infrastructure (FIST)'' [File No. SR/FST/MS-I/2019/41].

\bibliographystyle{utphys}
\bibliography{biblio}
\end{document}